\documentclass[longauth]{aa}
\usepackage{graphicx}
\usepackage{txfonts}
\usepackage{color}
\usepackage[dvipsnames]{xcolor}
\usepackage{subcaption}
\usepackage{float}
\usepackage{gensymb}
\usepackage{hyperref}
\hypersetup{
    colorlinks,
    linkcolor={blue},
    citecolor={blue},
    urlcolor={blue}
}

\graphicspath{{./figures/}}
\begin{document}

\title{Mass distribution in the core of MACS~J1206}

\subtitle{Robust modeling from an exceptionally large sample of central multiple images}

\titlerunning{Mass distribution in the core of MACS~J1206 from a large sample of central multiple images}
\authorrunning{G.~B.~Caminha et al} 

\author{G.~B.~Caminha       \inst{\ref{unife},\,\ref{inafbologna}}
                            \thanks{e-mail address: \href{mailto:gbcaminha@fe.infn.it}{gbcaminha@fe.infn.it}}
        C.~Grillo           \inst{\ref{unimilano},\,\ref{dark}}         \and
        P.~Rosati           \inst{\ref{unife},\,\ref{inafbologna}}      \and
        M.~Meneghetti       \inst{\ref{inafbologna},\,\ref{infnbologna}}\and
        A.~Mercurio         \inst{\ref{inafcapo}}                       \and
        S.~Ettori           \inst{\ref{inafbologna},\,\ref{infnbologna}}\and
        I.~Balestra         \inst{\ref{obs_munich},\,\ref{inaftrieste}} \and
        A.~Biviano          \inst{\ref{inaftrieste}}                    \and
        K.~Umetsu           \inst{\ref{sinica}}                         \and
        E.~Vanzella         \inst{\ref{inafbologna}}                    \and
        M.~Annunziatella    \inst{\ref{inaftrieste}}                    \and 
        M.~Bonamigo         \inst{\ref{dark}}                           \and
        C.~Delgado-Correal  \inst{\ref{unife}}                          \and
        M.~Girardi          \inst{\ref{inaftrieste},\,\ref{unitrieste}} \and
        M.~Lombardi         \inst{\ref{unimilano}}                      \and
        M.~Nonino           \inst{\ref{inaftrieste}}                    \and
        B.~Sartoris         \inst{\ref{unitrieste}}                     \and
        P.~Tozzi            \inst{\ref{inafflorence}}                   \and
        M.~Bartelmann       \inst{\ref{heidelberg}}                     \and  
        L.~Bradley          \inst{\ref{stsi}}                           \and
        K.~I.~Caputi        \inst{\ref{Kapteyn}}                        \and
        D.~Coe              \inst{\ref{stsi}}                           \and
        H.~Ford             \inst{\ref{hopk}}                           \and
        A.~Fritz            \inst{\ref{inafmilano}}                     \and
        R.~Gobat            \inst{\ref{school_korea}}                   \and
        M.~Postman          \inst{\ref{stsi}}                           \and
        S.~Seitz            \inst{\ref{obs_munich},\,\ref{mpi}}         \and
        A.~Zitrin           \inst{\ref{BenGurion}}              
        }
\institute{
Dipartimento di Fisica e Scienze della Terra, Universit\`a degli Studi di Ferrara, Via Saragat 1, I-44122 Ferrara, Italy\label{unife}\and
INAF - Osservatorio Astronomico di Bologna, via Gobetti 93/3, 40129 Bologna, Italy\label{inafbologna}\and
Dipartimento di Fisica, Universit\`a  degli Studi di Milano, via Celoria 16, I-20133 Milano, Italy\label{unimilano} \and
Dark Cosmology Centre, Niels Bohr Institute, University of Copenhagen, Juliane Maries Vej 30, DK-2100 Copenhagen, Denmark\label{dark}\and
INFN - Sezione di Bologna, viale Berti Pichat 6/2, 40127 Bologna, Italy\label{infnbologna} \and
INAF - Osservatorio Astronomico di Capodimonte, Via Moiariello 16, I-80131 Napoli, Italy\label{inafcapo} \and
University Observatory Munich, Scheinerstrasse 1, 81679 Munich, Germany\label{obs_munich}\and
INAF - Osservatorio Astronomico di Trieste, via G. B. Tiepolo 11, I-34143, Trieste, Italy\label{inaftrieste}\and
Institute of Astronomy and Astrophysics, Academia Sinica, P.O.Box 23-141, Taipei 10617, Taiwan \label{sinica} \and
Dipartimento di Fisica, Universit\`a  degli Studi di Trieste, via G. B. Tiepolo 11, I-34143 Trieste, Italy\label{unitrieste}\and
INAF - Osservatorio Astrofisico di Arcetri, Largo E. Fermi, I-50125, Firenze, Italy\label{inafflorence} \and
Universität Heidelberg, Zentrum für Astronomie, Institut für Theoretische Astrophysik, Philosophenweg 12, 69120 Heidelberg,
Germany \label{heidelberg} \and
Space Telescope Science Institute, 3700 San Martin Drive, Baltimore, MD 21208, USA\label{stsi} \and
Kapteyn Astronomical Institute, University of Groningen, Postbus 800, 9700 AV Groningen, The Netherlands \label{Kapteyn} \and
Department of Physics and Astronomy, The Johns Hopkins University, 3400 North Charles Street, Baltimore, MD 21218, USA \label{hopk} \and
INAF - Istituto di Astrofisica Spaziale e Fisica Cosmica (IASF) Milano, via E. Bassini 15, 20133, Milano, Italy \label{inafmilano} \and
School of Physics, Korea Institute for Advanced Study, Hoegiro 85, Dongdaemun-gu, 02455 Seoul, Republic of Korea\label{school_korea} \and
Max Planck Institute for Extraterrestrial Physics, Giessenbachstrasse, 85748 Garching, Germany\label{mpi}\and
Physics Department, Ben-Gurion University of the Negev, P.O. Box 653, Be'er-Sheva 84105, Israel \label{BenGurion}
}

\abstract{We present a new strong lensing analysis of the galaxy cluster MACS~J1206.2$-$0847 (MACS~1206), at $z=0.44$, using deep spectroscopy from CLASH-VLT and VLT/MUSE archival data in combination with imaging from the Cluster Lensing and Supernova survey with Hubble.
MUSE observations enable the spectroscopic identification of 23 new multiply imaged sources, extending the previous compilations by a factor of approximately five.
In total, we use the positional measurements of 82 spectroscopic multiple images belonging to 27 families at $z=1.0-6.1$ to reconstruct the projected total mass distribution of MACS~1206. Remarkably, 11 multiple images are found within $50\,\rm kpc$ of the brightest cluster galaxy, making this an unprecedented set of constraints for the innermost projected mass distribution of a galaxy cluster.
We thus find that, although dynamically relaxed, the smooth matter component (dark matter plus hot gas) of MACS~1206 shows a significant asymmetry, which closely follows the asymmetric distribution of the stellar component (galaxy members and intracluster light).  We determine the value of the innermost logarithmic slope of the projected total mass density profile and find it to be close to the canonical Navarro-Frenk-White value. We demonstrate that this quantity is very robust against different parametrizations of the diffuse mass component; however, this is not the case when only one central image is used in the mass reconstruction. 
We also show that the mass density profile from our new strong lensing model is in very good agreement with dynamical and X-ray measurements at larger radii, where they overlap.}


\maketitle

\section{Introduction}

\begin{figure*}
  \centering
   \includegraphics[width = 0.9\textwidth]{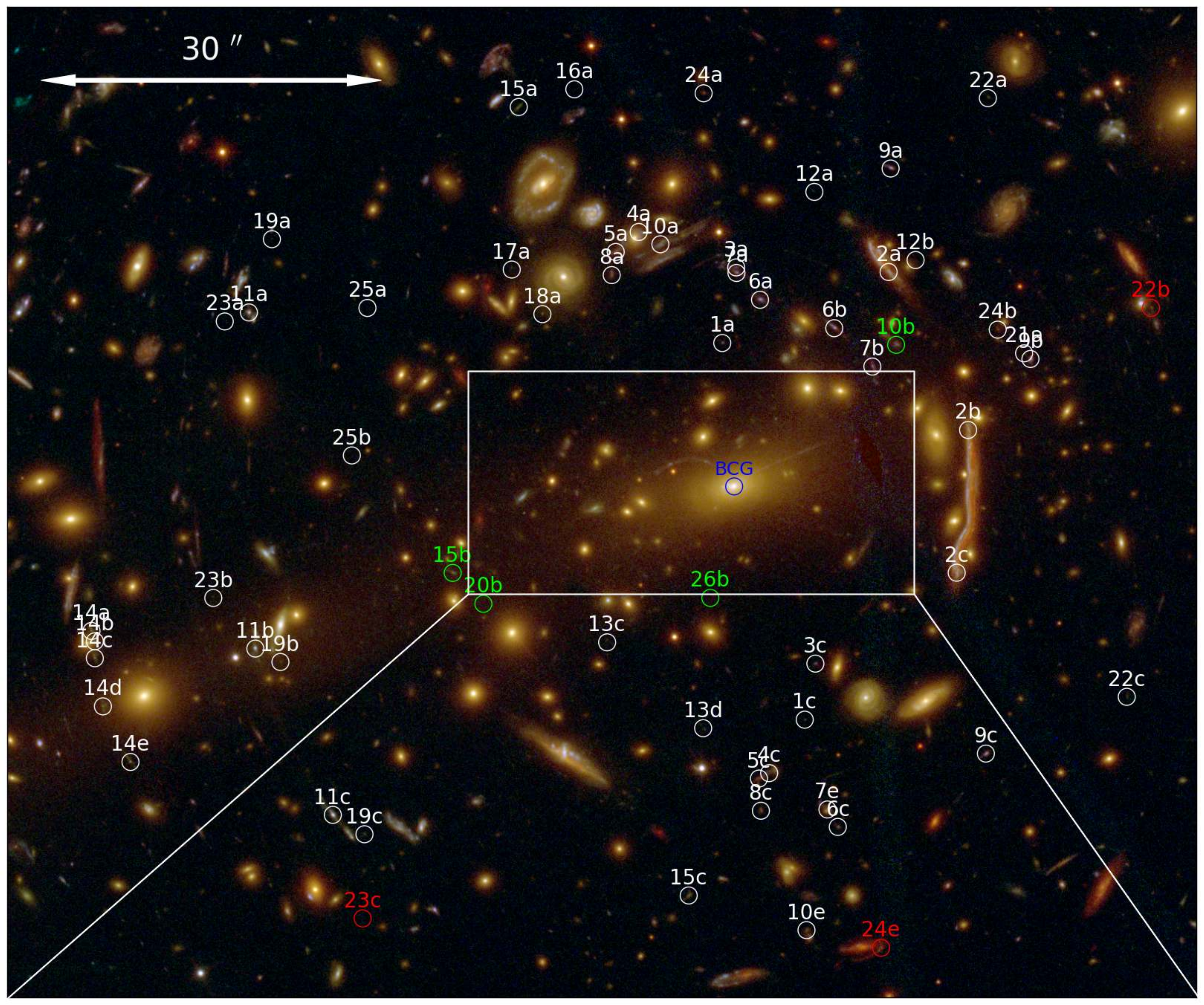}
   \includegraphics[width = 0.9\textwidth]{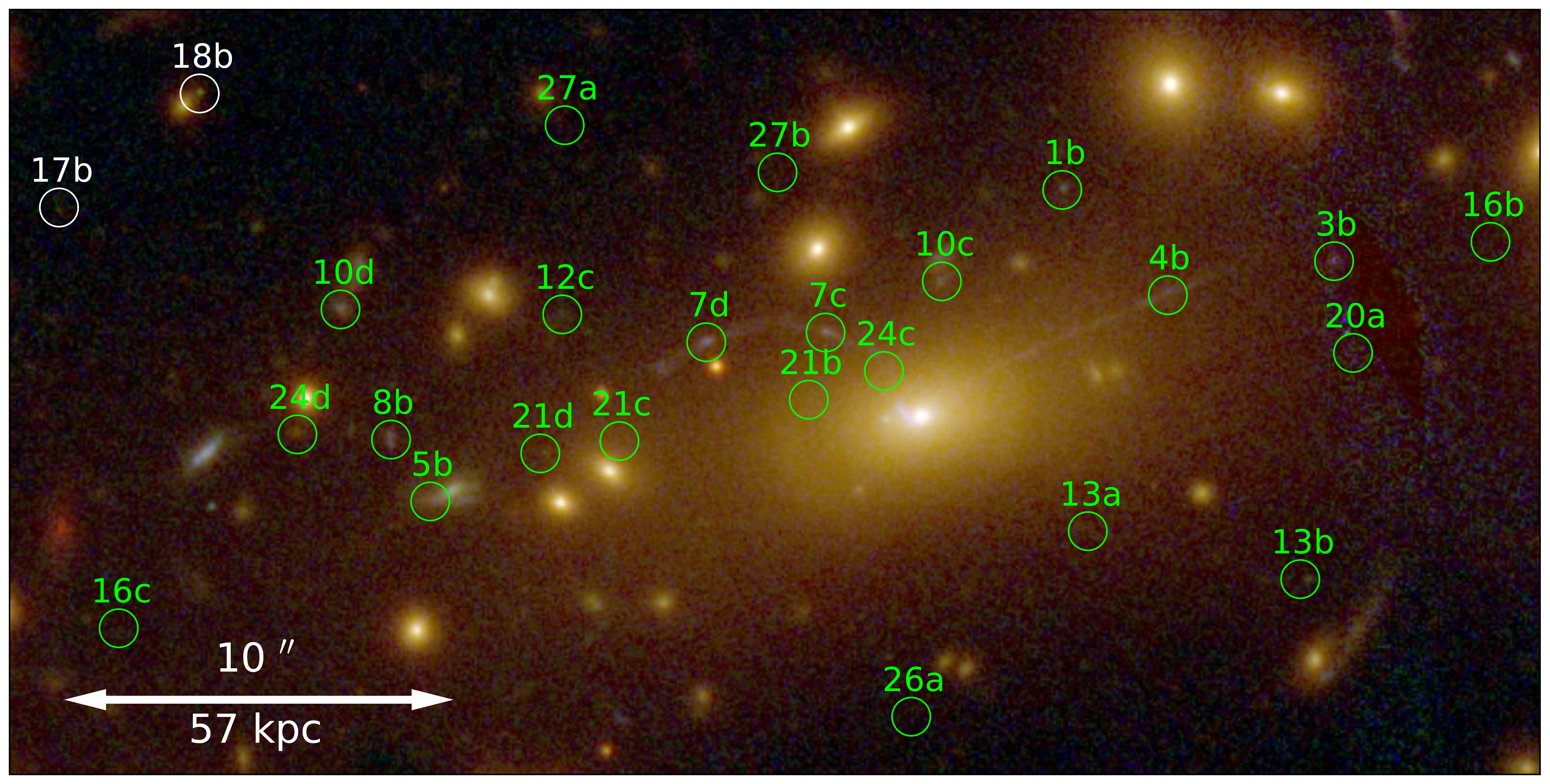}
   \caption{Spectroscopically confirmed families of multiple images in MACS~1206 overlaid on a color composite image based on 12 CLASH filters (from optical to near-infrared wavelengths). White and green circles indicate, respectively, the tangential ($\mu_{tan} > \mu_{rad}$) and radial ($\mu_{tan} < \mu_{rad}$) multiple images (see Eqs. (\ref{eq:mag_tan}) and (\ref{eq:mag_rad})) used in our strong lensing model (see Table \ref{tab:multiple_images}). The three red circles indicate multiple images that are excluded from our models because they might be significantly deflected by massive and angularly close early-type galaxies at distances between those of the galaxy cluster and the sources.}
  \label{fig:arcs}
\end{figure*}

In recent years, significant efforts have been devoted to
observational and theoretical studies of the internal mass structure
of galaxy clusters, which host the largest bound dark matter
halos. These studies, particularly those focusing on the central
high-density regions, provide stringent tests of the structure
formation paradigm, in which cold dark matter (CDM) drives the
hierarchical assembly and shape the density profiles of DM halos via
dynamical processes over a wide range of scales. Cosmological $N$-body high-resolution simulations, where particles are treated
as a collisionless fluid, consistently find that the central slope of
the 3D mass density profile is $\rho_{\rm DM}(r)\propto
r^{-\gamma}$, with $\gamma\simeq 1$ \citep[][]{1996ApJ...462..563N, 1998ApJ...499L...5M, 2012MNRAS.425.2169G}, and where there is no evidence of a central core.  However, the
presence of baryons, which eventually become dominant in real massive
clusters at small radii where the brightest cluster galaxy (BCG)
resides, complicates this scenario. Baryonic processes, such as
adiabatic cooling or heating due to star formation or accretion onto a
supermassive black hole, can significantly alter the gravitational
potential in the inner regions; as a result the DM mass distribution
will dynamically adjust to it. The observational evidence that DM
halos on cluster scales seem to have shallow inner slopes from a first
systematic study using stellar kinematics and lensing in the strong
and weak regime \citep{2004ApJ...604...88S, 2011ApJ...728L..39N, 2013ApJ...765...25N, 2013ApJ...765...24N} has
stimulated a number of theoretical investigations on the role of the
baryons in shaping cluster cores. Despite recent progress in
complex hydrodynamical simulations of clusters, there is still no
consensus  on the net effect of such feedback mechanisms on the inner density slope \citep[see, e.g.,][]{2004ApJ...616...16G, 2012MNRAS.422.3081M, 2015MNRAS.452..343S}.
In addition, dynamical processes due to 
infalling galaxies may also play a relevant role in producing a flat
central slope \citep{2004MNRAS.355.1119N, 2012MNRAS.424...38D}.

Interestingly, the central logarithmic slope $\gamma$ is also quite
sensitive to the physical properties of the DM particles. For example,
a  nonvanishing self-interaction cross section of DM particles would
lead to the formation of a core in the highest density regions,
approximately within the central 50~kpc \citep[$\gamma\simeq 0$; e.g.,][]{2000PhRvL..84.3760S, 2000MNRAS.315L..29F, 2012MNRAS.423.3740V, 2012MNRAS.424.1105M}.
This would have significant implications on the strong lensing cross section of galaxy clusters \citep[see, e.g.,][]{2001MNRAS.325..435M}.
Clearly, an understanding of the 
baryonic effects is critical in order  to turn observational
constraints of the innermost DM density profiles into a powerful probe
of the nature of DM particles in combination with constraints on the
self-interaction cross section from bullet-like cluster mergers
\citep[e.g.,][]{2004ApJ...606..819M, 2015Sci...347.1462H}. In this regard, the
extension of these observational studies to galaxy scale systems is particularly interesting. $N$-body
simulations predict similarly cuspy profiles given the approximate
self-similarity of density profiles across halo masses; however,
baryonic effects are expected to vary across the mass spectrum of DM
halos.
Observational studies of the inner density profiles of
galaxy systems have focused on DM dominated dwarf galaxies, which have
consistently revealed a flattening of the central density \citep{1995ApJ...447L..25B, 2003ApJ...583..732S, 2014ApJ...789...63A} and
field elliptical galaxies, which are instead found to have isothermal total mass 
profiles  (i.e., $\rho_{\rm TOT}(r)\propto r^-2$) using kinematic and lensing
methods, implying a lack of cores for $\rho_{\rm DM}(r)$ \citep{2012ApJ...747L..15G,2012ApJ...752..163S}.

In general, the best observational constraints on the cluster density profiles can
only be obtained by combining all possible tracers of the {total}
cluster mass distribution, as each tracer is most sensitive in a
different radial range and is affected by different astrophysical
systematics. These methods are stellar kinematics of the BCG and
strong gravitational lensing in the central regions \cite[e.g.,][]{2011ApJ...728L..39N}; hydrostatic equilibrium of the hot X-ray gas out to $\sim
R_{500}$; weak gravitational lensing out to the virial radius,
$R_{200}$; and galaxy dynamics at $R\gtrsim 100$ kpc to well beyond
the virial radius, which has recently become feasible with extensive
spectroscopic surveys \citep{2013A&A...558A...1B}. The quality and
homogeneity of gravitational lensing data on large samples of
clusters have increased dramatically in recent years thanks to
dedicated programs with the \emph{Hubble} Space Telescope (HST), particularly with the  Cluster Lensing and Supernova survey with \emph{Hubble} \citep[CLASH,][]{2012ApJS..199...25P}, later enhanced with the \emph{Hubble} Frontier Fields program \citep[HFF,][]{2017ApJ...837...97L},
supplemented by ground-based panoramic high-quality imaging data, especially with
Subaru/SupCam \citep{2014ApJ...795..163U, 2014MNRAS.439....2V}.

These studies have shown that the overall total mass density profiles of
massive clusters is on average well reproduced by the Navarro-Frenk-White parametrization \citep[NFW,][]{1996ApJ...462..563N, 1997ApJ...490..493N}, down to
$R\sim 100$ kpc. At lower radii, possibly down to a few kpc,
separating the luminous and dark matter contributions becomes increasingly
difficult. The most promising approach is indeed the combination of a
highly precise strong lensing model, based on a large number of
spectroscopically confirmed multiple images, and spatially resolved
kinematics of the BCG. The latter can probe the inner cluster
potential out to 2-3 effective radii ($R_e$) in clusters at $z\sim 0.4$
or approximately 50~kpc, whereas strong lensing can robustly probe
these small radii ($R<50$~kpc) provided that an adequate number of internal
multiple images are present, which is the case of the cluster
MACS~J1206.2$-$0847 (hereafter MACS~1206), the subject of this study.

The combination of strong lensing and internal kinematics is also not
straightforward;  a dynamical model for the mass profile from the
projected velocity dispersion and density profile requires an estimate
of the mass-to-light ratio (e.g., a knowledge the BCG star formation history and
the stellar IMF), as well as an ansatz on the isotropy of the
stellar orbits. Therefore, a simple dynamical model may not capture the
complexity of the 3D phase space stellar distribution in the inner core
and is prone to systematics or degeneracies.

In this paper we present a new, significantly enhanced measurement
of the inner total mass distribution of MACS~1206 based on an
exceptionally large number of central multiple images identified with
CLASH multi-band imaging; this measurement is possible thanks to the unique
sensitivity of the Multi Unit Spectroscopic Explorer (MUSE) on the
Very Large Telescope.  We describe the new spectroscopic confirmation
of 73 multiple images combined with previous studies, which
provide a sample of 27 background sources multiply lensed into 82
multiple images. A good fraction of these images is found in the inner 100
kpc, enabling a very robust determination of the total mass density
profile down to a few kpc, thus extending and improving previous
determinations of the density profiles based on weak lensing, X-ray
observations, and galaxy dynamics. In a forthcoming paper we will use
these new strong lensing measurements of the total mass to separate
the luminous and dark matter contributions, also taking advantage of
the internal kinematics of the BCG.

The galaxy cluster MACS~1206, discovered in the ROSAT All Sky Survey \citep[RXC~J1206.2$-$0848, ][]{2001A&A...369..826B}, at $z=0.44$,
has long been known as a powerful gravitational lens \citep{2001Natur.409...39B, 2009MNRAS.395.1213E}.
Its mass distribution has been studied
as part of the CLASH project, using a wide range of techniques.
A strong lensing model exploiting a large number of multiple images (mainly photometrically identified) was first developed by \citet{2012ApJ...749...97Z}, and later revisited by \citet{2013ApJ...774..124E}.
The overall cluster mass distribution was further studied through a combination of weak and strong lensing
\citep{2012ApJ...755...56U} and galaxy dynamics based on a large redshift sample of cluster members from the CLASH-VLT project
\citep{2013A&A...558A...1B, 2014Msngr.158...48R}.  Further analysis of
its gravitational potential and mass distribution have included a
combination of methods and  multiwavelength data sets
\citep{2015A&A...584A..63S, 2017MNRAS.467.3801S}.  The stellar content
of MACS~1206 has also been the subject of a number of studies using
spectrophotometric and kinematical information from the CLASH-VLT
data set: \citet{2014A&A...565A.126P} has studied the color and
morphological properties of its prominent  intracluster light (ICL),
detailed investigations of its galaxy populations, and internal
structure have been presented in \citet{2014A&A...571A..80A} and
\citet{2015A&A...579A...4G}, and more recently in \citet{2017arXiv170503839K}.

In this paper, we adopt the standard $\Lambda$CDM cosmological model, with $\Omega_m=0.3$ with vanishing curvature and $H_{0} = 70$km/s/kpc.
Adopting this cosmology, at the cluster redshift ($z=0.439$) one arcsec corresponds to 5.68~kpc.
All the images are oriented with north at top and east to the left, and
the angles are measured from the west and oriented counterclockwise.

\section{VLT and HST observations of MACS~1206}
\label{sec:data}

HST imaging of the core of the galaxy cluster MACS~1206 was obtained as part of the Cluster Lensing And Supernova survey with Hubble \citep[CLASH,][]{2012ApJS..199...25P}, from the UV to the near-IR (16 filters).
In Fig. \ref{fig:arcs}, we show the color composite image obtained from the combination of 12 HST filters.

As part of the CLASH-VLT Large Programme  \citep[][Rosati et al. in prep.]{2014Msngr.158...48R}, VLT/VIMOS was used to measure redshifts for
$\sim\! 2700$ sources over an area of ~400 arcmin$^2$,
spectroscopically confirming ~600 galaxy members
\citep{2013A&A...558A...1B}. The spectroscopic catalog was publicly
released in March 2014.

\subsection{MUSE observations}
\label{sec:muse_observations}

MACS~1206 was also observed with the integral field spectrograph
MUSE on the VLT \citep{2014Msngr.157...13B} under the GTO programs
095.A-0181(A) and 097.A-0269(A) (P.I. J. Richard) in April-May 2015
and April 2016.  The MUSE survey area includes three pointings, one
 centered on the BCG with a rotation of $\approx 20\degree$, and two
with an offset of $\approx 35 \arcsec$ towards the east and west. A total of 25 exposures of 1800 seconds each were
obtained covering an  effective area of 2.63 arcmin$^2$, of which 0.5 arcmin$^2$ have an exposure of 8.5
hours and the remaining area $\approx$ 4 hours. An
offset of $\approx 0\arcsec.5$ and a rotation angle of $90\degree$ were applied  in different exposures
to improve the removal of instrumental signatures and sky subtraction.

The raw data were reduced using the standard calibrations provided by the MUSE pipeline
\citep{2014ASPC..485..451W}, version 1.6.2, following the same
procedures described in \citet{C2017}.
For each of the 25 observations we checked the calibrated data-cube and wavelength-collapsed images, but did not find large variations on the observational conditions and final calibration.
We then combined all the MUSE {\tt PIXELTABLE}s, taking into account the positional offset of each observation, into one single data-cube.
The final data-cube has a field of view of 2.63 arcmin$^2$, with a spatial sampling of $0\arcsec.2$ in the wavelength range [$4750\,\AA - 9350\,\AA$].
The seeing, measured from stars in the wavelength-collapsed image of the final data-cube in regions with different exposure times, has a fairly constant value of $\approx 0\arcsec.6$.
To reduce the sky residuals, we applied the Zurich Atmosphere Purge \citep{2016MNRAS.458.3210S} using a SExtractor \citep{1996A&AS..117..393B} segmentation map to define sky regions.

In the next section we  describe the identification process of the multiply lensed images needed for the strong lensing model.

\subsection{Multiple image identification}
\label{sec:multiple_images_identification}

In keeping with our previous studies, we consider only families of multiple images with secure spectroscopic confirmation in order to avoid any image misidentification and to remove possible degeneracies between the values of the source redshifts and that of the cluster total mass.
Incorrect or incomplete information about the multiple images can bias the estimate of the total mass of a lens and suppress the statistical uncertainties \citep{2015ApJ...800...38G, C2016_2248, 2016ApJ...832...82J} of the strong lensing modeling. 

We combine the spectroscopic measurements from CLASH-VLT \citep{2013A&A...558A...1B} and MUSE (Delgado-Correal, in prep.) to find the largest possible number of families of multiple images based on their redshift values (see Sec. \ref{sec:muse_observations}).
In this remarkable data set, we identify 27 different background sources in the redshift range from $1.01$ to $6.06$ that are lensed into 85 multiple images.
Only four of these images lack a spectroscopic redshift, but their colors, shape, and image parity observed in the HST images ensure their correct association with two spectroscopically confirmed families.
The redshift distribution of the multiple images presents a clear overdensity at $z \approx 1.4$, containing five galaxies within $\Delta z = 0.0011$ (see Sec. \ref{sec:the_mass_distribution_of_macs1609}).

\begin{figure}
  \centering
  \includegraphics[width = 1.0\columnwidth]{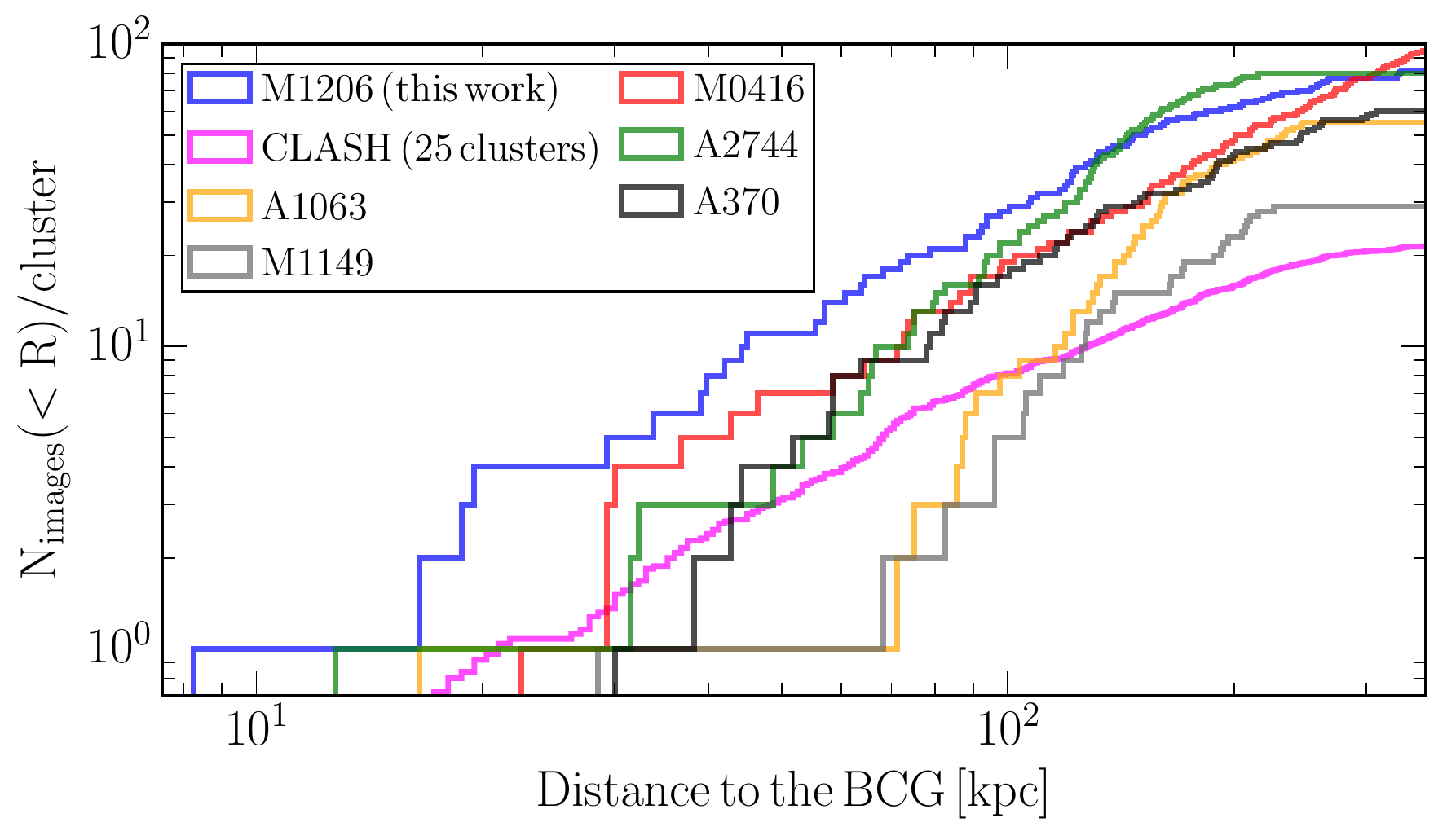}
  \caption{Cumulative distribution of the number of multiple images belonging to spectroscopically confirmed families of MACS~1206 and other clusters with similar MUSE and HST data. For the CLASH study by \citet{2015ApJ...801...44Z}, we consider all multiple images (also those with no spectroscopic information) and  the values are renormalized by the number of clusters in that work,  only 7 of the 25 clusters present spectroscopically confirmed multiple images within the inner 20 kpc.}
  \label{fig:images_dist}
\end{figure}
In Fig. \ref{fig:arcs}, we show the positions of the 85 multiple images, highlighting radial multiple images, i.e., those that according to our reference model (see Appendix \hyperref[ap:multiple_images]{A}) have magnification values that are greater in the radial than in the tangential direction, in very good agreement with the observed shapes of the images.
To illustrate this exceptionally large number of central multiple images, we show in Fig. \ref{fig:images_dist} the histogram of the cumulative number of  spectroscopically confirmed multiple images of MACS~1206, and other galaxy clusters with similar spectroscopic and photometric data, as a function of their projected distance from the BCG.
We also show the overall distribution of the CLASH sample from \citet{2015ApJ...801...44Z}, which also includes many photometrically selected multiple images.
Within the inner 20~kpc, MACS~1206 has strong lensing constraints from four multiple images, while at most one is usually present in other clusters.
At larger radii, MACS~1206 still contains the largest number of multiple image constraints out to $\approx 200$~kpc.
We also note that Abell~1063, which is known to be a very regular and relaxed cluster, presents only one central multiple image, while the merging clusters MACS~J0416.1$-$2403 \citep[hereafter MACS~0416,][]{2016ApJS..224...33B} and Abell~2744 have more multiple images in their inner regions.

Interestingly, 10 of the 27 multiply imaged sources are Lyman-$\alpha$ emitters with no detection in the CLASH imaging data (limiting AB mag in F814W $=27.7$), illustrating the high sensitivity of MUSE to  emission lines.
We note that Lyman-$\alpha$ emitters with no imaging counterpart have been found with MUSE even in deeper HST imaging, such as in the Hubble Frontier Fields \citep[see, e.g.,][]{C2017, 2017arXiv170206962M}.
Although they are very  faint, the images of these sources are clearly detected in the MUSE data-cube and their spectroscopic redshifts are secure (see Fig. \ref{fig:specs}).
We note that the Lyman-$\alpha$ profiles of these images are slightly narrower than those of the objects with clear HST detection, indicating a low level of scattering of the Lyman-$\alpha$ photons.

For each multiple image with HST detection, we carefully measure its position as the luminosity peak in the single band HST images.
For each of the families detected only by MUSE, we create a continuum subtracted data-cube and collapse this cube in the wavelength interval of the Lyman-$\alpha$ emission.
Then we use the peaks of the emission of the multiple images to measure their positions.
In the last column of Table \ref{tab:multiple_images} we indicate for each image whether the position was measured on the HST images and in which filter, or whether the measurement was performed on the MUSE data.
In order to reduce the noise in these pseudo-narrowband images, we apply a boxcar smoothing with kernel radius equal to one pixel (i.e., $0\arcsec.2$).
The positions of all detected multiple images are shown in Table \ref{tab:multiple_images} and the spectra and small cutouts of the HST color image in Fig. \ref{fig:specs}.
Here, as in previous works, we conservatively adopt a positional error of $\sigma^{\rm obs} = 0\arcsec.5$ for the images with HST detection and $\sigma^{\rm obs} = 1\arcsec$ for images detected only in the MUSE data because of the lower spatial resolution, noisier detection, and extended nature of the Lyman-$\alpha$ emission in some cases.
This should also account for large-scale mass perturbations along the line of sight and the limitations of parametric mass models \citep{2010Sci...329..924J, 2012MNRAS.420L..18H}.

We noticed that three multiple images, i.e., 22b, 23c, and 24e, are located very close in projection to massive spectroscopically confirmed background galaxies (see Fig. \ref{fig:arcs}), which can introduce significant deflections in addition to that associated with the cluster. 
Since the employed lensing software cannot take into account multiplane lensing effects, we decided not to include these multiple images in the reconstruction of the cluster total mass.
In the end, our strong lensing models are optimized on the observed positions of 82 multiple images from 27 distinct background sources.

\begin{figure}
  \centering
  \includegraphics[width = 1.0\columnwidth]{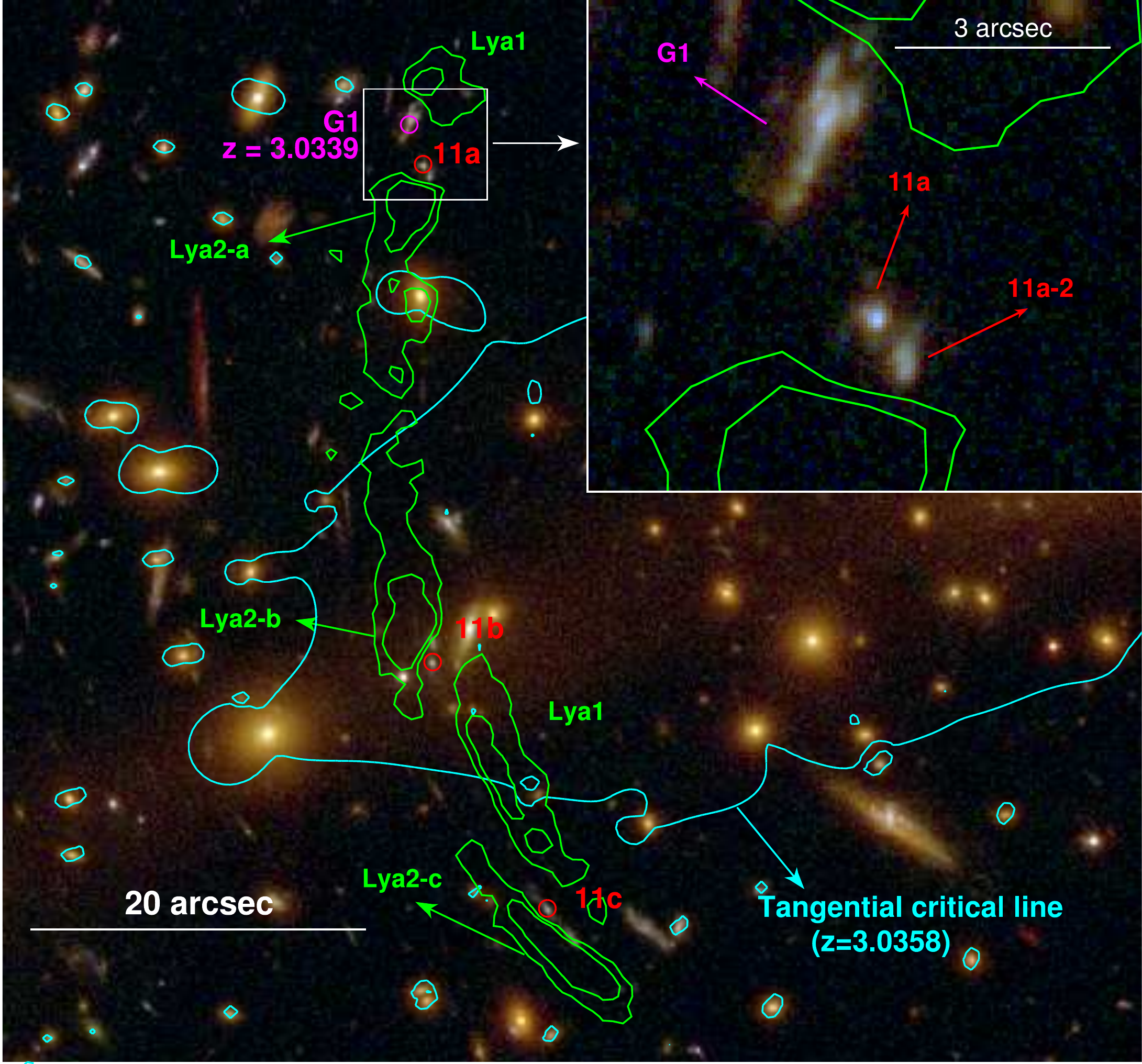}
  \includegraphics[width = 1.01\columnwidth]{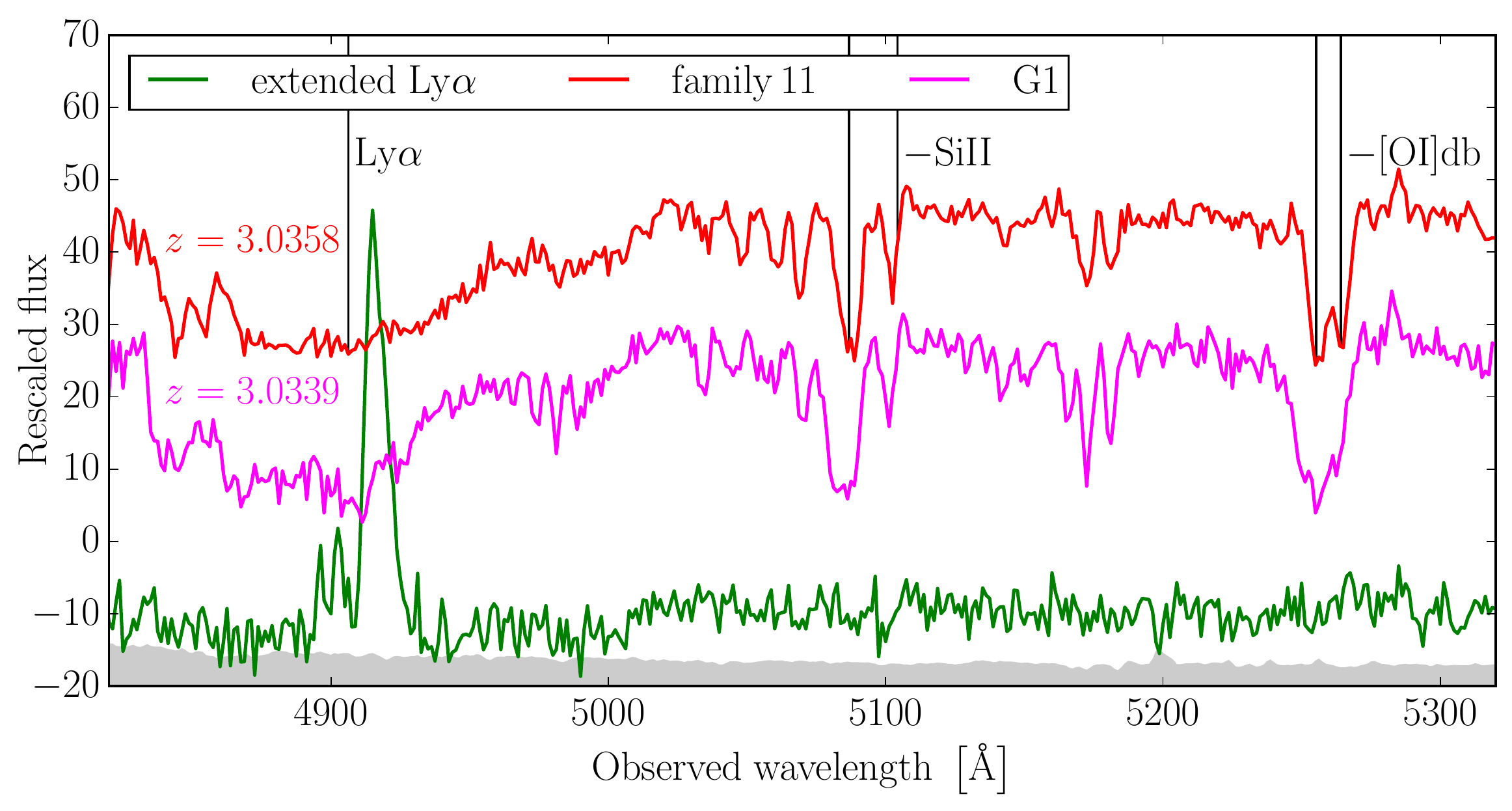}
  \caption{Top panel: Lyman-$\alpha$ extended emission (green contours) associated with  the multiple images (red circles) of family 11. The cyan lines show the tangential critical curves at the redshift  of family 11 ($z=3.0358$). A galaxy very close angularly and in redshift ($z=3.0339$), but singly imaged, is indicated with a magenta circle. Bottom panel: Red and magenta lines show, respectively, the stacked spectrum of the multiple images of family 11, corrected for the magnification factor and  shifted upwards by a factor of $+25$ (flux units are $\rm 10^{-20}\, erg\, s^{-1}\, cm^{-2}\,\AA^{-1}$), and the spectrum of the singly imaged galaxy. The green line shows the extended Lyman-$\alpha$ continuum-subtracted (to reduce the contamination from ICL) emission, extracted from the innermost green apertures. The magenta and green spectra have been rescaled by arbitrary factors for better visualization. The gray region represents the rescaled variance obtained from the data reduction pipeline.}
  \label{fig:lyman_alpha_arc}
\end{figure}
We also briefly comment on a remarkable extended Lyman-$\alpha$ emission associated with  family 11 at $z\approx 3.0$ \citep[IDs 2 and 3 in][]{2012ApJ...749...97Z}.
In the upper panel of Fig. \ref{fig:lyman_alpha_arc}, we show the positions of the multiple images 11a/b/c and the contour levels of the extended Lyman-$\alpha$ emission.
This consists of two blobs (Ly$\alpha 1$ and Ly$\alpha 2$) separated by 18~kpc (at $z=3.0358$, on the source plane), with no continuum emission detected in the MUSE and broadband HST photometric data.
The zoom-in panel shows in detail the region between the two blobs where three compact sources have been detected in the HST imaging: a galaxy named G1 (not multiply lensed), the multiple image 11a with a substructure 11a-2.
We note that in this region the Lyman-$\alpha$ emission is suppressed.
Deficient emission along the direction of the stellar continuum has also been observed in a Lyman-$\alpha$ blob behind Abell~1063 \citep{2016A&A...595A.100C}.
The images 11a, 11a-2, and Ly$\alpha 2$-a are lensed in two additional counter-images located towards the southern region and indicated as 11b/c, 11b/c-2, and Ly$\alpha 2$-b/c.
Instead, the galaxy G1 is outside the region of multiple image formation, while the extended Ly$\alpha 1$ is only partially multiply lensed, and its counter-image appears crossing the tangential critical line between the images 11b and 11a.

In the bottom panel of Fig. \ref{fig:lyman_alpha_arc}, we show the stacked MUSE spectra of the multiple images 11a/b/c corrected for the magnification factor, and also the emission of the single image G1.
These two sources present strong absorption features, including Lyman-$\alpha$ revealing a high column-density along the line of sight \citep[see also][for the VIMOS spectra of 11a/b/c]{2012ApJ...749...97Z}.
The emissions of Lya$\alpha$1 ad Ly$\alpha$2 are similar and do not have strong spatial variation or continuum.
In the same panel, we show the stacked continuum subtracted spectrum of the innermost green regions.
Using the 5(3)-$\sigma$ limiting magnitude in the band F606W \citep[see][]{2012ApJS..199...25P}, we estimate a lower limit for the equivalent width of Lya$\alpha$1 and Ly$\alpha$2 of $200(400)\,\AA$, which -- in combination with the deficient emission around the compact sources -- implies that the Lyman-$\alpha$ emission is plausibly generated by fluorescence \citep{2012MNRAS.425.1992C}.
An offset between the ionizing sources and the Lyman-$\alpha$ emission, which in our case is 4.7~kpc between G1 and Ly$\alpha$1, and 4.1~kpc between 11a and Ly$\alpha$2, has been shown to be relatively common \citep[see, e.g.,][]{2011MNRAS.418.1115R,2017MNRAS.465.3803V}.
Other Lyman-$\alpha$ emitters behind galaxy clusters have been studied in detail \citep{2016MNRAS.456.4191P, 2017MNRAS.467.3306S}, showing the advantages of using the lensing magnification to explore high-$z$ sources.

\section{Strong lensing model}
\label{sec:strong_lens_model}

In this section, we describe the reconstruction of the total mass distribution of MACS~1206 through a strong lensing analysis. We adopt the same modeling strategy and software \citep[{\tt lenstool},][]{1996ApJ...471..643K, 2007NJPh....9..447J} described in \citet{C2016_2248, C2017}; we therefore provide a brief summary here and refer to these previous works for further details.

\subsection{Strong gravitational lensing definitions}
\label{sec:sl_definitions}

We use the observed positions of the multiple images as constraints to reconstruct the total mass distribution of MACS~1206. To optimize a model, we minimize the $\chi^2$ function, defined on the lens plane as
\begin{equation}
\label{eq:chi2_lens}
    \chi^2(\vec{\Pi})  :=  \sum_{j=1}^{N_{\rm fam}} \sum_{i=1}^{N^{j}_{\rm im}}\left( \frac{\left| \vec{\theta}^{\rm obs}_{i,j} -
              \vec{\theta}_{i,j}^{\rm pred}\left( \vec{\Pi}\right) \right|}{\sigma^{\rm obs}_{i,j}} \right)^2,
\end{equation}
where $\vec{\theta}^{\rm obs}$ and $\vec{\theta}^{\rm pred}$ are the multiple image observed and model-predicted positions, respectively; $N_{\rm fam}$ is the total number of families of multiple images; and $N^j_{\rm im}$ is the number of images associated with family $j$.
The parameters that characterize a model are given by the vector $\vec{\Pi}$, and $\sigma^{\rm obs}$ is the adopted uncertainty on the observed image positions.

The total magnification, $\mu$, of a point-like image located in \vec{\theta} is given by
\begin{equation}
\mu^{-1}(\vec{\theta}) = \left[1 - \kappa(\vec{\theta})\right]^2 - \gamma^2(\vec{\theta}),
\label{eq:mag_total}
\end{equation}
where $\kappa$ and $\gamma$ are the convergence and the modulus of the shear, respectively \citep[see, e.g.,][]{1992grle.book.....S}.
The shear can be represented as a complex quantity, $\vec{\gamma} = \gamma e^{2i\varphi}$, and the direction of distortion of an observed image is aligned with the phase $\varphi$ of the shear. The total magnification can be expressed as the product of a tangential ($\mu_{tan}$) and a radial ($\mu_{rad}$) term, defined as
\begin{eqnarray}
\mu_{tan}^{-1}(\vec{\theta}) &=& 1 - \kappa(\vec{\theta}) - \gamma(\vec{\theta}), \label{eq:mag_tan}\\
\mu_{rad}^{-1}(\vec{\theta}) &=& 1 - \kappa(\vec{\theta}) + \gamma(\vec{\theta}). \label{eq:mag_rad}
\end{eqnarray}

\subsection{Parametric total mass model components}

Using multiwavelength observations and different probes, \citet{2017MNRAS.467.3801S} have recently proved that MACS~1206 does not show strong deviations from thermal equilibrium and that we are probably observing this cluster in a face-on projection.
Given its relaxed state, it is expected that the smooth baryonic component traces  the total gravitational potential well, and consequently also traces the dark matter distribution.
Moreover, the distributions of the galaxy members and of the ICL suggest that MACS~1206 has a relatively high elongation along a sightly tilted east-west direction (see Fig. \ref{fig:arcs}).

Motivated by this regularity, we divide the cluster total mass distribution into two terms: 1) a smooth component accounting for the mass in the form of principally dark matter, but also hot-gas and ICL, and 2) the clumpy component of the total mass of the galaxy members.
Moreover, to account for the possible presence of massive structures in the outer regions of the cluster ($\gtrsim 300 \,\rm kpc$), we include an external shear term parametrized by its intensity $\gamma_{ext}$ and orientation $\theta_{ext}$.

\subsubsection{Selection of cluster members}
\label{sec:cluster_members_selection}

For the clumpy component of the total mass distribution, we first select the spectroscopically confirmed galaxy members.
They are defined as the galaxies that provide a velocity dispersion value lower than $\Delta_{\rm v} = 2919\, \rm km\, s^{-1}$ in the cluster rest frame \cite[$z=0.439$,][]{2015A&A...579A...4G} (this corresponds to a redshift range of $z = 0.425 - 0.453$).
Within the HST field of view, this criterion is satisfied by 114 galaxies with measured magnitudes brighter than 24 in the F160W filter.
In order to select galaxy members with no spectroscopic information, we employ the method described in \citet{2015ApJ...800...38G}, which uses the HST photometry in 13 different filters to define the color-space region where the spectroscopic members are located.
The distance from this region of each galaxy with no spectroscopic measurement is then converted into the probability of belonging (or not)  to the cluster.

\begin{figure}[th!]
  \centering
   \includegraphics[width = 1.\columnwidth]{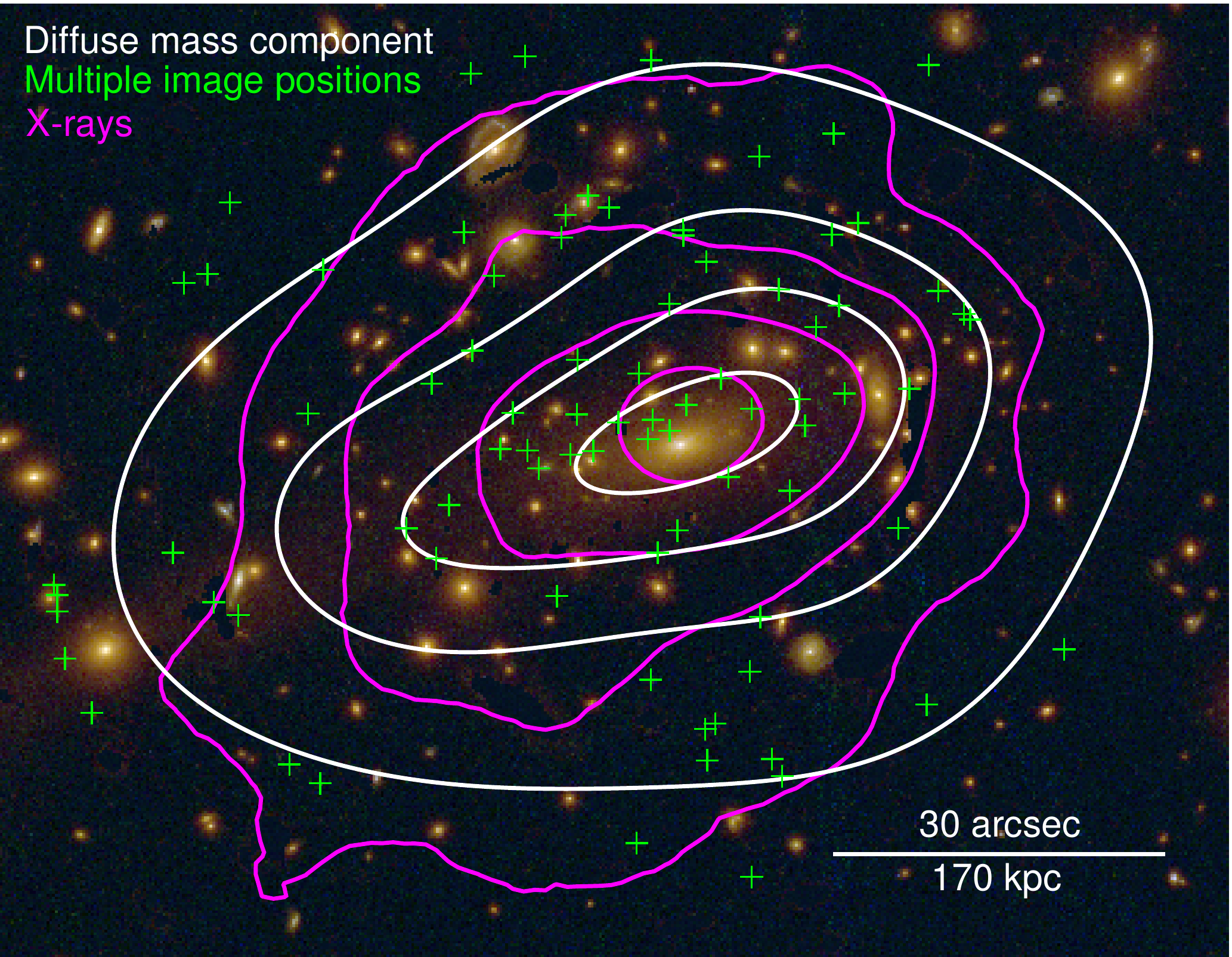}
   \caption{Cluster members of MACS~1206 used in the strong lensing models. Galaxies that do not satisfy the selection criteria described in Sec. \ref{sec:cluster_members_selection} have been masked out of the original HST image using the SExtractor segmentation map. The white lines show the contour levels of the reconstructed surface mass density of the diffuse mass components only (obtained by removing  the surface mass density of the cluster member halos from
that of the  total), corresponding to the values of $\rm [1.0, 1.5, 2.0, 3.0] \times 10^{15}\, M_{\odot}\,Mpc^{-2}$. The magenta lines show the isocontours of the $0.5-7$~keV \emph{Chandra} X-ray detection. The green crosses indicate the positions of the multiple images shown with white and green circles in Figure \ref{fig:arcs}.}
  \label{fig:objects_members}
\end{figure}
From our previous analyses of the clusters MACS~0416 \citep[][]{2015ApJ...800...38G, C2017}, MACS~J1149.5$+$2223 \citep{2016ApJ...822...78G}, and Abell~S1063 \citep{C2016_2248} and from tests performed in MACS~1206 where we compared the cluster membership based on the first VIMOS-only measurements with the additional MUSE data, we can conclude that this methodology results in a low contamination of $\approx5\%$ and a completeness of $\approx95\%$.
Using this method, we selected 147 additional photometric members in the HST/WFC3 field of view, with F160W magnitudes brighter than 24.
Moreover, to account for bright and massive galaxies that do not satisfy the strict dynamical selection criterion, we relaxed the velocity dispersion limit to $\rm \Delta_{v} = 4000\,km\,s^{-1}$ for galaxies with stellar mass values higher than $10^{10}\rm\, M_{\odot}$ (following the scaling relation $\rm \log_{10}\left(M_{\star}/M_{\odot}\right) =  20.16 - 0.475\times F160W$ derived from the SED fitting analysis presented in Delgado-Correal in prep.), thus adding four extra spectroscopically confirmed members.
In conclusion, the final cluster member catalog used in our strong lens model contains 265 galaxies, $\approx 45\%$ of which have spectroscopic confirmation.
Cluster members included in this sample are shown in the background HST image of Fig. \ref{fig:objects_members} by masking out all the other detected sources.

As in \citet{C2016_2248},  we adopt for each cluster member a truncated dual pseudo-isothermal elliptical mass distribution with vanishing ellipticity and core radius.
This model is described by two parameters, the central velocity dispersion ($\sigma_{v,i}^{gals}$) and a truncation radius ($r_{cut,i}^{gals}$) \citep[see, e.g.,][]{2007arXiv0710.5636E, 2010A&A...524A..94S}.
In order to reduce the number of free parameters describing the 265 cluster members, we use a constant total mass-to-light ratio given by
\begin{equation}
\sigma_{v,i}^{gals} = \sigma_v^{gals}\left(\frac{L_i}{L_0} \right)^{0.25}, r_{cut,i}^{gals} = r_{cut}^{gals}\left(\frac{L_i}{L_0} \right)^{0.5}
,\label{eq:member_scale}
\end{equation}
where $L_0$ is a reference luminosity, that we associate with  the BCG ($\rm mag_{F160W} = 17.2$), while the two normalizations $\sigma_v^{gals}$ and $r_{cut}^{gals}$ are free to vary in the modeling and describe the cluster members.

\subsubsection{Diffuse mass distribution}
\label{sec:total_mass_smooth_component}

In order to probe systematics originated from the modeling assumptions on the cluster total mass parametrization, we consider two different models to describe the smooth dark matter component.
Firstly, we adopt a parametric model given by a pseudo-isothermal elliptical mass distribution \citep[PIEMD,][]{1993ApJ...417..450K}.
The projected mass distribution of this model is parametrized in terms of its center ($x_0$ and $y_0$), an effective velocity dispersion $\sigma_v$, a core radius $r_{core}$, an ellipticity $\varepsilon \equiv (a^2 - b^2)/(a^2 + b^2)$ (where $a$ and $b$ are the major and minor axes), and its orientation $\theta$.
The expression of the projected mass density $\Sigma$ is given by
\begin{equation}
\Sigma \left(R\right) = \frac{\sigma_v^2}{2 G}\left( R^2(\varepsilon)+ r_{core}^2 \right)^{-1/2}.
\label{eq:sigma_dPIEMD}
\end{equation}
The parameter $r_{core}$ accounts for a finite value of the density $\Sigma$ at the origin, i.e., if $r_{core} \neq 0$, it follows that $\Sigma$ does not change for small values of $R$.
Moreover, if $r_{core}$ is sufficiently large, the profile will be flat in its inner regions, departing from a cuspy mass distribution.

We also use the well-established NFW profile \citep[][]{1996ApJ...462..563N, 1997ApJ...490..493N} to model the smooth mass component.
$N$-body simulations show that this model describes well the spherically averaged dark matter distribution of halos in equilibrium. 
The NFW profile is parametrized by a characteristic density ($\rho_s$) and a scale radius ($r_s$).
In this model, the concentration parameter is defined as $c_{200} \equiv r_{200}/r_s$ , where $r_{200}$ is the radius of the sphere inside which the mean cluster density is $200$ times the critical density of the Universe at the cluster redshift.
The expression for the 2D projected density and the derivation of the lensing quantities can be found in \citet{1996A&A...313..697B}, \citet{2000ApJ...534...34W}, \citet{2002A&A...390..821G}, or \citet{2003MNRAS.346...67M}.

In order to account for variations in the central slope, an extension of the 3D NFW density profile, known as the generalized NFW model \citep[see, e.g.,][]{1996MNRAS.278..488Z, 2000ApJ...529L..69J, 2001ApJ...555..504W}, is defined as
\begin{equation}
\rho_{\rm gNFW}\left(r\right) =  \dfrac{\rho_s}{\left(r/r_{s}\right)^{\gamma_{\rm gNFW}} \left( 1 + r/r_s\right)^{3 - \gamma_{\rm gNFW}}},
\label{eq:rho_gnfw}
\end{equation}
where if $\gamma_{\rm gNFW}=1$ the original NFW model is recovered, and $\gamma_{\rm gNFW}<1$ or $\gamma_{\rm gNFW}>1$ correspond, respectively, to shallower or steeper 3D profiles in the core.

In the {\tt lenstool} code, the ellipticity of the NFW and gNFW models is introduced in the lens potential ($\varepsilon_{pot}$) and not in  the projected mass.
This approximation allows a fast computation of the deflection angle across the image plane; however, it presents some limitations for relatively high values of the ellipticity.
Basically, for ellipticity higher than 0.5, the associated projected mass has a nonphysical dumbbell shape and negative values along the minor axis \cite[see, e.g.,][]{2002A&A...390..821G, 2012A&A...544A..83D}.
Indeed, in our strong lensing models that use the NFW and gNFW, $\varepsilon_{pot}$ reaches values on the order of 0.5.
However, we use these parametrizations only for the sake of comparison with our reference model based on the PIEMD model, where the ellipticity is introduced in the projected mass distribution.

\subsection{Optimizing the strong lensing model}
\label{sec:final_sl_model}

\begin{figure}
  \includegraphics[width = 1.0\columnwidth]{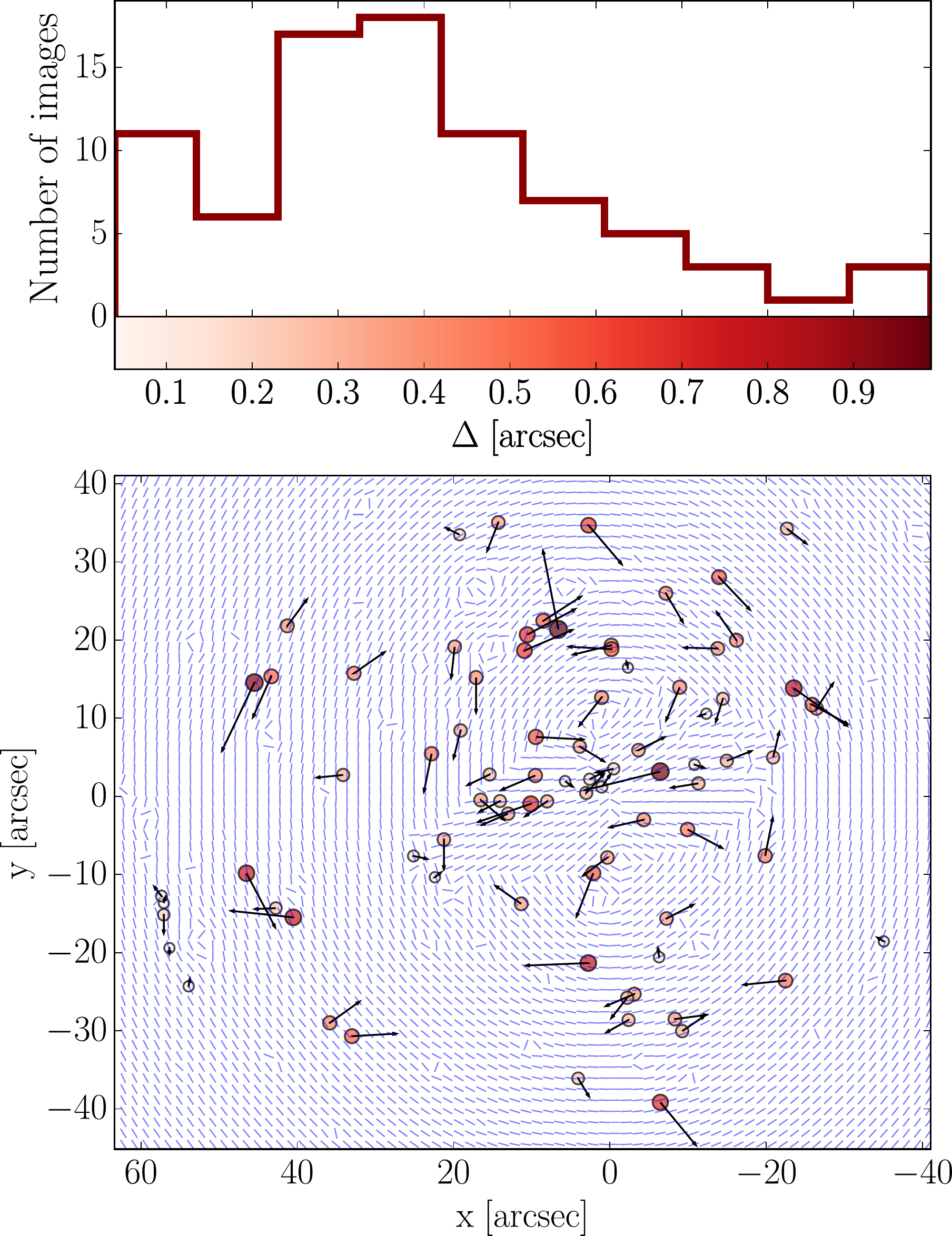}

  \caption{Offset between the observed and model-predicted (model ID P3$\varepsilon$ in Table \ref{tab:summary_bf}) positions of the multiple images in MACS~1206. Darker hues of red correspond to larger offsets. In the top panel, the histogram shows the distribution of the absolute values of the differences ($\Delta$). In  the bottom panel, the circles indicate the observed positions of the multiple images, and the arrows point towards the model-predicted positions. For the sake of clarity, the arrow lengths are ten times larger than the values of the corresponding offsets. The blue sticks represent  the shear orientation on a regular
grid (see Sect. \ref{sec:sl_definitions}) for a source at $z_{src}=1.4$.}
  \label{fig:offset_map}
\end{figure}

In order to find the parametrization of the total mass distribution that best reproduces the multiple image positions, we begin by adopting one single PIEMD profile to represent the diffuse component, as discussed in Section \ref{sec:total_mass_smooth_component}, which has six free parameters.
In addition to that, the galaxy members and the external shear add two more free parameters each, for a total of ten free parameters.
We note that this parametrization is very similar to that adopted in \citet{2013ApJ...774..124E} (see Sect. \ref{sec:comparison_models}), who used a very different set of multiple images,  many of which were based  on photometric information alone.
The results from this first model, named P1, are shown in Table \ref{tab:summary_bf}.
We adopt a flat prior for all free parameters in the lens models presented in this work.
The model P1 yields a root mean square ($\rm rms$) difference ($\Delta_{\rm rms}$) between the model-predicted and observed positions of $2\arcsec.04$.
Although this value is comparable to previous strong lensing models of MACS~J1206 \citep[][who used a significantly smaller set of spectroscopically confirmed families]{2012ApJ...749...97Z, 2013ApJ...774..124E}, it cannot reproduce the observed multiplicity and positional configuration of families 12, 17, 18, 21, and 26.
Moreover, we find that multiple images located in the northern region of the cluster have a slightly larger offset between the model-predicted and observed positions ($\Delta$), suggesting an underlying asymmetry in the mass distribution.

We also note that the exceptionally large number of radially elongated images (27 out of 85) is suggestive of a projected mass distribution with high-order asymmetries, i.e., more complex than an elliptically symmetric distribution.
Using numerical simulations, \citet{2004MNRAS.349..476T} have shown that following the evolution of a cluster in cosmic time, the cross-section of radial arc formation can be boosted by a factor of 10 in the presence of asymmetries in the cluster core (see their Figures 5 and 11).
This is in agreement with recent studies of FF clusters based on deep MUSE spectroscopy, e.g., MACS~0416 \citep{C2017}, Abell~2744 \citep{2017arXiv170206962M}, MACS~1149 \citep{2016ApJ...822...78G, 2016ApJ...817...60T}, and Abell~370 \citep{2016arXiv161101513L}.
\citet{2013A&A...558A...1B} and \citet{2017MNRAS.467.3801S} have shown that MACS~1206 is a fairly relaxed cluster; however, the remarkably asymmetric shape of its ICL \citep{2014A&A...565A.126P} and the distribution of the galaxy members (see Fig. \ref{fig:objects_members}) reveal a large degree of asymmetry of the overall mass distribution.

\begin{table*}[!]
\centering
\small
\caption{Summary of the best fit models of MACS 1206 and comparison with previous works on other clusters using only spectroscopically confirmed families of multiple images.}
\begin{tabular}{l c c c c c c l} \hline \hline
Model ID & N. par. & DOF & $\Delta_{\rm rms} ['']$ & $\chi^2_{min}$  & BIC &AIC & Description \\
\hline
P1               & 10 &100& 2.04 & 819   & 1023 & 996 &One elliptical PIEMD halo, 265 members and external shear \\
P2               & 14 & 96& 1.37 & 394   &  616 & 579 &Same as P1, but with a second circular PIEMD halo\\
P3               & 18 & 92& 0.56 & 62.6  &  304 & 256 &Same as P2, but with a third circular PIEMD halo\\
P3$\varepsilon$  & 22 & 88& 0.44 & 41.0  &  301 & 242 & Same as P3, but $\varepsilon$ and $\theta$ values of all halos are free to vary (reference)\\
\hline
N3$\varepsilon$  & 22 & 85& 0.40 & 33.3  &  294 & 235 & Same as P3$\varepsilon$, but the central halo is a pseudo-elliptical NFW\\
G3$\varepsilon$  & 23 & 84& 0.40 & 33.3  &  299 & 237 & Same as P3$\varepsilon$, but the central halo is a pseudo-elliptical gNFW\\
\hline
P4               & 22 & 88& 0.53 & 52.4  &  313 & 254 &Same as P3, but with a fourth circular PIEMD halo\\
P3$\varepsilon$BCG& 25 & 82& 0.44 & 40.0 & 315 & 247 &Same as P3, but with BCG parameters free to vary\\
\hline
Abell~S1063                  & 17 & 41 & 0.32 & 16.8 & ---& --- & Updated model from \citet{C2016_2248} and \citet{Karman_2017} \\
MACS~J0416                   & 26 & 104& 0.59 & 143  & --- & ---& Model by \citet{C2017} \\
Abell~2744                   & 30 & 78 & 0.67 & 133  &--- & ---& Model with the ``gold constraints'' by \citet{2017arXiv170206962M} \\ 
\hline \hline
\end{tabular}
\label{tab:summary_bf}
\tablefoot{
Summary of the strong lensing models and their global results. Columns show the model IDs, the number of optimized parameters (N. par.) and degrees of freedom (DOF), the best fit positional $\Delta_{\rm rms}$, the value of the reference $\chi^2_{min}$, the Bayesian Information Criterion \citep{schwarz1978}, the Akaike information criterion \citep{1974ITAC...19..716A}, and a short description of each model. For the definitions of these quantities, we refer to \citet{C2016_2248}.
}
\end{table*}

In  an attempt to correctly reproduce the multiplicity of all multiple images, we then consider another model including an extra circular pseudo-isothermal profile whose center and core radius are left free to vary across the field of view.
This model has now 14 free parameters and is named P2 (see Table \ref{tab:summary_bf}).
We find that the second diffuse halo makes the total mass distribution more elongated towards the north direction, while the main diffuse halo remains centered at $\approx 0\arcsec.7$ from the BCG.
This model has a $\Delta_{\rm rms}$ value of $1\arcsec.37$ and can reproduce the multiplicity of most of the multiple images, except for family 26.
However, this value of $\Delta_{\rm rms}$ is significantly higher then the typical subarcsecond average residuals obtained in recent studies using similar high-quality data sets, which combine HST and MUSE observations.

We therefore include a third circular halo in the parametrization of the total mass distribution, leading to 18 free parameters (model P3).
In this case, the value of $\Delta_{\rm rms}$ reduces significantly to $0\arcsec.56$ and the multiplicity of all multiple images can be reproduced correctly.
We note that the multi-mass components introduced here should not be associated with extra dark matter halos, but rather to extra asymmetries or high-order multipoles that a single parametric profile cannot account for.
Numerical simulations have shown that asymmetries in the mass distribution have a significant impact on the strong lensing properties of galaxy clusters \citep{2003MNRAS.346...67M, 2007A&A...461...25M}.

We further investigate the complexity of the cluster mass distribution by considering a model P3$\varepsilon$ in which the ellipticity and orientation of the extra components are also free to vary, adding four more free parameters in the modeling.
This model reproduces the positions of the observed multiple images with a $\Delta_{\rm rms}$ value of $0\arcsec.44$, a reduction of $\approx 27\%$ when compared to P3.
In the top panel of Fig. \ref{fig:offset_map}, we show the distribution of the offset of each multiple image.
In the bottom image, the circles indicate the observed positions of the multiple images and the arrows point towards the model predicted positions with sizes ten times the offset values.
The blue lines indicate the direction of the distortion (see Sec. \ref{sec:sl_definitions}), which is in excellent agreement with the observed orientations of the gravitational arcs (tangential and radial) in Fig. \ref{fig:arcs}.
As expected, there is a mild correlation between the directions of the offsets and the distortion.
Since the model P3$\varepsilon$ can reproduce very well the positions of all multiple images with a relatively low number of free parameters, given the number of constraints, we use it as our reference model to study the total mass distribution of MACS~1206.
The best fit parameters and confidence levels are quoted in Table \ref{tab:model_params}.
The inspection of the covariance between the parameters shows all the expected degeneracies in parametric lens models.
Particularly, the external shear parameters show a mild degeneracy with the ellipticity and orientation of the second and third diffuse halos, however not with the central one.

In order to explore the sensitivity of our modeling to specific parametrizations, we also use the NFW and gNFW models to describe the central diffuse component, N3$\varepsilon$ and G3$\varepsilon$ respectively, where $\gamma_{\rm gNFW}$ is fixed to one in the former and left free to vary in the latter.
In these models we keep the same PIEMD parametrization for the second and third diffuse halos as in the reference model P3$\varepsilon$.
The slope value $\gamma_{\rm gNFW}$ of the central diffuse component is not necessarily similar to the slope of the cluster total mass distribution ($\gamma^{\rm total}_{\rm gNFW}$), represented in our parametrization as the sum of the diffuse components (generally more than one) and the galaxy members.
We find that these two best fit models have similar mass distributions, yielding a $\Delta_{\rm rms}$ of $0\arcsec.40$.
Indeed, the reconstructed best fit values are very similar and $\gamma_{\rm gNFW}^{\rm halo}$ is $1.06$, making the two models virtually the same.
Regarding the limitation of the pseudo-elliptical NFW model, the estimated values from the MCMC sampling are $\varepsilon_{\rm NFW} = 0.44_{-0.07}^{+0.09}$ and $\varepsilon_{\rm gNFW} = 0.45_{-0.08}^{+0.08}$ (95\% confidence level), reaching the limit of $\approx 0.5$ mentioned in Sec. \ref{sec:total_mass_smooth_component}.
These two models are used only for the sake of comparison with our reference P3$\varepsilon$ model, since the high values of $\varepsilon_{NFW}$ might lead to nonphysical projected mass distributions.

We investigated whether further increasing the complexity of the cluster total mass distribution reduces the value of $\Delta_{\rm rms}$.
We did that by letting the velocity dispersion and cut-radius of the BCG free to vary, i.e., not following the other member total mass-to-light scaling realation.
Despite these extra free parameters, we find that the $\Delta_{\rm rms}$ does not change and all other best fit parameters remain well within 68\% confidence levels, thus not justifying this extra complexity in the mass modeling.
We also find that by including a fourth diffuse component, whose best fitting position is at $\approx 30 \arcsec$ southeast from the BCG, we cannot improve the overall fit of the multiple image positions.
Similarly, negligible differences are found if we adopt nonzero ellipticities for the cluster members, fixing them to the values measured from the F160W band.
Finally, we verify that by including the three multiple images 22b, 23c, and 24e (marked in red in Fig. \ref{fig:arcs}), located near massive background galaxies, the value of $\Delta_{\rm rms}$ of the reference model increases to $0\arcsec.71$.

\section{Mass distribution of MACS1206}
\label{sec:the_mass_distribution_of_macs1609}

In Fig. \ref{fig:objects_members}, we show the surface mass density contours associated with the diffuse component of the mass distribution of our reference model.
Owing to the inclusion of the three diffuse halos, the overall distribution is clearly asymmetric. However, this distribution is smooth and presents only one pronounced peak close to the BCG.
We also note that the diffuse 2D mass component resembles the spatial distribution of the stellar component (ICL and galaxy members), as well as the hot X-ray gas.
Interestingly, the simulated galaxy cluster \emph{Hera} \citep{2016arXiv160604548M}, extracted from high-resolution $N$-body simulations \citep{2014MNRAS.438..195P}, presents a morphology similar to that of MACS~1206 \citep[see Fig. 1 in][]{2016arXiv160604548M}, which indicates that such asymmetric structures might not be rare in the population of massive clusters.

\begin{figure*}
  \centering
  \includegraphics[width = 0.9\columnwidth]{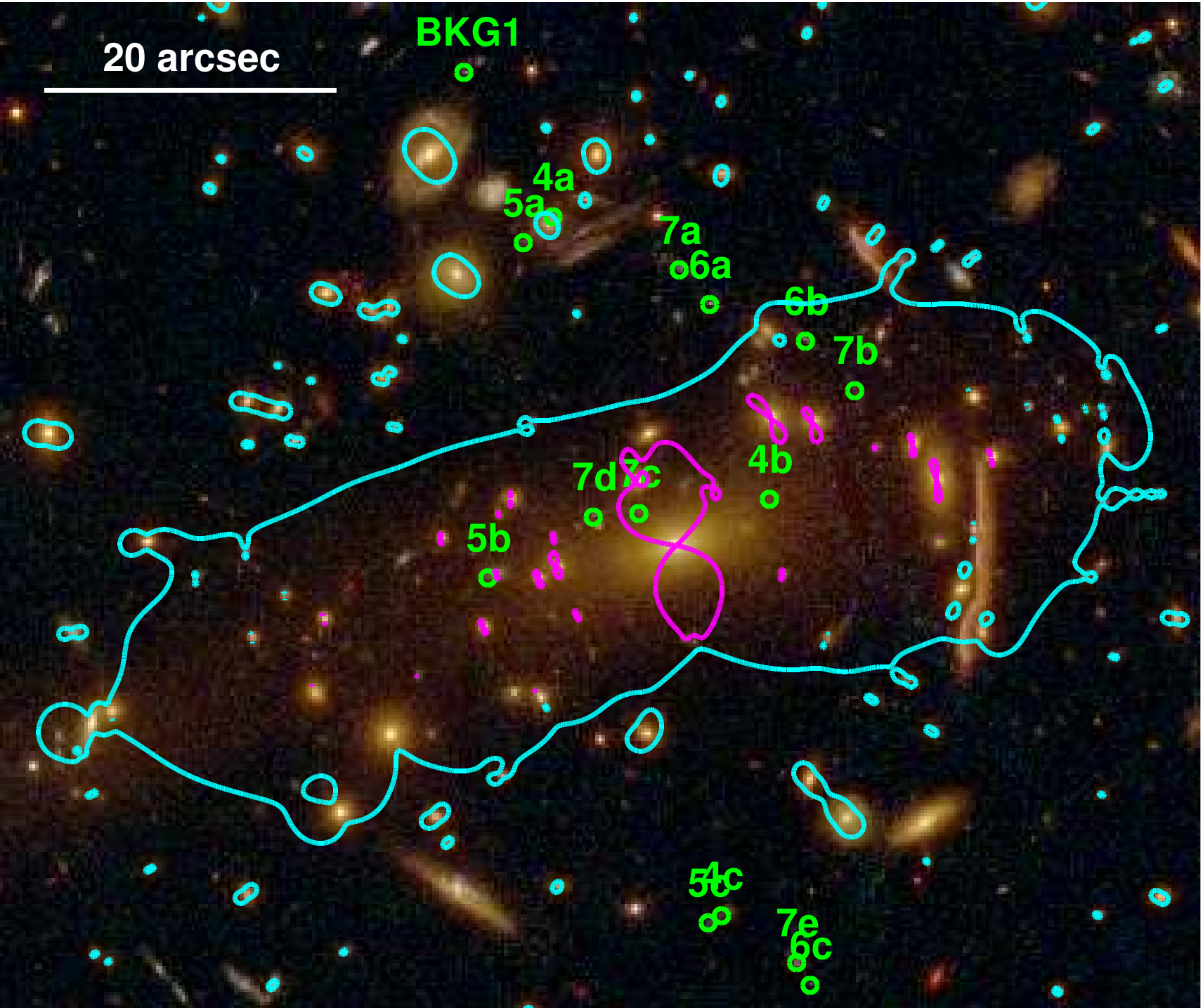}
  \includegraphics[width = 1.1\columnwidth]{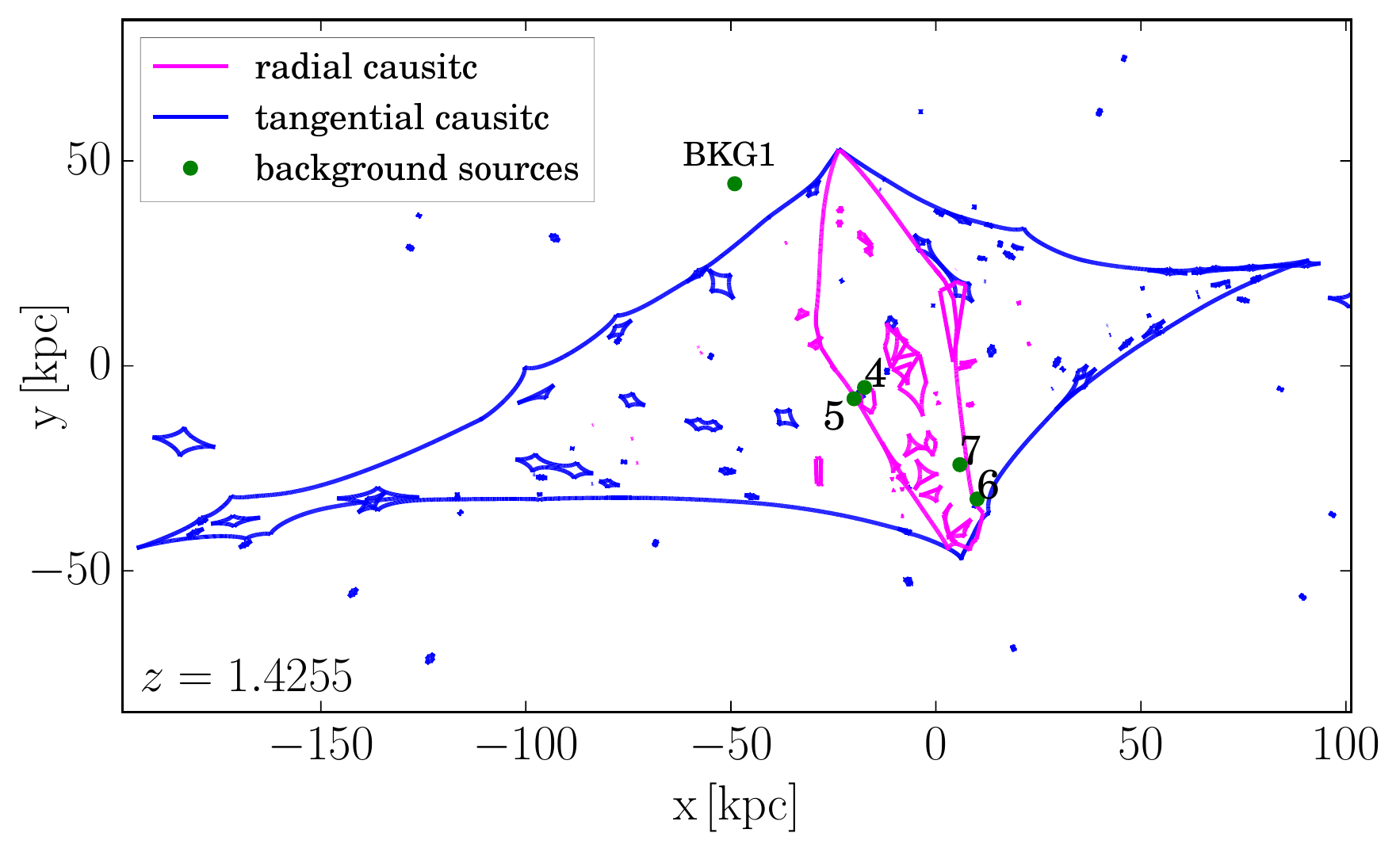}
  \caption{Critical curves and caustics of the reference model P3$\varepsilon$ for a source at $z_{src} = 1.4255$ (the mean redshift values of the sources). Left panel: Tangential (cyan) and radial (magenta) critical lines on the image plane. The green circles show the observed positions of the multiple images belonging to the four families within $\Delta z \leq 0.0011$. BKG1 is a background galaxy not multiply lensed by MACS~1206. Right panel: Tangential (cyan) and radial (magenta) caustics on the source plane, and the reconstructed positions of the background sources.}
  \label{fig:critic_lines}
\end{figure*}
As mentioned above, numerical simulations have suggested that such asymmetric mass distributions increase the likelihood that radial arcs will form and  large radial caustics will be created.
To illustrate this, we show in Fig. \ref{fig:critic_lines} the radial critical lines for sources at $z=1.4255$ (corresponding to the aforementioned background source overdensity) and the respective caustics on the source plane from our reference strong lensing model.
We also indicate the positions of the five background sources located in the narrow redshift range $[1.4248 - 1.4259]$.
Source BKG1 is not multiply lensed, while the other sources have different multiple-image configurations on the image plane.
These five sources are relatively close in projection (within $\approx 100$~kpc) and might be associated with a group in an early stage of formation at $z \approx 1.4$.

\begin{figure*}
  \includegraphics[width = 1.0\columnwidth]{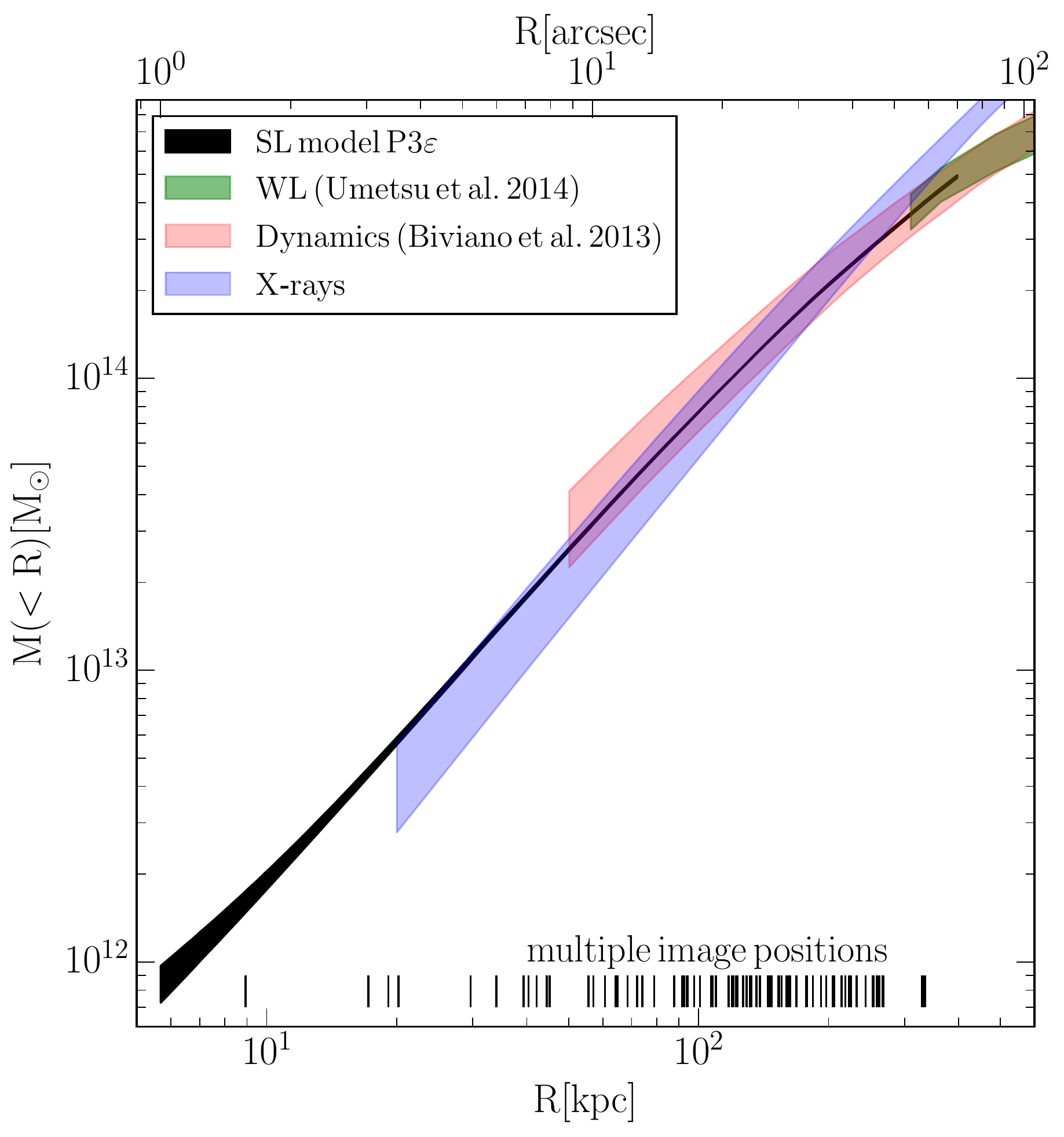}
  \includegraphics[width = 1.0\columnwidth]{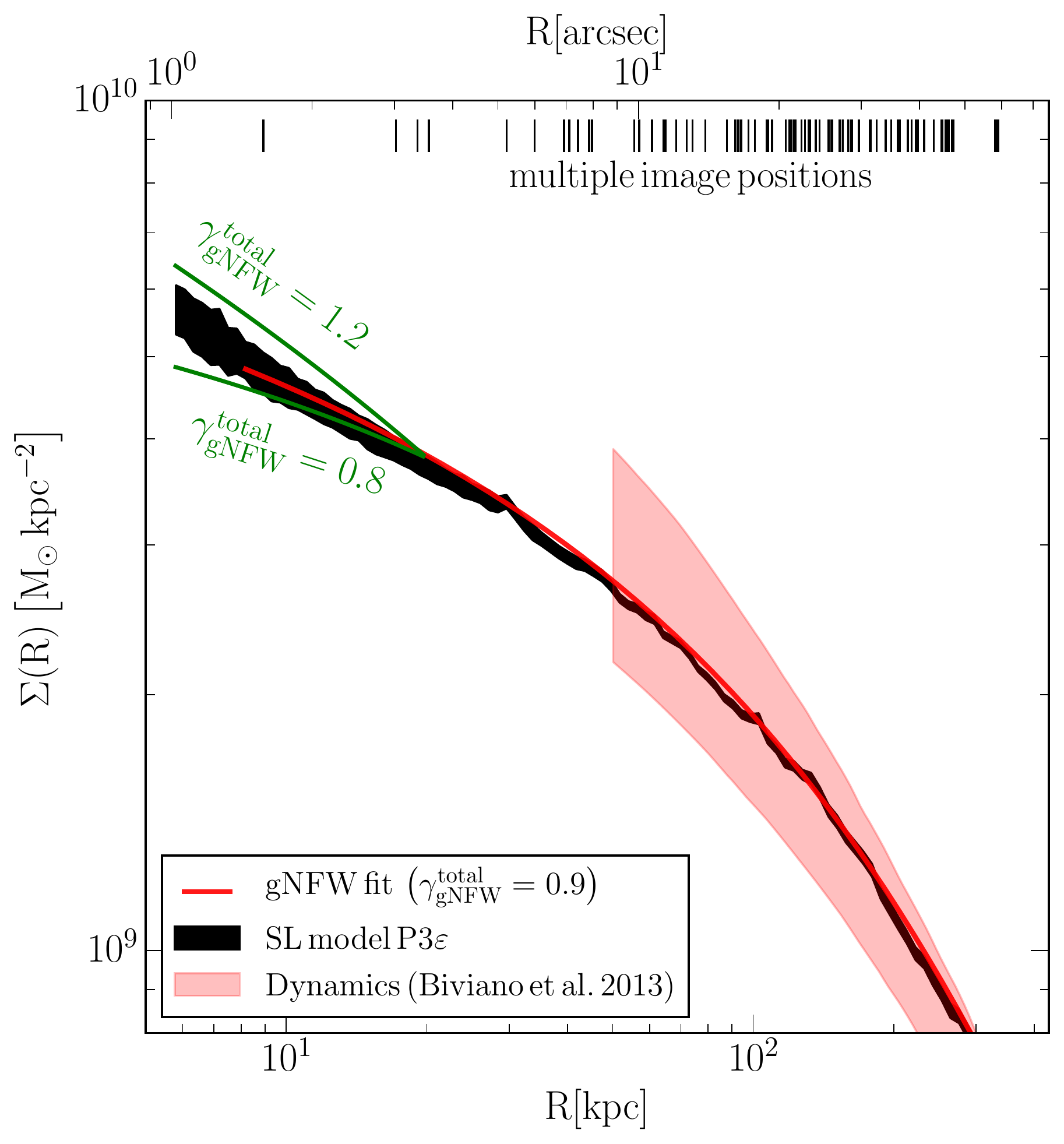}
  \caption{Comparison of different independent projected total mass determinations in MACS~1206. The black region represents the results at the 95\% confidence level of our strong lensing analysis (reference model P3$\varepsilon$). At the 68\% confidence level, the dynamical \citep[][]{2013A&A...558A...1B}, weak-lensing \citep[][]{2014ApJ...795..163U}, and X-ray (see Sect.~\ref{sec:xray}) total mass estimates are shown in red, green, and blue, respectively. The vertical lines indicate the projected radial distances from the cluster center of the multiple images presented in this work. The cumulative projected total mass profile is shown in the left panel. In the right panel we show the projected total surface mass density profile and the red line is a fit using a gNFW model. The green lines show two different gNFW profiles, with values of $\rm \gamma_{gNFW}^{\rm total}$ equal to 0.8 and 1.2, normalized at $\rm R=20\,kpc$.}
  \label{fig:mass_profile}
\end{figure*}
In Figure \ref{fig:mass_profile}, we show the cumulative total mass profile of our reference model P3$\varepsilon$.
To compute the 95\% confidence level region, we extract 200 random models from the MCMC and create total mass maps with a spatial resolution of $0\arcsec.05$.
For each map, we compute the mass within different radii and the 2.3rd and 97.7th percentiles from the distribution of the 200 measurements.
The vertical lines locate the positions of the multiple images, i.e., the region with strong lensing constraints, from $\rm R\simeq 9$~kpc out to $\rm R\simeq 300$~kpc from the BCG center.
The colored regions show the total mass profiles obtained from galaxy member dynamics \citep{2013A&A...558A...1B}, X-ray hydrostatic analysis (see Sec. \ref{sec:xray}), and weak gravitational lensing \citep[not combining with strong lensing,][]{2014ApJ...795..163U}.
Although these independent measurements are valid only at large radii and cannot be extrapolated to the inner regions ($\rm R<50$~kpc), the agreement with our strong lensing mass reconstruction in the overlapping radial range is remarkable.

In the right panel of Figure \ref{fig:mass_profile}, we show the total surface mass density profile (95\% confidence level) and compare it
with the results from \citet{2013A&A...558A...1B}.
We find that one single gNFW profile accurately represents the radial profile of the total mass distribution down to $\rm R = 8$~kpc (solid red curve), where the contribution of the BCG halo starts to become significant. 
The recovered gNFW parameters are $\rho_s = (1.9 \pm 0.3) \times 10^{6} \rm \, M_{\odot}\,kpc^{-3}$, $r_s = (300 \pm 3)\, \rm kpc,$ and $\gamma_{\rm gNFW}^{\rm total} = 0.91 \pm 0.04$ (at 68\% confidence level), corresponding to a concentration of $c \approx 5.1\pm 0.2$.
These values are in very good agreement with those obtained from the dynamical analysis \citep[see Table 3 of ][]{2013A&A...558A...1B}, and show that the radial total mass distribution does not strongly deviate from a NFW profile, since $\gamma_{gNFW}^{\rm total}$ is close to one.
In the same panel, we illustrate two gNFW profiles with slopes of 1.2 and 0.8, with the same $r_s$ value and $\rho_s$ rescaled in order to have the same $\Sigma \rm (R)$ at 20~kpc.
We remark that the measurement of the total mass density profile in projection, within circular apertures, and superposing several diffuse components might also be consistent with profiles different from a gNFW profile. 

\begin{figure}
  \centering
  \includegraphics[width = 1.0\columnwidth]{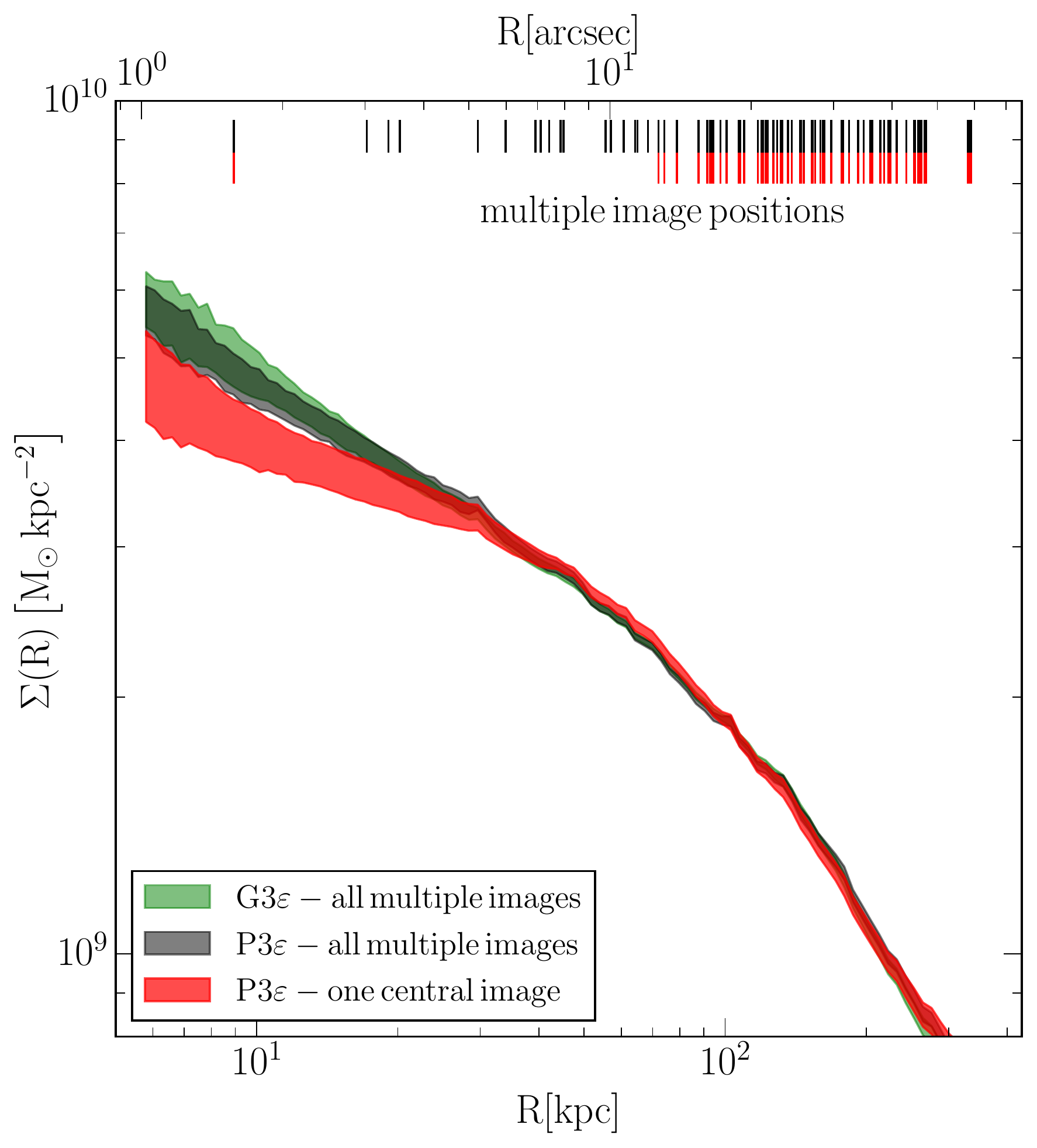}

  \includegraphics[width = 1.0\columnwidth]{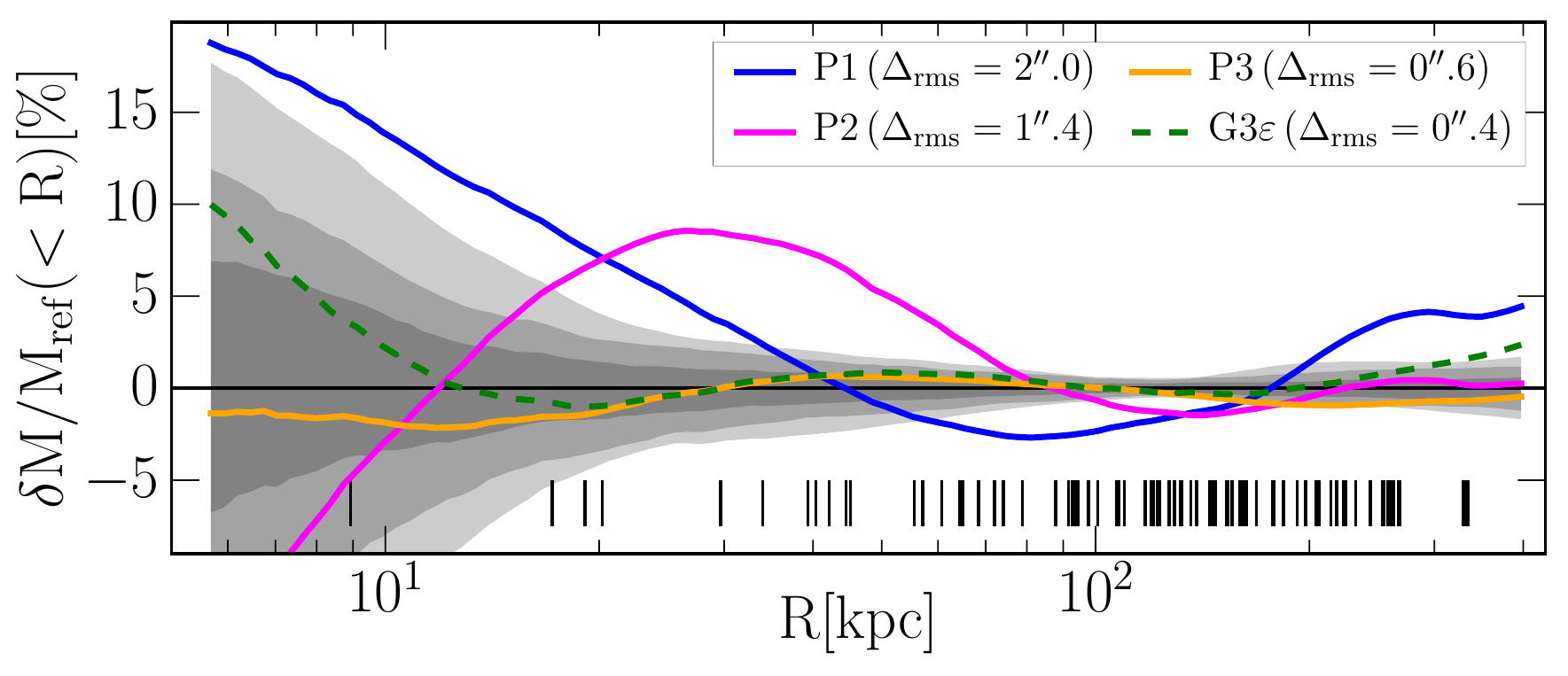}
  \caption{Comparison of the total mass density profiles of MACS~1206 obtained from strong lensing models with different cluster mass parametrizations and sets of multiple images. The vertical lines indicate the projected radial distances from the cluster center of the multiple images presented in this work (in black) and a subset of them (in red). Upper panel: Gray and green regions show, respectively, the profiles for the reference model P3$\varepsilon$ and model G3$\varepsilon$. The red region shows the profile with the same cluster mass parametrization as for the reference model, obtained by excluding all multiple images within 70~kpc in projection of the cluster center except the innermost one. The errors correspond to the 95\% confidence level and are estimated from 200 different models extracted from the corresponding MCMCs. Bottom panel: Radial profiles of the relative difference in the cumulative projected total mass values obtained from the reference model P3$\varepsilon$ and other models with different cluster mass parametrizations ($\rm \delta M = M_{ref} - M_{i}$), all employing the full set of multiple images. The gray regions show the statistical errors at the 68\%, 95\% and 99.8\% confidence levels of the reference model.}

  \label{fig:mass_clusters}
\end{figure}
We also performed a test to investigate possible systematic effects on
the determination of the slope of the total projected mass profile
when only a limited number of strong lensing constraints is used, or
available, in the inner regions.  In Fig. \ref{fig:mass_clusters}, we
compare the density profile obtained with the P3$\varepsilon$ and
G3$\varepsilon$ models using the full set of 82 multiple images with
that obtained with the P3$\varepsilon$ model when only one multiple
image within 70~kpc (in red) is considered.  As discussed above, the
latter configuration reproduces the most common set of constraints
available on the CLASH or Frontier Field clusters, e.g., Abell~1063 and
MACS~J1149 (see Fig. \ref{fig:images_dist}).  We note that the
determination of the density profile using the new large set of
central multiple images in MACS~1206 is very robust against different
models (P3$\varepsilon$ and G3$\varepsilon$ are shown), whereas the
model based on only one central image underestimates the inner slope
of the total density profile.

\subsection{Comparison with previous strong lensing analyses}
\label{sec:comparison_models}

Strong lensing analyses of MACS~1206 have been presented in
\citet{2012ApJ...749...97Z} and \citet{2013ApJ...774..124E}.  The two
studies used a similar set of approximately 50 multiple images
belonging to 12 single background sources.  This provides a relatively
large number of constraints; however, their identification and
distances were based largely on CLASH photometric redshifts, since
only three multiply lensed sources had a spectroscopic redshifts from
CLASH-VLT at that time.  As discussed in our previous strong lensing
analysis of the Frontier Field clusters, which take advantage of large sets of
spectroscopically confirmed multiple images \citep{C2016_2248,
  2016ApJ...822...78G, 2016arXiv161101513L, C2017,
  2017arXiv170206962M}, the use of multiple image systems
identified on the basis of their photometric properties is prone to
intrinsic systematics in the reconstruction of the internal structure of
the mass distribution and in magnification measurements.  Nonetheless,
the use of a large number of photometric multiple images is often
enough to recover the circularized mass density profiles.

Indeed, previous analyses obtained a relatively high value for
$\Delta_{\rm rms}$: \citet{2012ApJ...749...97Z} found
$\Delta_{\rm rms}=1\arcsec.8$, while \citet{2013ApJ...774..124E} found
$\Delta_{\rm rms}=0\arcsec.85$, after removing two misidentified multiple images belonging to two families.
Both these studies used only one
elliptical profile to describe the smooth mass component.
On the other hand, the total projected mass within $\rm 100\,kpc$ of
$(\rm 8 \pm 1)\times 10^{13}M_{\odot}$ reported in
\citealt{2012ApJ...749...97Z} and $\rm (7.11 \pm 0.04) \times
10^{13}M_{\odot}$ in \citet{2013ApJ...774..124E} are in very good
agreement (given the statistical and systematic errors) with the value of the study presented
here of $\rm (7.25 \pm 0.02 )\times 10^{13}M_{\odot}$. The large set
of spectroscopic multiple images used in our new analysis enables
the determination of the asymmetric shape of the smooth mass component
(as shown in Fig. \ref{fig:objects_members}), but also a robust
determination of the mass componenent due to the subhalo population.

We note that the 27 multiple image families spectroscopically confirmed by MUSE contains four with previously known redshifts and seven families for which only photometric information was available (see Table 2 in \citealt{2013ApJ...774..124E}, which updated the list from \citealt{2012ApJ...749...97Z}).
We were not able to confirm only one photometric family from \citet{2013ApJ...774..124E} (their ID~5).
When comparing the new spectroscopic measurements with their model predicted redshifts, we find an error of $\approx
10\%$, which is larger
than the photometric redshift error ($\approx 5\%$). We
also note that the purity of the cluster member catalog adopted in our
new analysis, based on the highly complete sample of MUSE redshifts,
is significantly higher than previous studies where member
galaxies were selected from the cluster red sequence. This further
contributes to the accuracy and precision of our new model.

\subsection{The hydrostatic mass distribution of MACS1206}
\label{sec:xray}

The hydrostatic mass profile shown in Fig. \ref{fig:mass_profile} is based on an archival
Chandra ACIS-I observation (ObsId~3277). We have reprocessed it with a standard pipeline based on CIAO
4.9 \citep{2006SPIE.6270E..1VF} and CALDB 4.7.4 to create a new
events-2 file which includes filtering for grade, status, bad pixels,
and time intervals for anomalous background levels. We obtain a
cumulative good time interval of 22.9 ksec.  To recover the
hydrostatic mass profile, we assume that the hydrostatic equilibrium
holds between the intracluster medium and the gravitational potential,
and that both have a spherically symmetric distribution
\citep[e.g.,][]{2013SSRv..177..119E}. Then we combine the gas density,
temperature, and a parametrized dark matter distribution following the
backward method described in \cite{2013SSRv..177..119E}. The gas
density profile has been recovered from the geometrical deprojection
of the X-ray surface brightness profile extracted from the
exposure-corrected image in the 0.7--2 keV energy range (see Fig. \ref{fig:objects_members}). This energy
range has been chosen to maximize the signal-to-noise ratio. All the
point sources detected with the CIAO routine wavdetect have been
masked. A background defined locally in the same exposure by selecting
three circular regions of $2'$ radius, and between $6'$ and $8'$
from the peak of the cluster's emission, has been used to correct both
the surface brightness profile and the spectra extracted in four
independent spatial bins, by requiring $\sim$3000 net counts in the
0.6--7 keV band in each bin. The spectral fitting has been performed
with Xspec 12.9 \citep{1996ASPC..101...17A}, using a thermal component (apec)
with 3 degrees of freedom (normalization, temperature and metallcity),
absorbed by a Galactic column density $n_H$ fixed to the local value
of $4.35 \times 10^{20}$ cm$^{-2}$ \citep{2005A&A...440..775K}, and at
the nominal redshift of 0.439. A NFW functional form
for the gravitational potential is adopted with two free parameters
(normalization and $R_{200}$). These parameters are constrained by
performing a grid-based search for a minimum in the distribution of
the $\chi^2$ evaluated by comparing the observed spectral temperature
profile (also considering the relative errors) and the one predicted
from the inversion of the hydrostatic equilibrium equation in which
the observed gas density profile is used \citep[for further details
see][]{2010A&A...524A..68E}.
At each radius, we associate a symmetric
error on the mass profile that represents the range of values allowed
from the $1 \sigma$ statistical uncertainties on the two free parameters
(i.e., as defined from the region enclosed within $\Delta
\chi^2=2.3$). The projected cumulative X-ray mass profile is shown in
Fig.~\ref{fig:mass_profile} and found to be in general good agreement with
other determinations. It is important to note that this comparison
should be limited to the radial range in which the ICM properties are
properly constrained. In the present case, the profile of the hydrostatic mass density profile can be extended down to $ 4\arcsec = 23$~kpc radius, where the temperature can be determined.
Moreover, a clear tension between {\it Chandra} ACIS
and {\it XMM-Newton} EPIC spectral measurements is observed in
MACS~1206 at $r >$ 500 kpc, as shown in \cite{2014ApJ...794..136D}
(see their Fig.~3). This systematic difference is partially due to
residual mismatches in the cross-calibration of the
two instruments \citep[see][]{2015A&A...575A..30S} and has been
observed to be particularly significant ($\sim$20\% and above) in high-temperature ($T>10$ keV) systems such as MACS~1206. As a result of these
significant uncertainties in the X-ray spectral analysis, we conclude that the
hydrostatic mass in MACS~1206 is presently well constrained only over
the radial range between $\sim$ 20~kpc and 500 kpc, with the relatively shallow archival \emph{Chandra} data.

\section{Conclusions}
\label{sec:conclusions}

In this work, we have presented a new strong lensing analysis of the galaxy cluster MACS~1206, based on the identification of 23 new spectroscopically confirmed multiply lensed sources, using deep MUSE archival observations in combination with CLASH imaging.
With the new measurements, we have confirmed seven multiple image families which were previously identified with photometric information alone, and four with previously CLASH-VLT (VIMOS) redshift measurements.
In total, we therefore used 27 spectroscopic families at $z=1.0-6.1$, lensed into 82 multiple images as constraints for our strong lensing model.
As in previous studies exploiting MUSE and HST observations, these sources are largely low-luminosity Lyman-$\alpha$ emitters at $z>3$, including extended Lyman-$\alpha$ halos.
Remarkably, this study has revealed an exceptional number of central multiple images, which, within the inner 20~kpc, is approximately a factor of 4 larger than in any other known cluster to date.
As a result, this spectroscopic data set has allowed us to obtain a very robust estimate of the inner slope of the projected total mass density profile and a detailed reconstruction of the inner projected mass distribution, which has been shown to be challenging when using photometrically selected multiple images \citep{2016A&A...588A..99L}.
We summarize our results as follows:
\begin{itemize}

\item We model the total mass distribution with a clumpy component associated with the galaxy member halos, which are reliably identified with MUSE spectroscopy, and a smooth component tracing the cluster contribution at larger radial scales.
We find that the model that best fits this large number of multiple images requires significant asymmetry, departing from the elliptical parametrization usually adopted for the diffuse component of a cluster mass distribution.
To reproduce this more complex mass distribution, we add extra components in the parametrization of the total mass distribution, significantly improving the overall fit when compared to a model with a single diffuse elliptical halo.
With this optimized model, we are able to reproduce the observed image positions with a $\Delta_{\rm rms}$ of $0\arcsec.44$, in keeping with previous high precision lens models based on CLASH/HFF data supplemented with MUSE spectroscopy. The resulting projected mass distribution closely follows the asymmetric distribution of the galaxy members and the ICL, as well as the intra-cluster gas (see Fig. \ref{fig:objects_members}).

\item We believe that this exceptional number of central multiple images is due to such an asymmetric inner mass distribution, as corroborated by numerical simulations of forming massive clusters.
The direct inspection of the radial caustics in MACS~1206 reveals indeed that they are unusually wide, thus favoring the formation of a large number of radial images.
Based on their radial and tangential magnifications from our reference model, we find that $\sim$30\% of the 82 multiple images are classified as ``radial arcs''.

\item The large number of constraints in the inner 50~kpc leads to a determination of the projected total mass density profile that is very robust against different parametrizations of the diffuse mass component.
We find that systematic errors due to the adopted parametrization are comparable with the statistical errors (a few percent at $\rm R \gtrsim 50$~kpc), while the accuracy in the mass determination remains within 10\% (at 95\% confidence level) down to $R=10$~kpc.
We fit the azimuthally averaged projected total density profile with a gNFW model and find a central slope of $\gamma_{gNFW}^{\rm total} = 0.91 \pm 0.04$, close to the canonical NFW.

\item Using the same reference model parametrization, we test the robustness of the inner slope determination by keeping only one central image within $R<70$~kpc, which is the typical case in other clusters.
Interestingly, we find that in this case the best fit inner profile is no longer stable against different parametrizations and might lead to significant biases on the innermost slope measurements, especially in lenses with irregular mass distribution in the core.

\end{itemize}

The determination of the total mass distribution with unprecedented accuracy in the inner core of MACS~1206 opens the opportunity to accurately measure the dark matter profile by separating the baryonic contribution \citep[see, e.g.,][]{2017arXiv170510322B}.
We will address this central issue in a forthcoming paper, taking advantage of the internal kinematics of the BCG, which provides a strong independent constraint on the gravitational potential in the innermost regions.
We also plan to investigate the conditions of the formation of radial arcs with the state-of-the-art $N$-body and hydrodynamical simulations of clusters.
Moreover, the increased number of strong lensing constraints presented in this work makes MACS~1206 an interesting study case for strong lensing cosmography \citep{2010Sci...329..924J, C2016_2248, 2017arXiv170405380A}.
The strong lensing maps (lens convergence, shear, and magnification), as well as the configuration files of the {\tt lenstool} software, are publicly available\footnote{All the files can be found at the link: \url{http://www.fe.infn.it/u/gbcaminha/}.}.

\begin{acknowledgements}
  The authors thank the anonymous referee for the useful comments on the manuscript.
  We acknowledge financial support from PRIN-INAF 2014 1.05.01.94.02.
  C.G. acknowledges support from VILLUM FONDEN Young Investigator Programme grant 10123.
  C.D.C. is supported by the Erasmus Mundus Joint Doctorate Program by the grant number 2014-0707 from the EACEA of the European Commission.
  This work made use of the CHE cluster, managed and funded by ICRA/CBPF/MCTI, with financial support from FINEP (grant 01.07.0515.00 from CT-INFRA - 01/2006) and FAPERJ (grants E-26/171.206/2006 and E-26/110.516/2012).
  This research made use of Astropy, a community-developed core Python package for Astronomy \citep{2013A&A...558A..33A}.
  This work made use of data taken under the ESO program IDs 095.A-0181(A), 097.A-0269(A), and 186.A-0798(A).
\end{acknowledgements}

\bibliographystyle{aa}
\bibliography{references}


\newpage
\noindent {\bf \large Appendix A: Multiple images}
\label{ap:multiple_images}

In Table \ref{tab:multiple_images}, we present relevant information about the multiple images used in our strong lensing model.
In Fig. \ref{fig:specs}, we show the MUSE spectra around relevant spectral features used to determine the redshift of multiple images and the corresponding image cutouts for multiple images.
The median values and the 68\%, 95\%, and 99.7\% confidence levels of the free parameters of the model P3$\varepsilon$, as determined from the MCMC analyses, are shown in Table \ref{tab:model_params}.

\renewcommand{\thetable}{A.\arabic{table}}
\renewcommand{\thefigure}{A.1}

\longtab{
\setcounter{table}{0}
\begin{longtable}{lccccrlcc}
\caption{\label{tab:multiple_images} Information on spectroscopically identified multiple images in MACS~1206.}\\
\hline\hline
 ID & RA & DEC & $z_{\rm MUSE}$ & $z_{\rm previous}^a$ & $\mu_{tot}$ & $\mu_{tan}/\mu_{rad}$ &  Comments\\
\hline
\endfirsthead
\caption{continued.}\\
\hline\hline
 ID & RA & DEC & $z_{\rm MUSE}$ & $z_{\rm previous}^a$ & $\mu_{tot}$ & $\mu_{tan}/\mu_{rad}$ &  Comments\\
\hline
\endhead
\hline
\endfoot
1a  & 181.550916 & $-$8.797422  & 1.0121 & ---    & $10.2_{-0.7}^{+0.6}$ & $1.48_{-0.06}^{+0.05}$ & HST/F606W \\
1b  & 181.549604 & $-$8.799294  & 1.0121 & ---    &  $6.5_{-0.6}^{+0.7}$ & $0.67_{-0.05}^{+0.06}$ & HST/F606W \\
1c  & 181.548870 & $-$8.806655  & 1.0121 & ---    &  $3.4_{-0.1}^{+0.1}$ & $1.27_{-0.02}^{+0.02}$ & HST/F606W \\
\hline
2a  & 181.546790 & $-$8.795680  & 1.0369 & 1.0348 &  $6.9_{-0.2}^{+0.2}$ & $3.19_{-0.10}^{+0.10}$ & HST/F105W \\
2b  & 181.544819 & $-$8.799553  & 1.0369 & 1.0336 &  $3.0_{-0.2}^{+0.3}$ & $2.16_{-0.12}^{+0.15}$ & HST/F105W \\
2c  & 181.545102 & $-$8.803052  & 1.0369 & 1.0336 &  $6.9_{-0.2}^{+0.2}$ & $2.18_{-0.12}^{+0.13}$ & HST/F105W \\
\hline
3a  & 181.550570 & $-$8.795568  & 1.0433 & ---    &  $5.2_{-0.2}^{+0.2}$ & $1.56_{-0.03}^{+0.03}$ & HST/F606W \\
3b  & 181.547611 & $-$8.799811  & 1.0433 & ---    &  $5.1_{-0.3}^{+0.3}$ & $0.77_{-0.05}^{+0.06}$ & HST/F606W \\
3c  & 181.548607 & $-$8.805281  & 1.0433 & ---    &  $6.1_{-0.3}^{+0.3}$ & $2.42_{-0.15}^{+0.16}$ & HST/F606W \\
\hline
4a  & 181.552987 & $-$8.794699  & 1.4248 & ---    &  $6.9_{-0.6}^{+0.7}$ & $2.31_{-0.27}^{+0.32}$  & MUSE \\
4b  & 181.548830 & $-$8.800057  & 1.4248 & ---    &  $7.4_{-1.2}^{+1.5}$ & $0.09_{-0.02}^{+0.02}$  & MUSE \\
4c  & 181.549752 & $-$8.807965  & 1.4248 & ---    &  $3.9_{-0.1}^{+0.2}$ & $1.46_{-0.03}^{+0.04}$  & MUSE \\
(4c)& 181.554126 & $-$8.801587  & ---    & ---    &  $2.2_{-0.2}^{+0.2}$ & $0.48_{-0.03}^{+0.04}$  & blended with gal. \\
\hline
5a  & 181.553557 & $-$8.795189  & 1.4254 & ---    &  $5.8_{-0.4}^{+0.4}$ & $1.46_{-0.12}^{+0.13}$  & HST/F110W \\
5b  & 181.554237 & $-$8.801552  & 1.4254 & ---    &  $2.0_{-0.1}^{+0.1}$ & $0.74_{-0.04}^{+0.03}$  & HST/F110W \\
5c  & 181.550005 & $-$8.808098  & 1.4254 & ---    &  $3.8_{-0.1}^{+0.1}$ & $1.45_{-0.04}^{+0.04}$  & HST/F110W \\
\hline
6a  & 181.549979 & $-$8.796362  & 1.4255 & ---    & $18.3_{-1.6}^{+2.1}$ & $2.60_{-0.13}^{+0.13}$  & HST/F606W \\
6b  & 181.548139 & $-$8.797058  & 1.4255 & 2.1743$^b$ & $27.5_{-2.4}^{+2.6}$ & $3.76_{-0.39}^{+0.37}$  & HST/F606W \\
6c  & 181.548050 & $-$8.809283  & 1.4255 & ---    & $3.2_{-0.1}^{+0.1}$ & $1.55_{-0.02}^{+0.03}$  & HST/F606W \\
\hline
7a  & 181.550563 & $-$8.795704  & 1.4257 & ---    &  $9.8_{-0.7}^{+0.7}$ & $2.04_{-0.06}^{+0.07}$ & HST/F606W \\
7b  & 181.547193 & $-$8.797998  & 1.4257 & ---    &  $6.2_{-0.3}^{+0.4}$ & $1.19_{-0.05}^{+0.05}$ & HST/F606W \\
7c  & 181.551339 & $-$8.800328  & 1.4257 & ---    &  $3.9_{-0.5}^{+0.6}$ & $0.29_{-0.03}^{+0.03}$ & HST/F606W \\
7d  & 181.552213 & $-$8.800398  & 1.4257 & ---    &  $5.3_{-0.5}^{+0.6}$ & $0.22_{-0.02}^{+0.02}$ & HST/F606W \\
7e  & 181.548307 & $-$8.808856  & 1.4257 & ---    &  $3.4_{-0.1}^{+0.1}$ & $1.57_{-0.03}^{+0.03}$ & HST/F606W \\
\hline
8a  & 181.553657 & $-$8.795756  & 1.4864 & ---    &  $6.0_{-0.4}^{+0.3}$ & $1.56_{-0.06}^{+0.07}$ & HST/F110W \\
8b  & 181.554524 & $-$8.801104  & 1.4864 & ---    &  $3.1_{-0.1}^{+0.2}$ & $0.87_{-0.05}^{+0.04}$ & HST/F110W \\
8c  & 181.549957 & $-$8.808887  & 1.4864 & ---    &  $3.5_{-0.1}^{+0.1}$ & $1.49_{-0.04}^{+0.05}$ & HST/F110W \\
\hline
9a  & 181.546741 & $-$8.793144  & 1.9600 & ---    &  $6.8_{-0.3}^{+0.3}$ & $3.42_{-0.11}^{+0.14}$ & HST/F606W \\
9b  & 181.543273 & $-$8.797812  & 1.9600 & ---    &  $6.4_{-0.4}^{+0.4}$ & $3.71_{-0.24}^{+0.29}$ & HST/F606W \\
9c  & 181.544378 & $-$8.807486  & 1.9600 & ---    &  $5.1_{-0.3}^{+0.4}$ & $2.71_{-0.07}^{+0.10}$ & HST/F606W \\
\hline
10a & 181.552450 & $-$8.795001  & 2.5393 & 2.5398 &  $9.4_{-0.7}^{+0.7}$ & $1.41_{-0.05}^{+0.06}$ & HST/F606W \\
10b & 181.546604 & $-$8.797465  & 2.5393 & ---    &  $5.9_{-0.3}^{+0.2}$ & $0.92_{-0.04}^{+0.04}$ & HST/F606W \\
10c & 181.550487 & $-$8.799957  & ---    & ---    &  $1.7_{-0.2}^{+0.2}$ & $0.39_{-0.02}^{+0.02}$ & HST/F606W \\
10d & 181.554894 & $-$8.800160  & ---    & ---    &  $4.9_{-0.3}^{+0.3}$ & $0.92_{-0.06}^{+0.06}$ & HST/F606W \\
10e & 181.548827 & $-$8.811813  & ---    & ---    &  $3.0_{-0.1}^{+0.1}$ & $1.61_{-0.04}^{+0.05}$ & HST/F606W \\
\hline
11a & 181.562654 & $-$8.796672  & 3.0358 & 3.0363 &  $4.6_{-0.2}^{+0.2}$ & $3.13_{-0.08}^{+0.12}$ & HST/F606W \\
11b & 181.562495 & $-$8.804911  & 3.0358 & 3.0371 &  $4.3_{-0.3}^{+0.4}$ & $4.10_{-0.23}^{+0.24}$ & HST/F606W \\
11c & 181.560573 & $-$8.808988  & 3.0358 & 3.0372 &  $7.5_{-0.4}^{+0.5}$ & $5.26_{-0.35}^{+0.37}$ & HST/F606W \\
\hline
12a & 181.548632 & $-$8.793717 & 3.3890 & ---     & $15.5_{-1.2}^{+1.1}$ & $4.32_{-0.28}^{+0.32}$  & MUSE \\
12b & 181.546121 & $-$8.795387 & 3.3890 & ---     &  $7.2_{-0.5}^{+0.5}$ & $3.13_{-0.20}^{+0.22}$  & MUSE \\
12c & 181.553268 & $-$8.800197 & 3.3890 & ---     &  $4.2_{-0.3}^{+0.3}$ & $0.28_{-0.02}^{+0.03}$  & MUSE \\
(12d)&181.551478 & $-$8.800081 & ---    & ---     &  $1.5_{-0.1}^{+0.1}$ & $0.58_{-0.04}^{+0.05}$  & blended with gal., low $\mu$ \\
(12e)&181.547097 & $-$8.812472 & ---    & ---     &  $3.1_{-0.1}^{+0.1}$ & $1.72_{-0.04}^{+0.04}$  & outside MUSE FoV \\
\hline
13a & 181.549416 & $-$8.801768  & 3.3961 & ---    &  $5.9_{-1.0}^{+1.7}$ & $0.18_{-0.03}^{+0.04}$  & MUSE \\
13b & 181.547859 & $-$8.802115  & 3.3961 & ---    & $14.4_{-2.0}^{+2.1}$ & $0.24_{-0.03}^{+0.04}$  & MUSE \\
13c & 181.553770 & $-$8.804759  & 3.3961 & ---    &  $6.8_{-0.6}^{+0.7}$ & $1.26_{-0.07}^{+0.08}$  & MUSE \\
13d & 181.551389 & $-$8.806864  & 3.3961 & ---    & $12.0_{-1.0}^{+1.1}$ & $2.05_{-0.13}^{+0.14}$  & MUSE \\
(13e)&181.554500 & $-$8.788383  & ---    & ---    &  $2.9_{-0.1}^{+0.1}$ & $1.67_{-0.03}^{+0.03}$  & outside MUSE FoV \\
\hline
14a & 181.566558 & $-$8.804480  & 3.7531 & ---    &  $6.2_{-0.6}^{+0.6}$ & $3.93_{-0.30}^{+0.34}$  & HST/F625W \\
14b & 181.566475 & $-$8.804733  & 3.7531 & ---    &  $3.2_{-0.4}^{+0.5}$ & $4.10_{-0.31}^{+0.34}$  & HST/F625W \\
14c & 181.566475 & $-$8.805147  & 3.7531 & ---    &  $7.5_{-0.7}^{+0.8}$ & $5.44_{-0.40}^{+0.42}$  & HST/F625W \\
14d & 181.566275 & $-$8.806328  & 3.7531 & ---    &  $6.5_{-0.8}^{+1.0}$ & $8.36_{-0.87}^{+0.99}$  & HST/F625W \\
14e & 181.565591 & $-$8.807690  & 3.7531 & ---    &  $7.2_{-0.4}^{+0.6}$ & $6.21_{-0.23}^{+0.25}$  & HST/F625W \\
\hline
15a & 181.555962 & $-$8.791635  & 3.7611 & ---    &  $4.2_{-0.1}^{+0.1}$ & $2.44_{-0.10}^{+0.10}$  & HST/F606W \\
15b & 181.557600 & $-$8.803056  & 3.7611 & ---    &  $4.5_{-0.4}^{+0.6}$ & $0.67_{-0.10}^{+0.13}$  & HST/F606W \\
15c & 181.551748 & $-$8.810964  & ---    & ---    &  $3.8_{-0.1}^{+0.1}$ & $1.70_{-0.07}^{+0.06}$  & HST/F606W \\
\hline
16a & 181.554584 & $-$8.791202  & 3.7617 & ---    &  $4.1_{-0.1}^{+0.1}$ & $1.99_{-0.05}^{+0.06}$  & MUSE \\
16b & 181.546465 & $-$8.799671  & 3.7617 & ---    &  $2.1_{-0.1}^{+0.1}$ & $0.35_{-0.02}^{+0.02}$  & MUSE \\
16c & 181.556520 & $-$8.802471  & 3.7617 & ---    &  $2.7_{-0.1}^{+0.2}$ & $0.61_{-0.05}^{+0.07}$  & MUSE \\
(16d)&181.550432 & $-$8.811056  & ---    & ---    &  $3.7_{-0.1}^{+0.1}$ & $1.60_{-0.05}^{+0.05}$  & MUSE edge \\
\hline
17a & 181.556136 & $-$8.795620  & 3.8224 & ---    &  $8.3_{-0.6}^{+0.7}$ & $2.22_{-0.14}^{+0.18}$  & MUSE \\
17b & 181.556958 & $-$8.799422  & 3.8224 & ---    &  $8.7_{-0.9}^{+0.9}$ & $2.44_{-0.19}^{+0.16}$  & MUSE \\
(17c)&181.547557 & $-$8.798385  & ---    & ---    &  $3.0_{-0.3}^{+0.3}$ & $0.30_{-0.03}^{+0.03}$  & low $\mu$ \\
(17d)&181.550041 & $-$8.813393  & ---    & ---    &  $2.8_{-0.1}^{+0.1}$ & $1.49_{-0.04}^{+0.05}$  & outside MUSE FoV \\
\hline
18a & 181.555376 & $-$8.796714  & 4.0400 & ---    & $17.0_{-2.0}^{+2.5}$ & $3.02_{-0.29}^{+0.39}$  & HST/F814W \\
18b & 181.555927 & $-$8.798595  & 4.0400 & ---    &  $3.8_{-0.4}^{+0.4}$ & $1.56_{-0.09}^{+0.11}$  & HST/F814W \\
(18c)&181.549137 & $-$8.798385  & ---    & ---    &  $2.8_{-0.1}^{+0.1}$ & $1.54_{-0.04}^{+0.04}$  & blended with gal., low $\mu$ \\
\hline
19a & 181.562084 & $-$8.794875  & 4.0520 & ---    &  $3.9_{-0.1}^{+0.1}$ & $2.58_{-0.06}^{+0.06}$  & HST/F814W \\
19b & 181.561873 & $-$8.805239  & 4.0520 & ---    &  $8.1_{-0.6}^{+0.7}$ & $5.27_{-0.61}^{+0.57}$  & HST/F814W \\
19c & 181.559788 & $-$8.809463  & 4.0520 & ---    &  $7.3_{-0.4}^{+0.5}$ & $4.82_{-0.31}^{+0.36}$  & HST/F814W \\
\hline
20a & 181.547472 & $-$8.800476  & 4.0553 & ---    &  $9.0_{-1.8}^{+2.7}$ & $0.08_{-0.02}^{+0.02}$  & MUSE \\
20b & 181.556839 & $-$8.803813  & 4.0553 & ---    &  $3.9_{-0.2}^{+0.3}$ & $0.65_{-0.09}^{+0.10}$  & MUSE \\
(20c)&181.555448 & $-$8.790286  & ---    & ---    &  $3.5_{-0.1}^{+0.1}$ & $1.54_{-0.04}^{+0.04}$  & MUSE edge \\
(20d)&181.551634 & $-$8.810099  & ---    & ---    &  $4.5_{-0.2}^{+0.2}$ & $1.96_{-0.05}^{+0.05}$  & MUSE edge \\
\hline
21a & 181.543431 & $-$8.797674  & 4.0718 & ---    &  $3.6_{-0.1}^{+0.2}$ & $1.88_{-0.10}^{+0.10}$  & MUSE \\
21b & 181.551462 & $-$8.800814  & 4.0718 & ---    &  $2.0_{-0.4}^{+0.8}$ & $0.10_{-0.03}^{+0.04}$  & MUSE \\
21c & 181.552850 & $-$8.801115  & 4.0718 & ---    &  $1.6_{-0.2}^{+0.2}$ & $0.28_{-0.02}^{+0.02}$  & MUSE \\
21d & 181.553430 & $-$8.801204  & 4.0718 & ---    &  $2.2_{-0.2}^{+0.2}$ & $0.23_{-0.01}^{+0.02}$  & MUSE \\
(21e)&181.550064 & $-$8.791270  & ---    & ---    &  $5.8_{-0.3}^{+0.3}$ & $2.60_{-0.07}^{+0.09}$  & outside MUSE FoV \\
\hline
22a & 181.544328 & $-$8.791418  & 4.2913 & ---    &  $6.6_{-0.3}^{+0.4}$ & $4.74_{-0.21}^{+0.23}$  & HST/F105W \\
22b$^\ast$& 181.540282 & $-$8.796562  & 4.2913 & ---    & --- & ---  & close to bkg. galaxy \\
22c & 181.540884 & $-$8.806094  & 4.2913 & ---    &  $7.0_{-0.4}^{+0.5}$ & $4.67_{-0.20}^{+0.24}$  & HST/F105W \\
\hline
23a & 181.563252 & $-$8.796893  & 4.7293 & ---    &  $5.8_{-0.3}^{+0.3}$ & $4.08_{-0.15}^{+0.24}$  & MUSE \\
23b & 181.563537 & $-$8.803670  & 4.7293 & ---    &  $6.3_{-0.4}^{+0.5}$ & $5.22_{-0.52}^{+0.61}$  & MUSE \\
23c$^\ast$& 181.559832 & $-$8.811526  & 4.7293 & ---    & --- & ---  & close to bkg. galaxy \\
\hline
24a & 181.551378 & $-$8.791300  & 5.6984 & ---    &  $6.4_{-0.3}^{+0.4}$ & $2.67_{-0.09}^{+0.10}$  & HST/F105W \\
24b & 181.544085 & $-$8.797094  & 5.6984 & ---    &  $3.7_{-0.2}^{+0.2}$ & $1.63_{-0.09}^{+0.09}$  & HST/F105W \\
24c & 181.550911 & $-$8.800607  & 5.6984 & ---    &  $0.39_{-0.04}^{+0.03}$ & $0.53_{-0.07}^{+0.04}$  & MUSE \\
24d & 181.555210 & $-$8.801066  & 5.6984 & ---    &  $4.4_{-0.3}^{+0.4}$ & $0.34_{-0.03}^{+0.02}$  & HST/F105W \\
24e$^\ast$& 181.546977 & $-$8.812247  & ---    & 5.7030 & --- & ---  & close to bkg. galaxy \\
\hline
25a & 181.559714 & $-$8.796562  & 5.7927 & ---    &  $8.8_{-0.5}^{+0.5}$ & $4.35_{-0.16}^{+0.19}$  & MUSE \\
25b & 181.560102 & $-$8.800177  & 5.7927 & ---    &  $9.0_{-0.7}^{+0.7}$ & $3.74_{-0.31}^{+0.34}$  & MUSE \\
(25c)&181.553429 & $-$8.813652  & ---    & ---    &  $3.2_{-0.1}^{+0.1}$ & $1.90_{-0.08}^{+0.09}$  & outside MUSE FoV \\
\hline
26a & 181.550711 & $-$8.803112  & 6.0106 & ---    & $25.2_{-5.2}^{+9.9}$ & $0.23_{-0.04}^{+0.04}$  & MUSE \\
26b & 181.551211 & $-$8.803668  & 6.0106 & ---    & $46.9_{-7.6}^{+11}$  & $0.22_{-0.05}^{+0.07}$  & MUSE \\
(26c)&181.553742 & $-$8.786102  & ---    & ---    &  $2.6_{-0.1}^{+0.1}$ & $1.54_{-0.03}^{+0.03}$  & outside MUSE FoV \\
\hline
27a & 181.553251 & $-$8.798825 & 6.0601 & ---     & $13.8_{-1.3}^{+1.7}$ & $0.43_{-0.05}^{+0.05}$  & MUSE \\
27b & 181.551691 & $-$8.799167 & 6.0601 & ---     &  $4.8_{-0.6}^{+0.8}$ & $0.71_{-0.06}^{+0.06}$  & MUSE \\
(27c)&181.546863 & $-$8.814885  & ---    & ---    &  $2.6_{-0.1}^{+0.1}$ & $1.43_{-0.04}^{+0.03}$  & outside MUSE FoV \\
\end{longtable}
\tablefoot{The magnification values, and their 68\% confidence level errors, are computed using the reference model P3$\varepsilon$. IDs in brackets are model predicted multiple images with no identification in the MUSE and HST data, therefore not used in the modeling.}
\tablefoottext{\textasteriskcentered}{Multiple images nearby background galaxies that are not used in the strong lensing models.}
\tablefoottext{a}{Spectroscopic redshifts from CLASH-VLT with VIMOS \citep{2012ApJ...749...97Z}.}
\tablefoottext{b}{The redshift of the multiple image 6b is reported with quality flag 2, i.e. reliability of 80\%, in the public catalogue by \citet{2013A&A...558A...1B}.}
}

\begin{figure*}
Family 1:

   \includegraphics[width = 0.666\columnwidth]{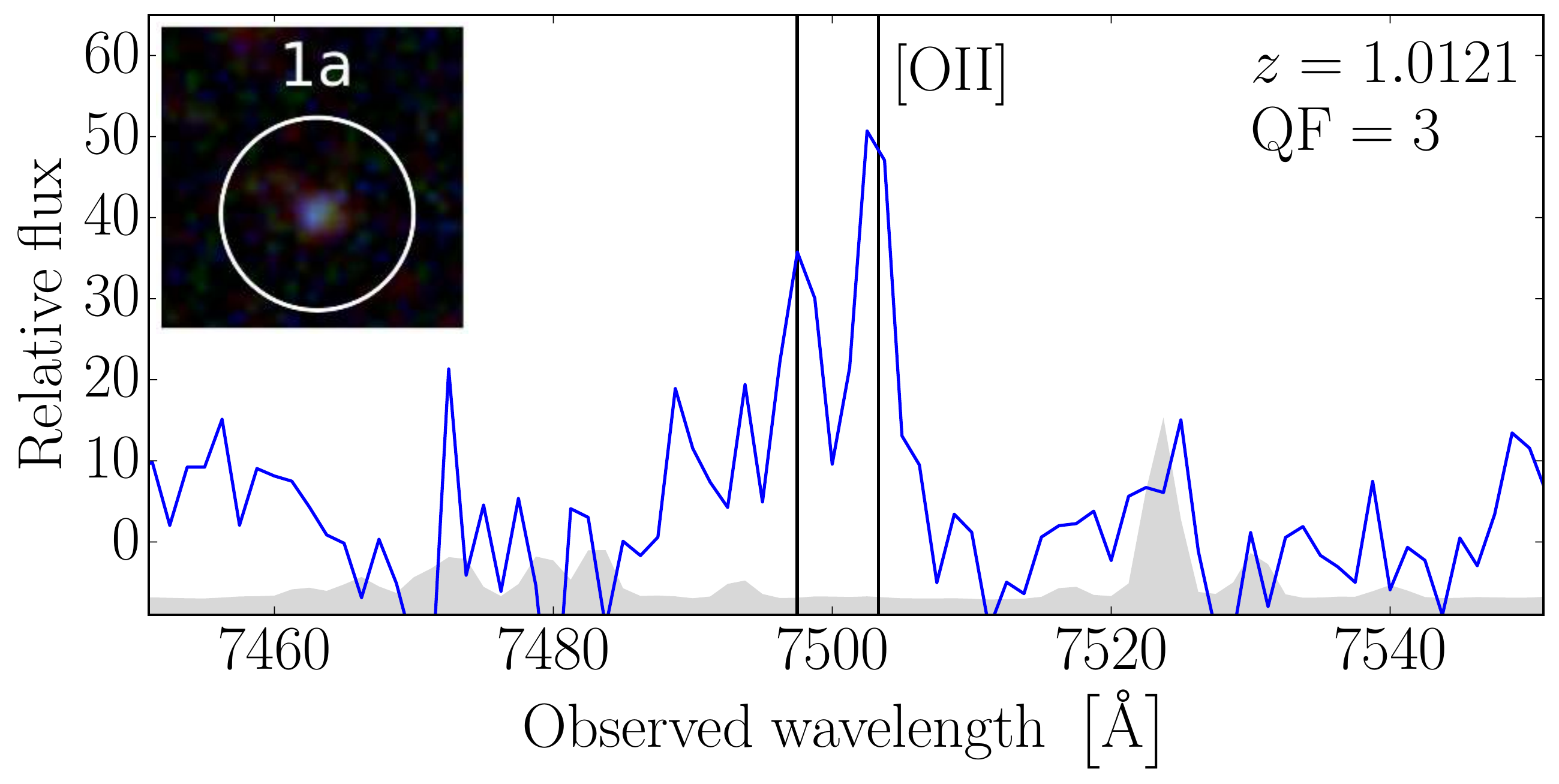}
   \includegraphics[width = 0.666\columnwidth]{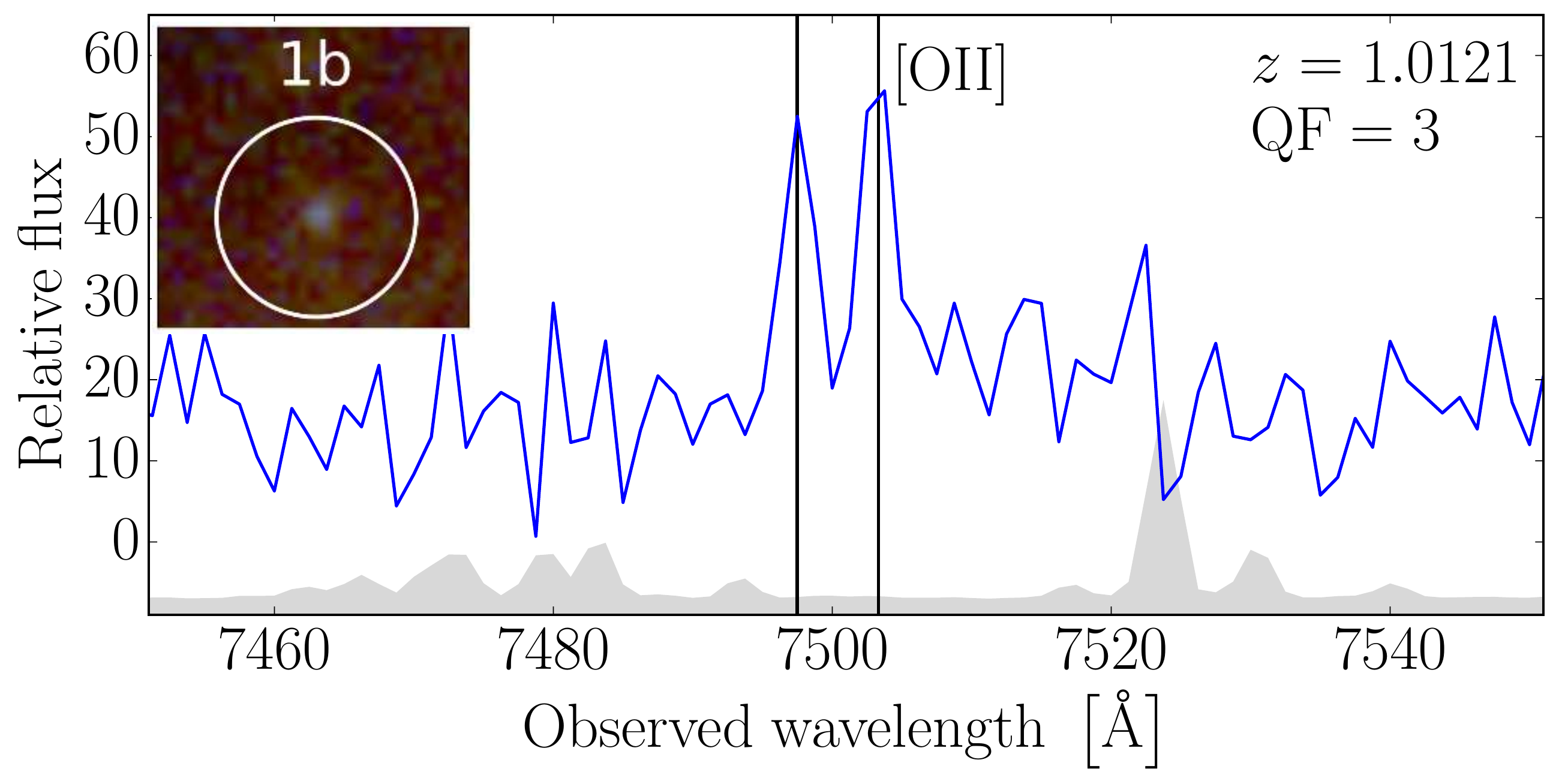}
   \includegraphics[width = 0.666\columnwidth]{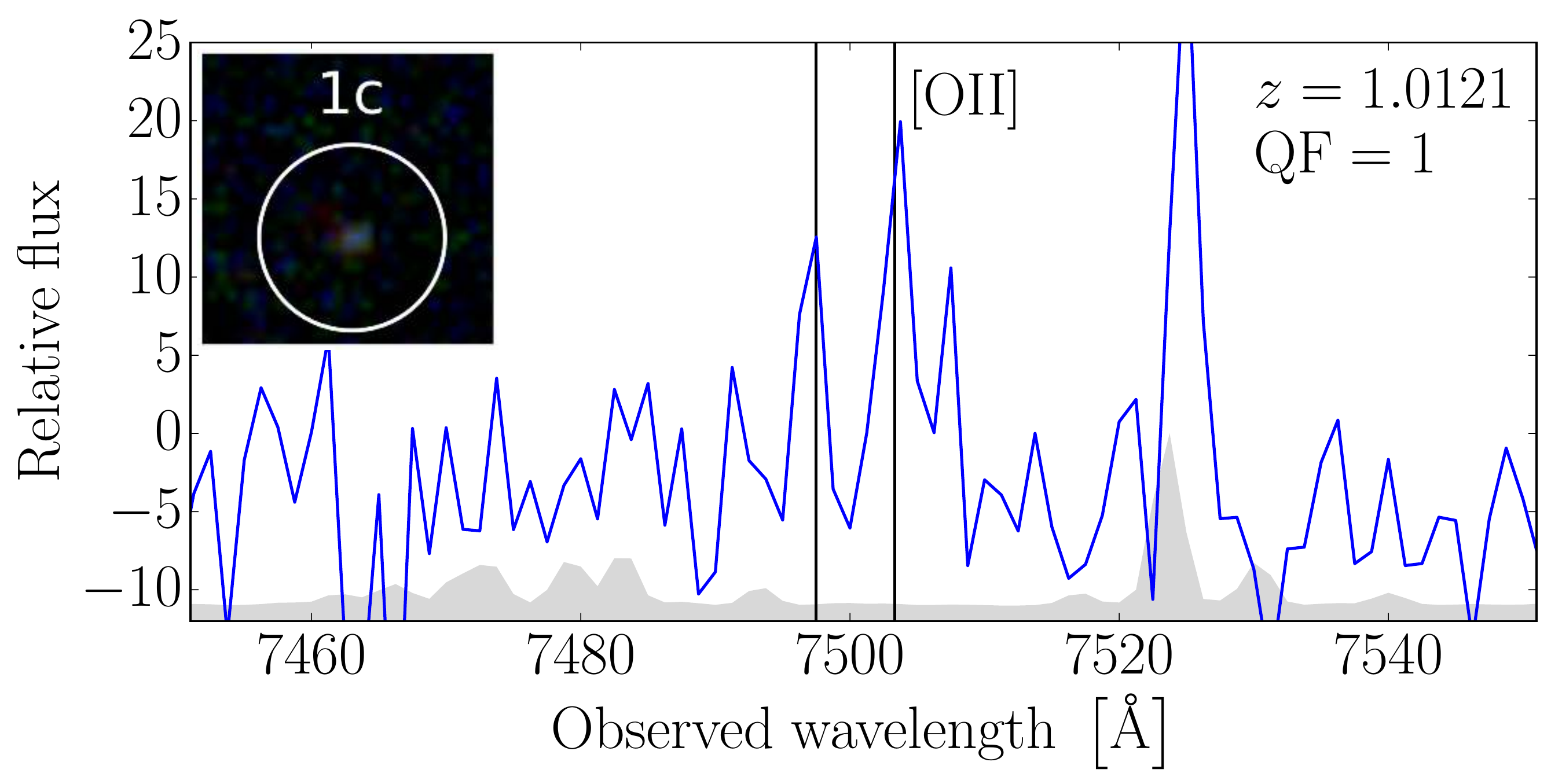}

Family 2:

   \includegraphics[width = 0.666\columnwidth]{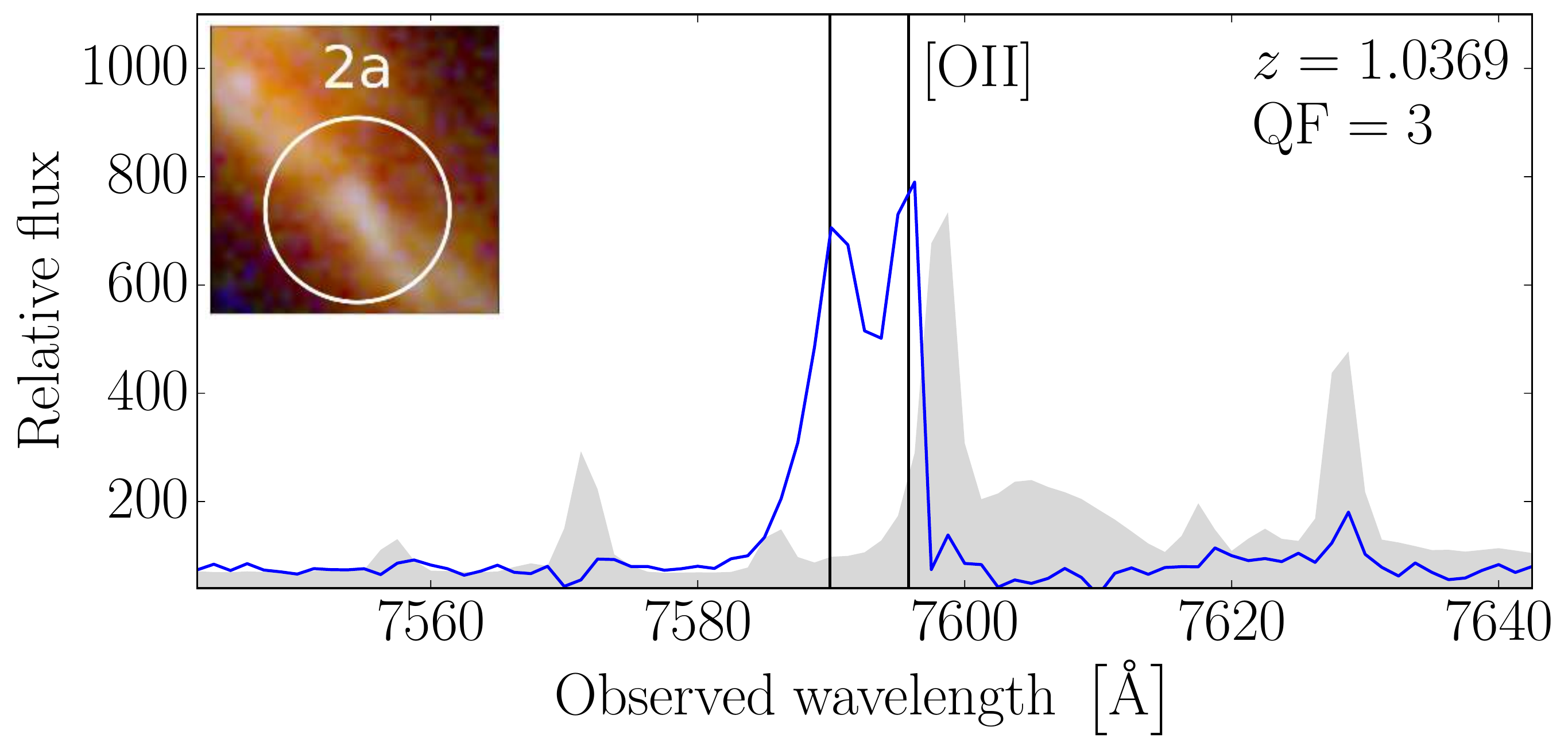}
   \includegraphics[width = 0.666\columnwidth]{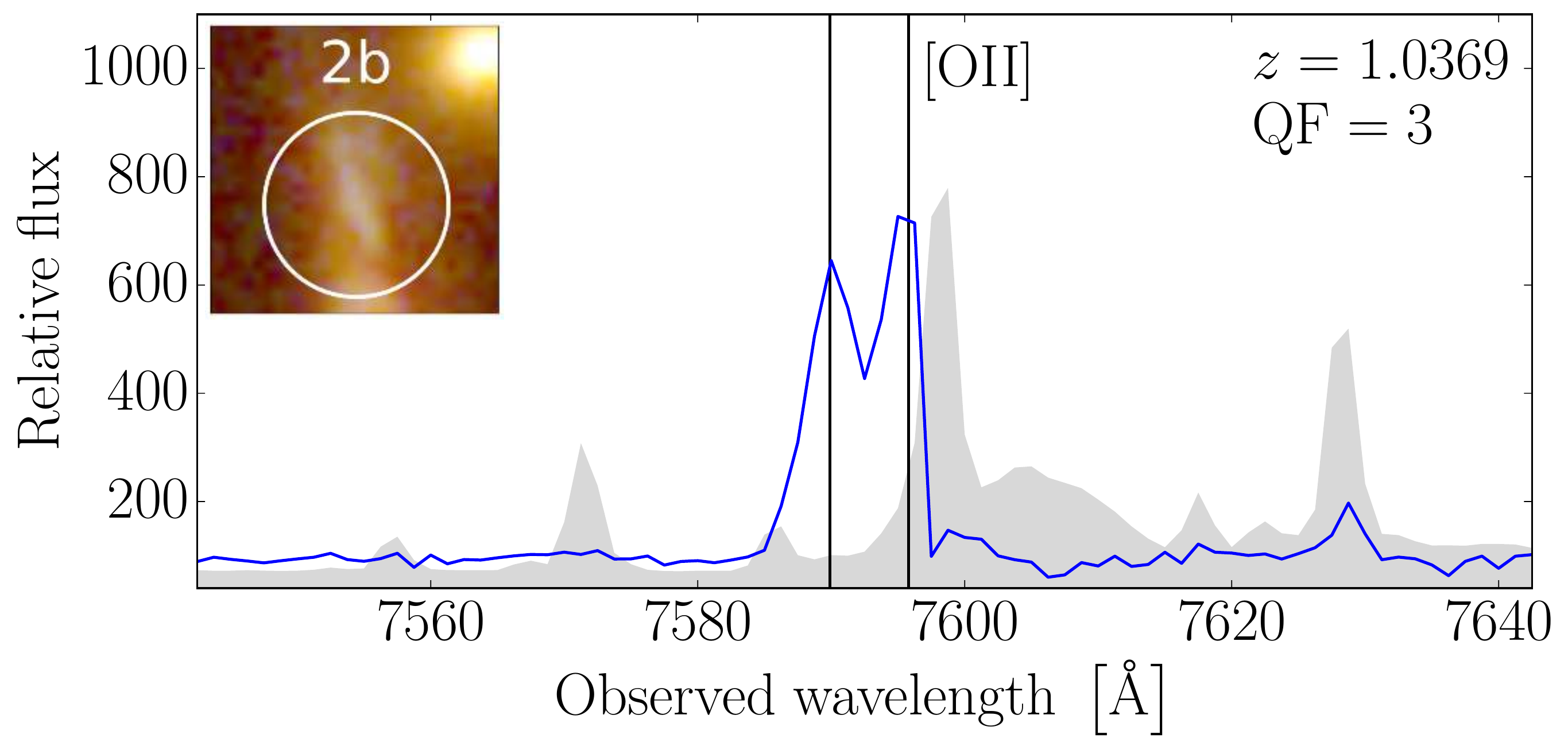}
   \includegraphics[width = 0.666\columnwidth]{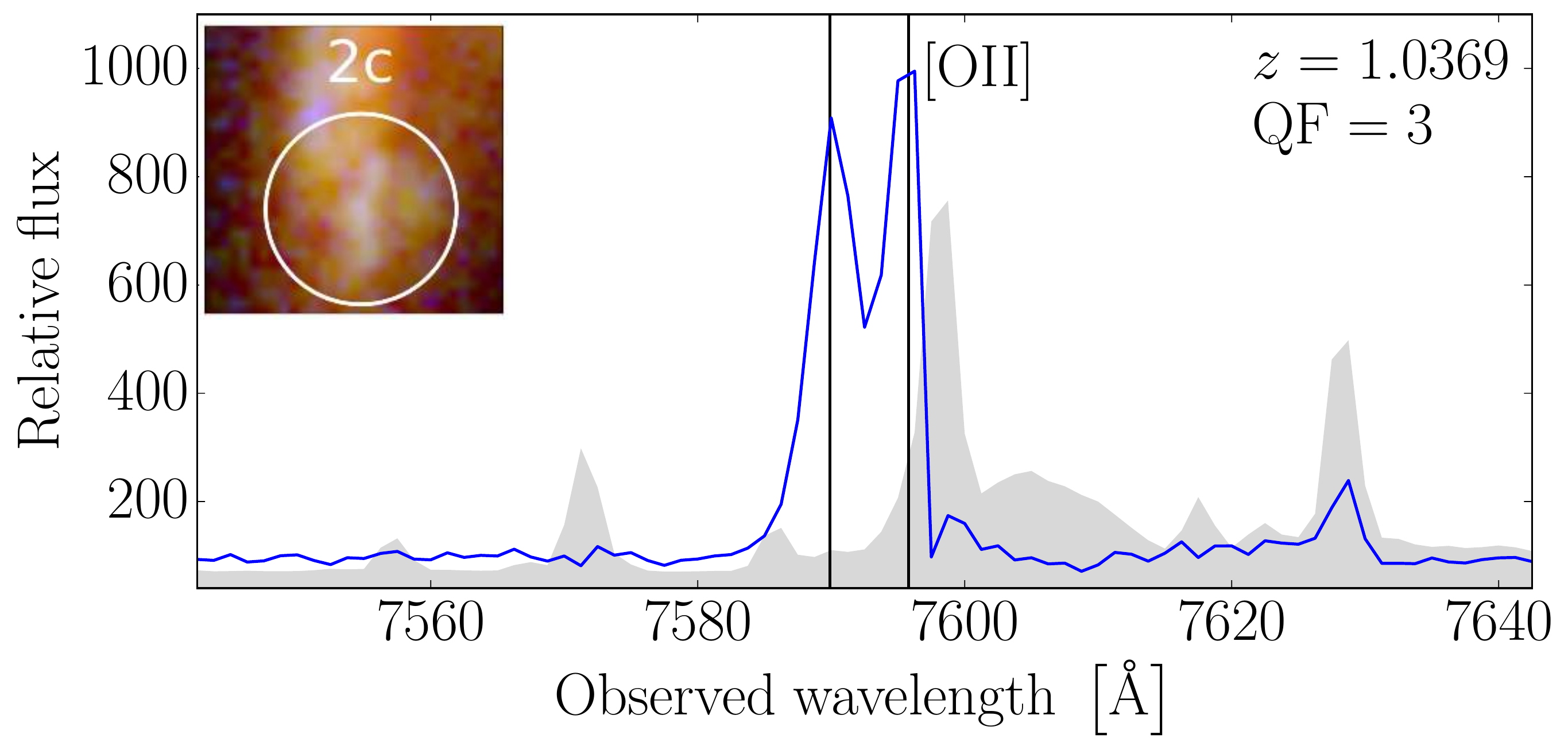}
Family 3:

   \includegraphics[width = 0.666\columnwidth]{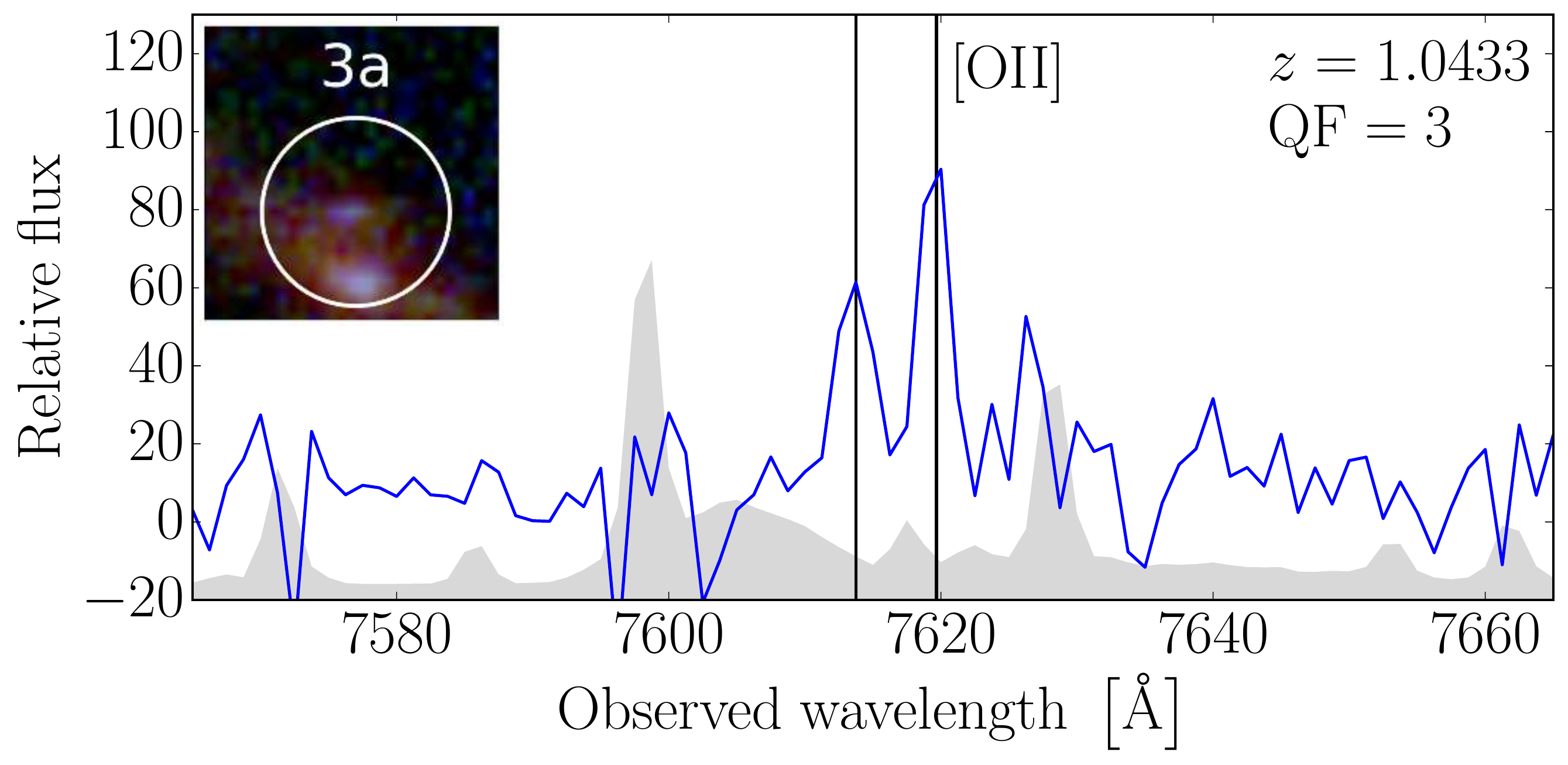}
   \includegraphics[width = 0.666\columnwidth]{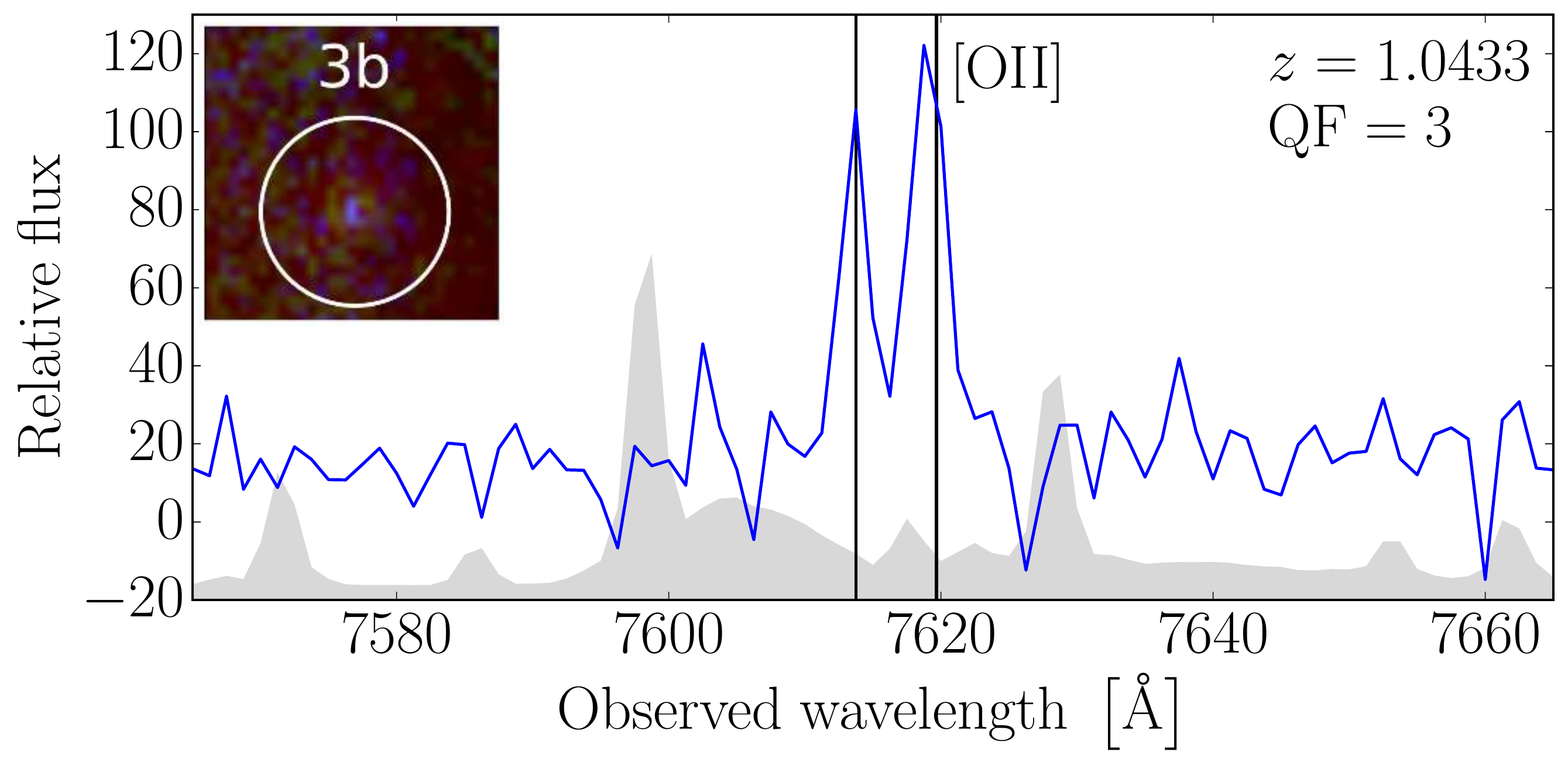}
   \includegraphics[width = 0.666\columnwidth]{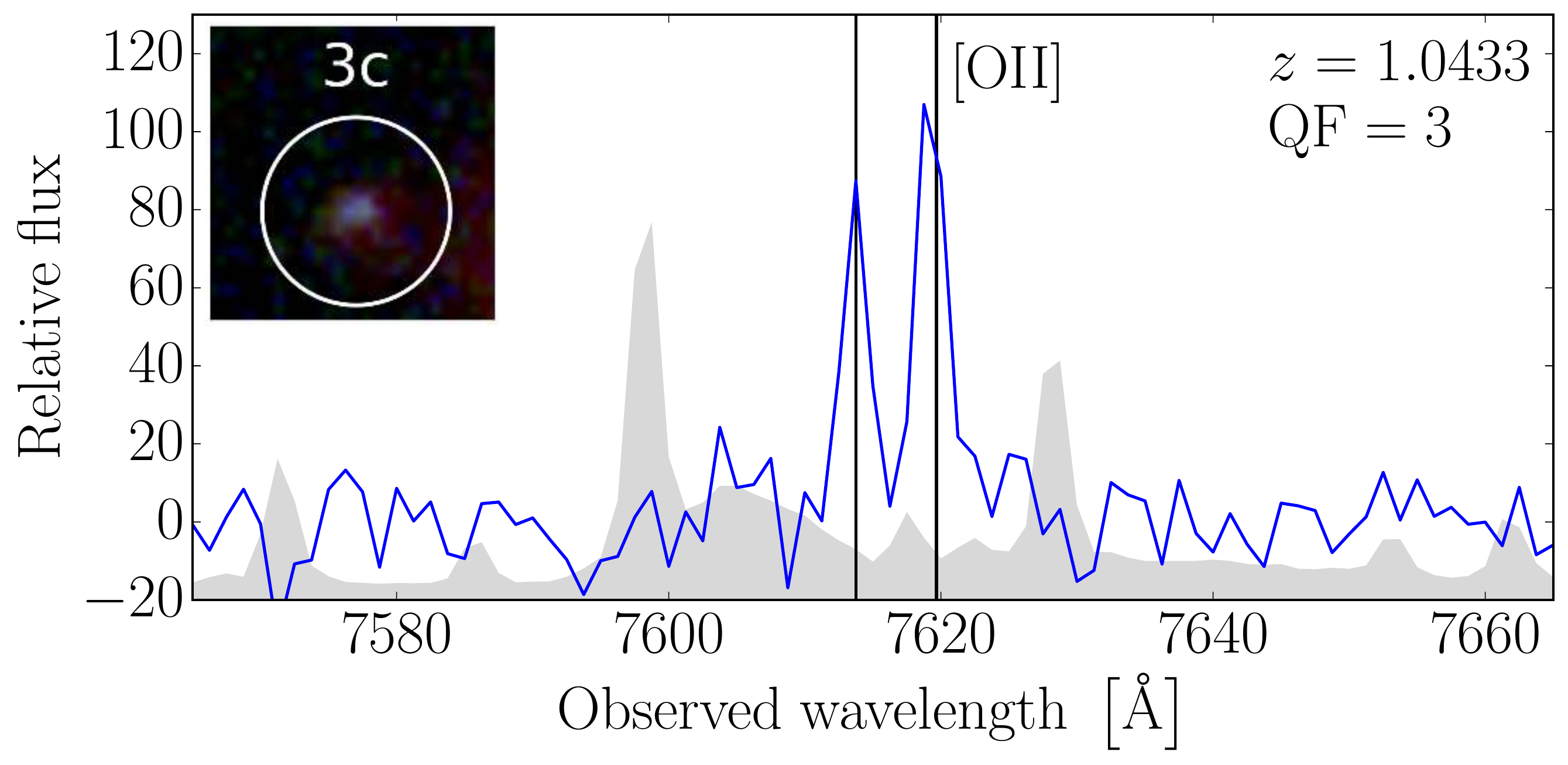}

Family 4:

   \includegraphics[width = 0.666\columnwidth]{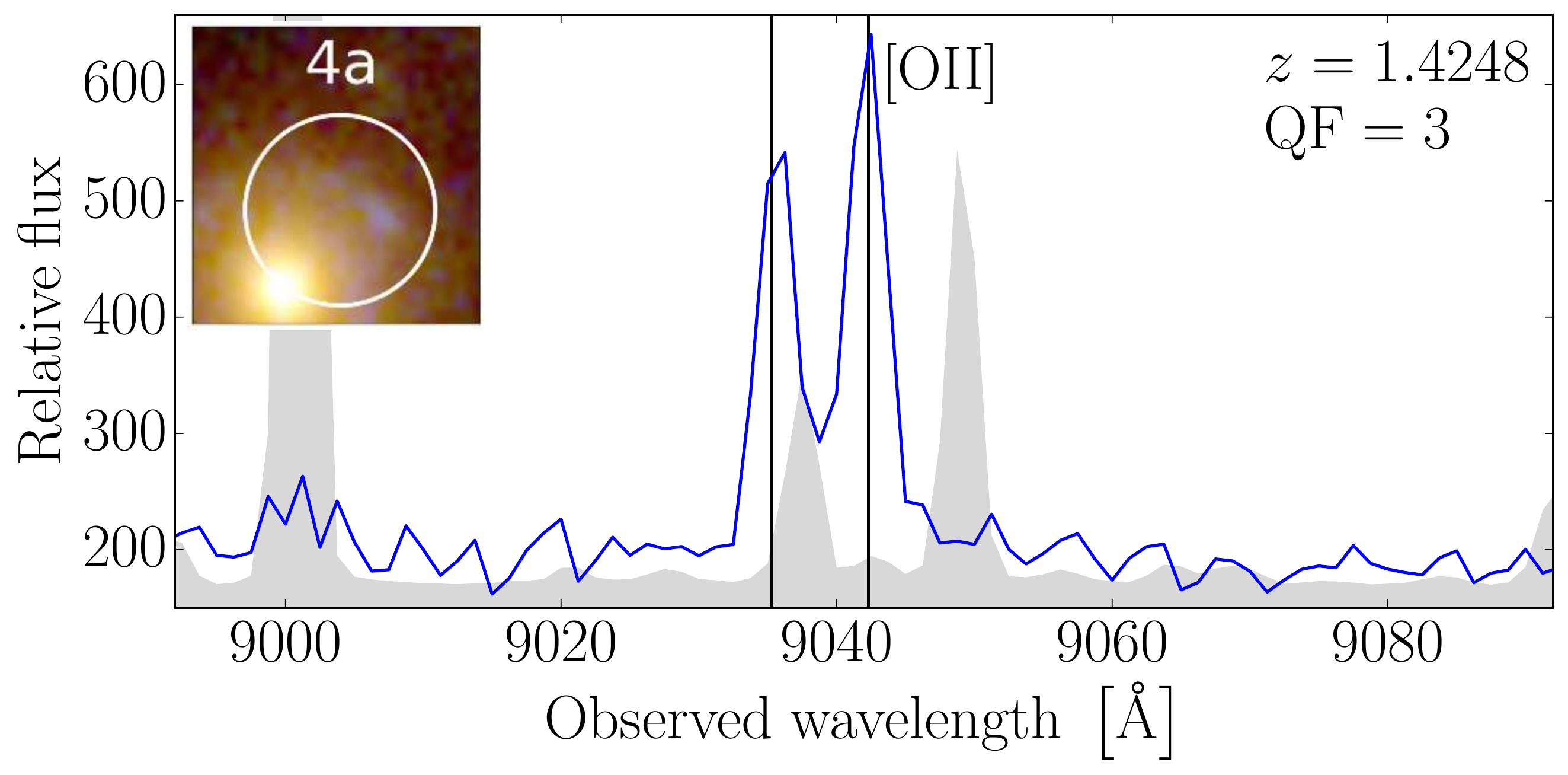}
   \includegraphics[width = 0.666\columnwidth]{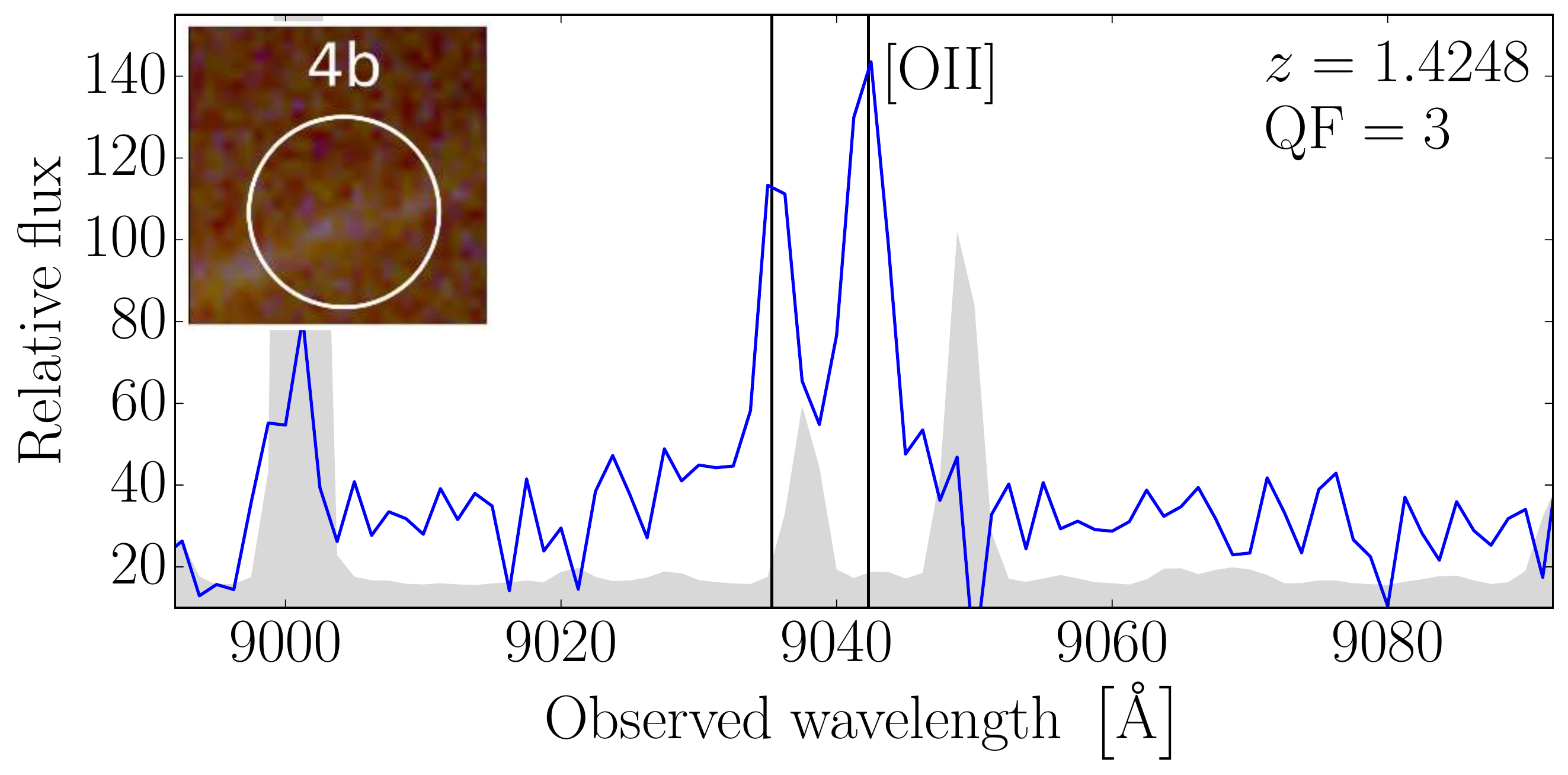}
   \includegraphics[width = 0.666\columnwidth]{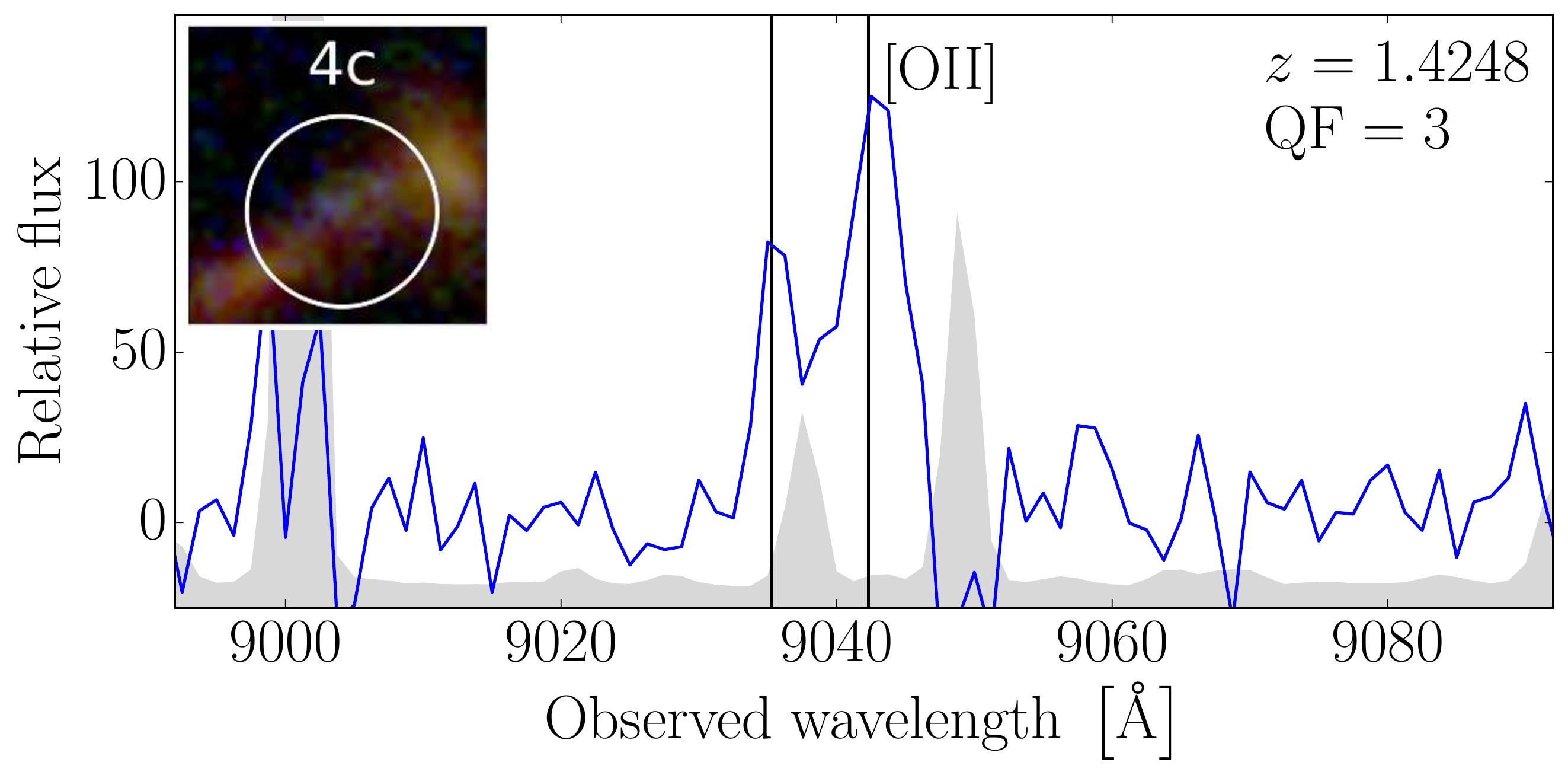}

Family 5:

   \includegraphics[width = 0.666\columnwidth]{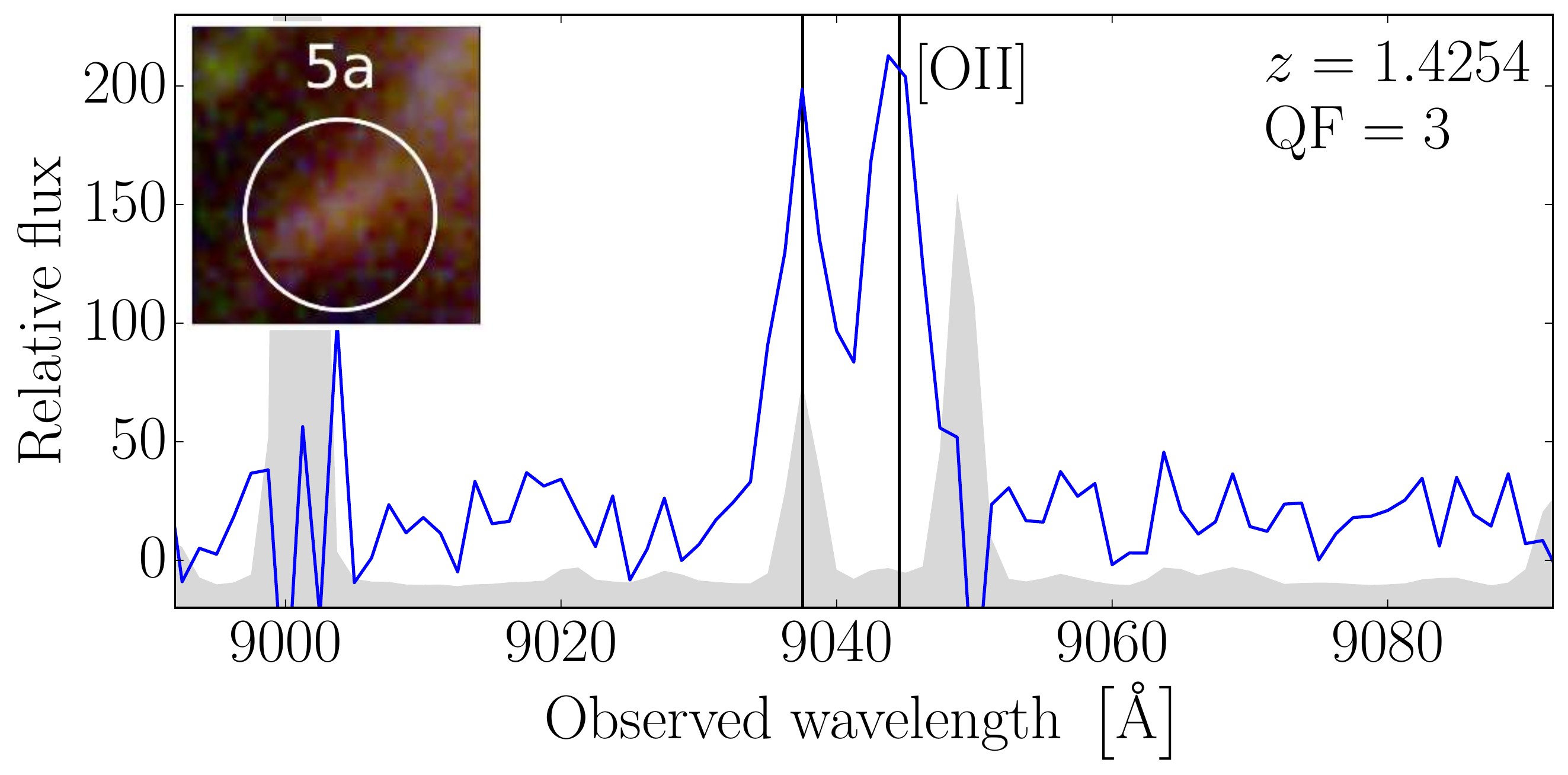}
   \includegraphics[width = 0.666\columnwidth]{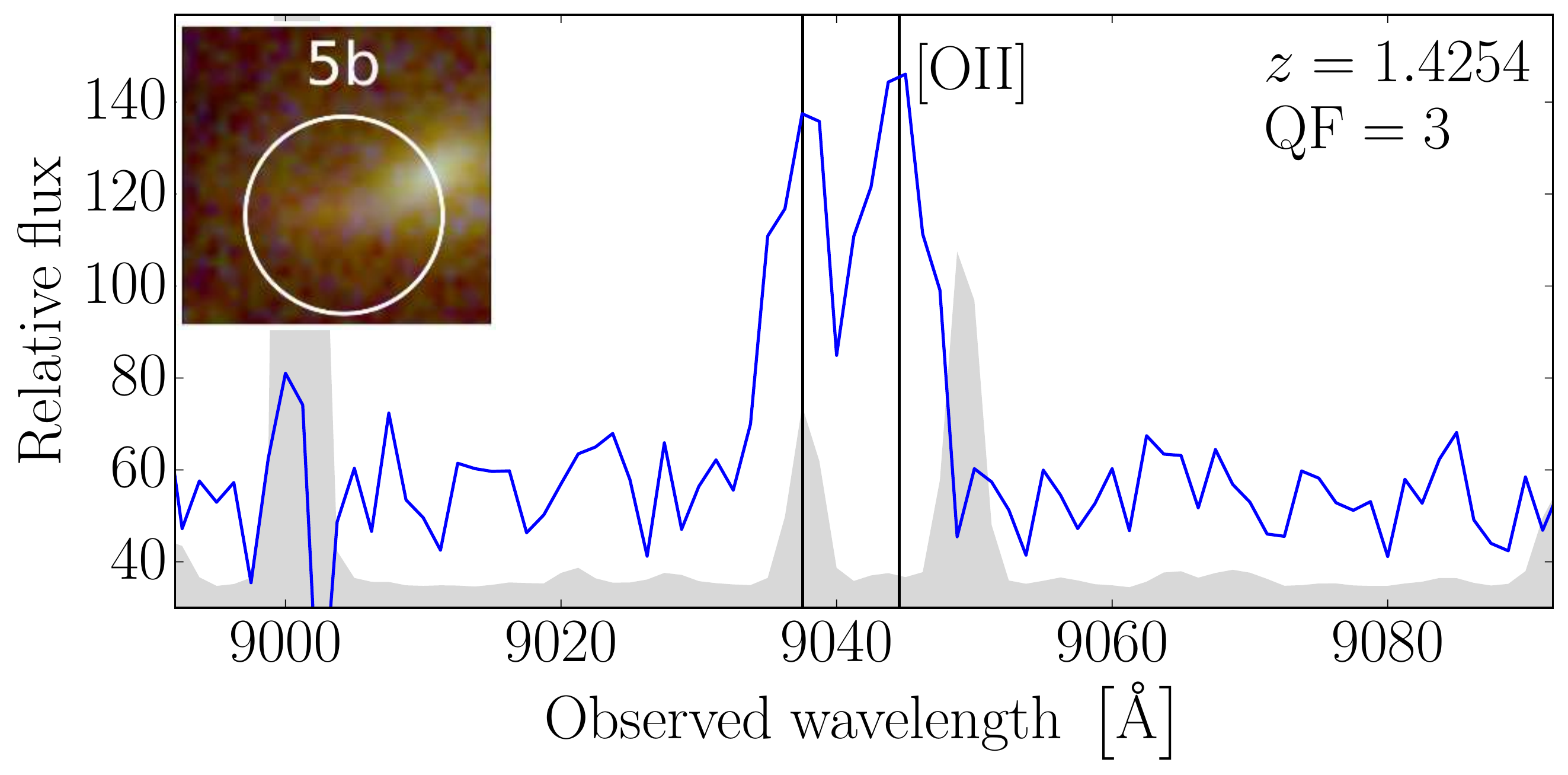}
   \includegraphics[width = 0.666\columnwidth]{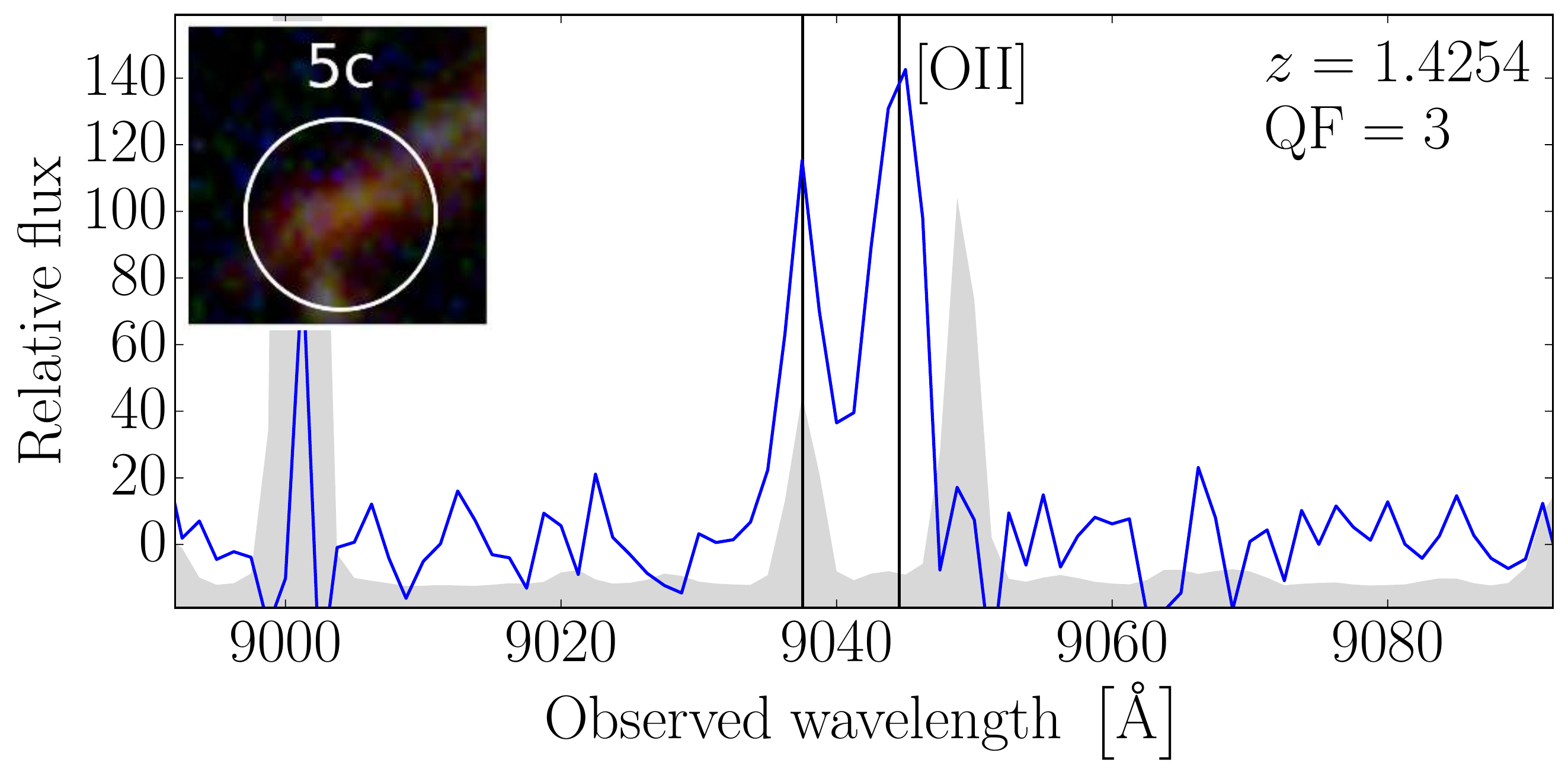}

Family 6:

   \includegraphics[width = 0.666\columnwidth]{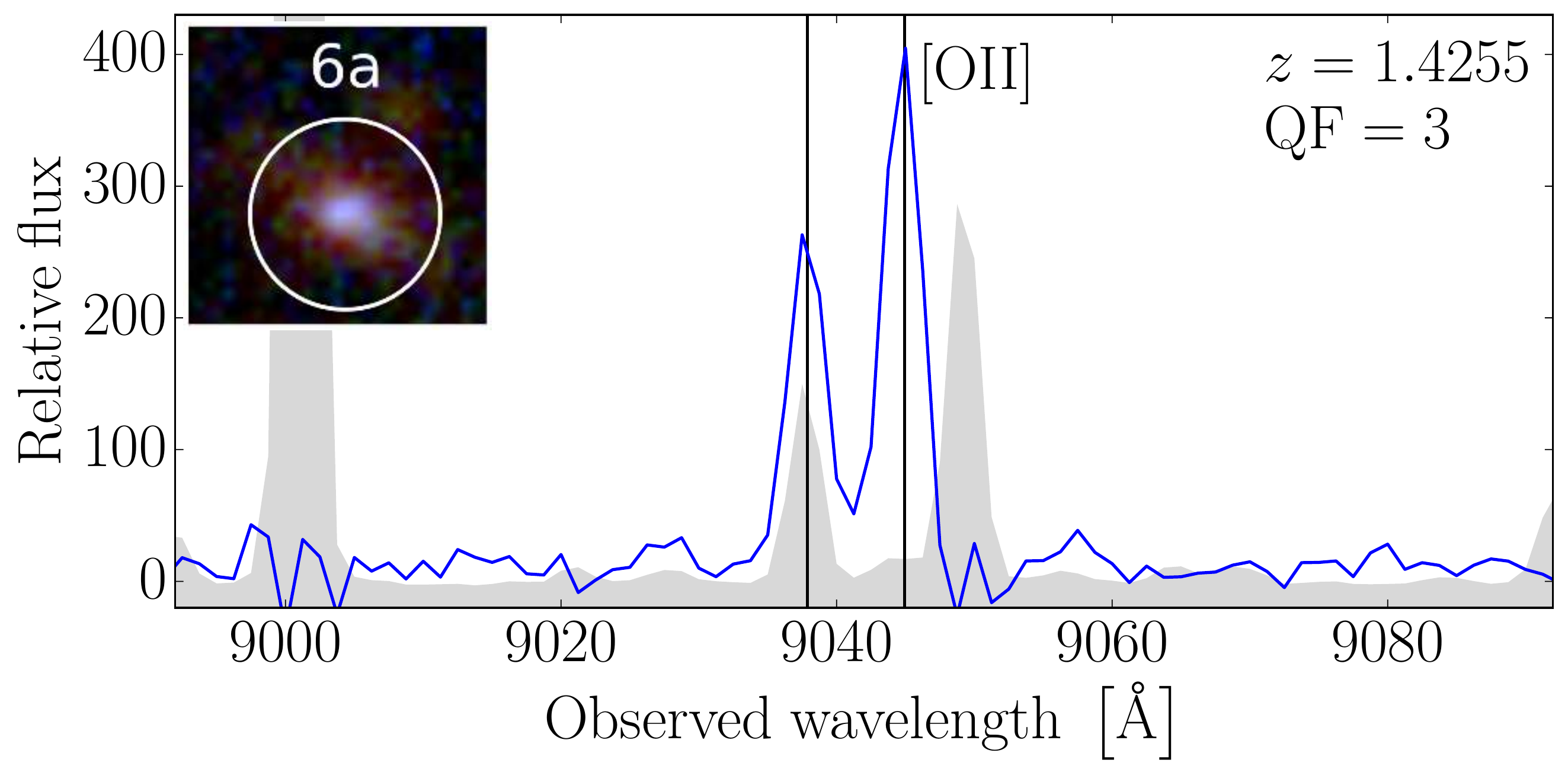}
   \includegraphics[width = 0.666\columnwidth]{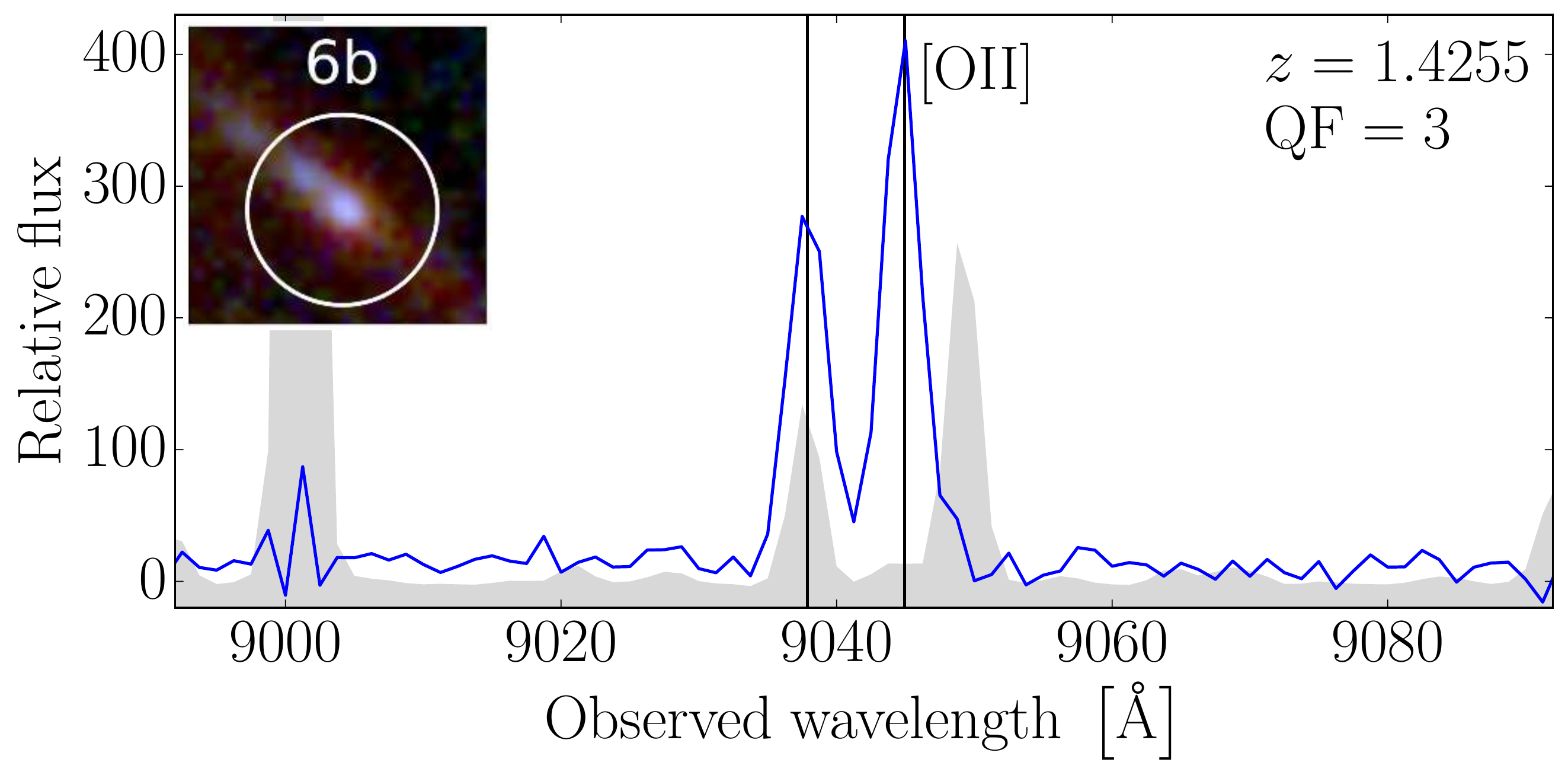}
   \includegraphics[width = 0.666\columnwidth]{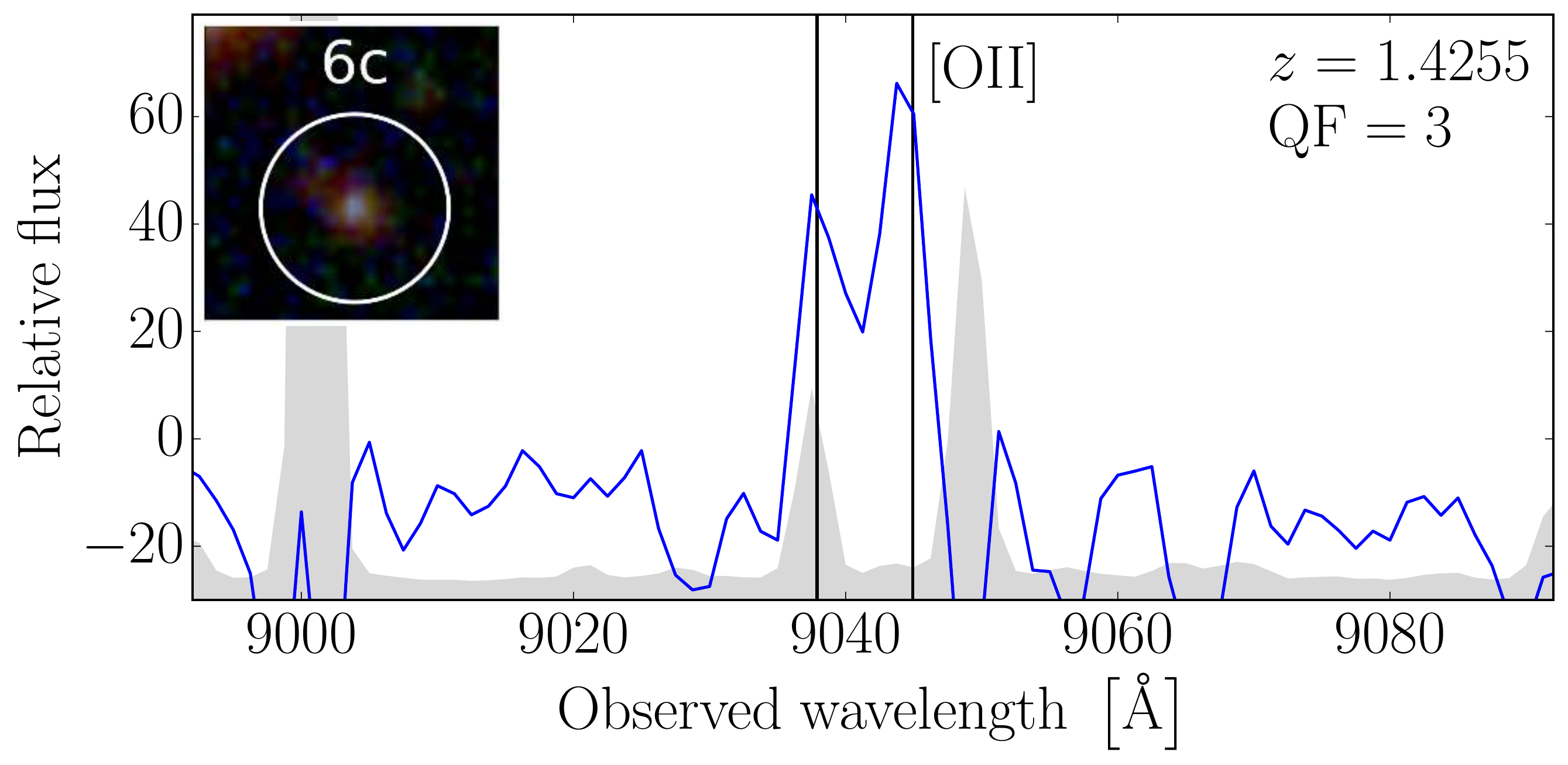}

   \caption{MUSE data of multiply lensed background sources of MACS~1206. The vertical black lines indicate the positions of the emission lines based on the best estimate of the systemic redshift. The gray area shows the rescaled variance obtained from the data reduction pipeline; the flux is given in units of $\rm 10^{-20}\, erg\, s^{-1}\, cm^{-2}\, \AA^{-1}$. The image cutouts in each panel have $2\arcsec$ across and are extracted from the CLASH color image. The white circles show the HST counterparts or are centered at the position of the MUSE emission in the cases of no apparent counterparts.}
  \label{fig:specs}
\end{figure*}

\begin{figure*}
\setcounter{figure}{\value{figure}-1}
Family 7:

   \includegraphics[width = 0.666\columnwidth]{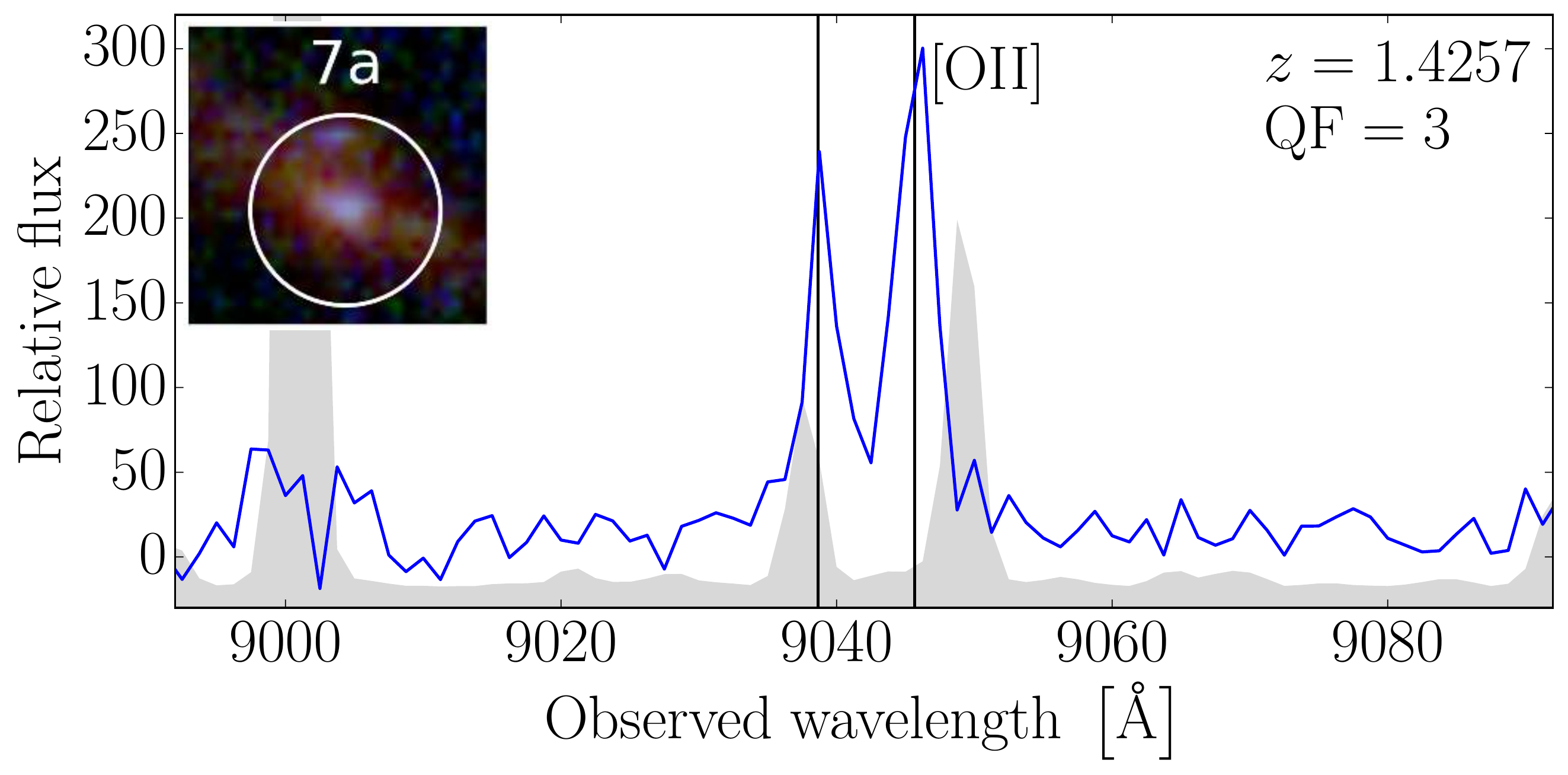}
   \includegraphics[width = 0.666\columnwidth]{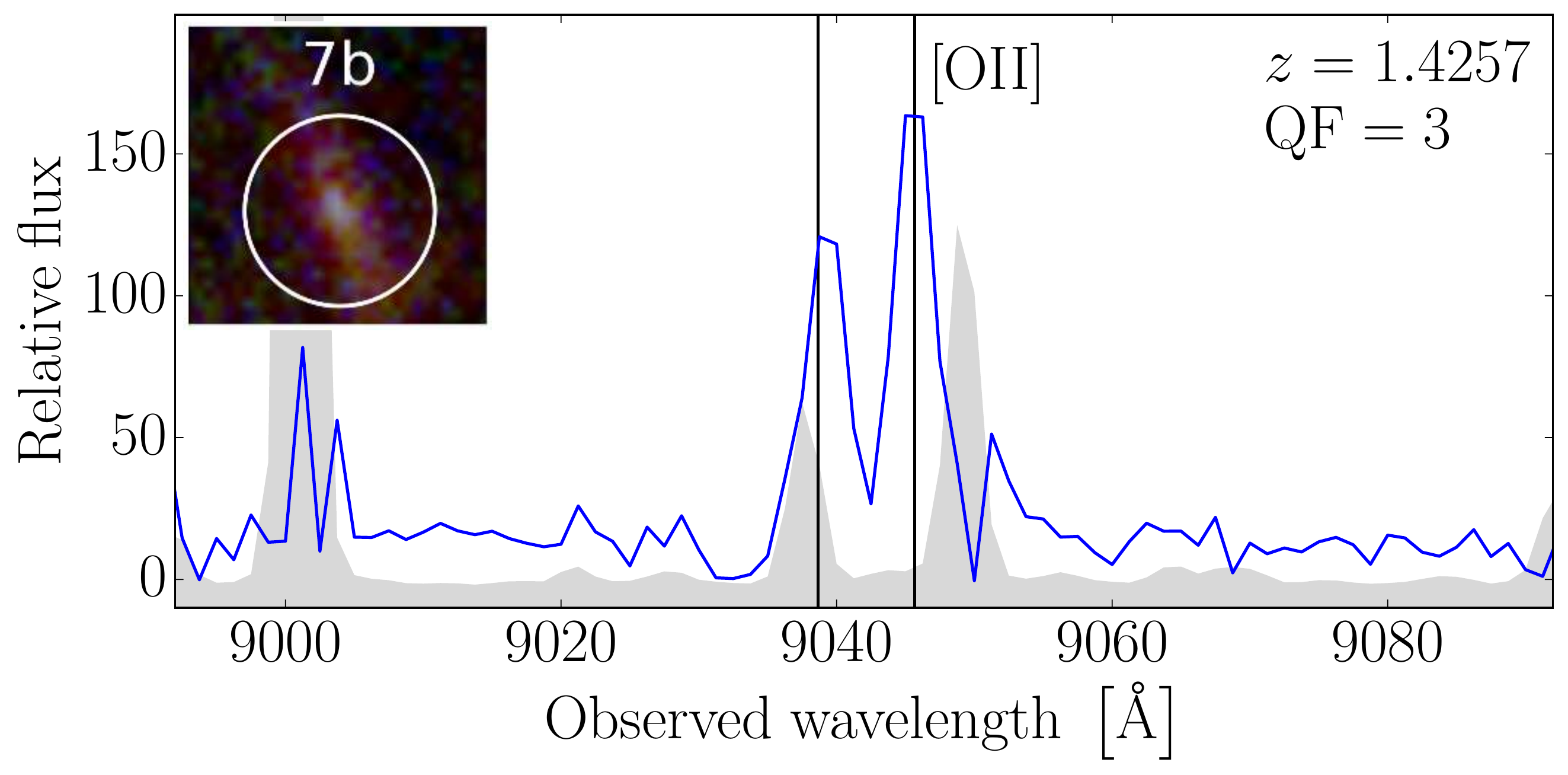}
   \includegraphics[width = 0.666\columnwidth]{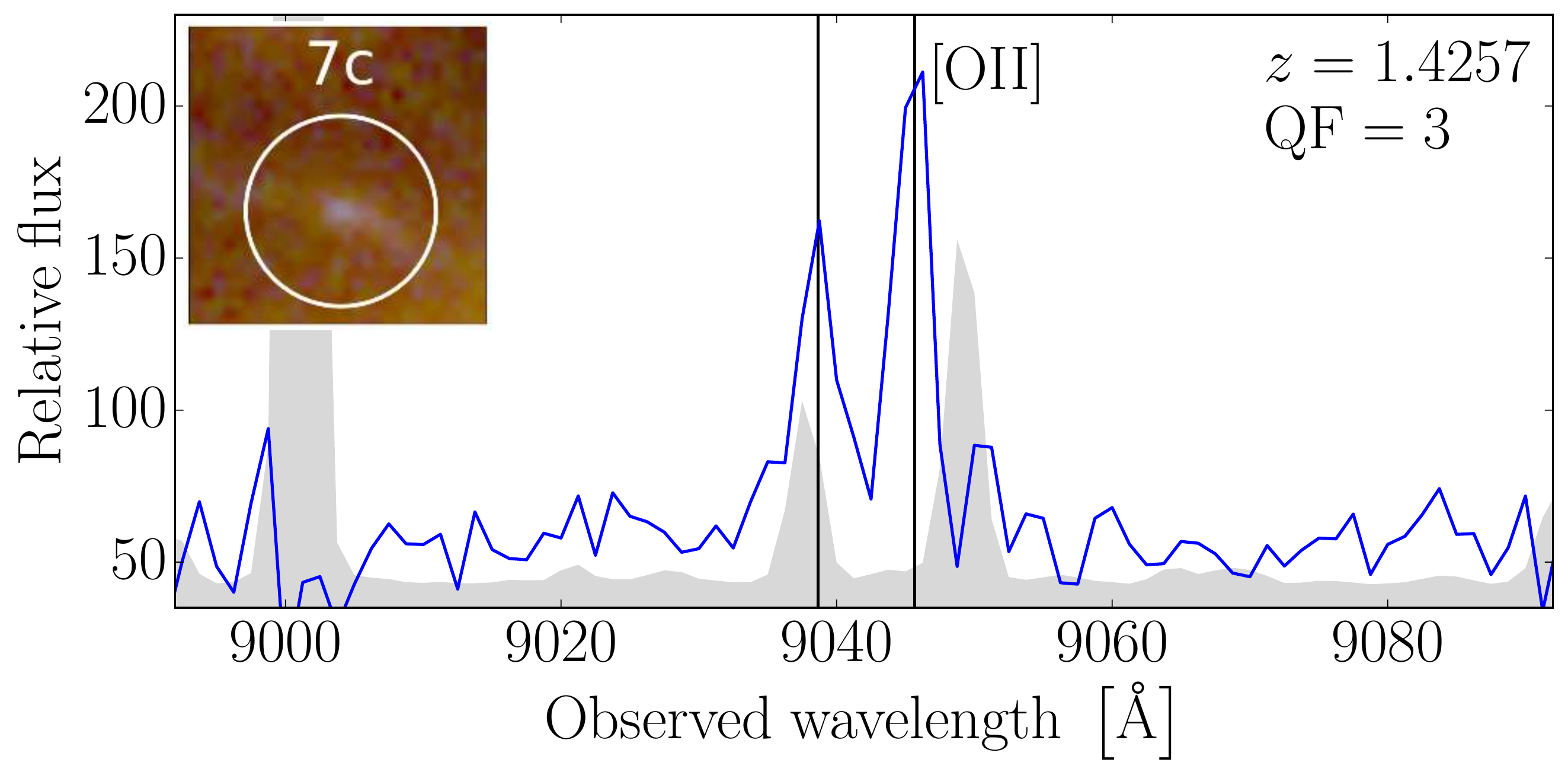}

   \includegraphics[width = 0.666\columnwidth]{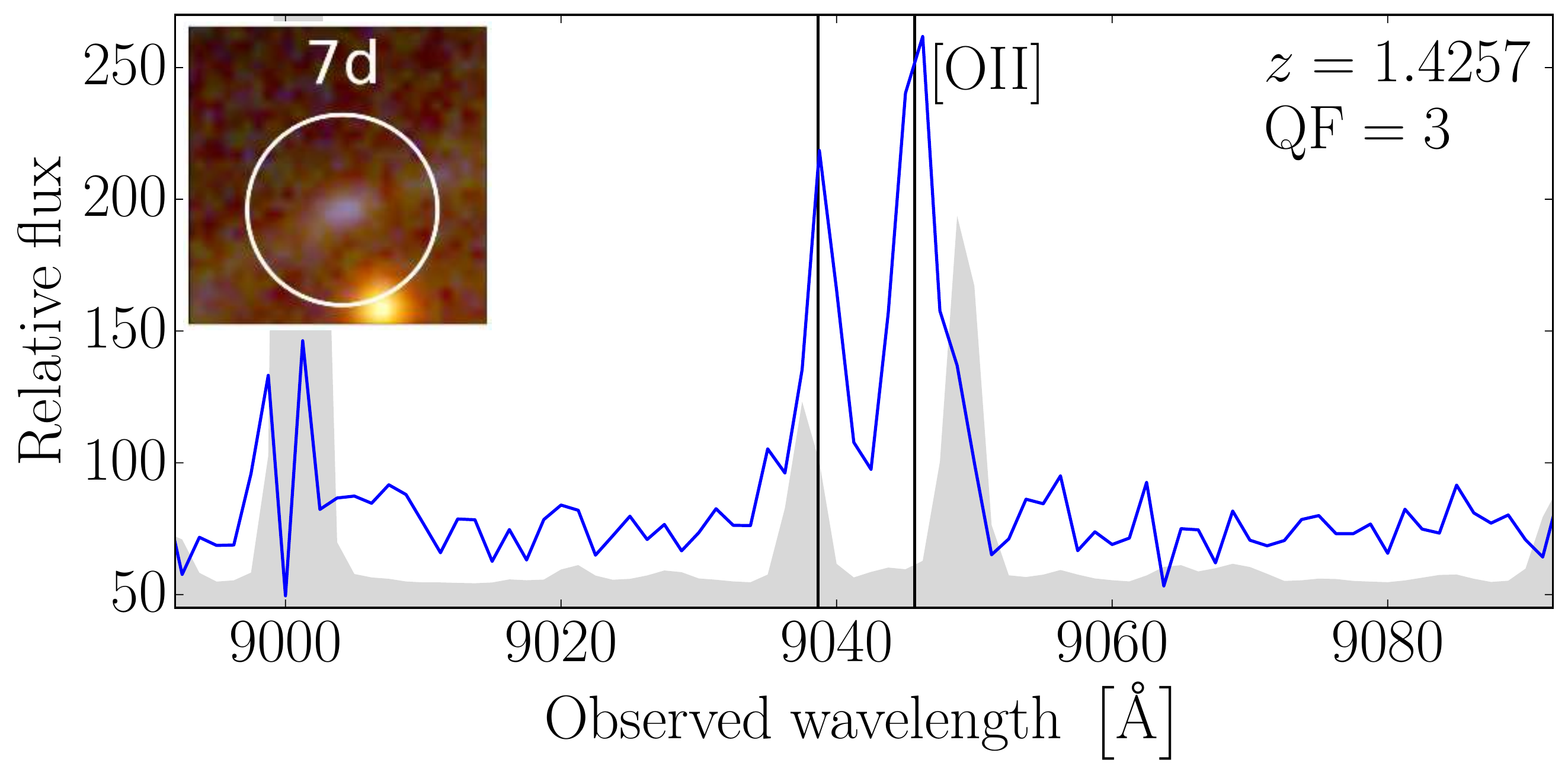}
   \includegraphics[width = 0.666\columnwidth]{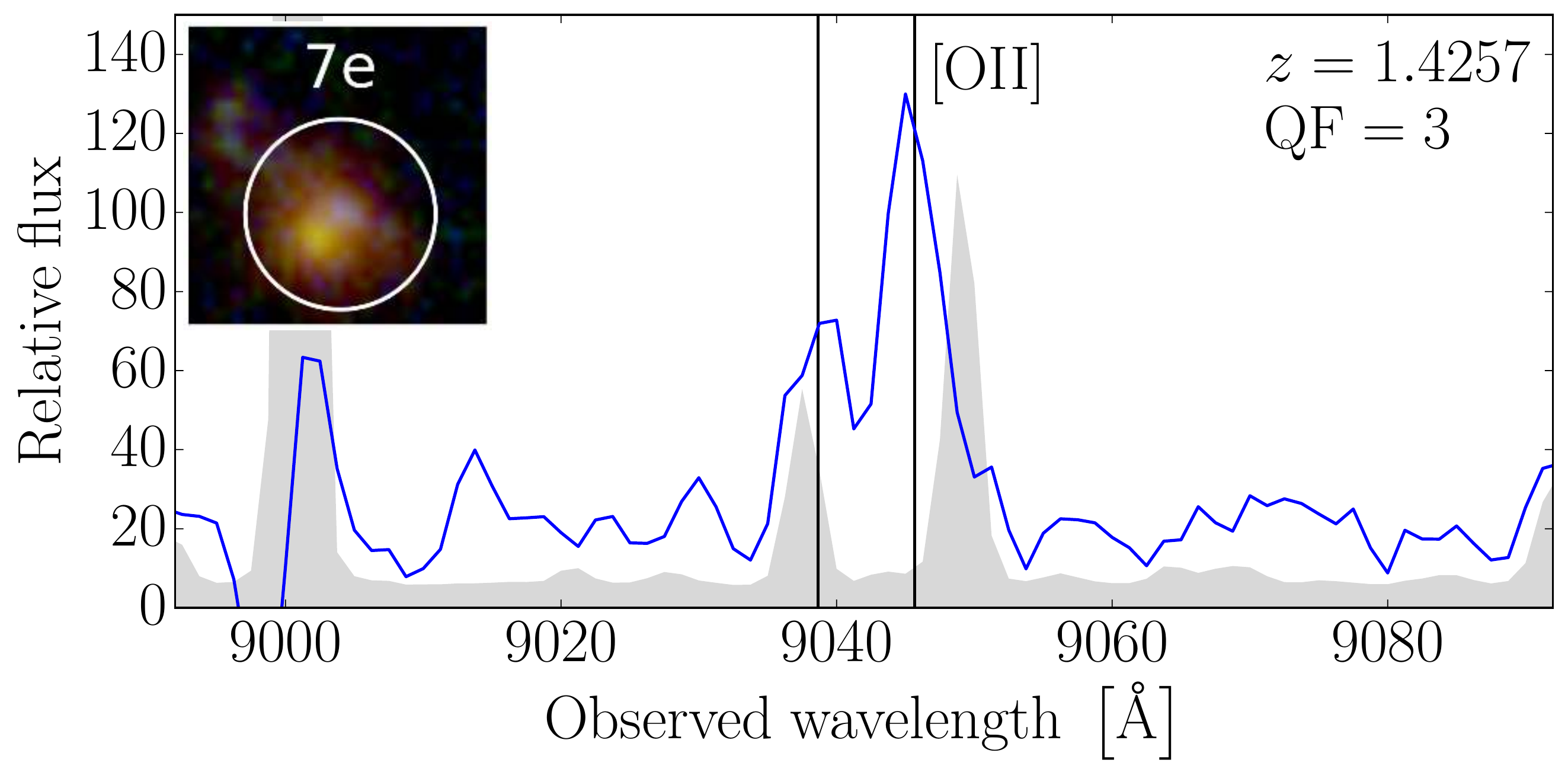}

Family 8:

   \includegraphics[width = 0.666\columnwidth]{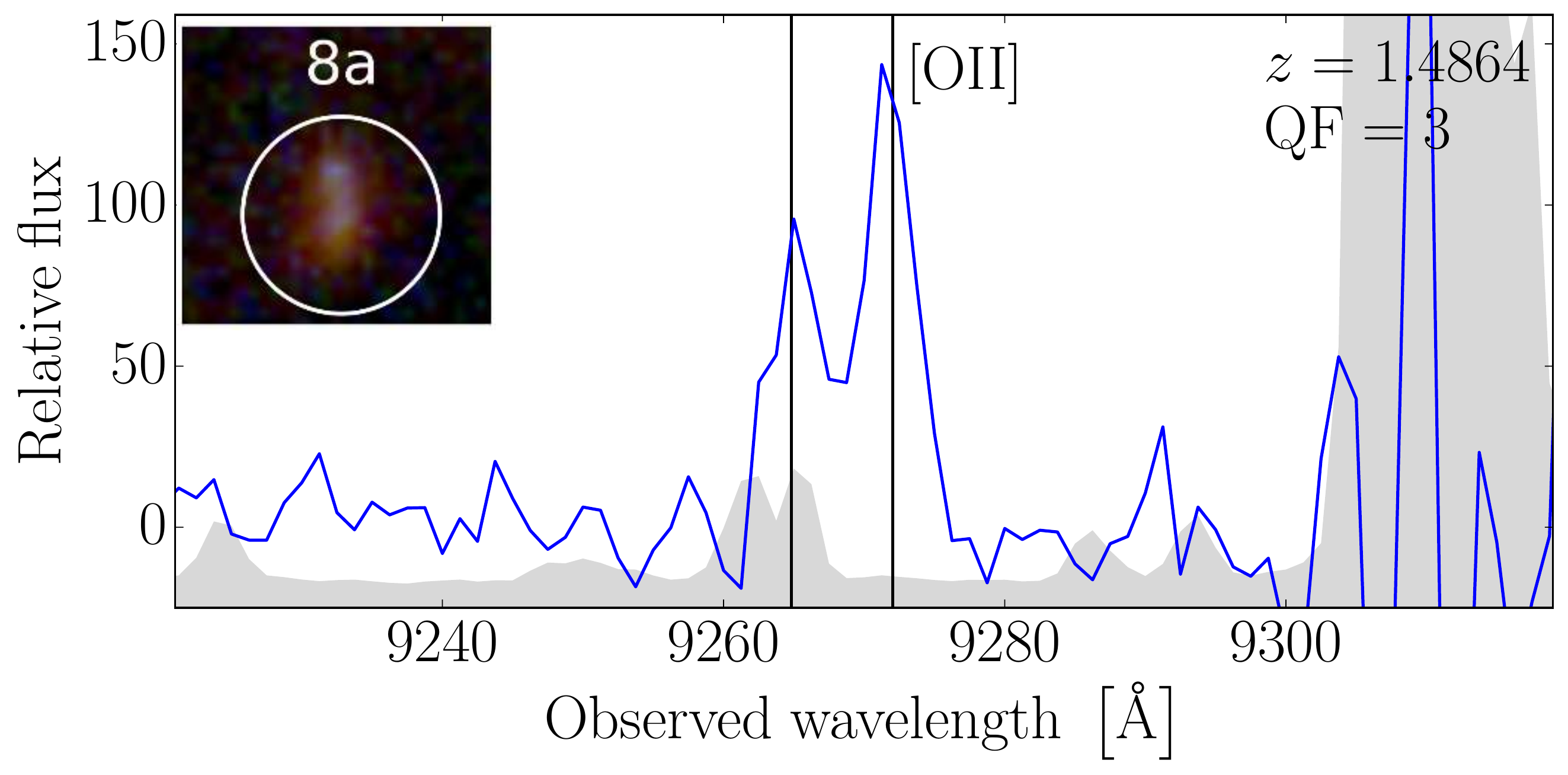}
   \includegraphics[width = 0.666\columnwidth]{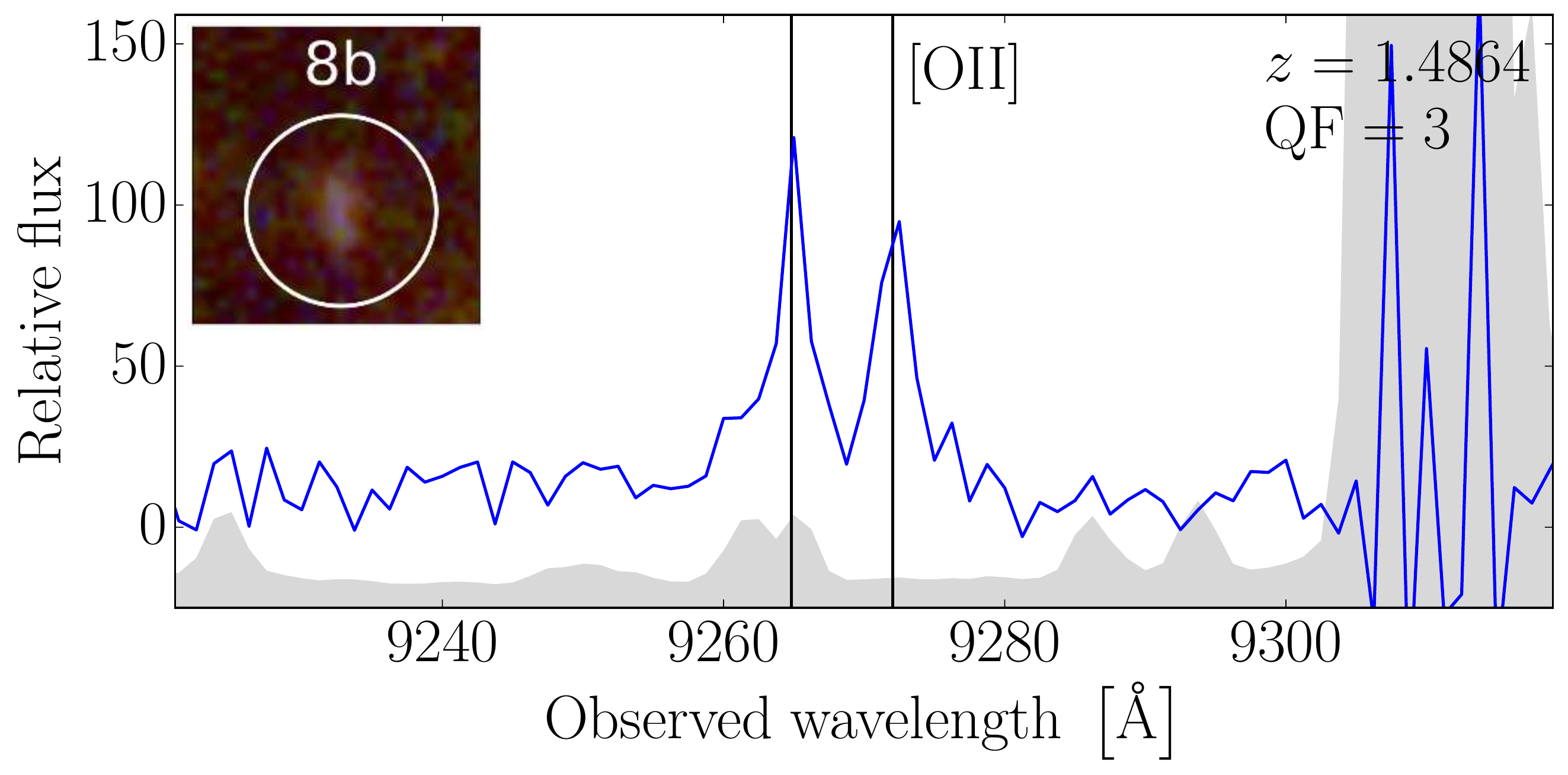}
   \includegraphics[width = 0.666\columnwidth]{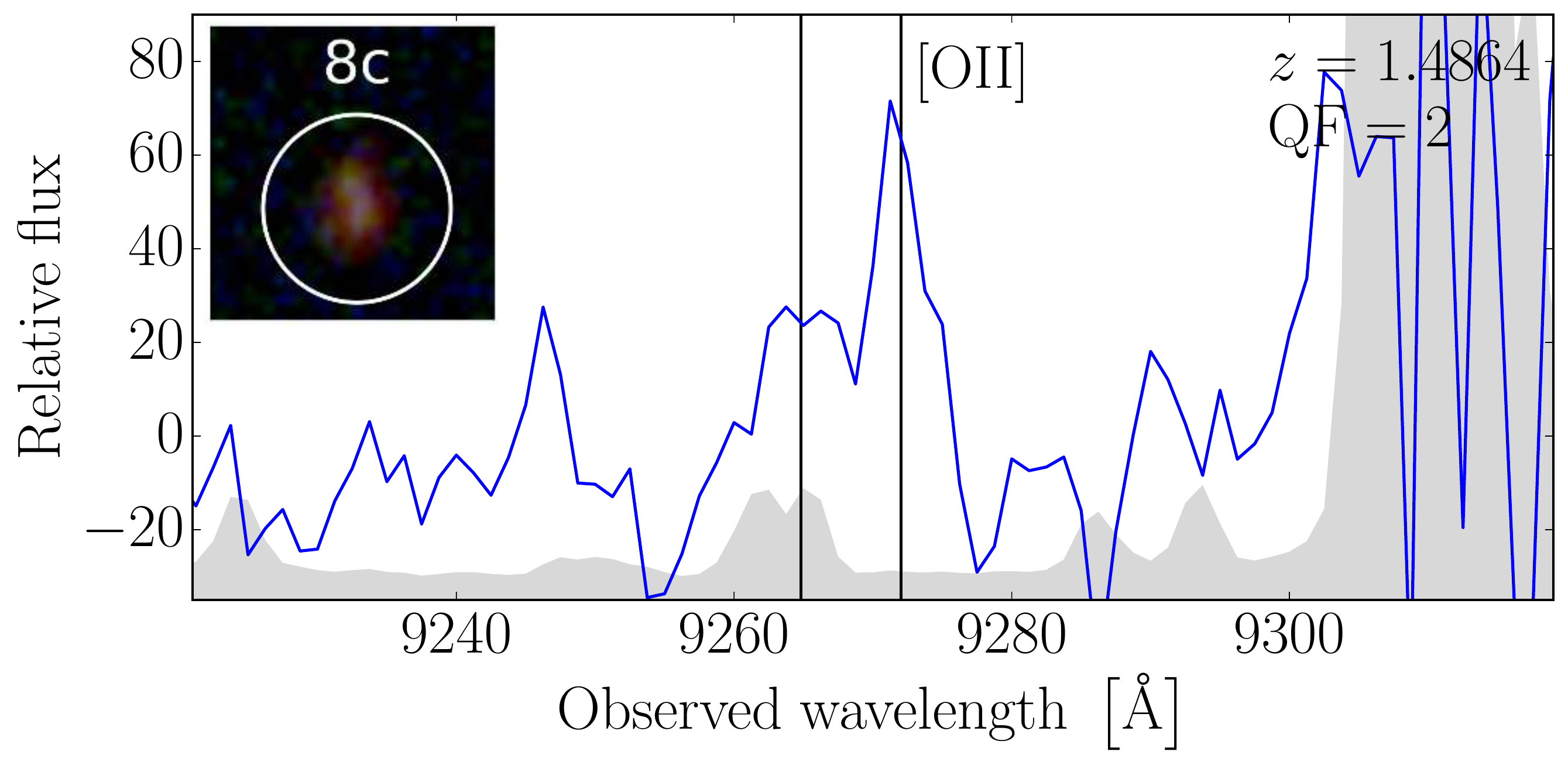}

Family 9:

   \includegraphics[width = 0.666\columnwidth]{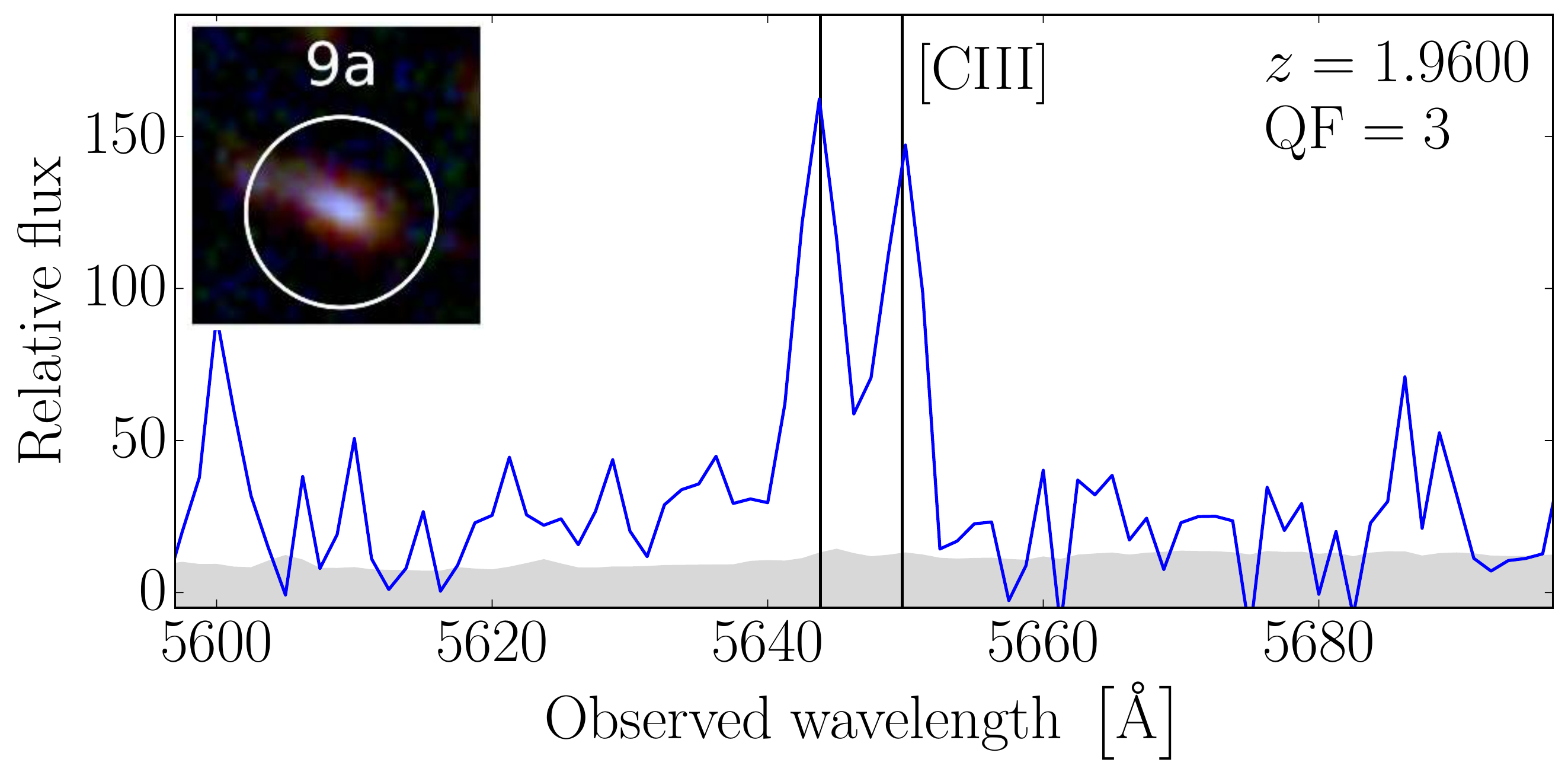}
   \includegraphics[width = 0.666\columnwidth]{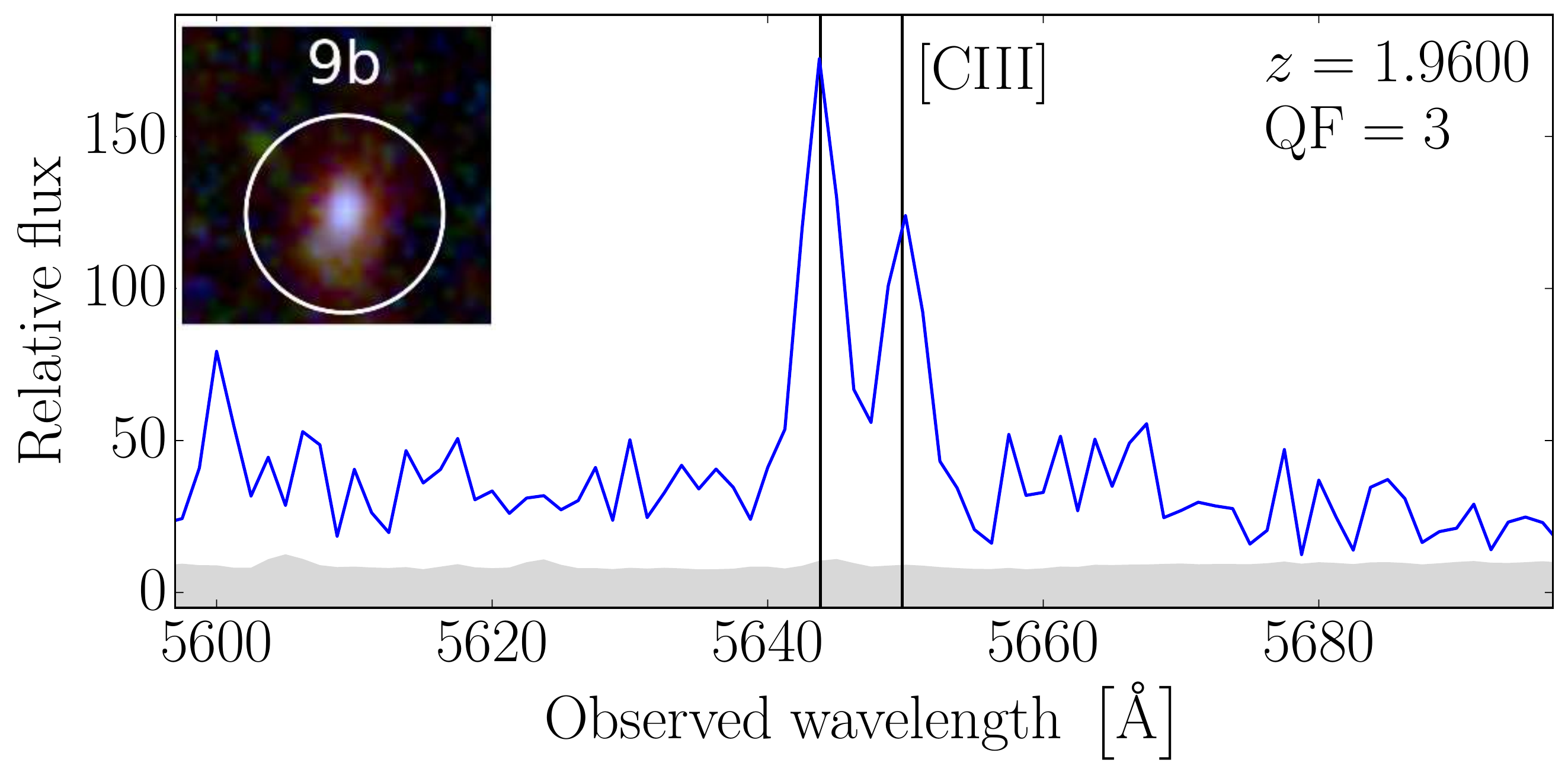}
   \includegraphics[width = 0.666\columnwidth]{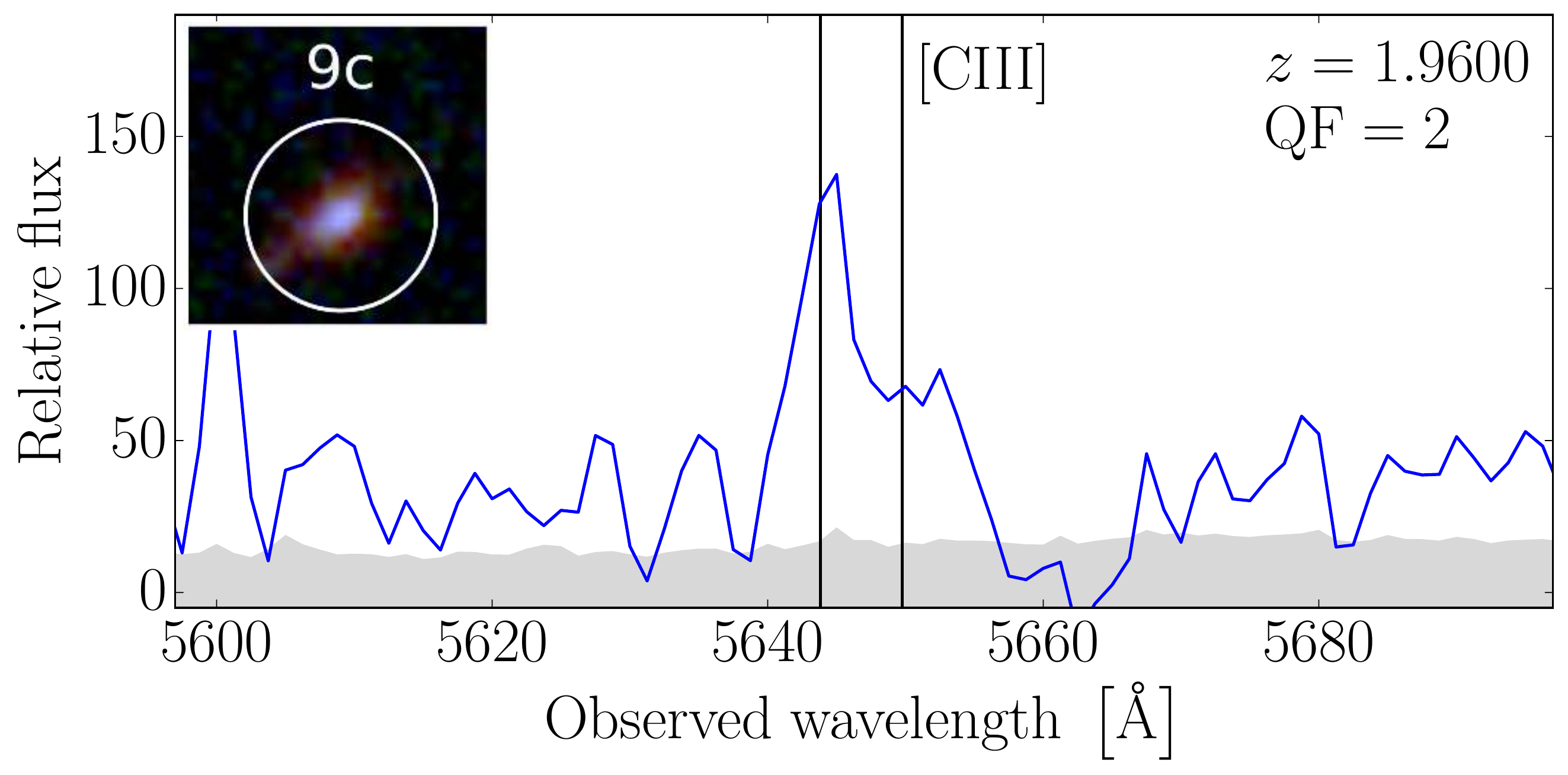}

Family 10:

   \includegraphics[width = 0.666\columnwidth]{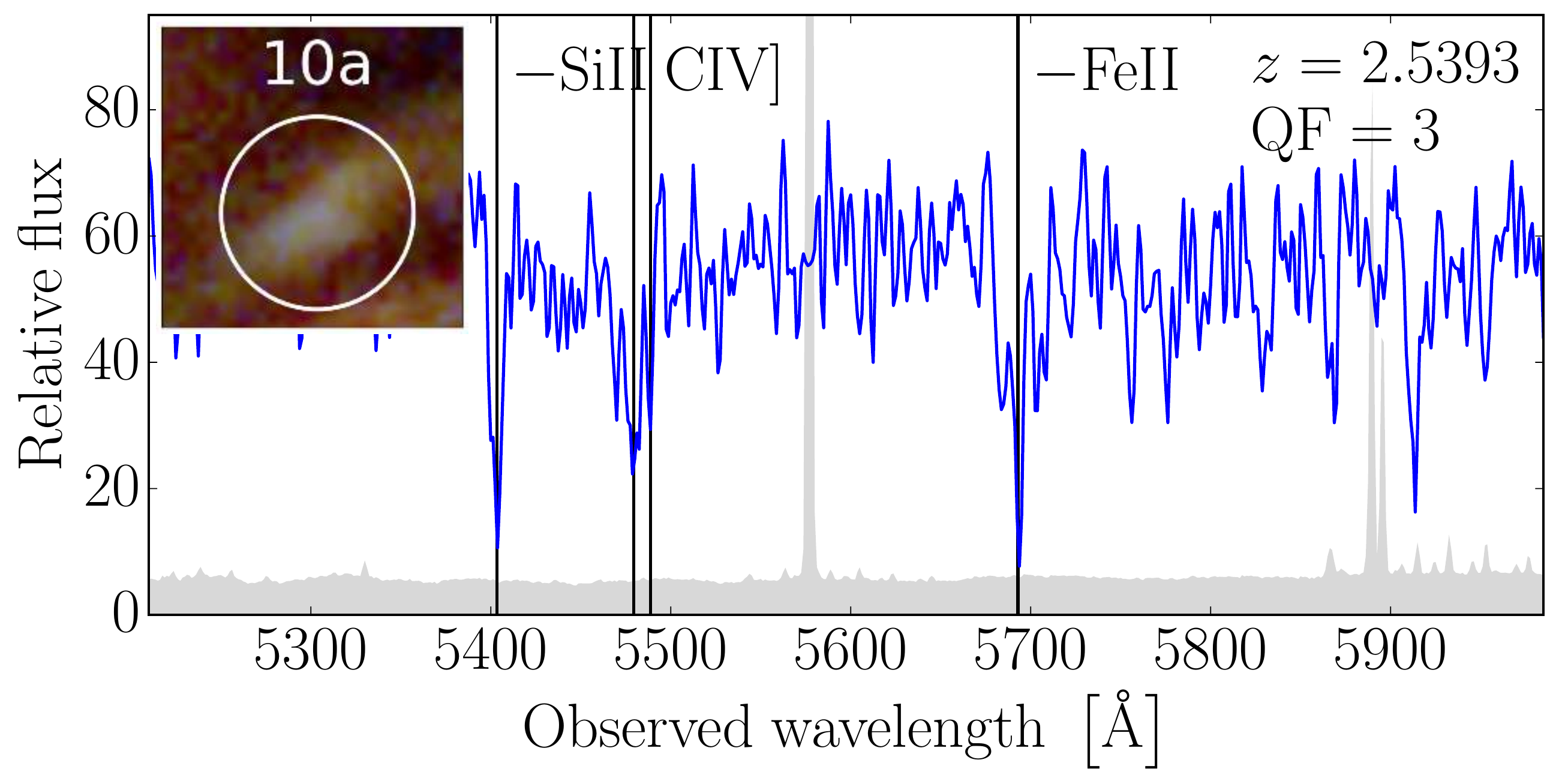}
   \includegraphics[width = 0.666\columnwidth]{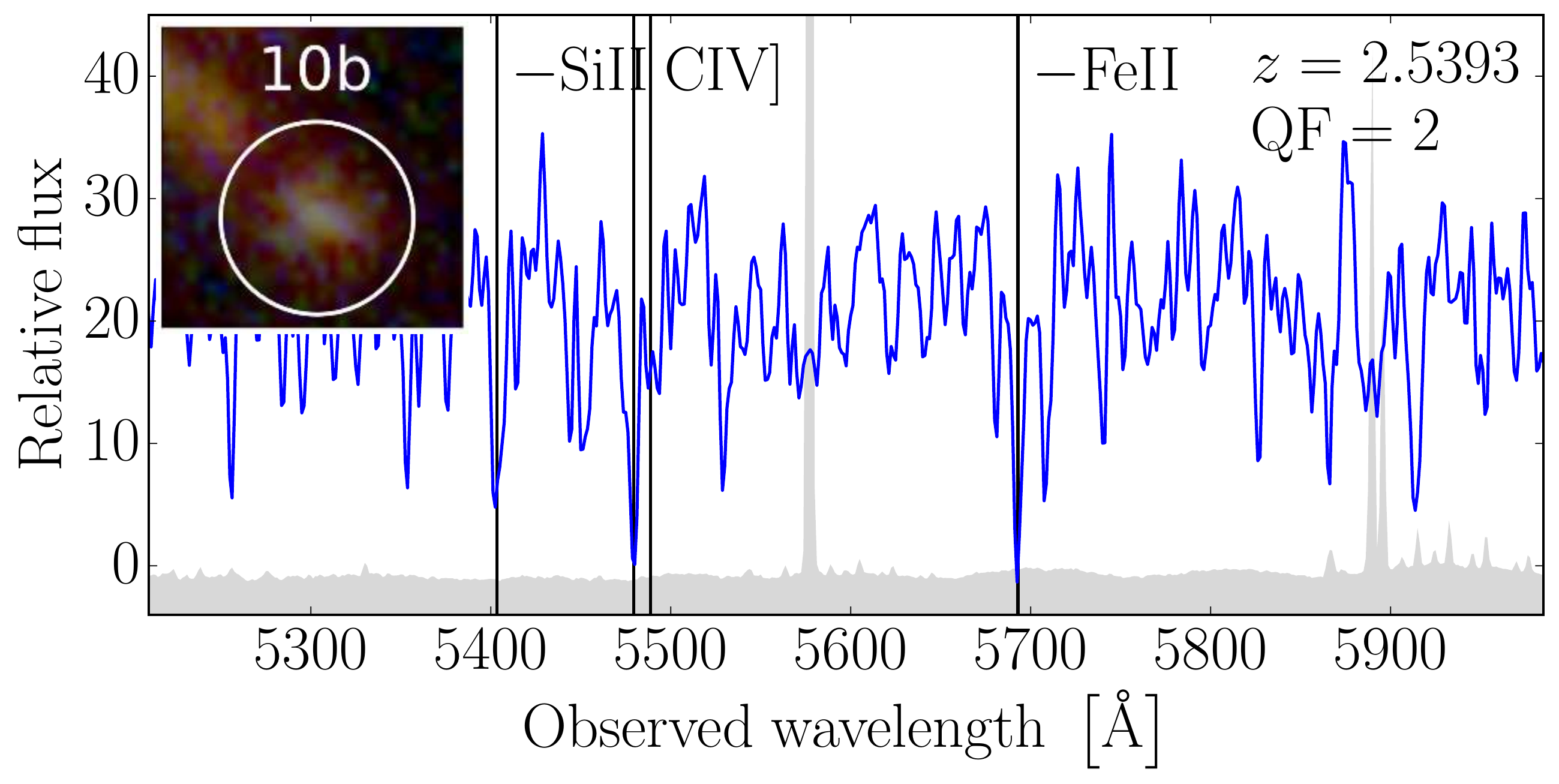}

Family 11:

   \includegraphics[width = 0.666\columnwidth]{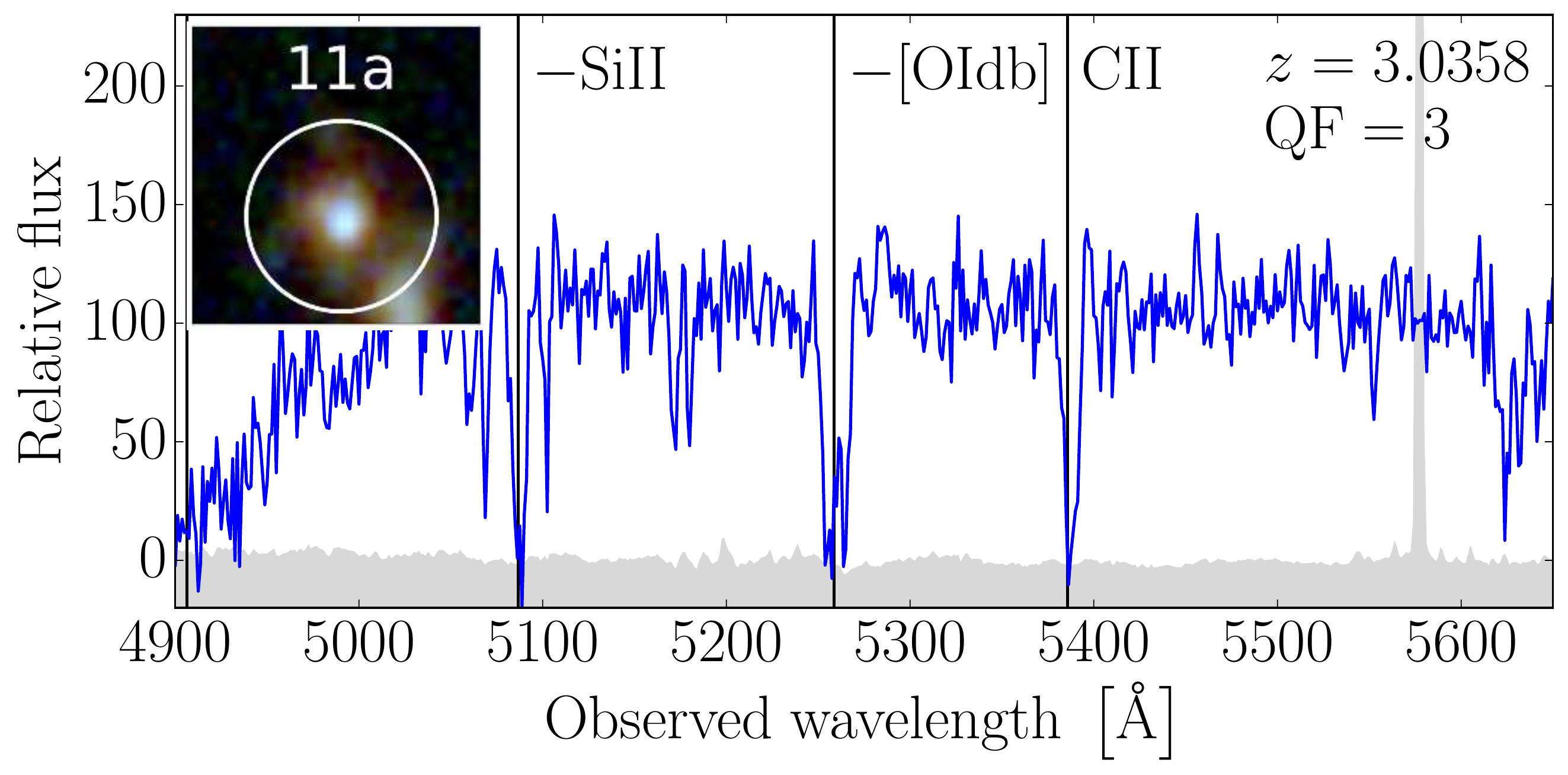}
   \includegraphics[width = 0.666\columnwidth]{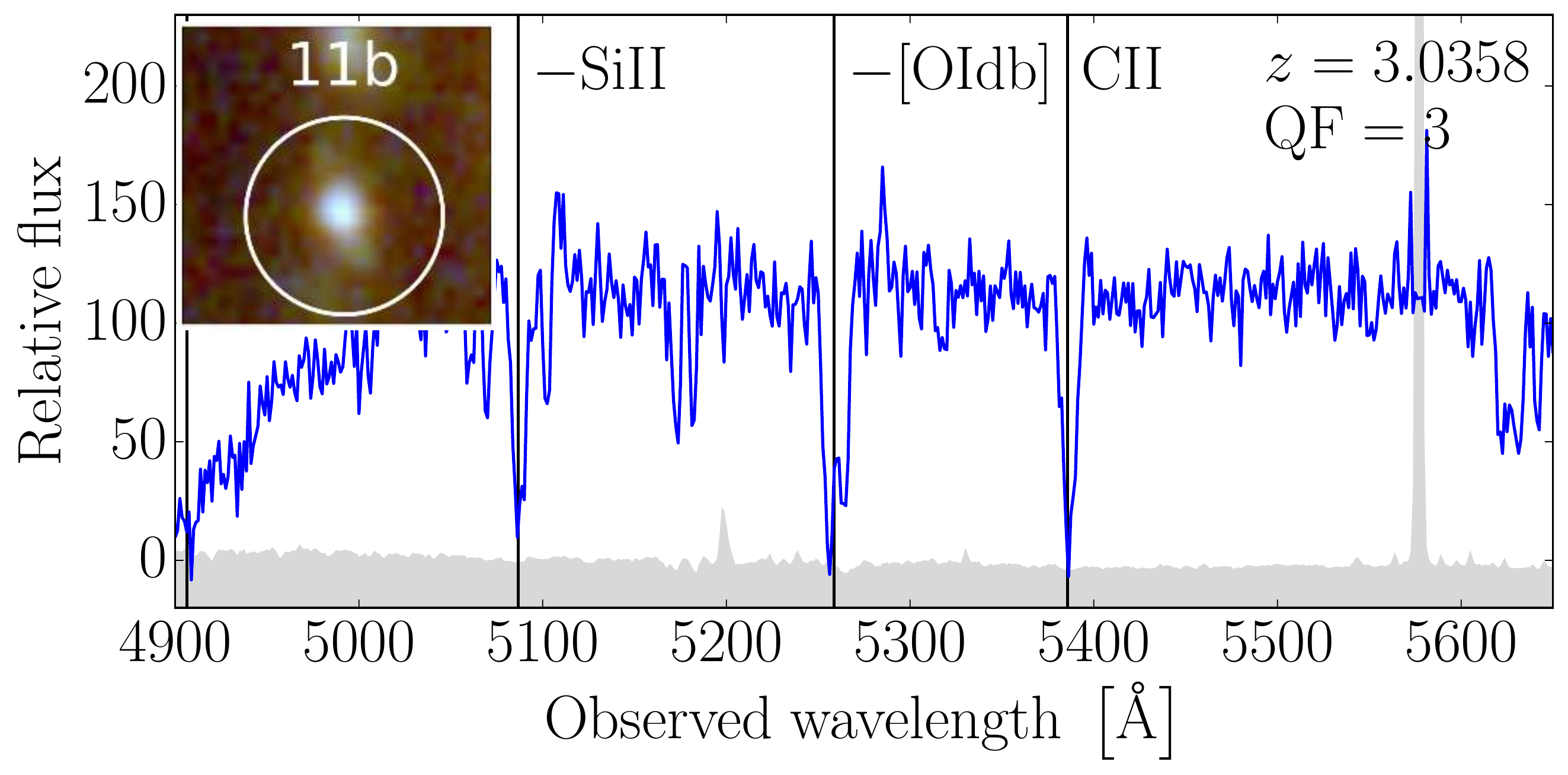}
   \includegraphics[width = 0.666\columnwidth]{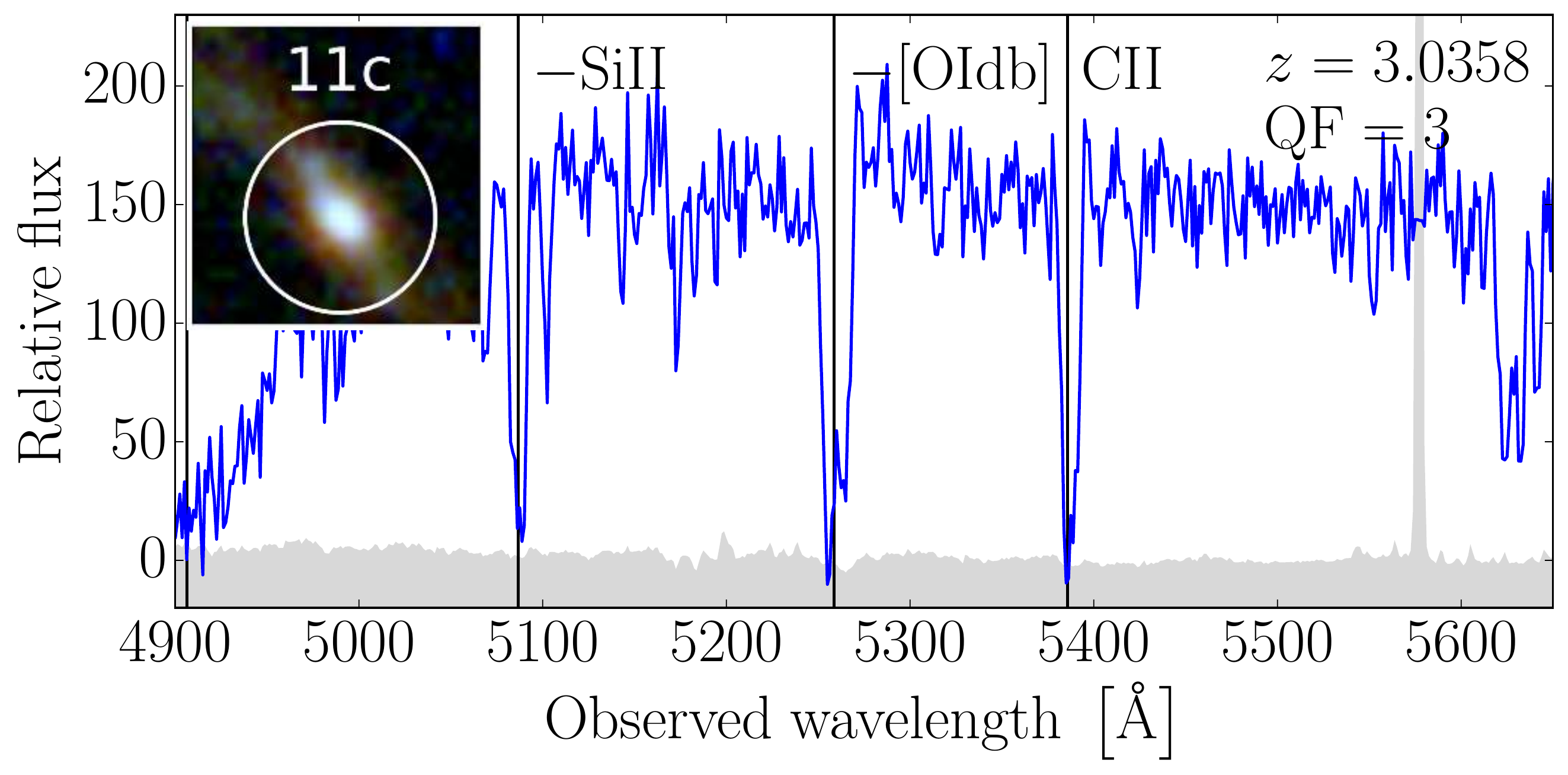}

Family 12:

   \includegraphics[width = 0.666\columnwidth]{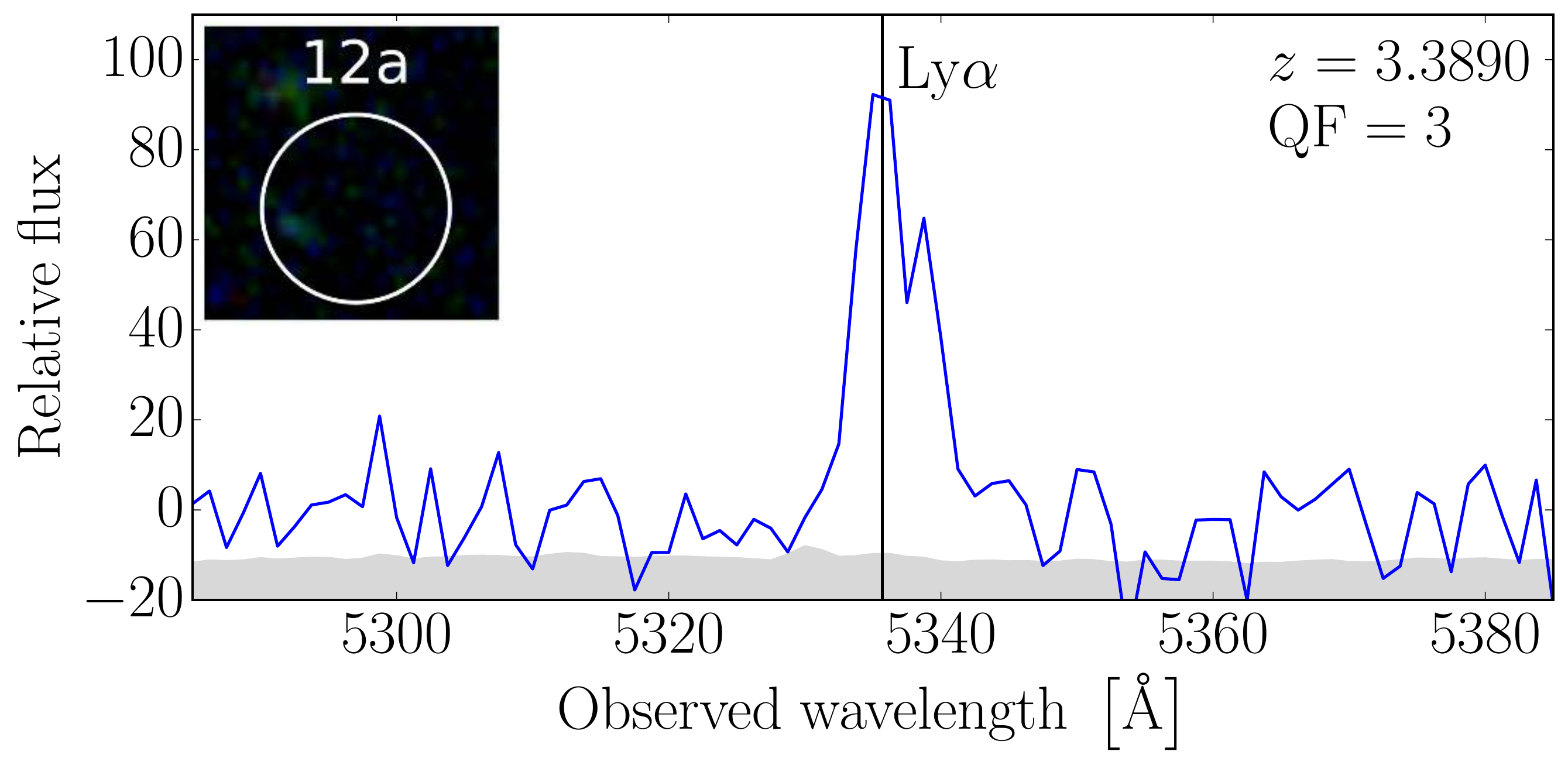}
   \includegraphics[width = 0.666\columnwidth]{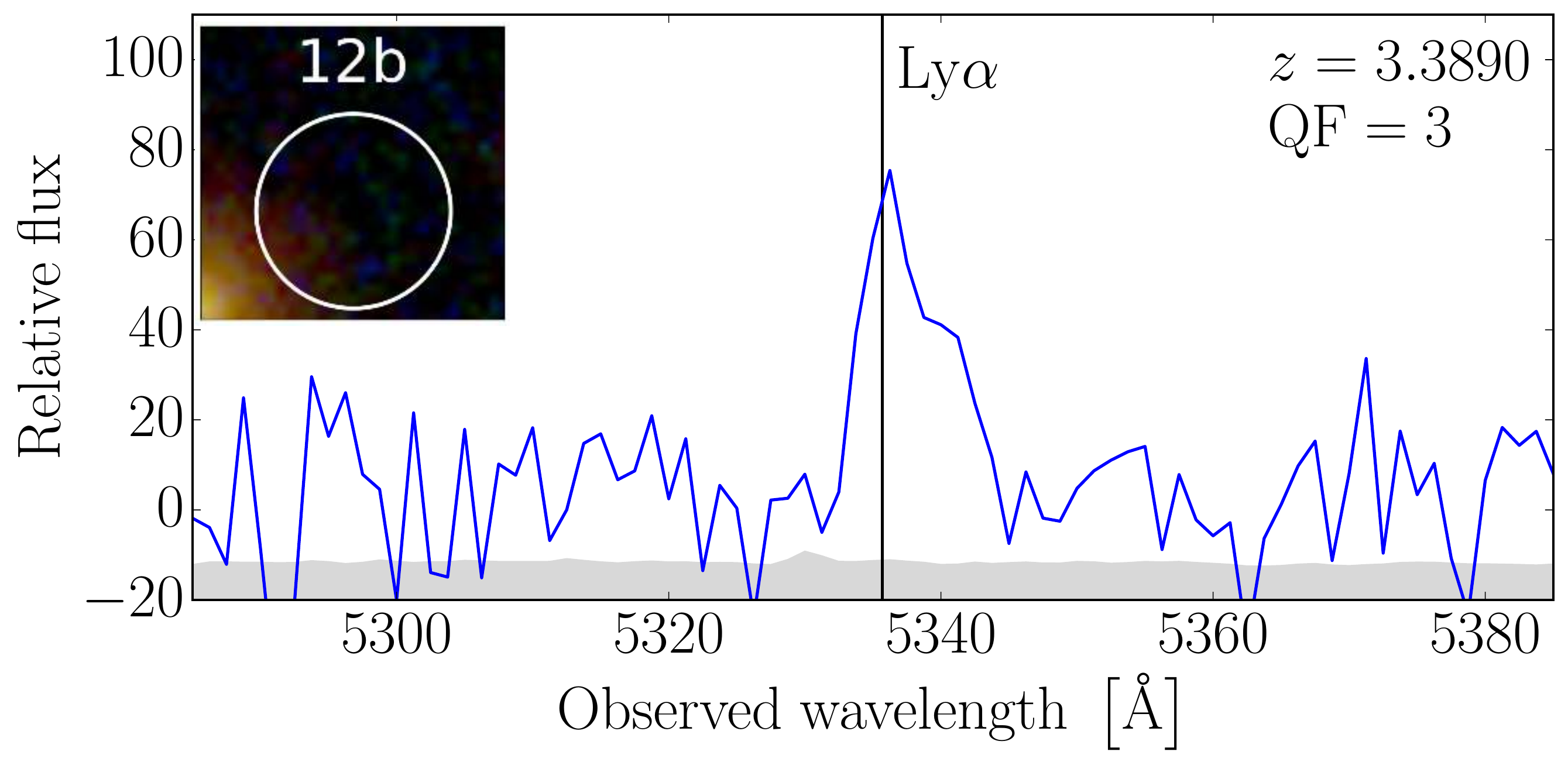}
   \includegraphics[width = 0.666\columnwidth]{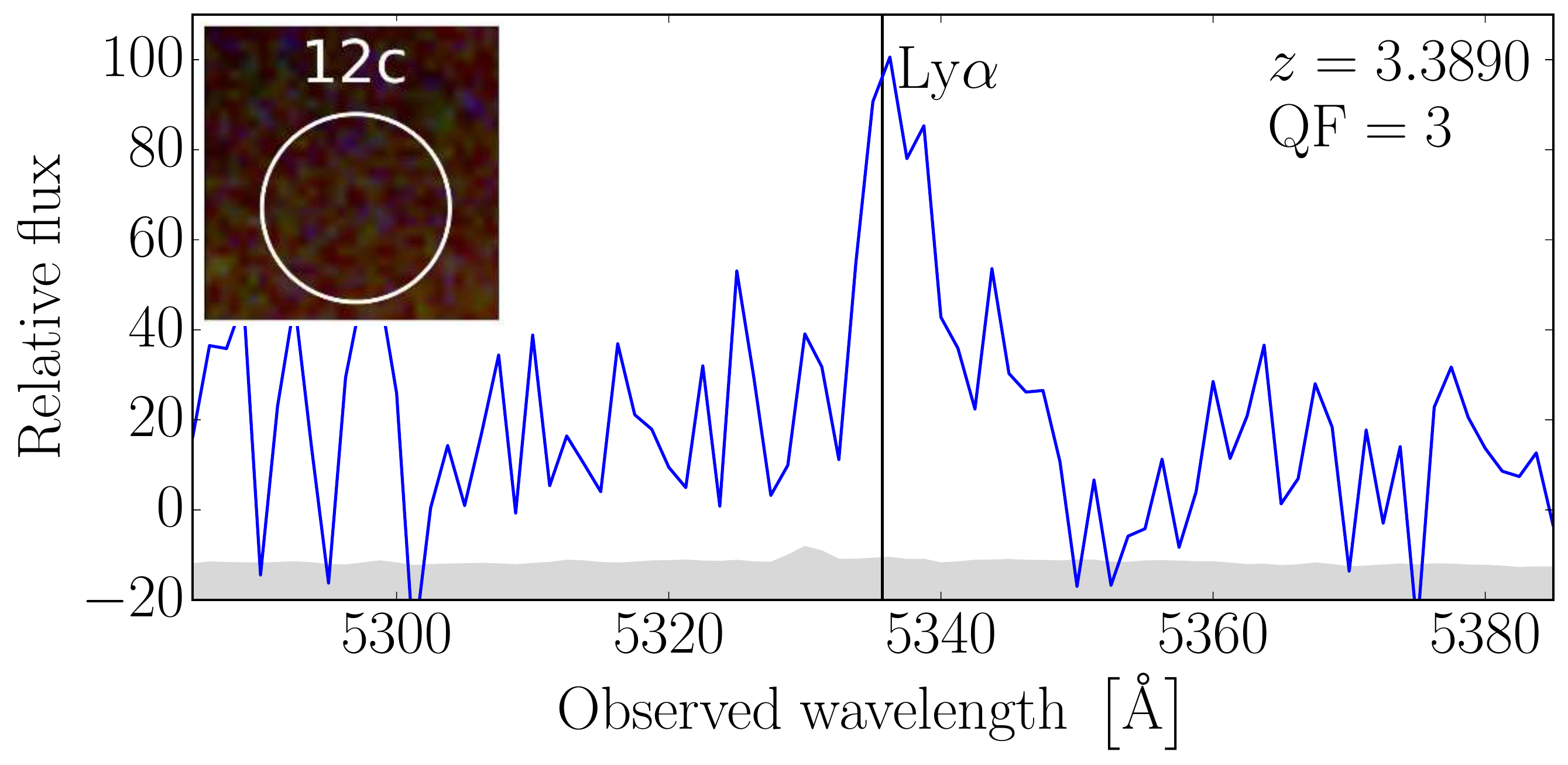}

  \caption{(Continued)}
  \label{fig:specs}
\end{figure*}

\begin{figure*}
\setcounter{figure}{\value{figure}-1}

Family 13:

   \includegraphics[width = 0.666\columnwidth]{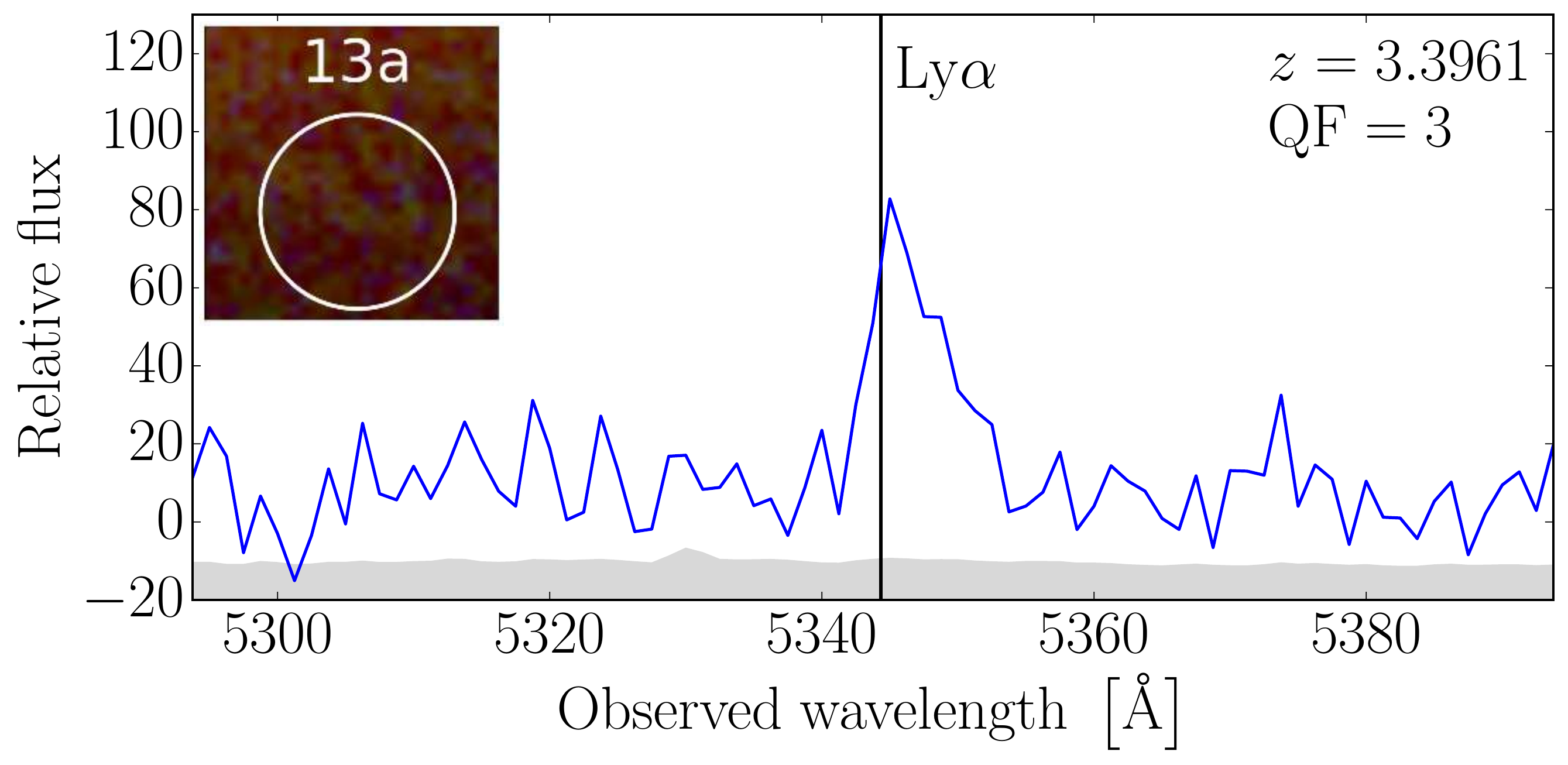}
   \includegraphics[width = 0.666\columnwidth]{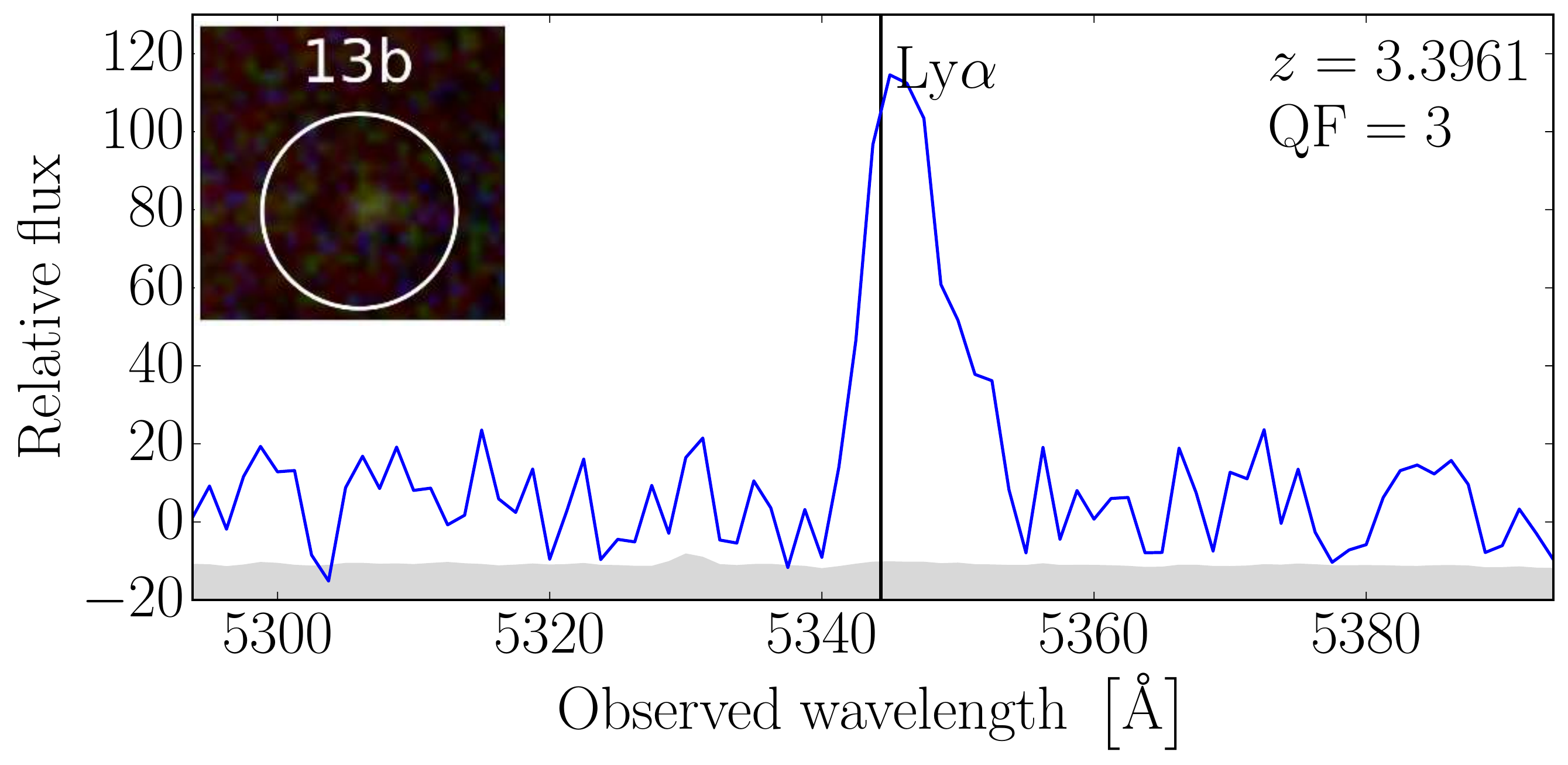}

   \includegraphics[width = 0.666\columnwidth]{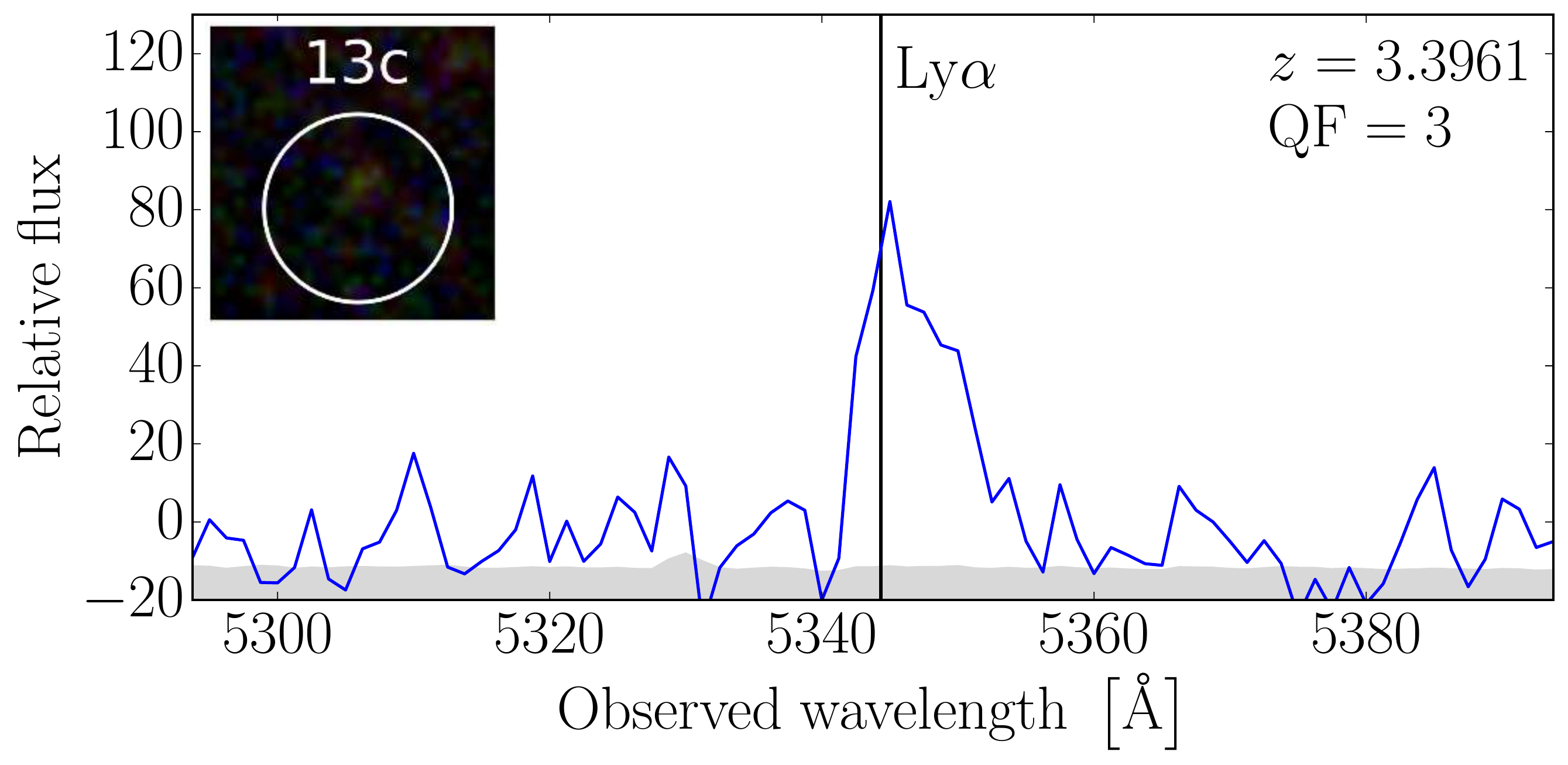}
   \includegraphics[width = 0.666\columnwidth]{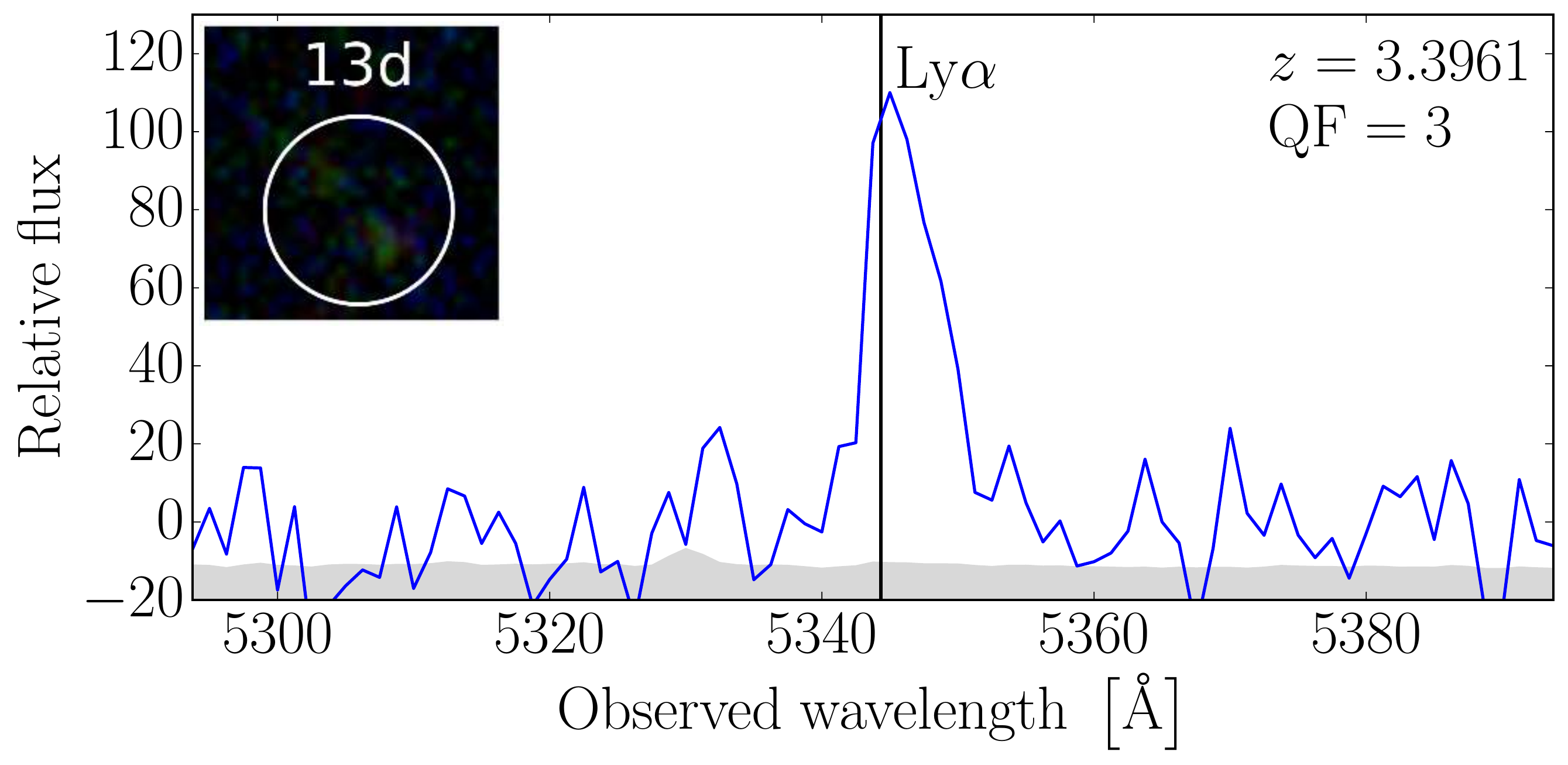}

Family 14:

   \includegraphics[width = 0.666\columnwidth]{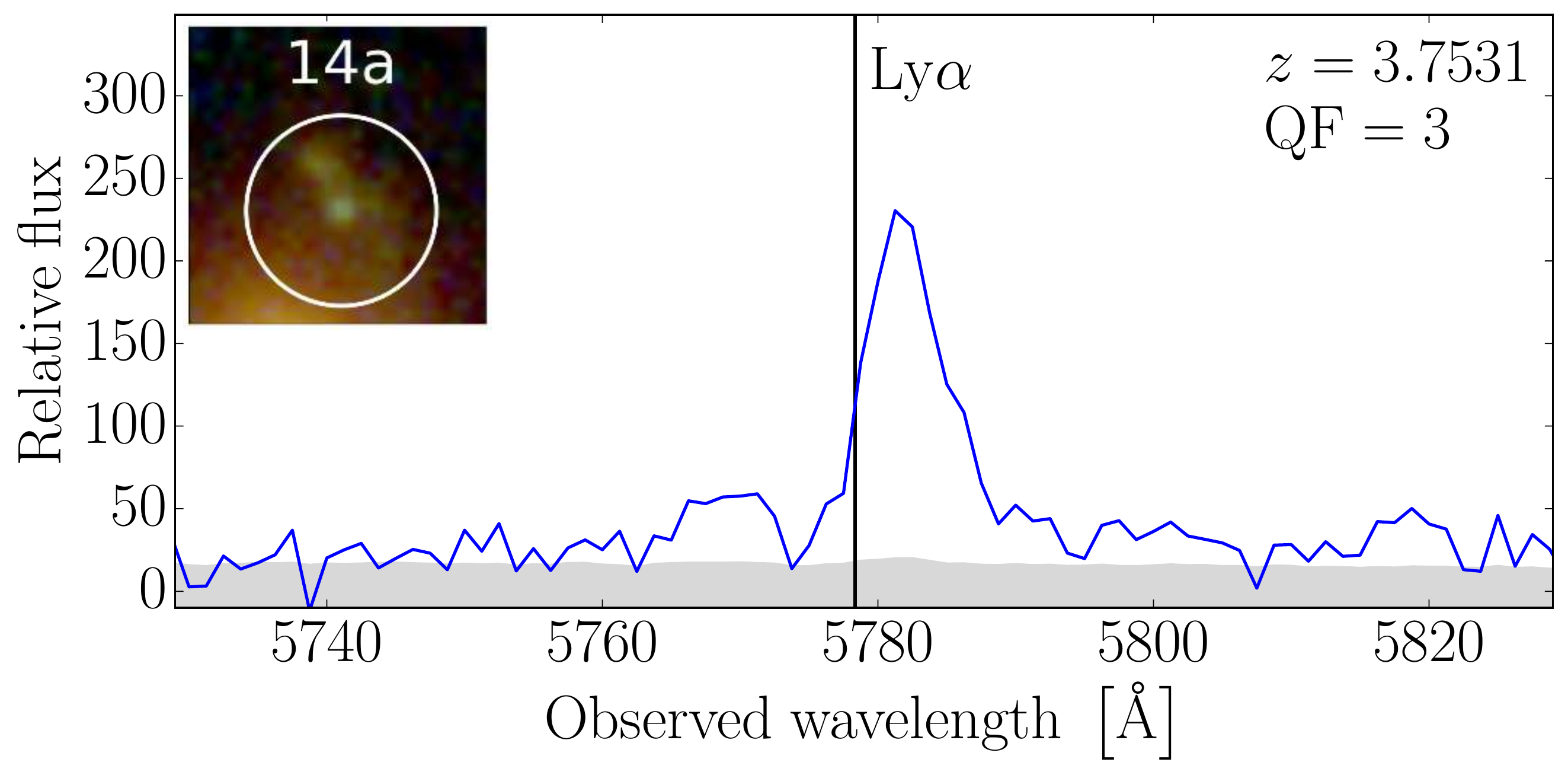}
   \includegraphics[width = 0.666\columnwidth]{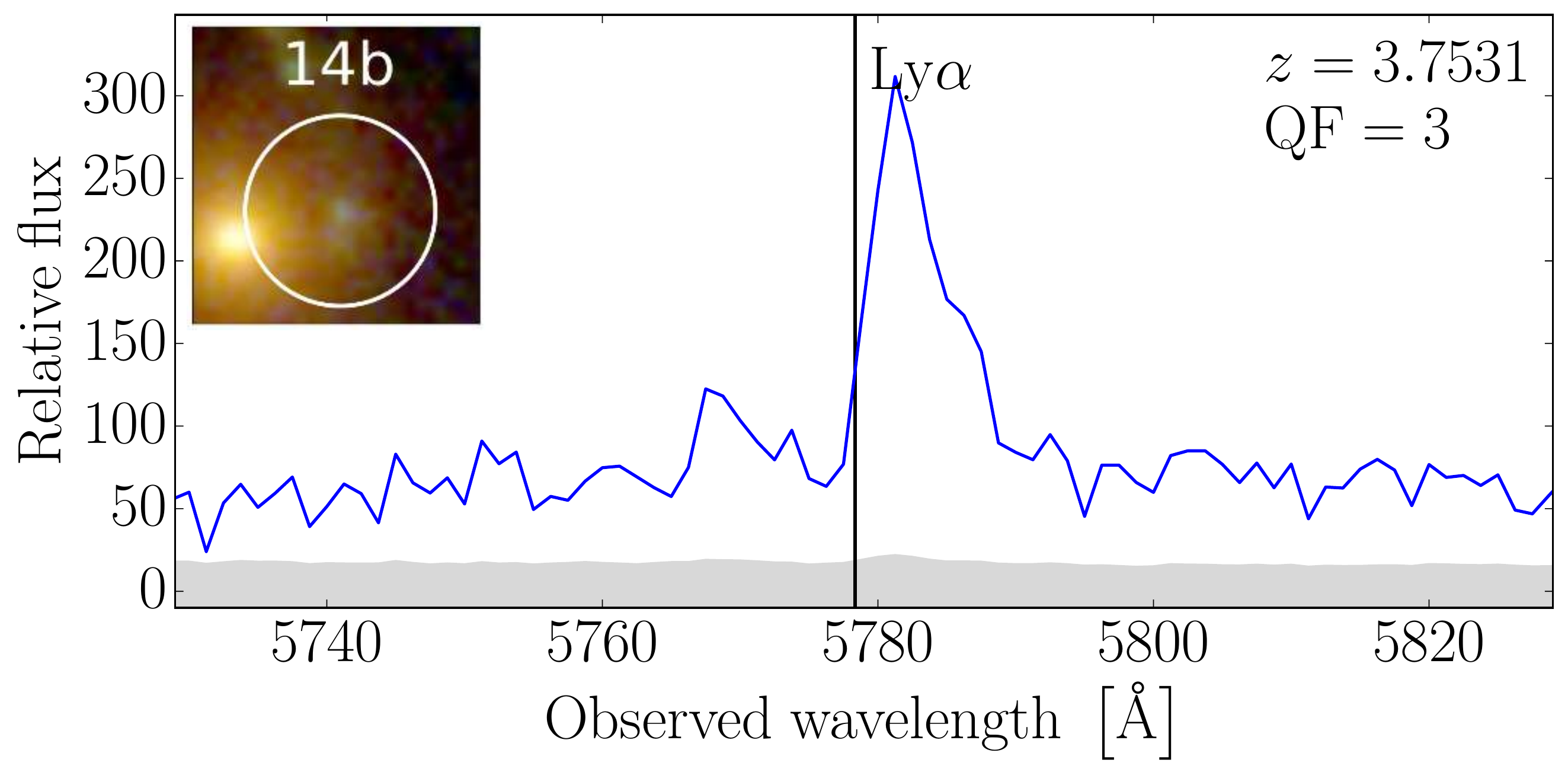}
   \includegraphics[width = 0.666\columnwidth]{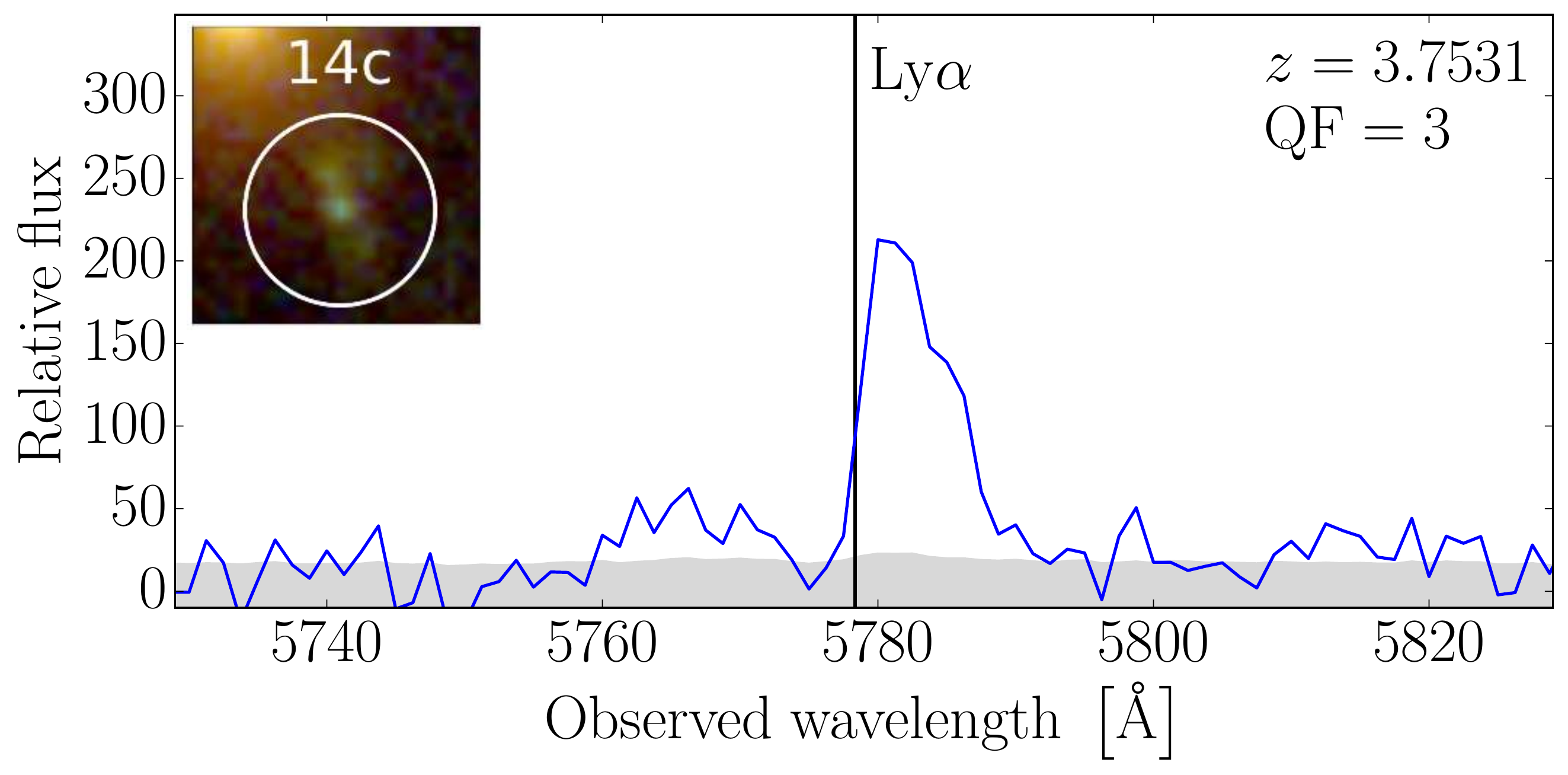}

   \includegraphics[width = 0.666\columnwidth]{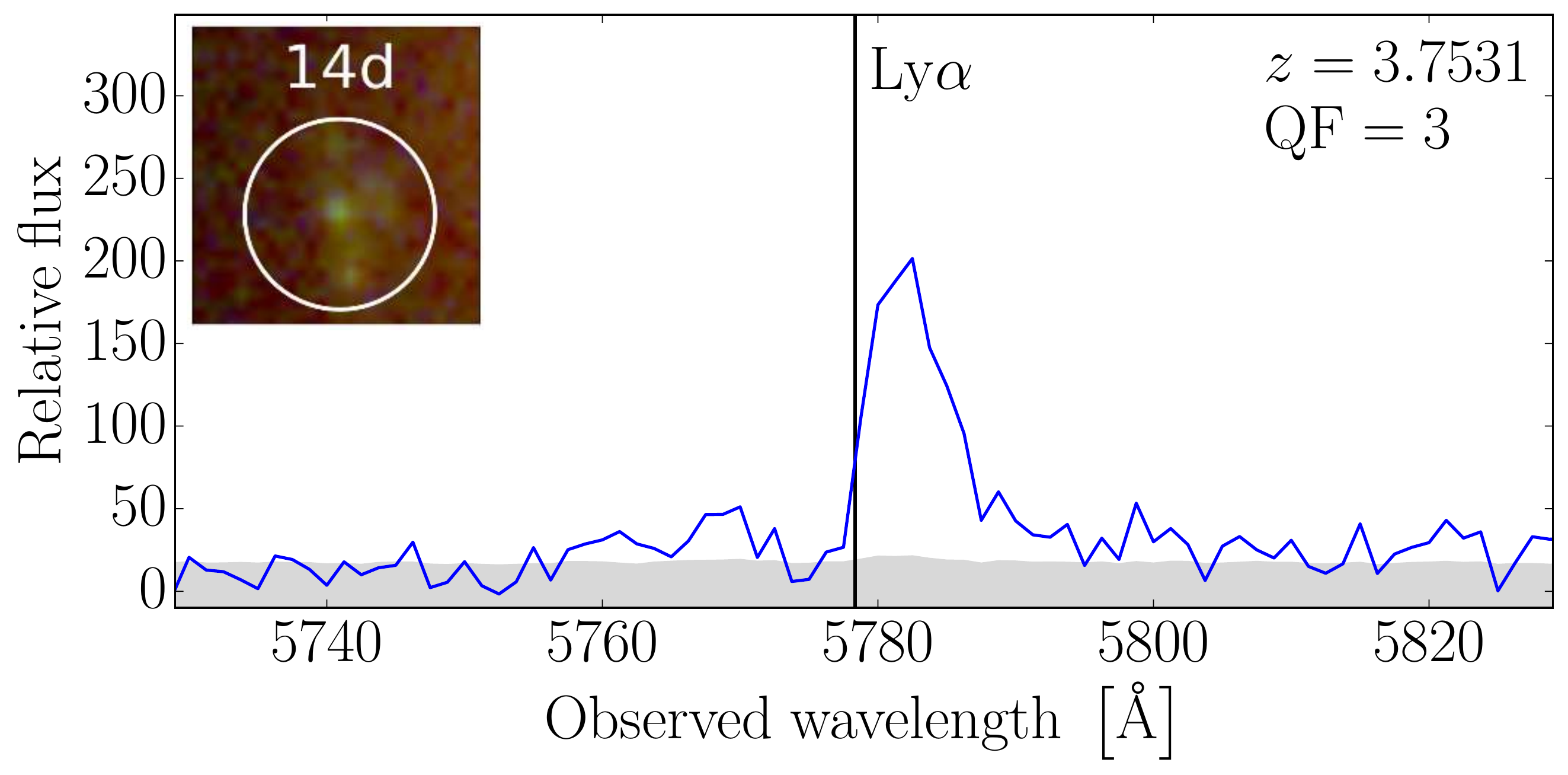}
   \includegraphics[width = 0.666\columnwidth]{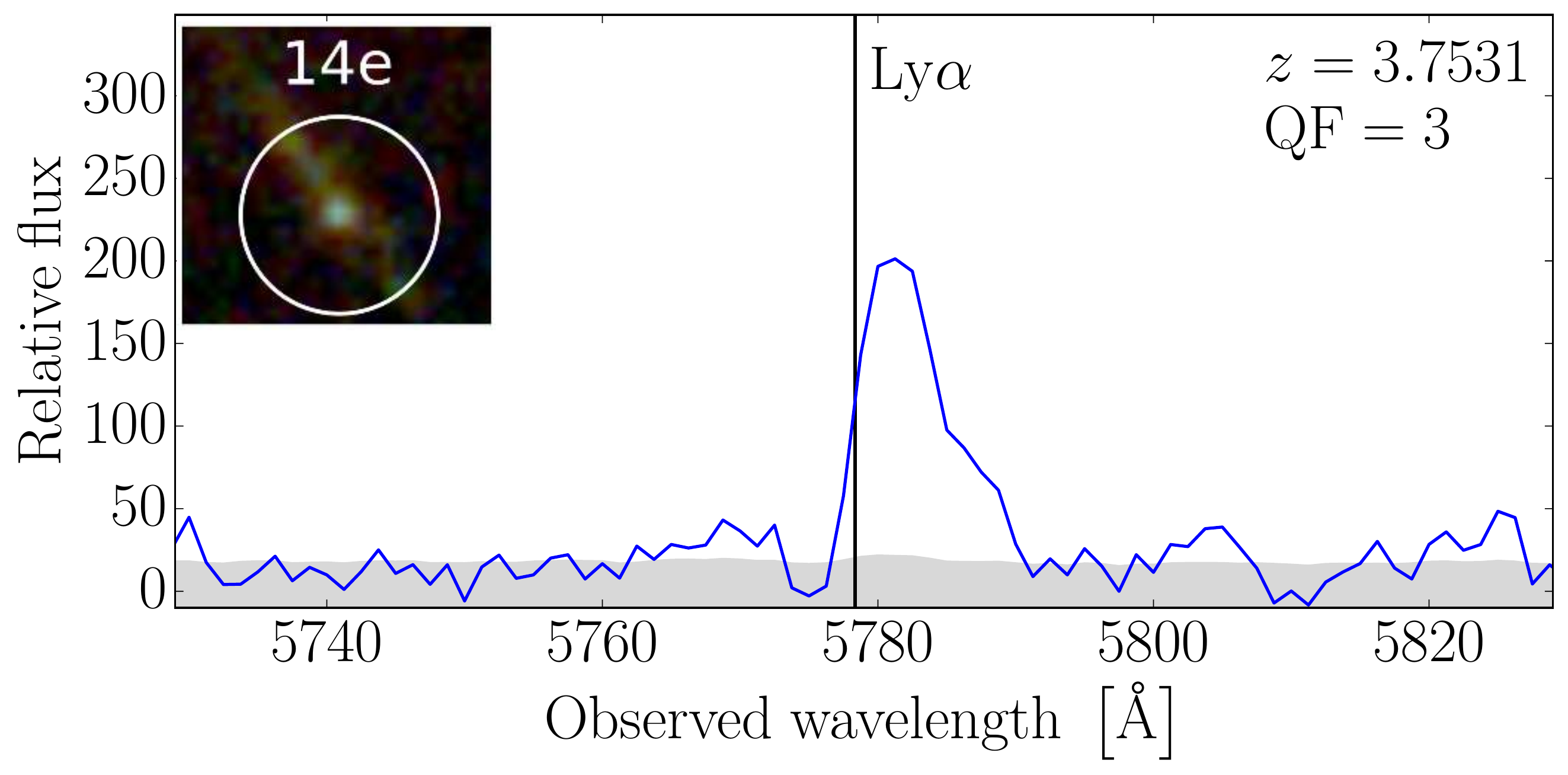}

Family 15:

   \includegraphics[width = 0.666\columnwidth]{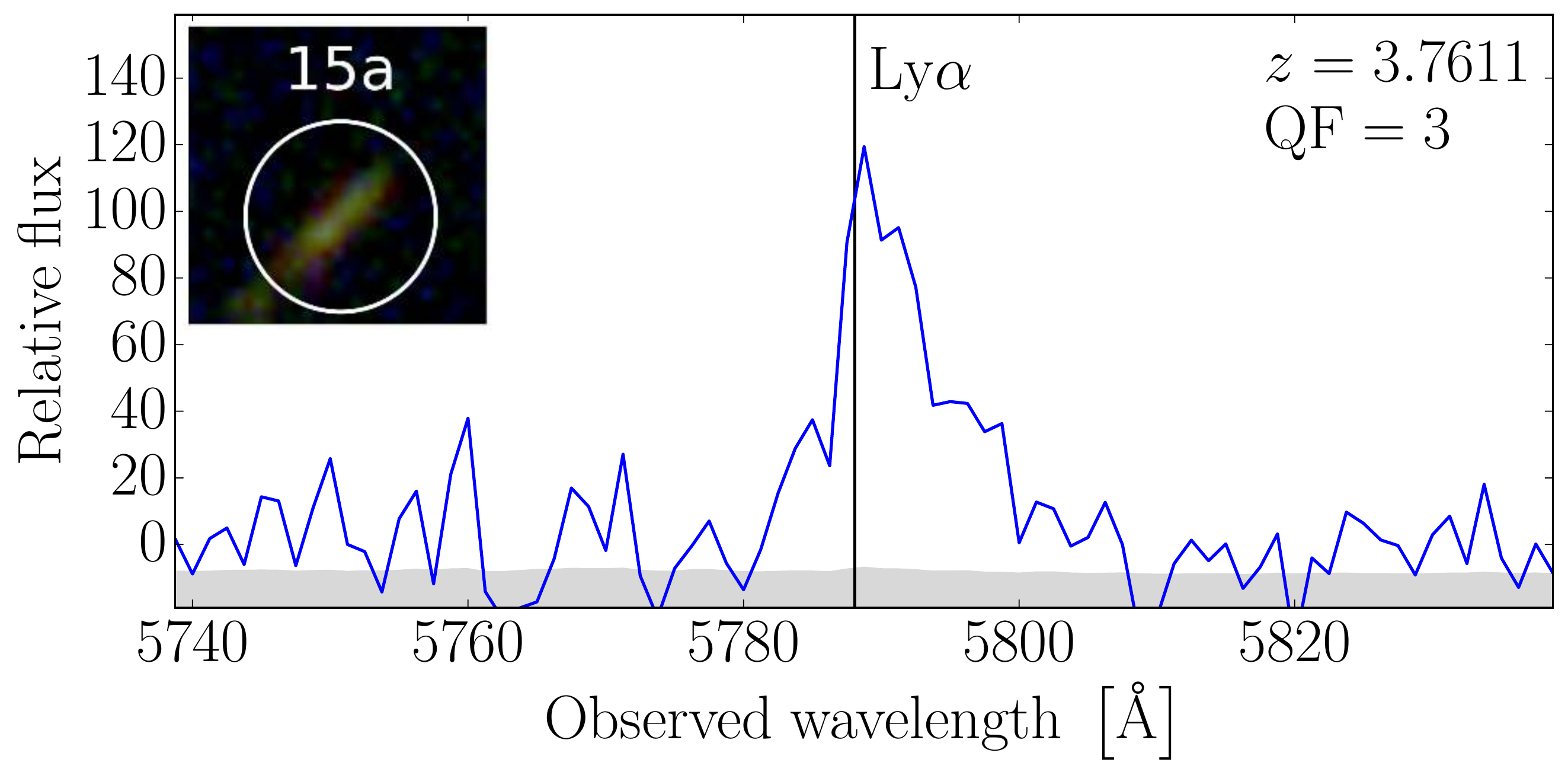}
   \includegraphics[width = 0.666\columnwidth]{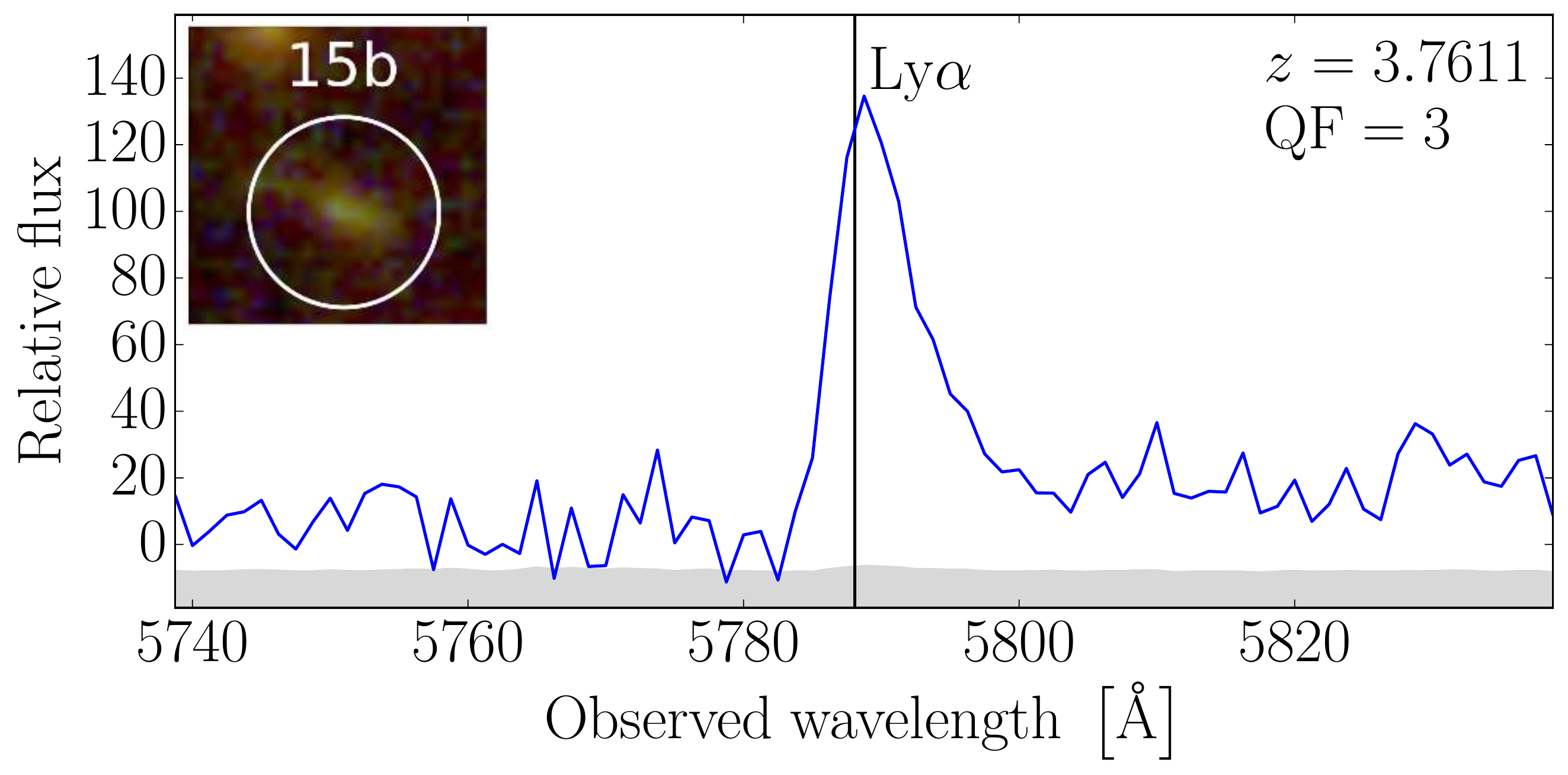}

Family 16:

   \includegraphics[width = 0.666\columnwidth]{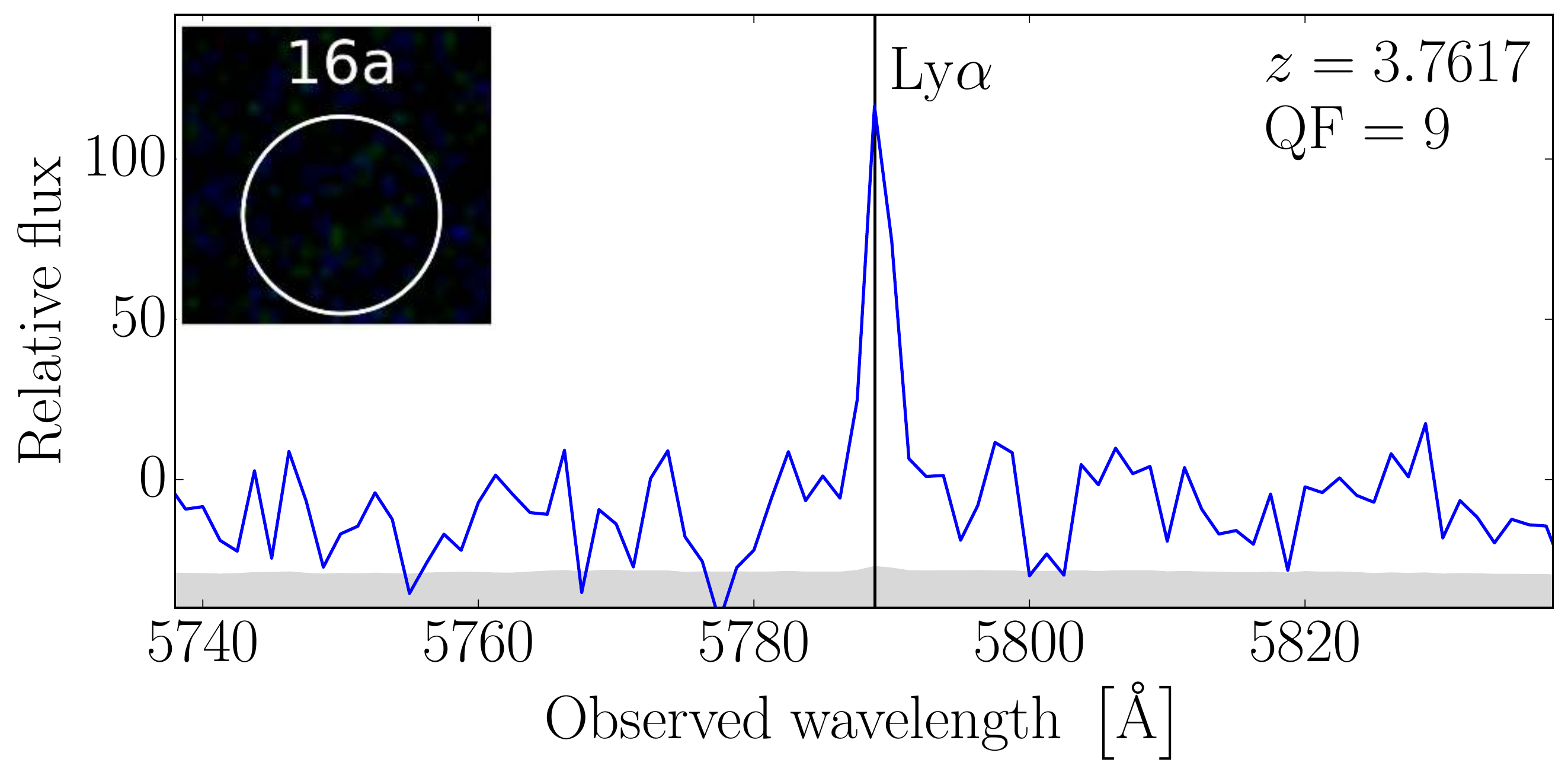}
   \includegraphics[width = 0.666\columnwidth]{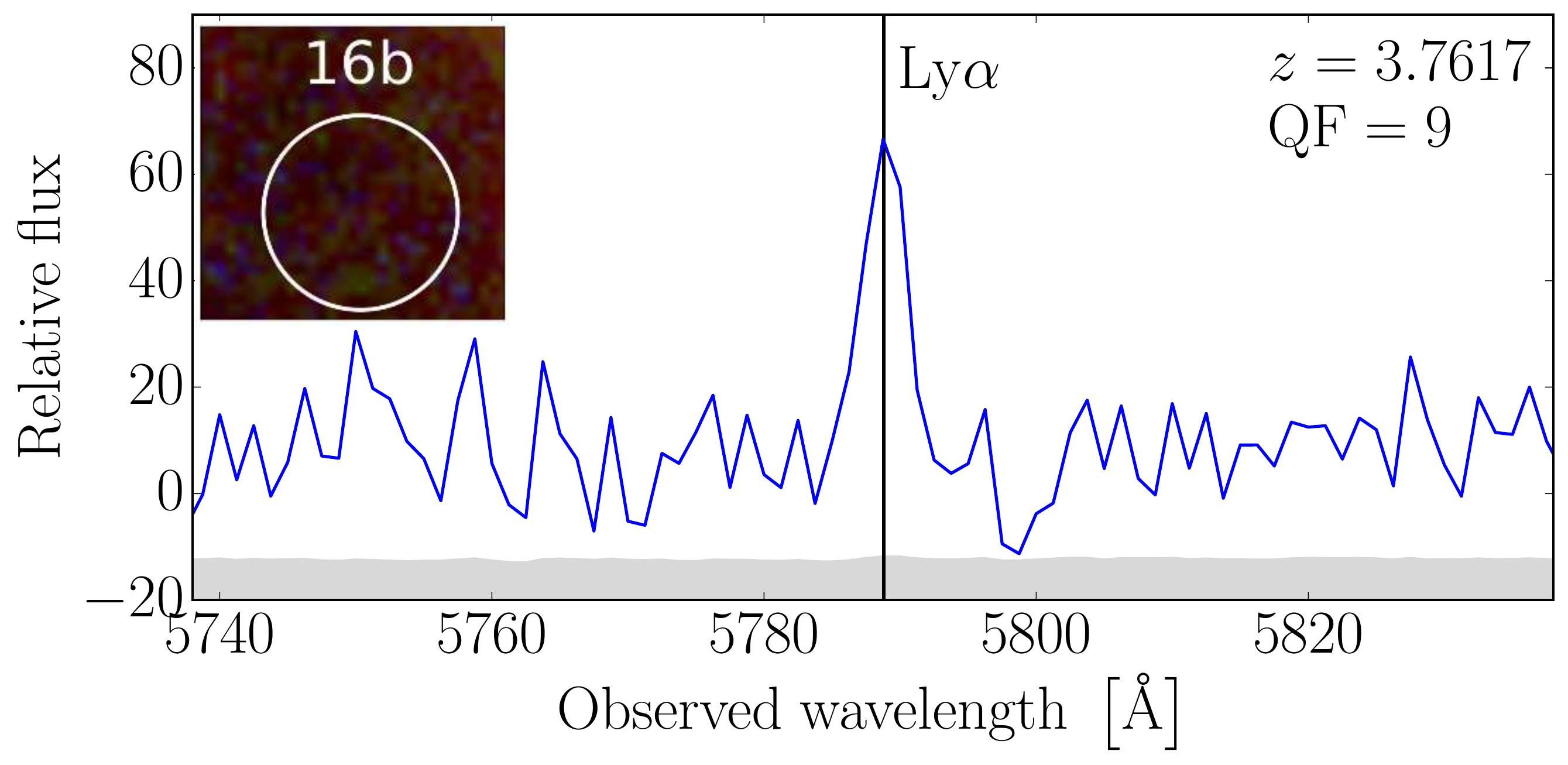}
   \includegraphics[width = 0.666\columnwidth]{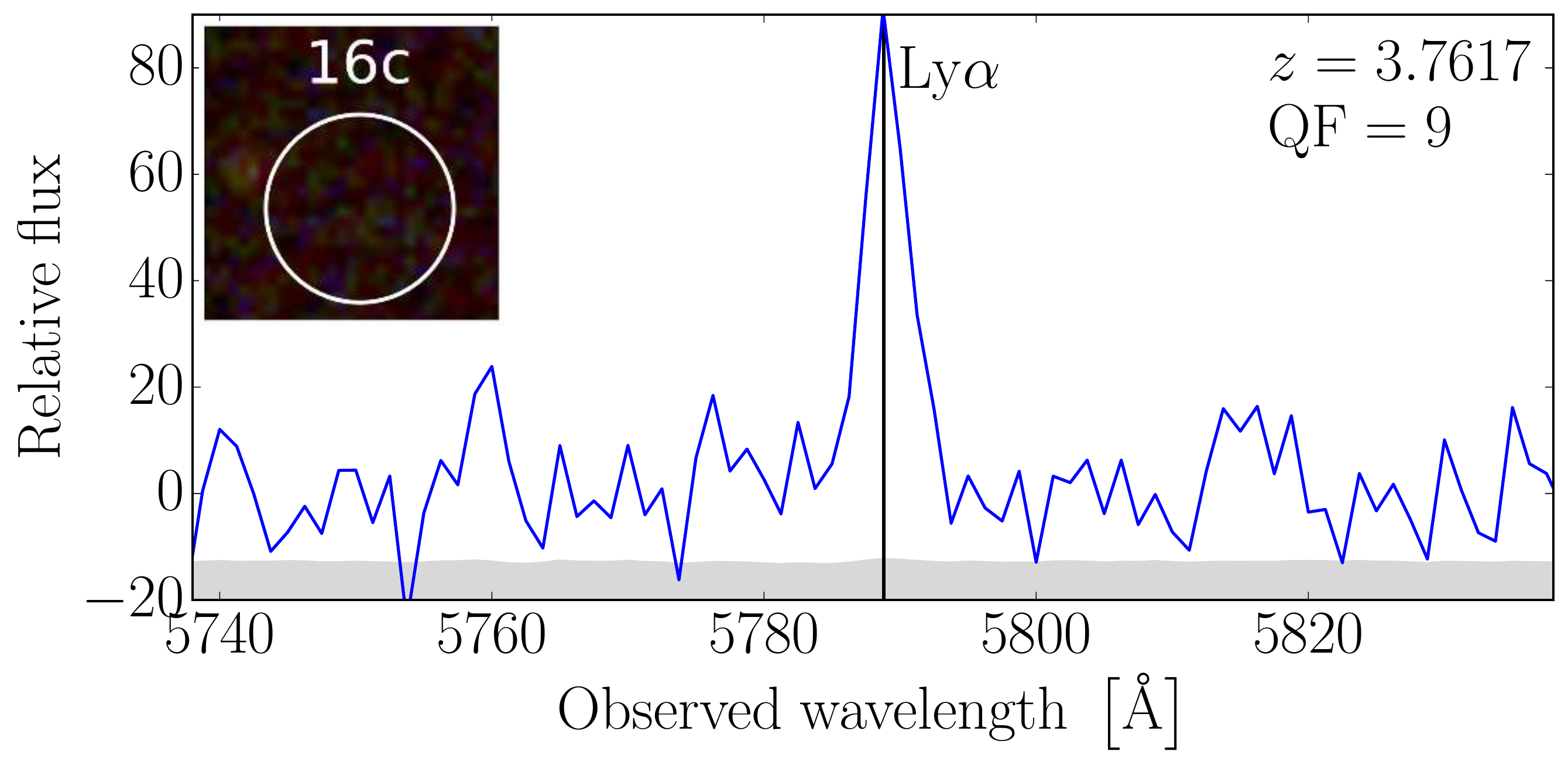}

Family 17:

   \includegraphics[width = 0.666\columnwidth]{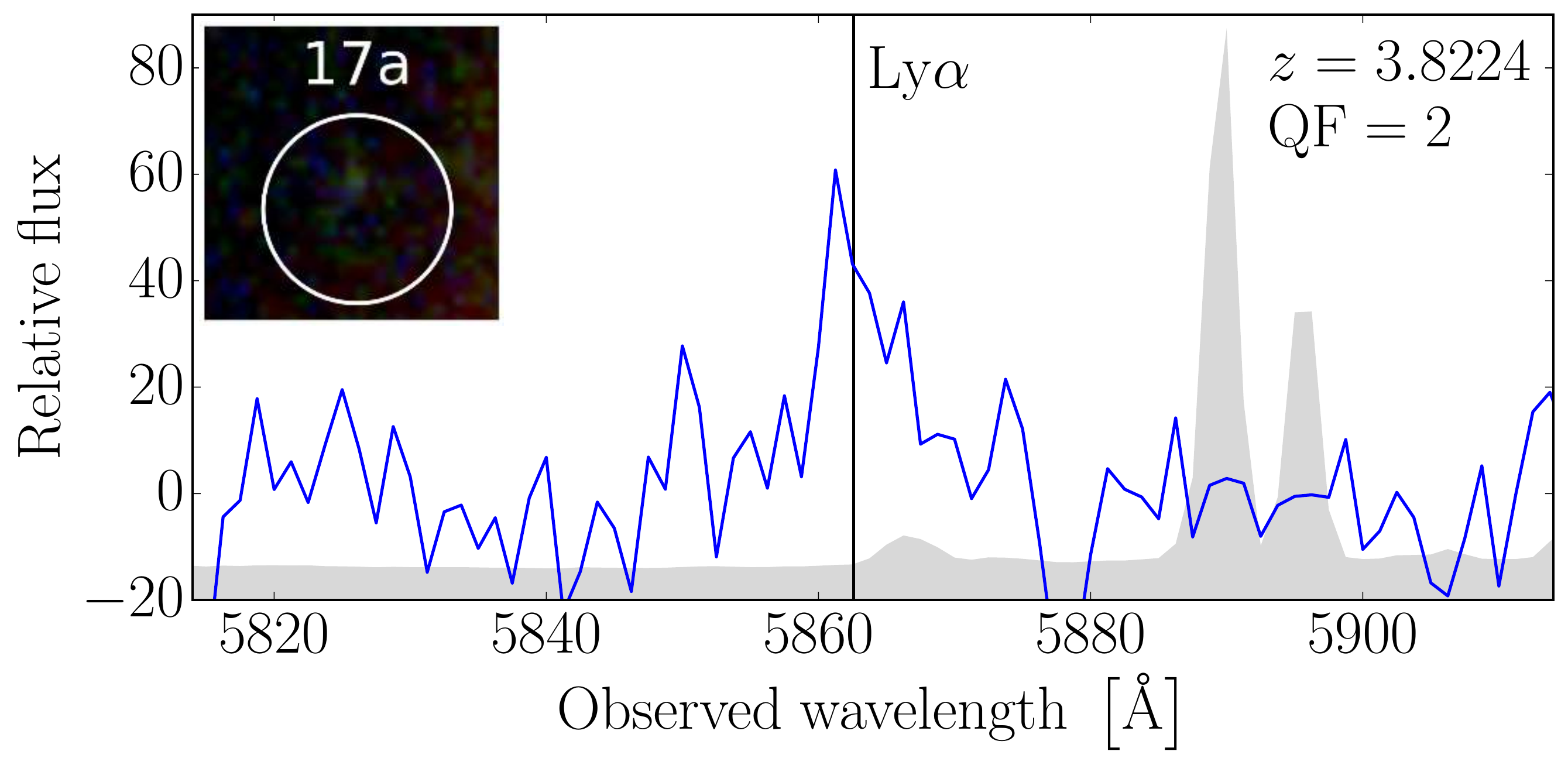}
   \includegraphics[width = 0.666\columnwidth]{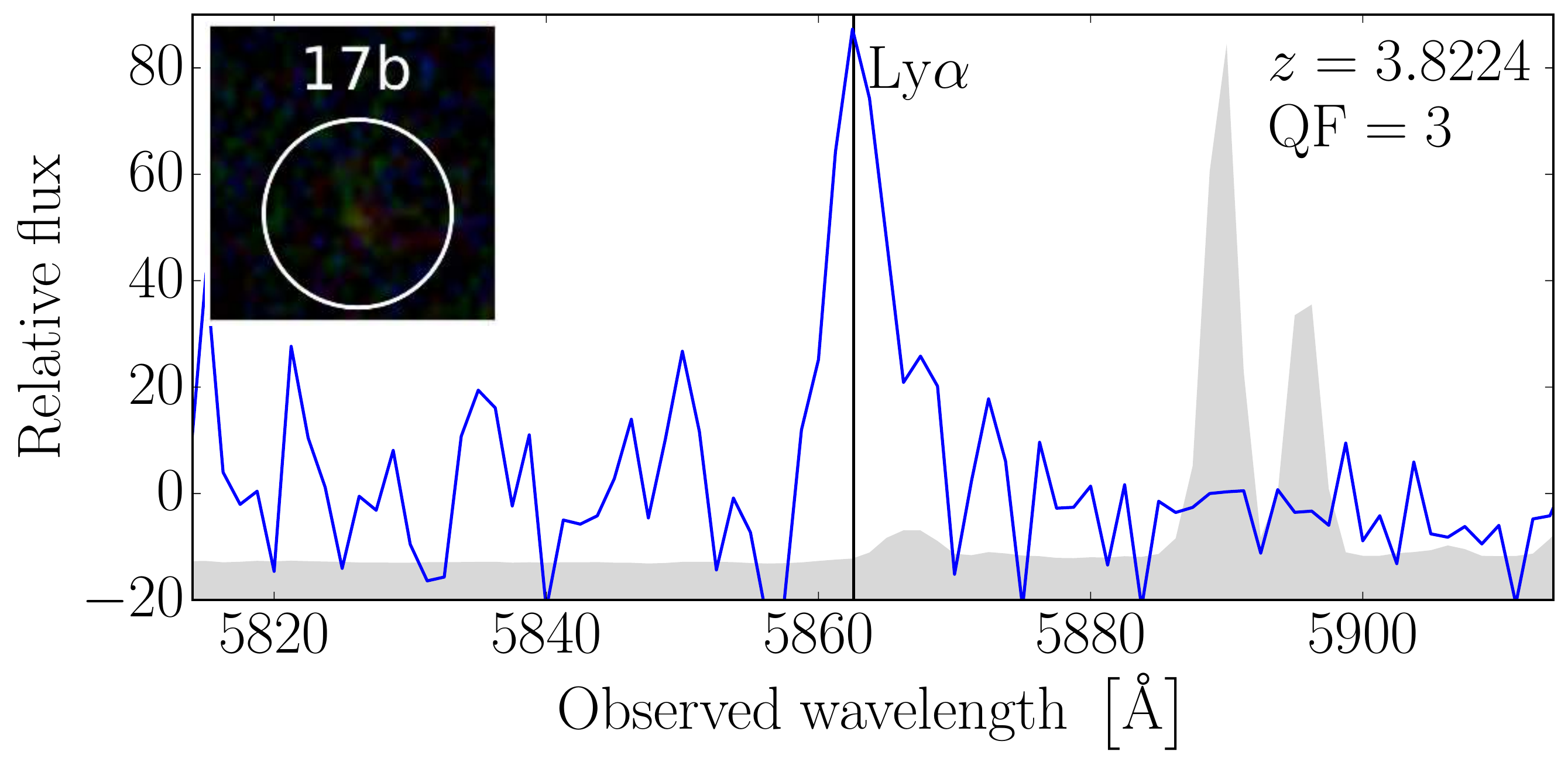}

  \caption{(Continued)}
  \label{fig:specs}
\end{figure*}

\begin{figure*}
\setcounter{figure}{\value{figure}-1}

Family 18:

   \includegraphics[width = 0.666\columnwidth]{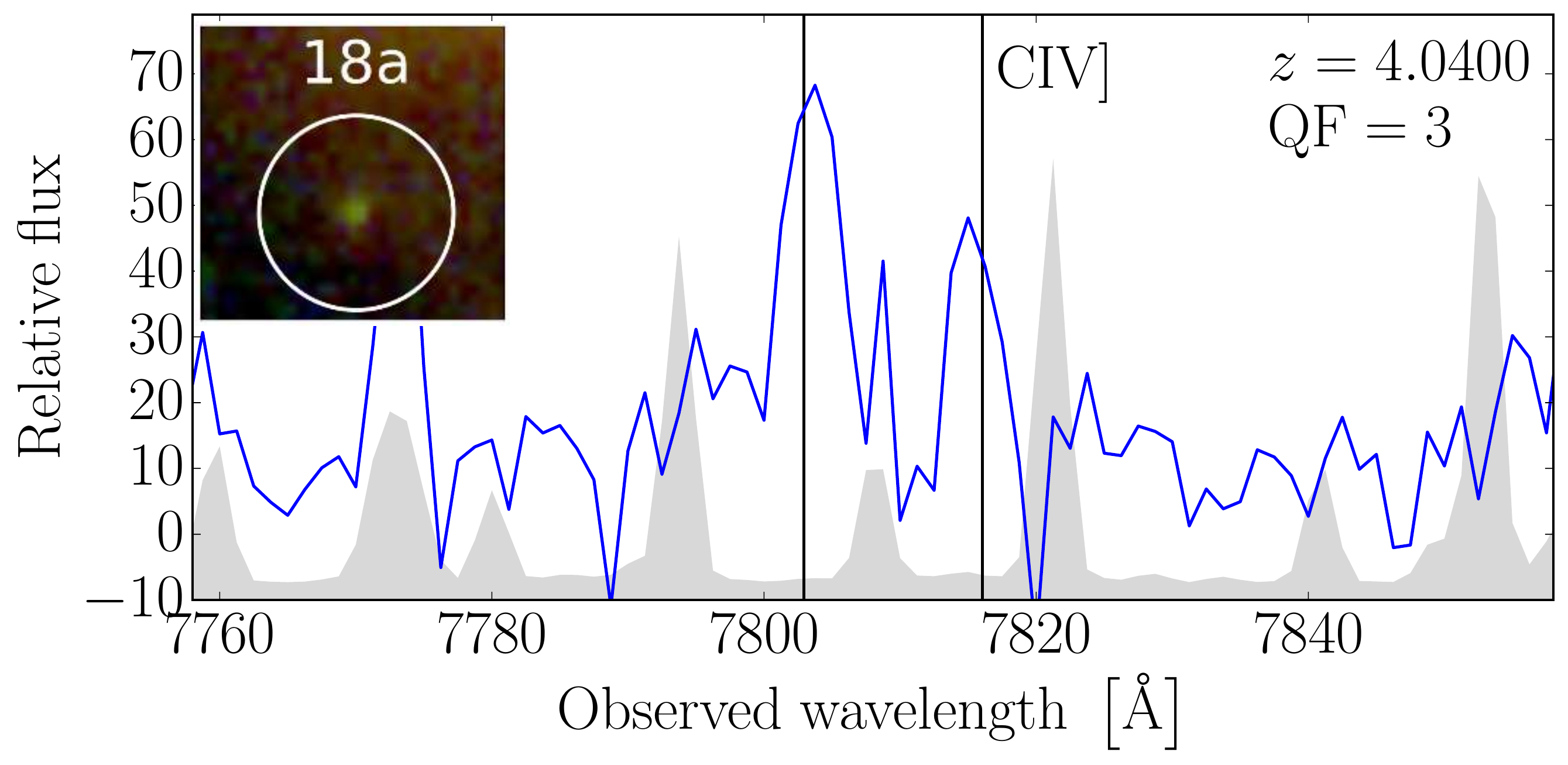}
   \includegraphics[width = 0.666\columnwidth]{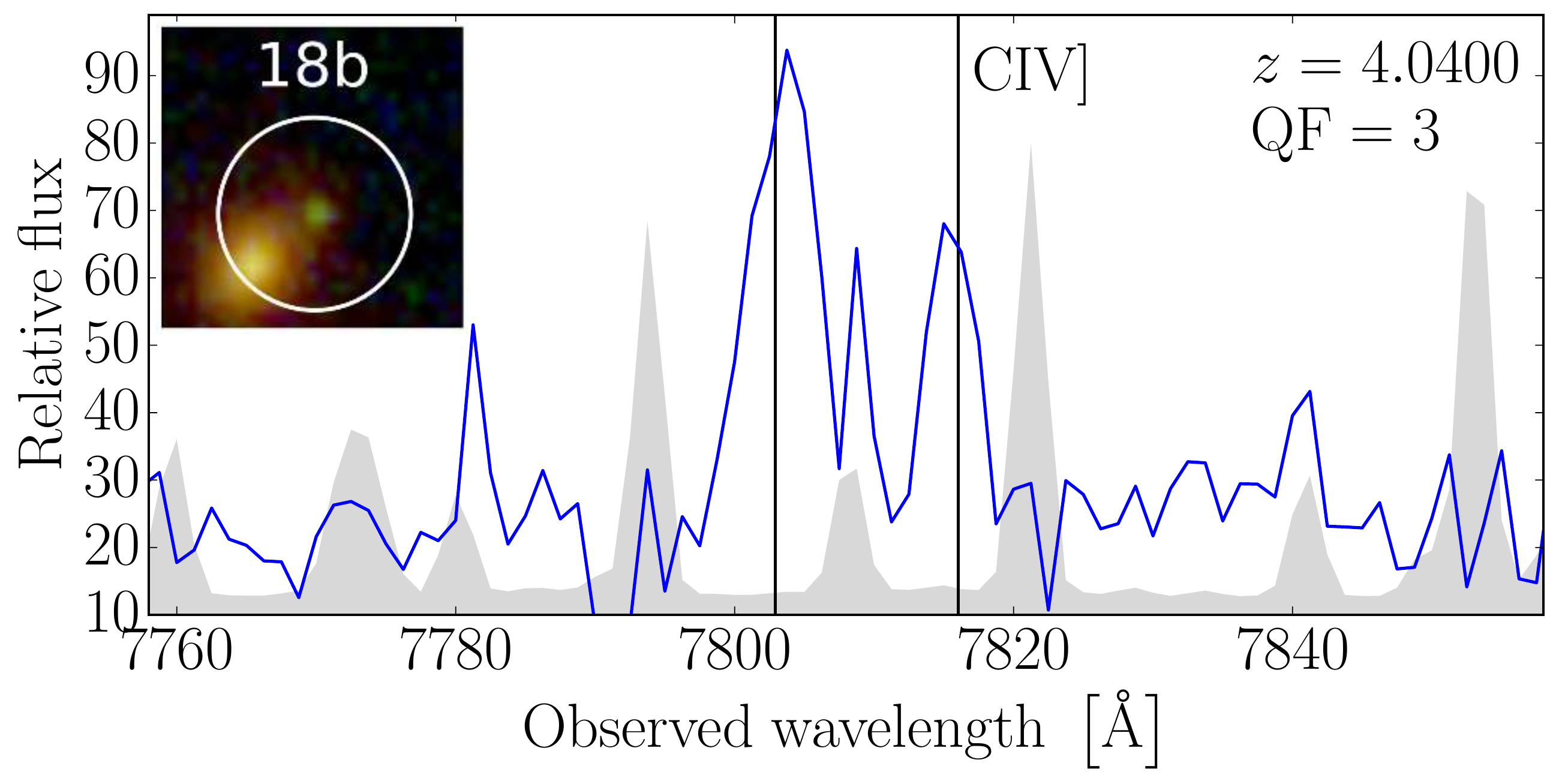}

Family 19:

   \includegraphics[width = 0.666\columnwidth]{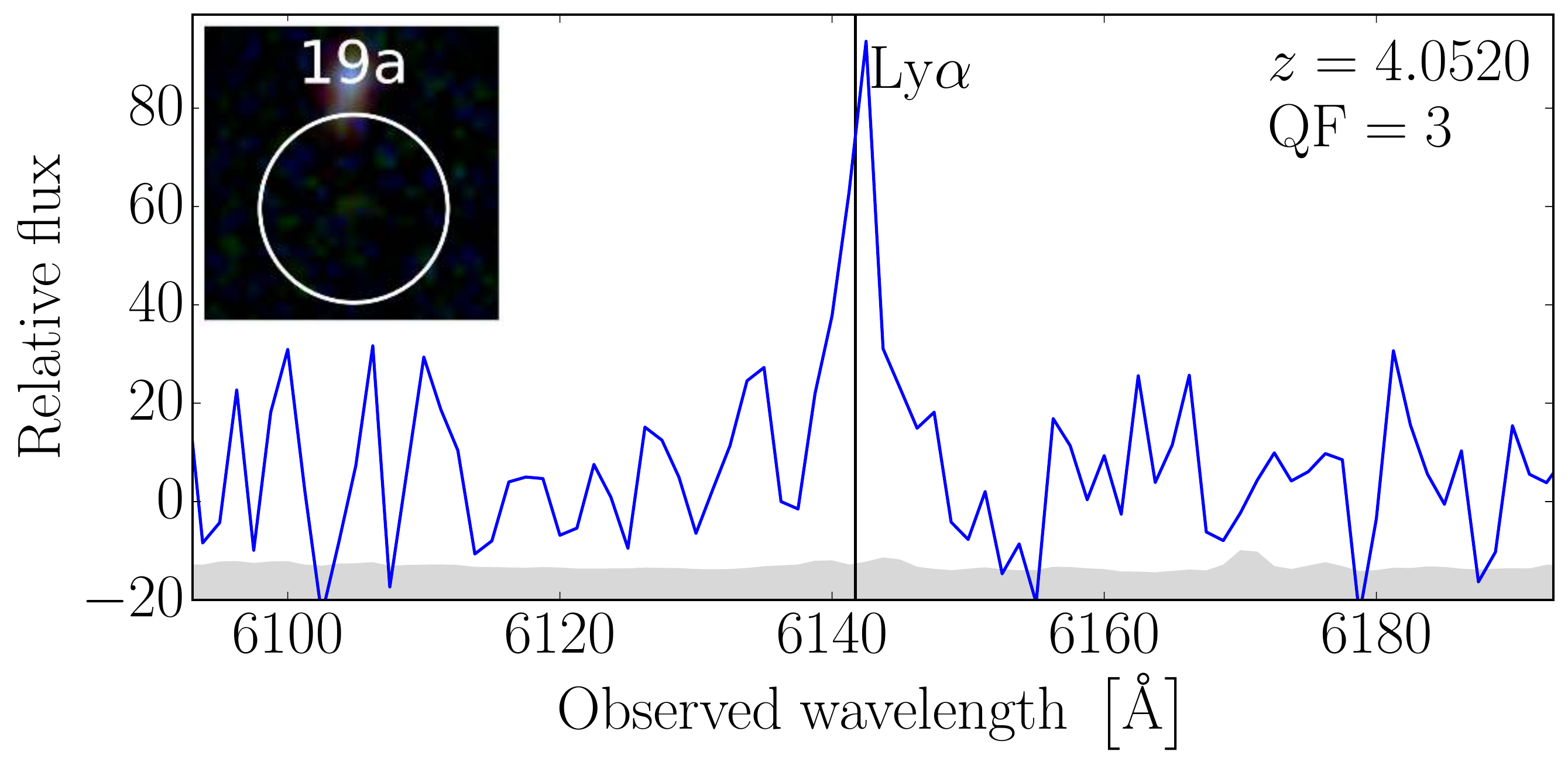}
   \includegraphics[width = 0.666\columnwidth]{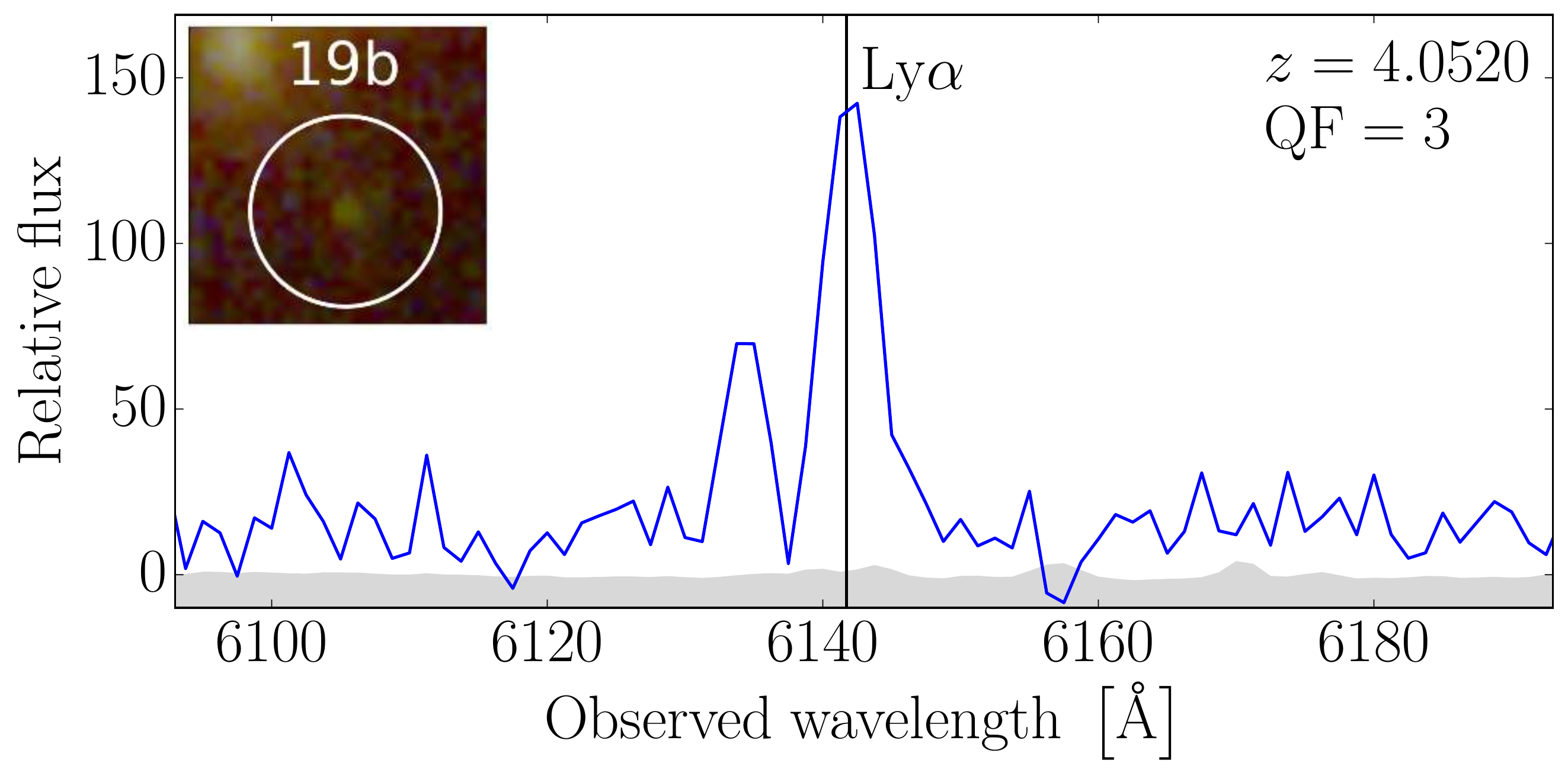}
   \includegraphics[width = 0.666\columnwidth]{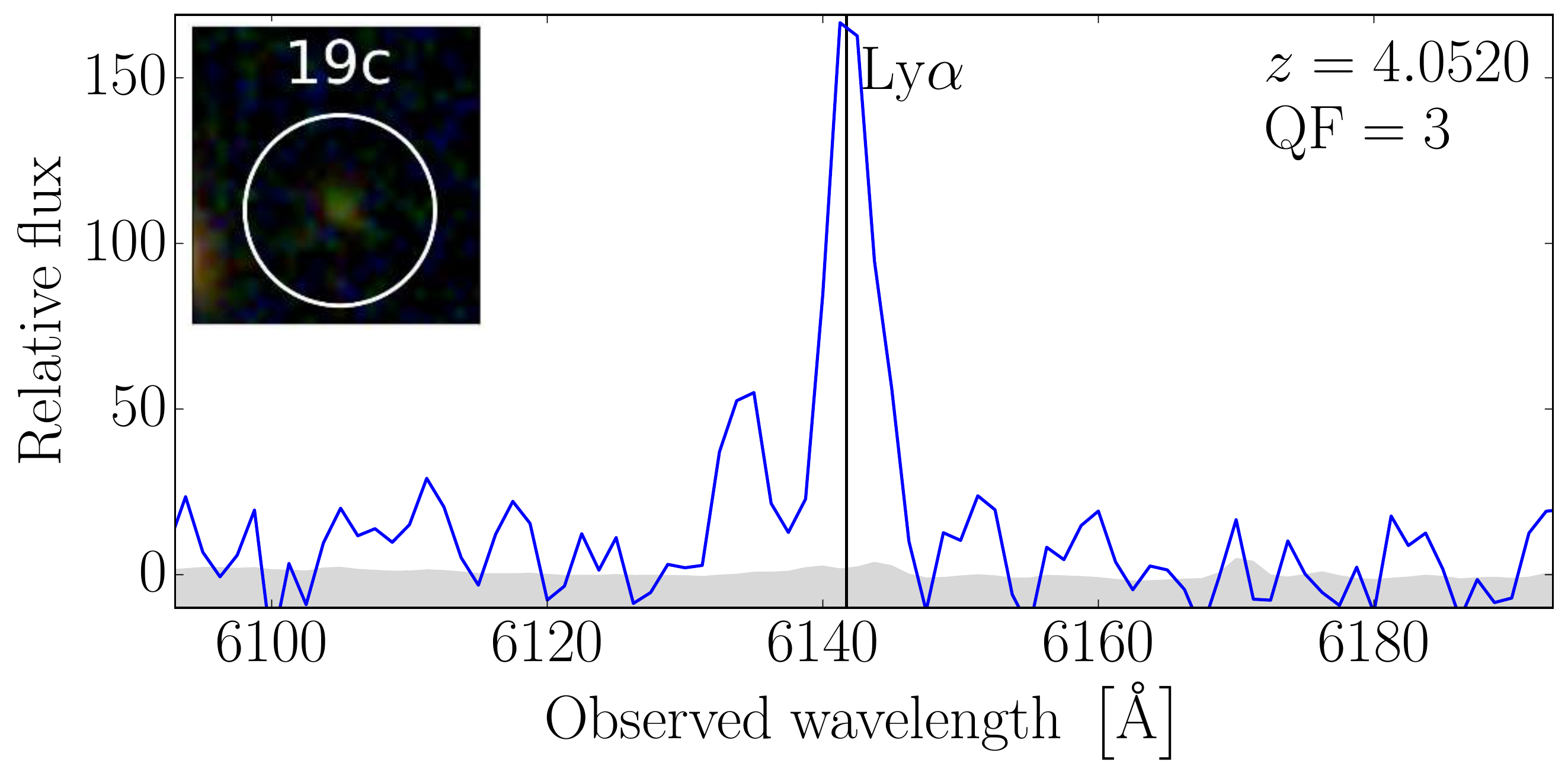}

Family 20:

   \includegraphics[width = 0.666\columnwidth]{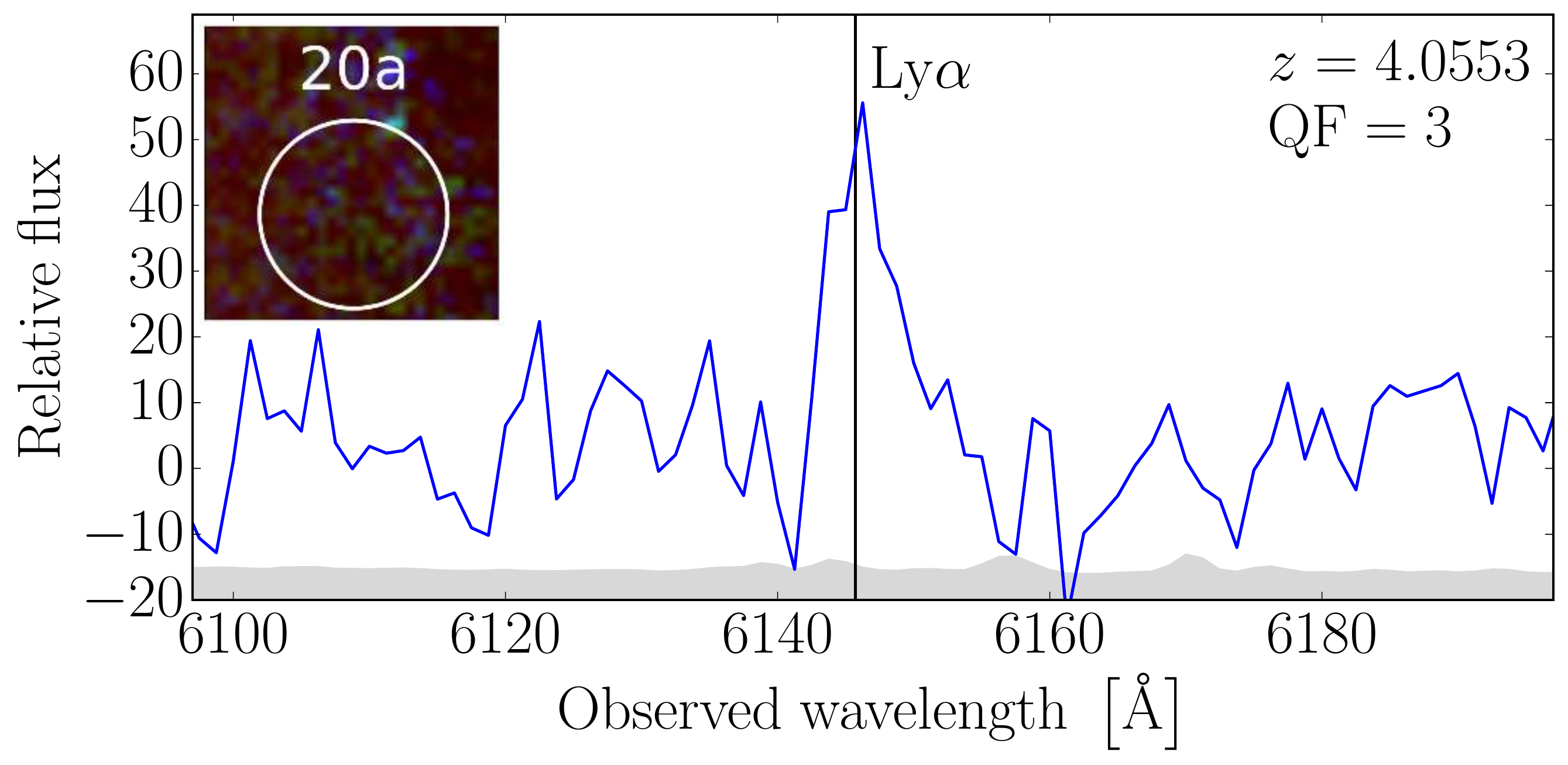}
   \includegraphics[width = 0.666\columnwidth]{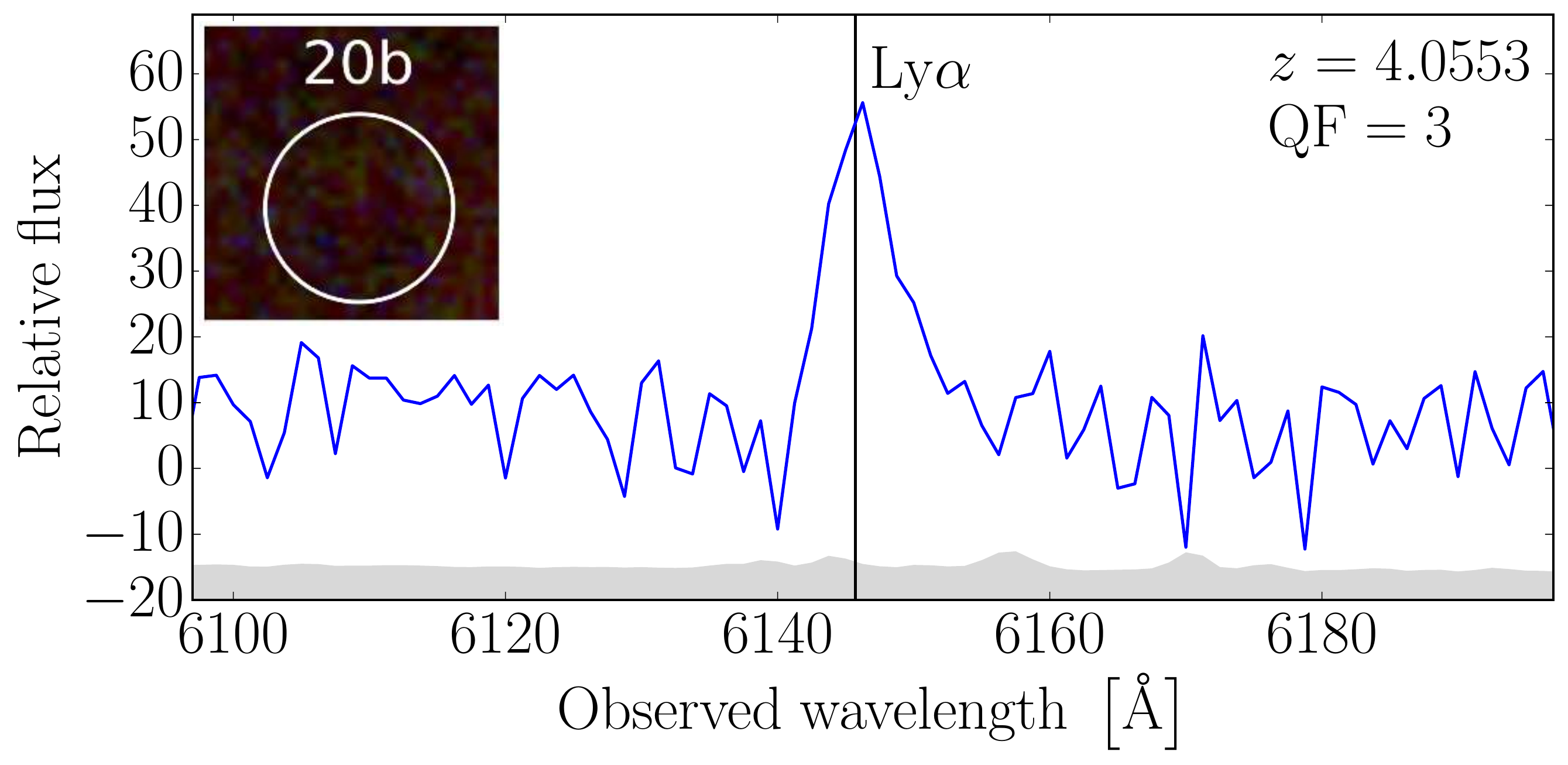}

Family 21:

   \includegraphics[width = 0.666\columnwidth]{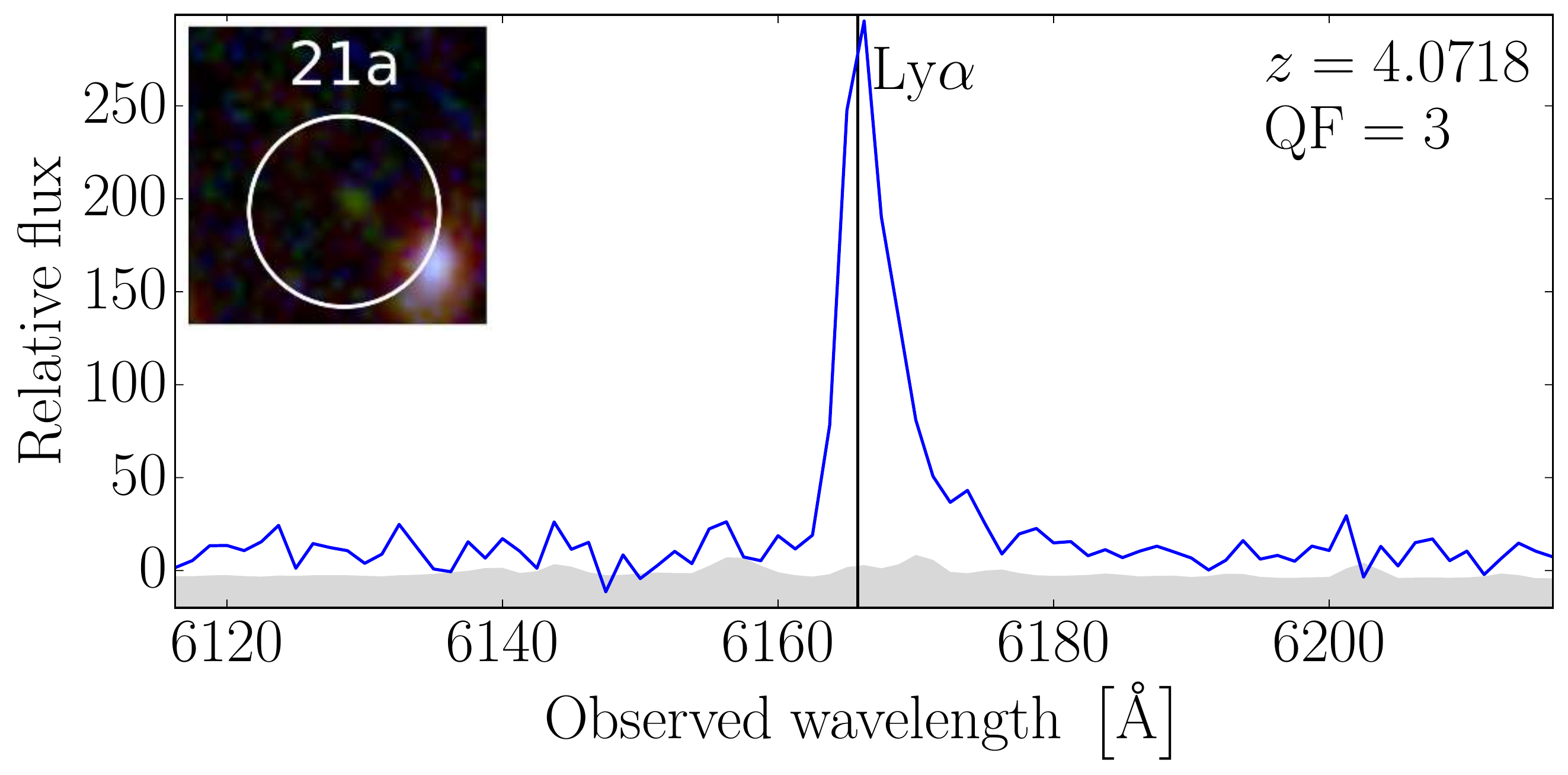}
   \includegraphics[width = 0.666\columnwidth]{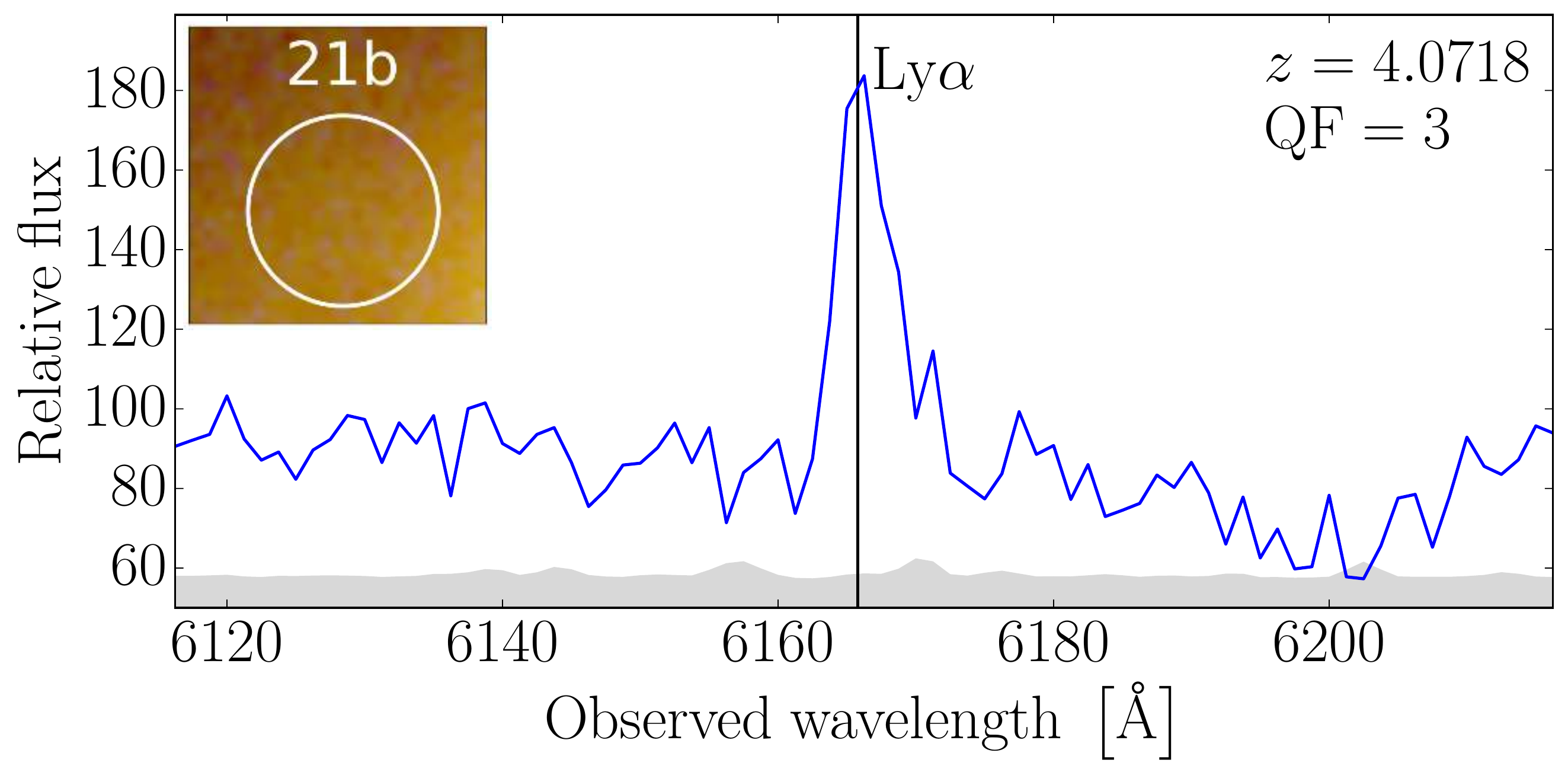}

   \includegraphics[width = 0.666\columnwidth]{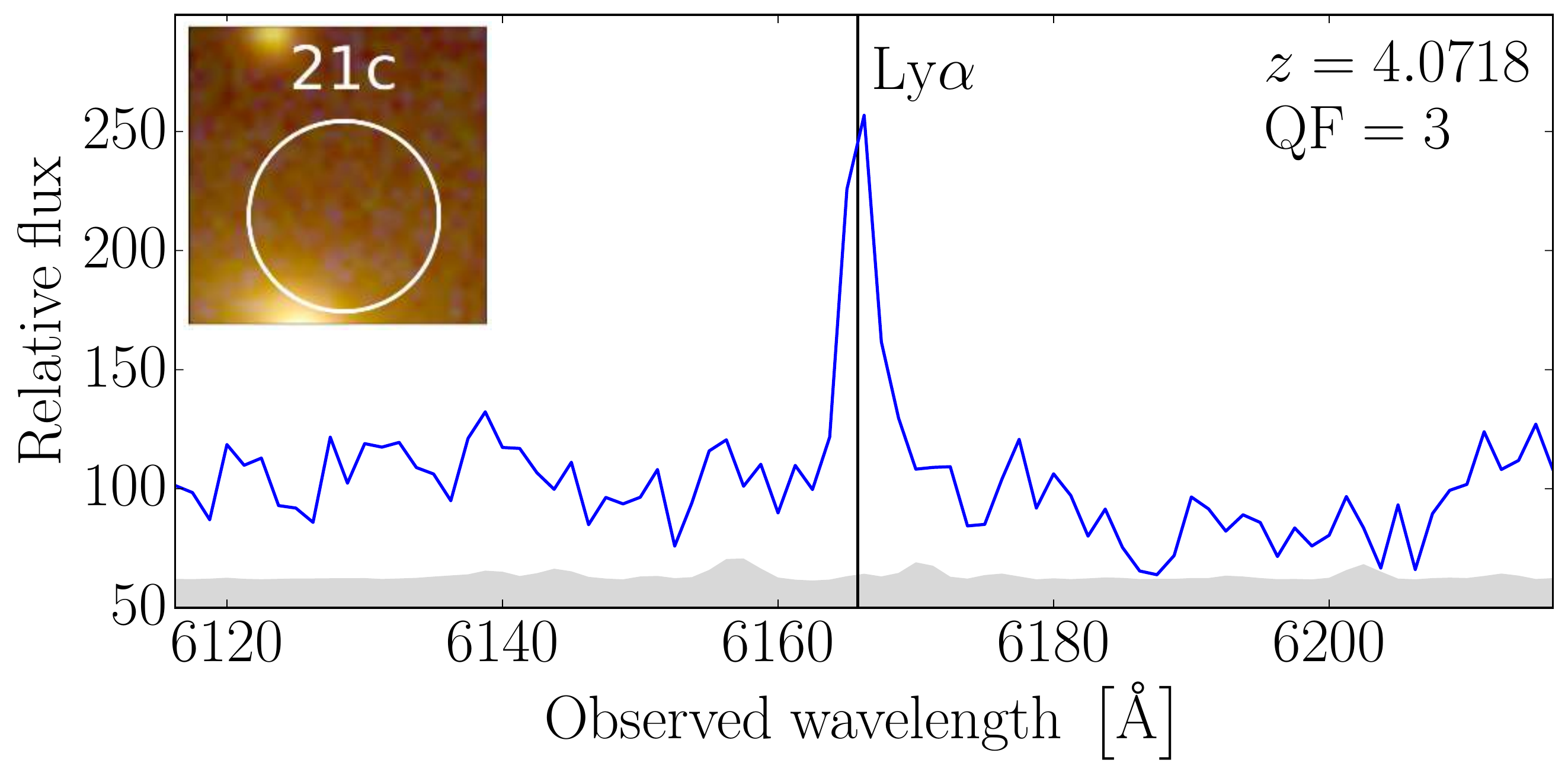}
   \includegraphics[width = 0.666\columnwidth]{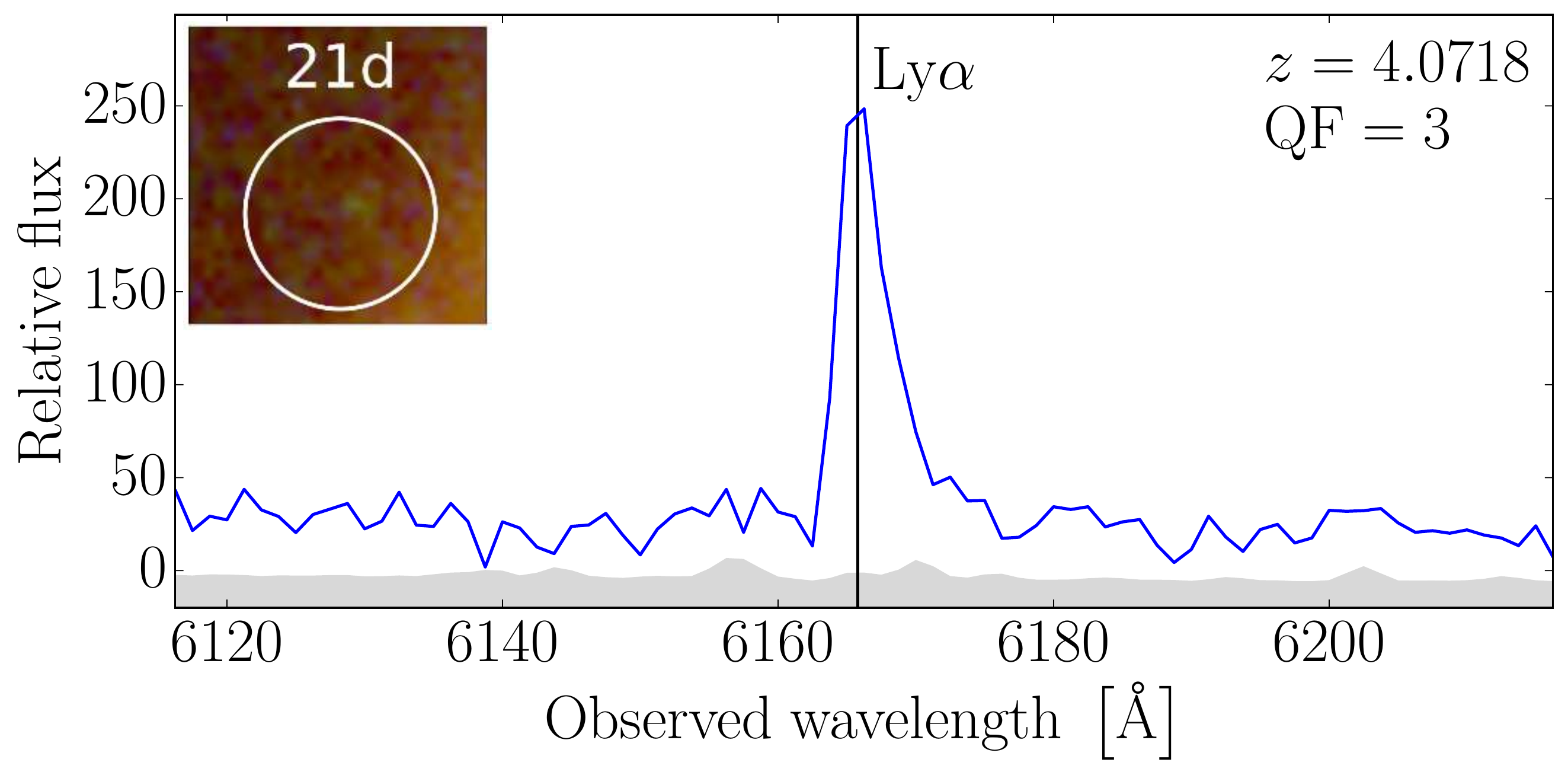}

Family 22:

   \includegraphics[width = 0.666\columnwidth]{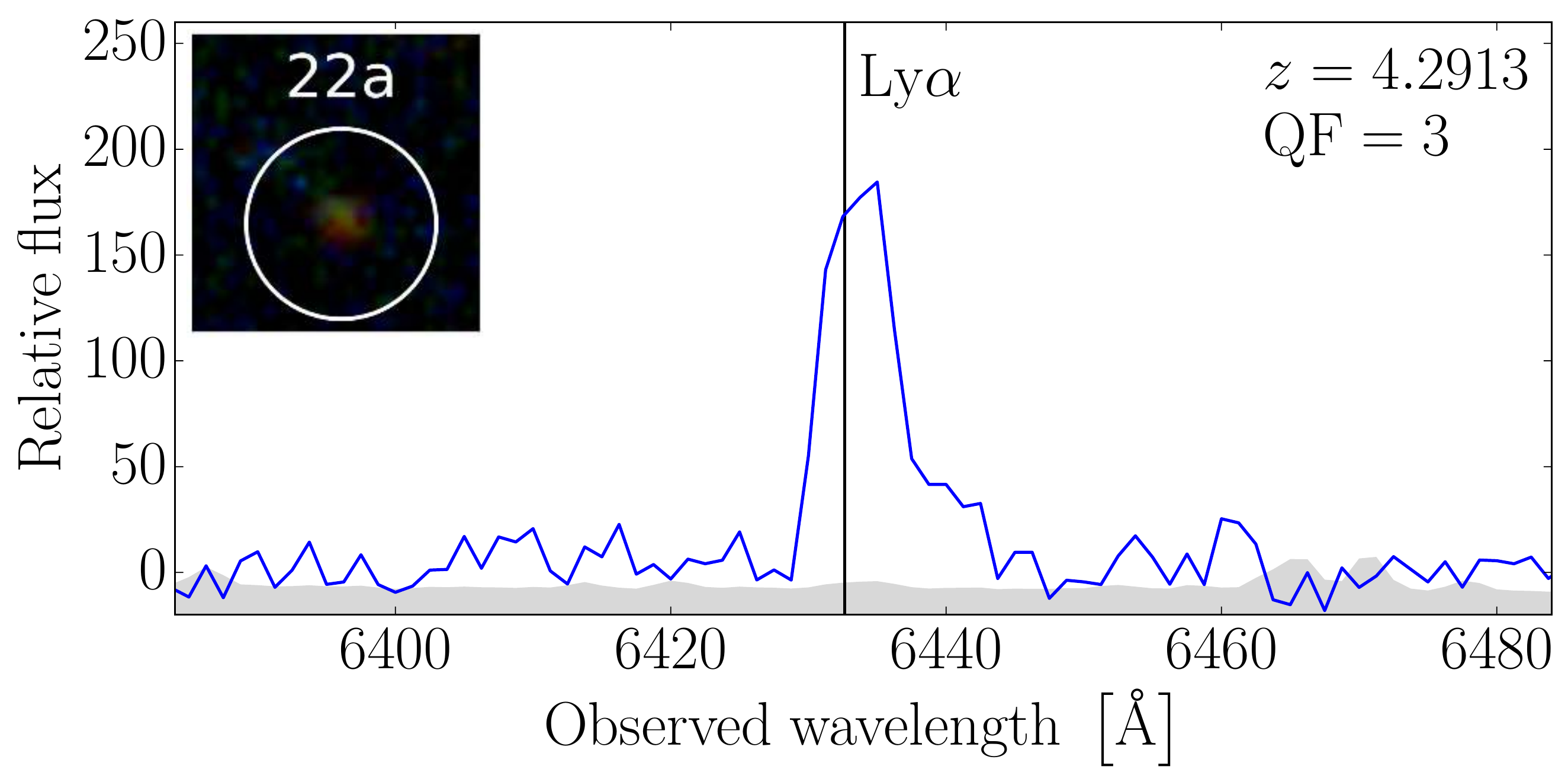}
   \includegraphics[width = 0.666\columnwidth]{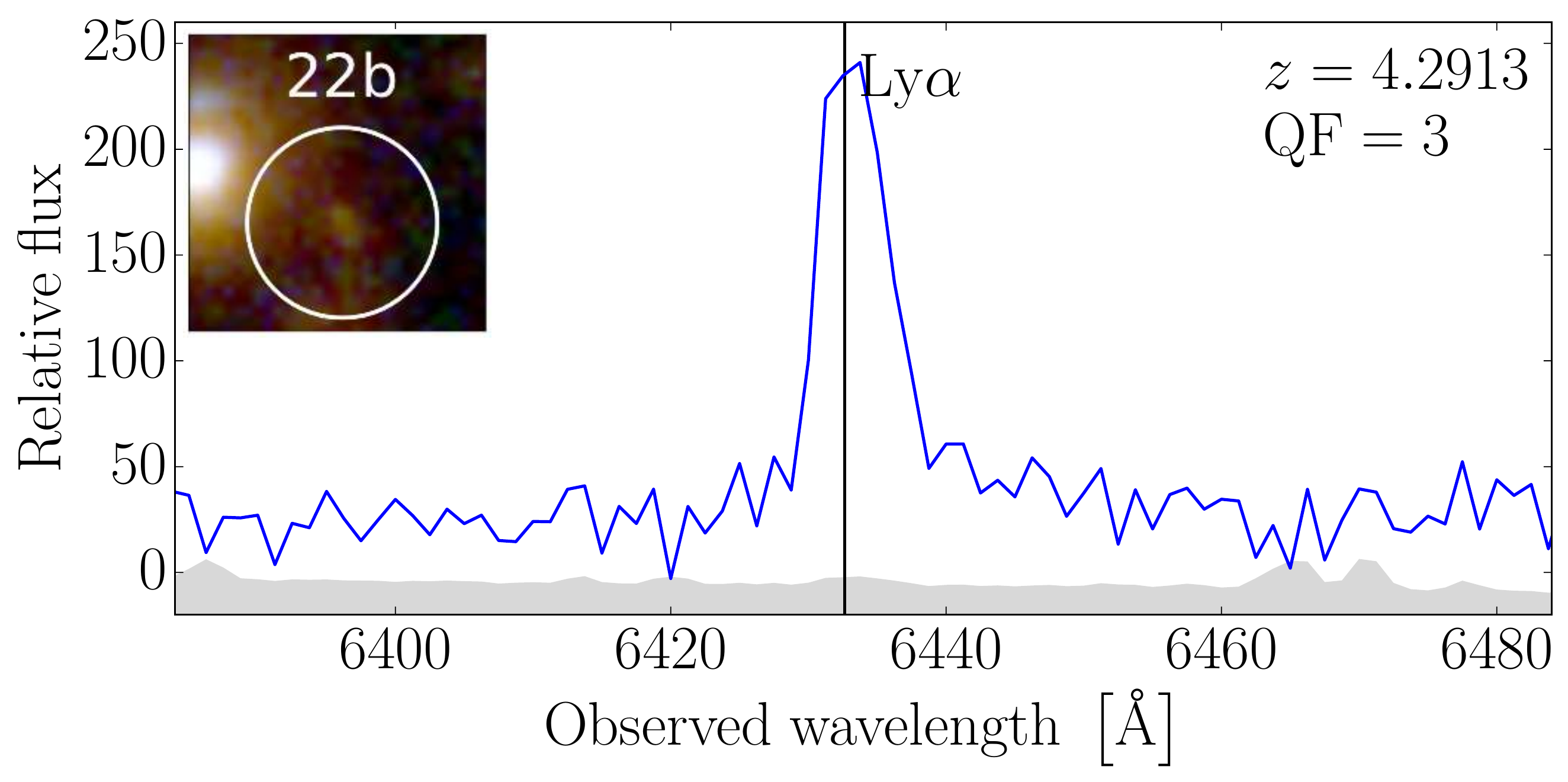}
   \includegraphics[width = 0.666\columnwidth]{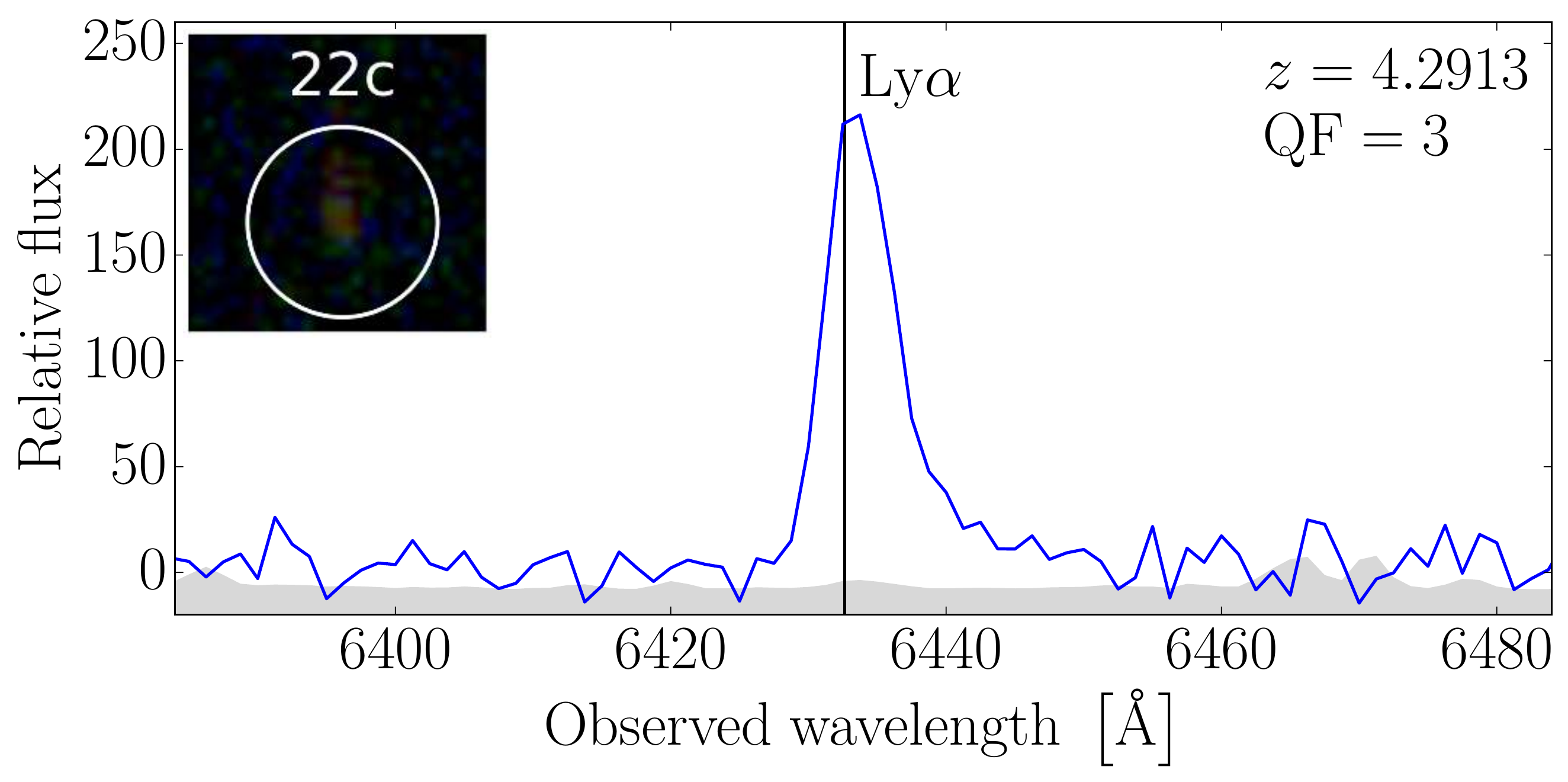}

Family 23:

   \includegraphics[width = 0.666\columnwidth]{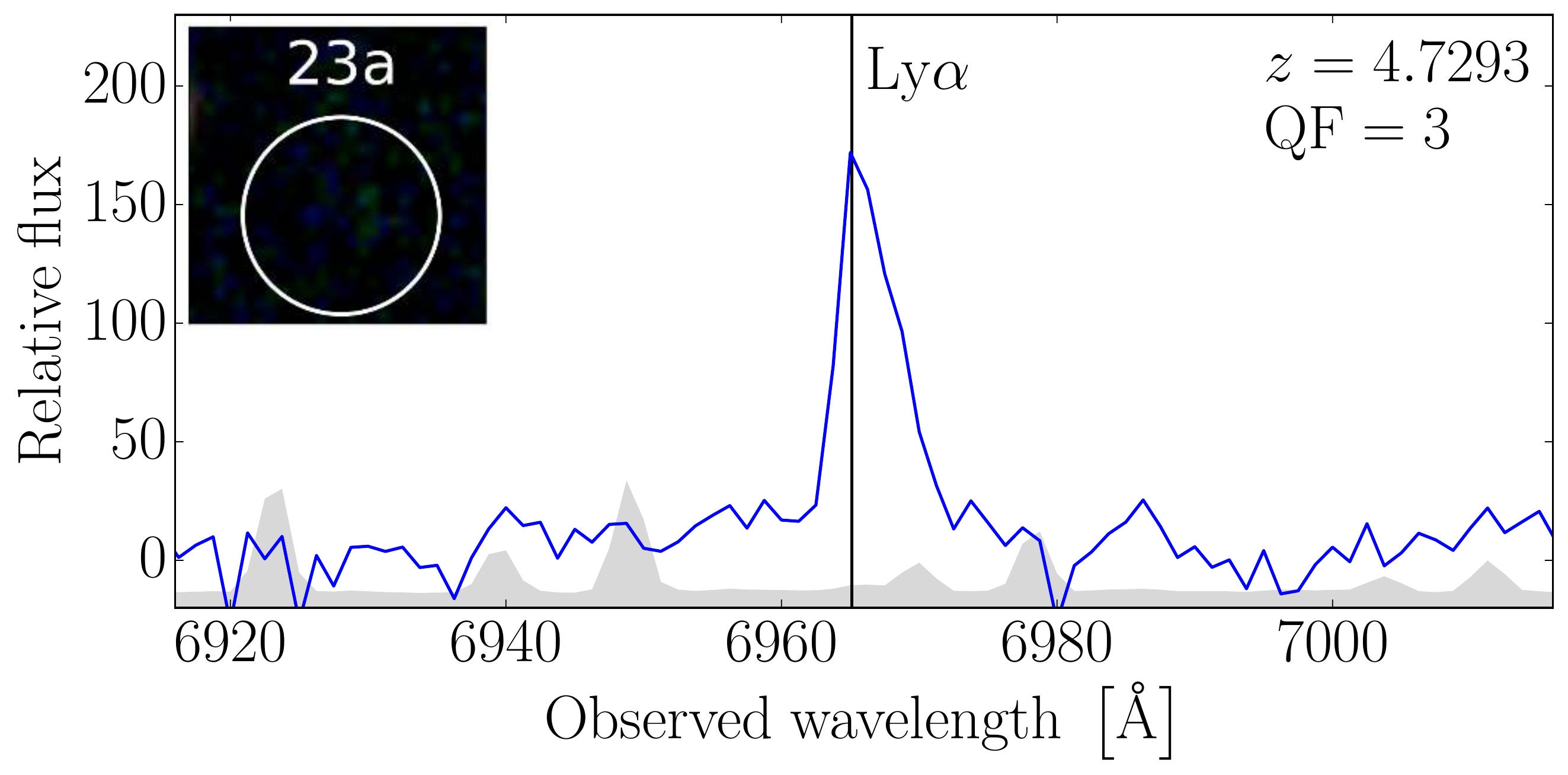}
   \includegraphics[width = 0.666\columnwidth]{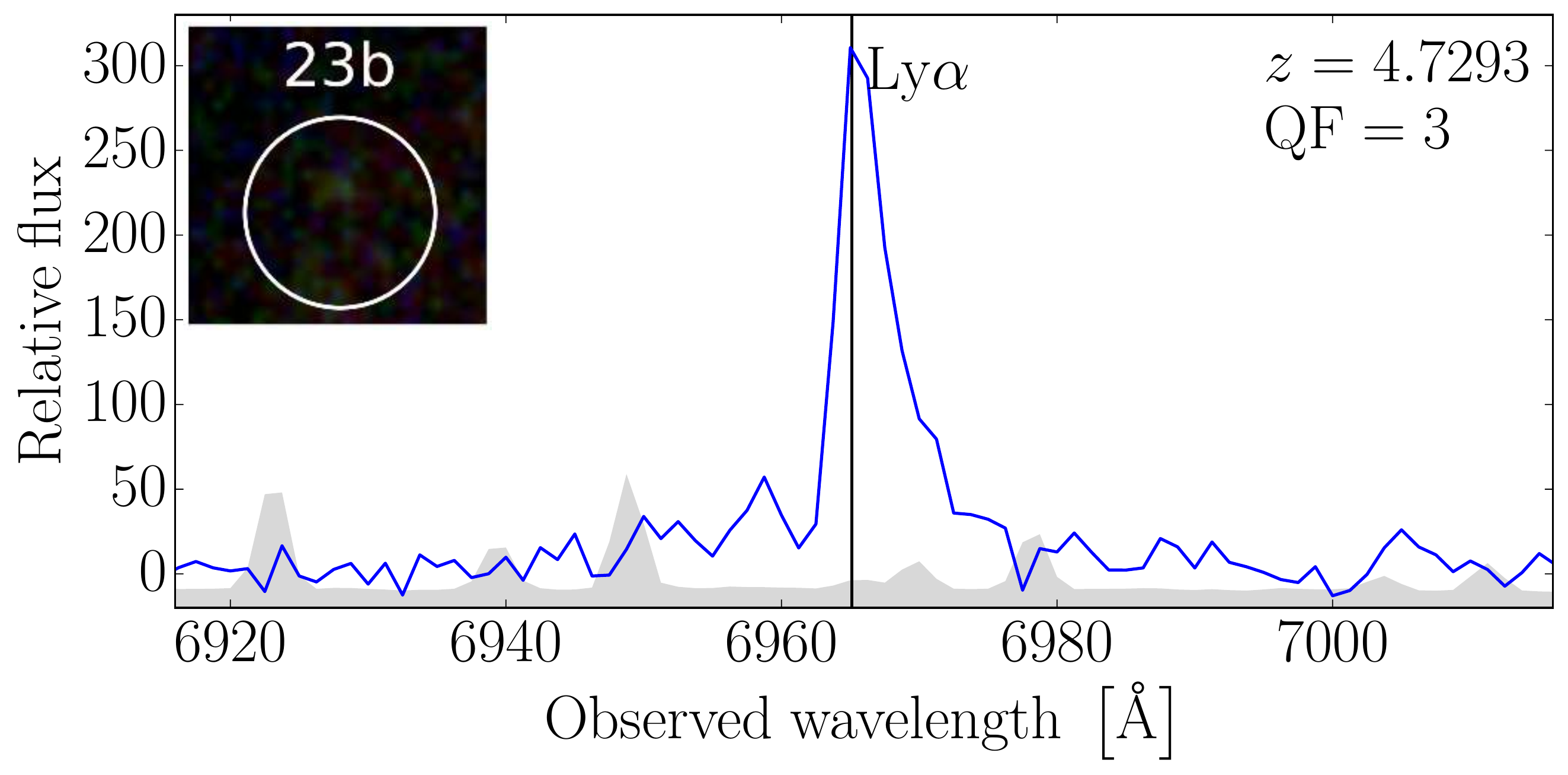}
   \includegraphics[width = 0.666\columnwidth]{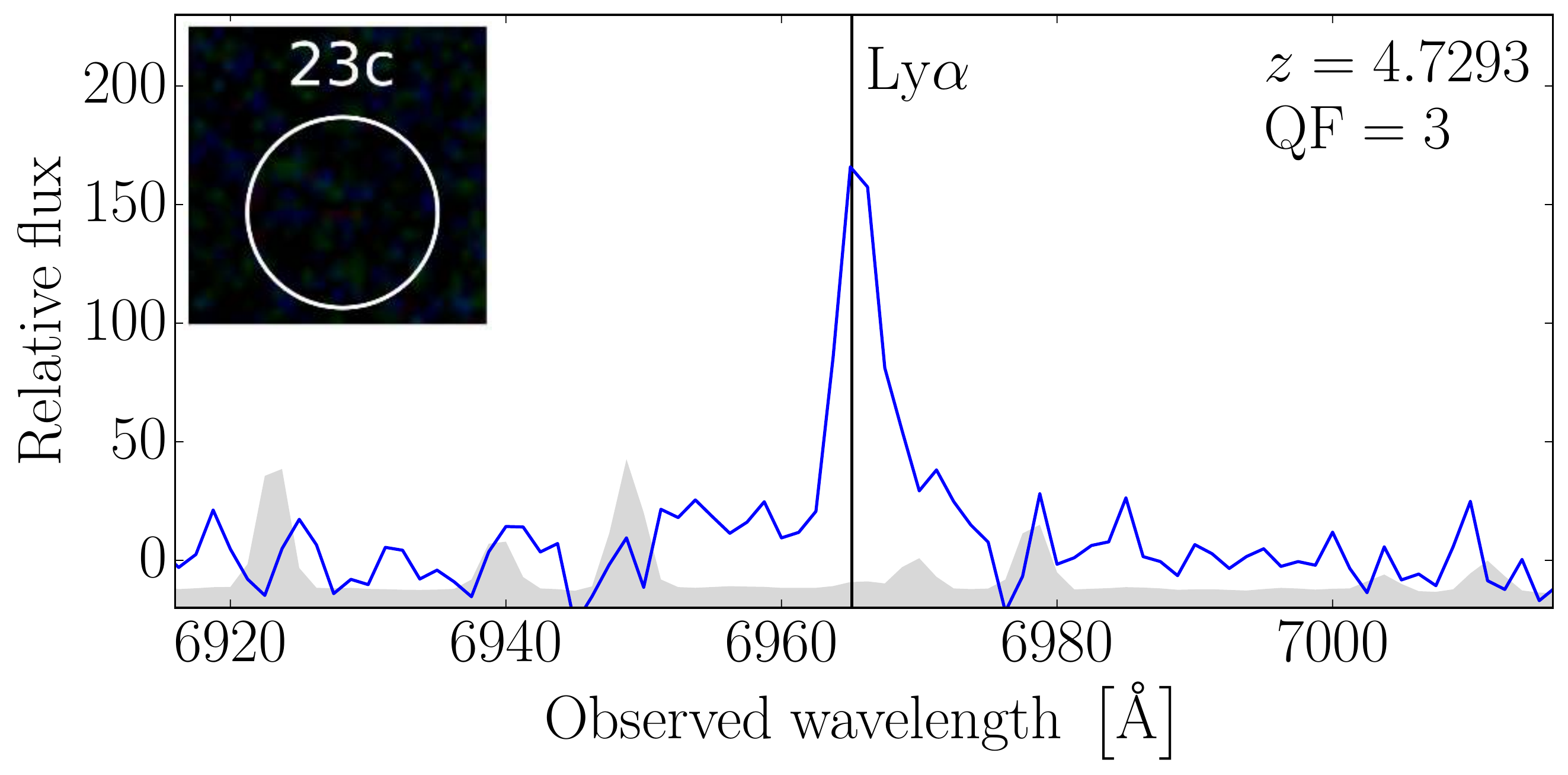}

  \caption{(Continued)}
  \label{fig:specs}
\end{figure*}

\begin{figure*}
\setcounter{figure}{\value{figure}-1}

Family 24:

   \includegraphics[width = 0.666\columnwidth]{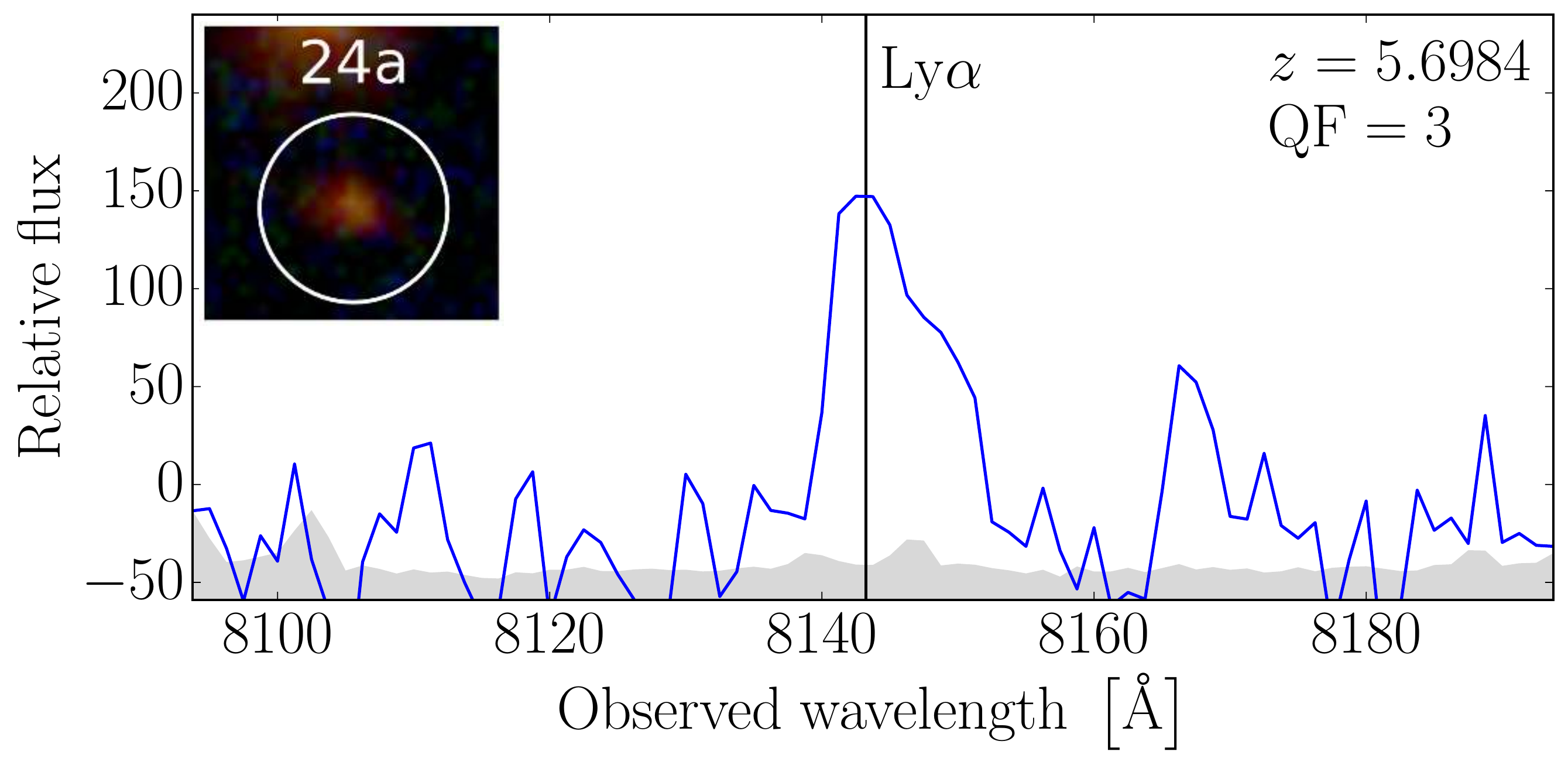}
   \includegraphics[width = 0.666\columnwidth]{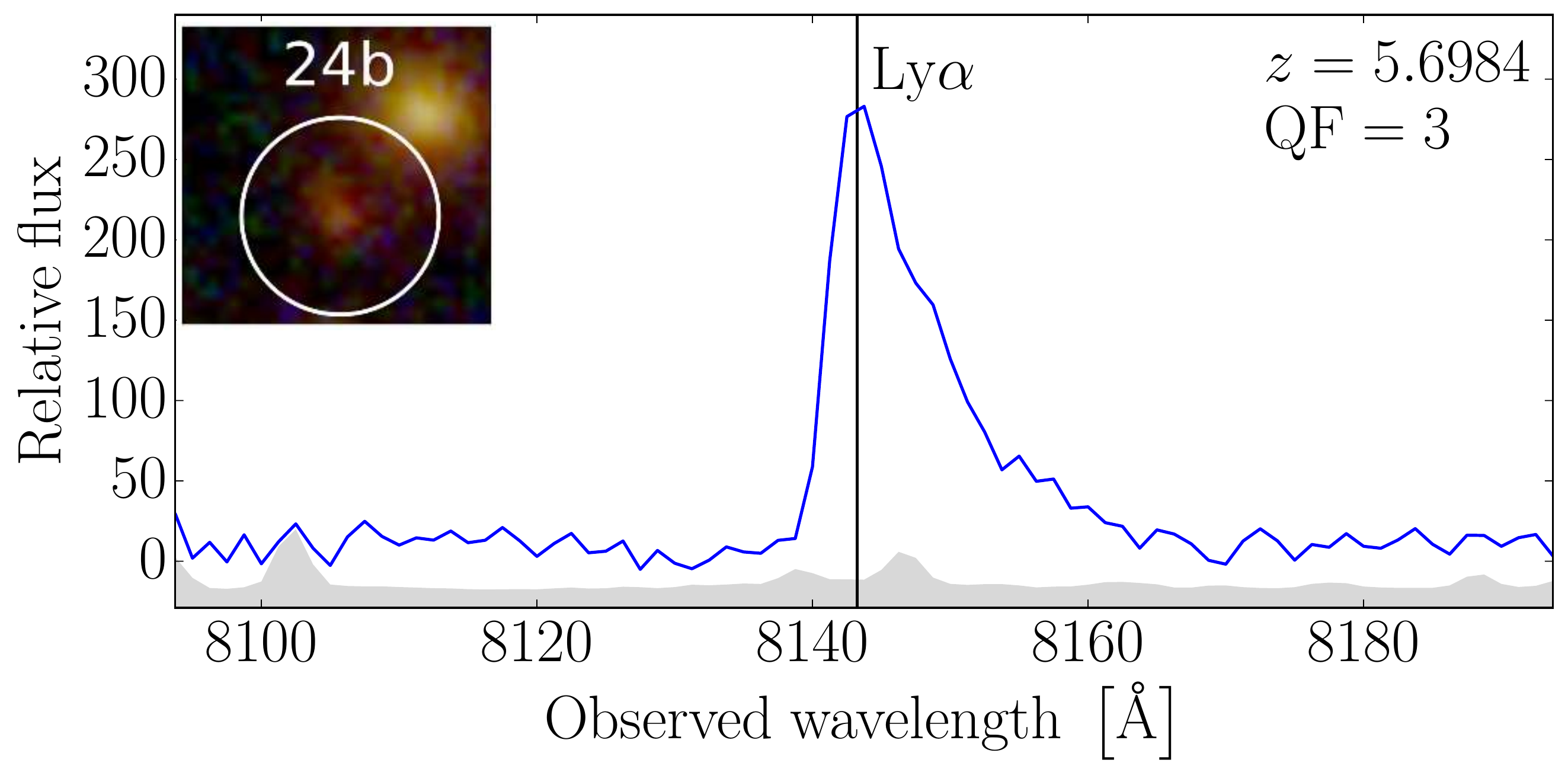}

   \includegraphics[width = 0.666\columnwidth]{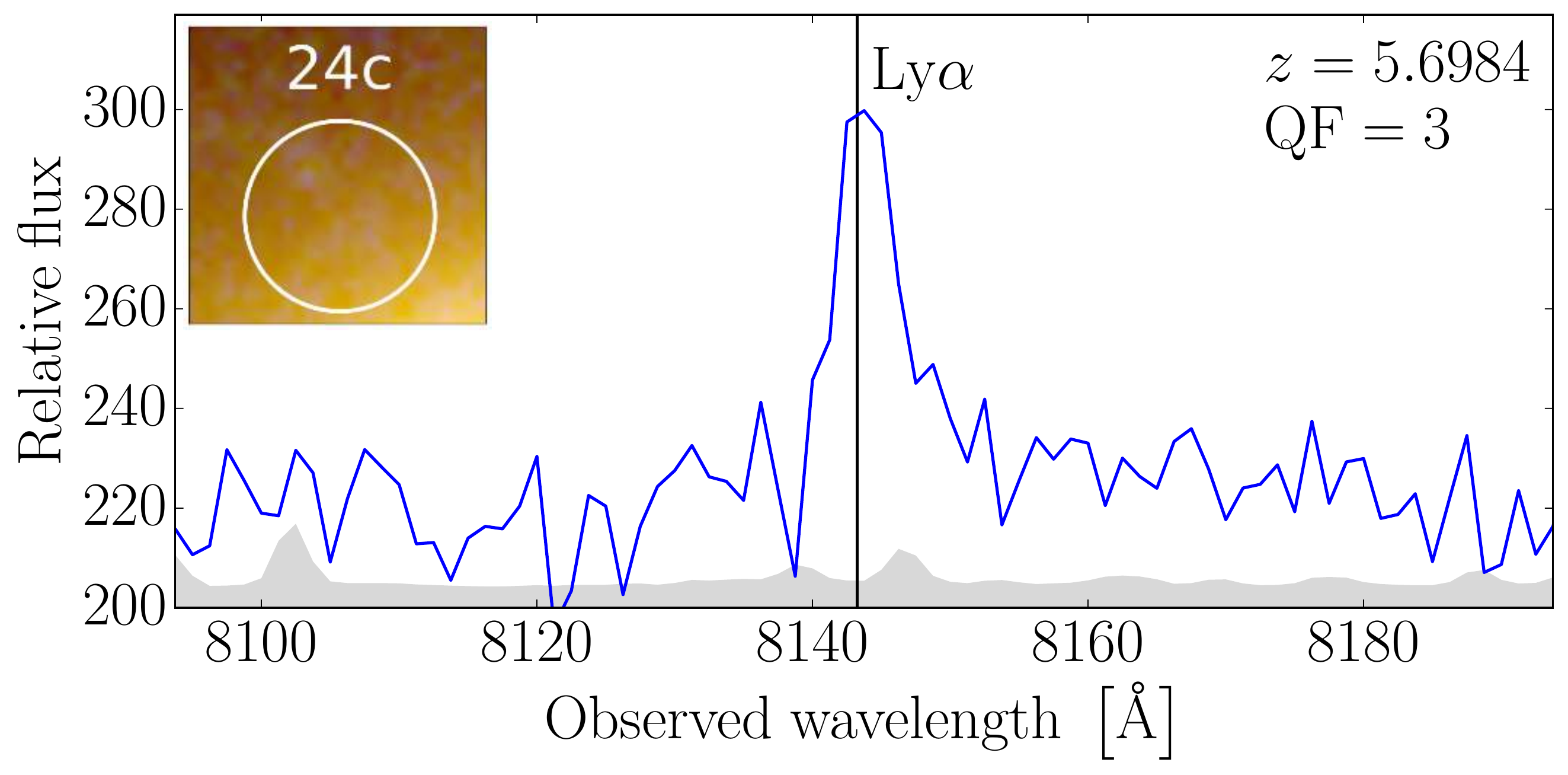}
   \includegraphics[width = 0.666\columnwidth]{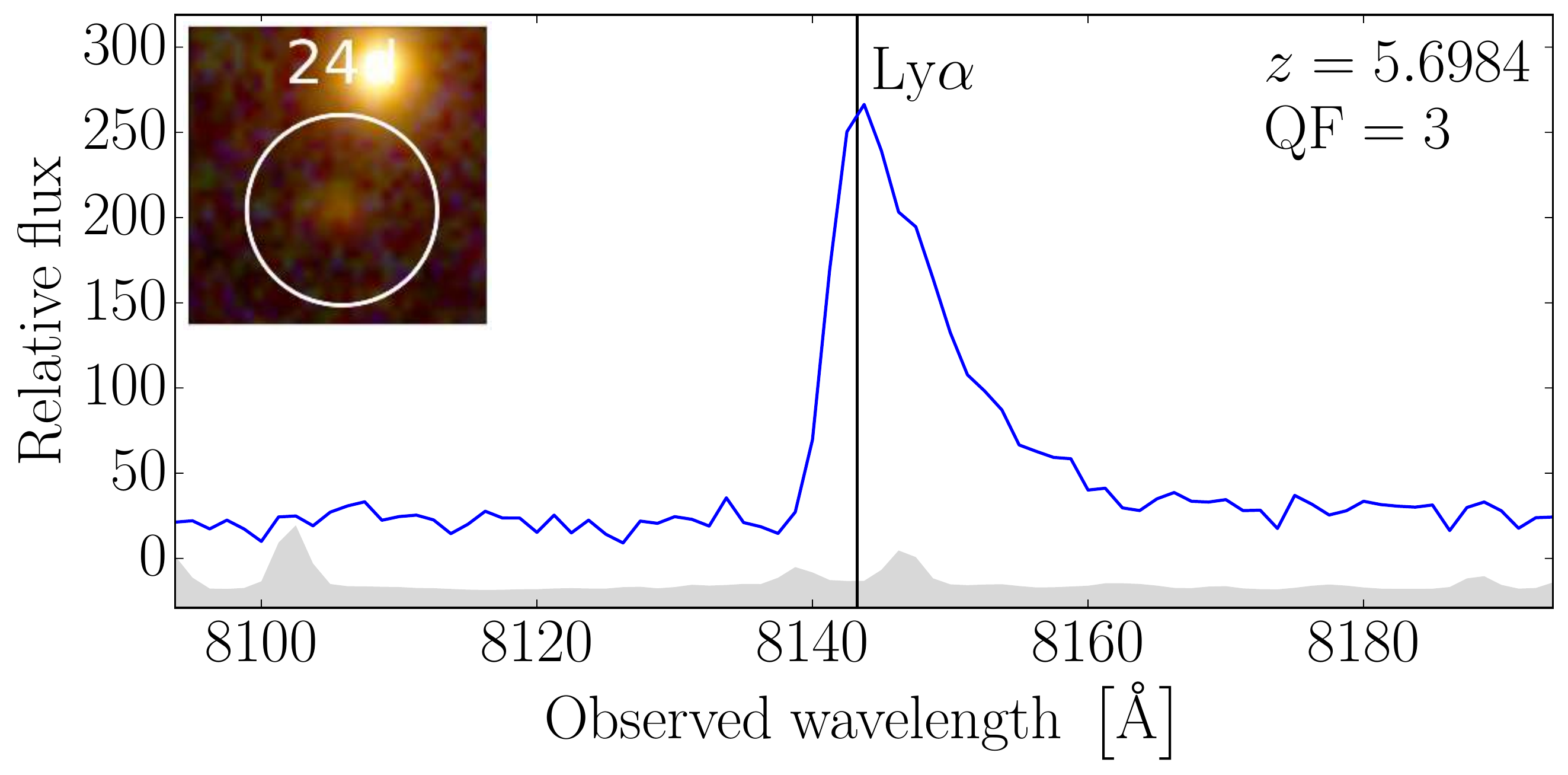}

Family 25:

   \includegraphics[width = 0.666\columnwidth]{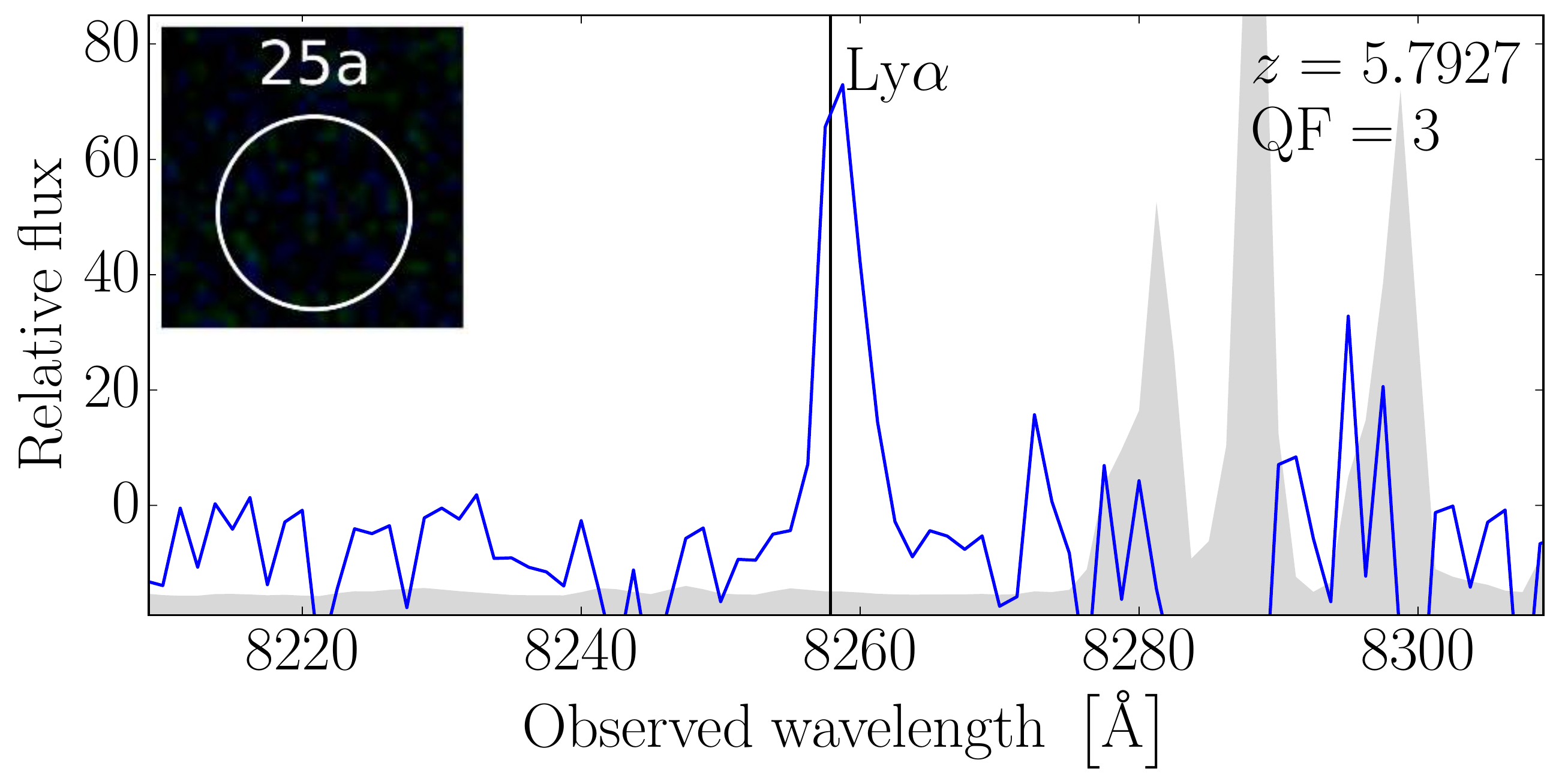}
   \includegraphics[width = 0.666\columnwidth]{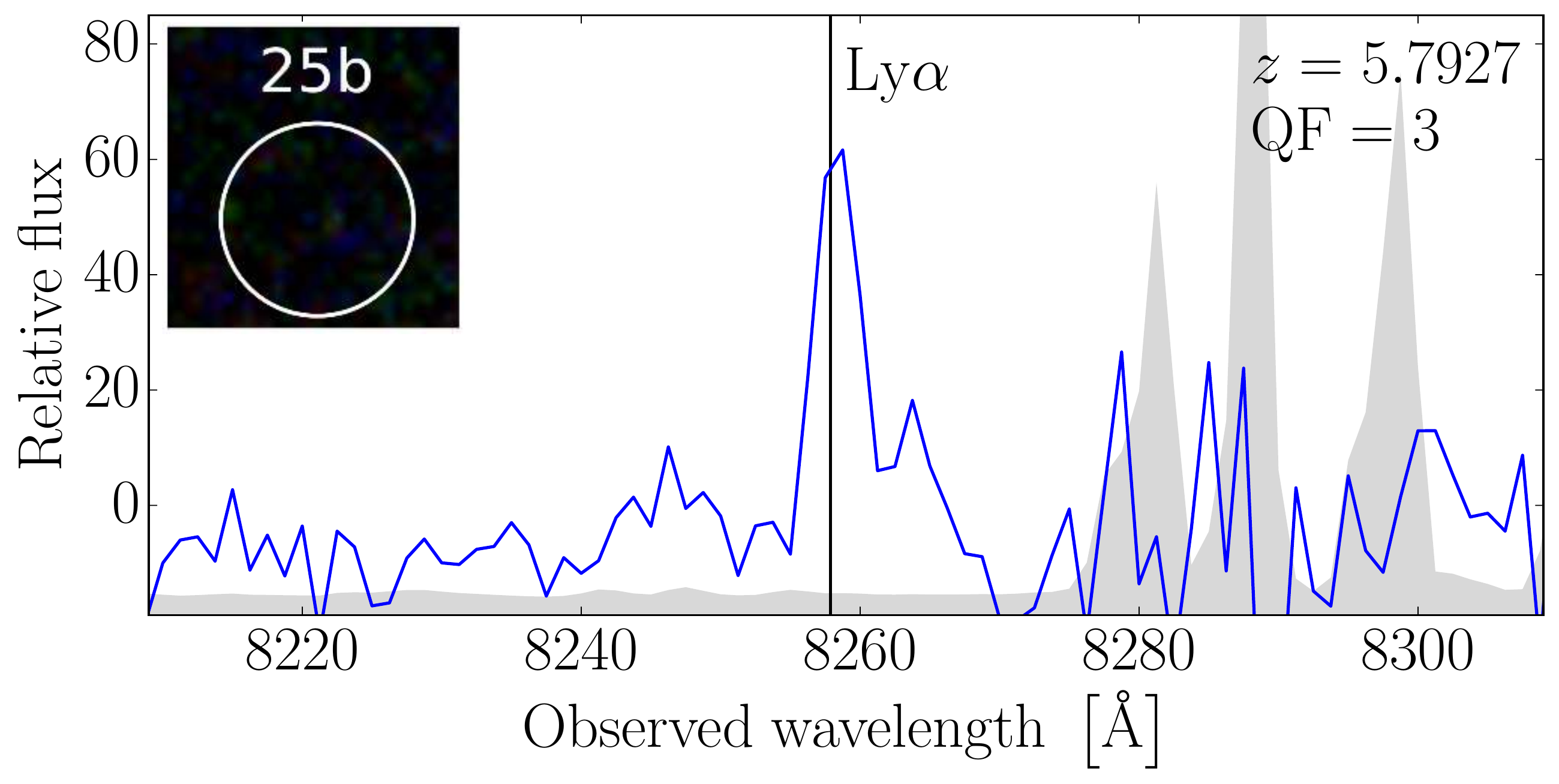}

Family 26:

   \includegraphics[width = 0.666\columnwidth]{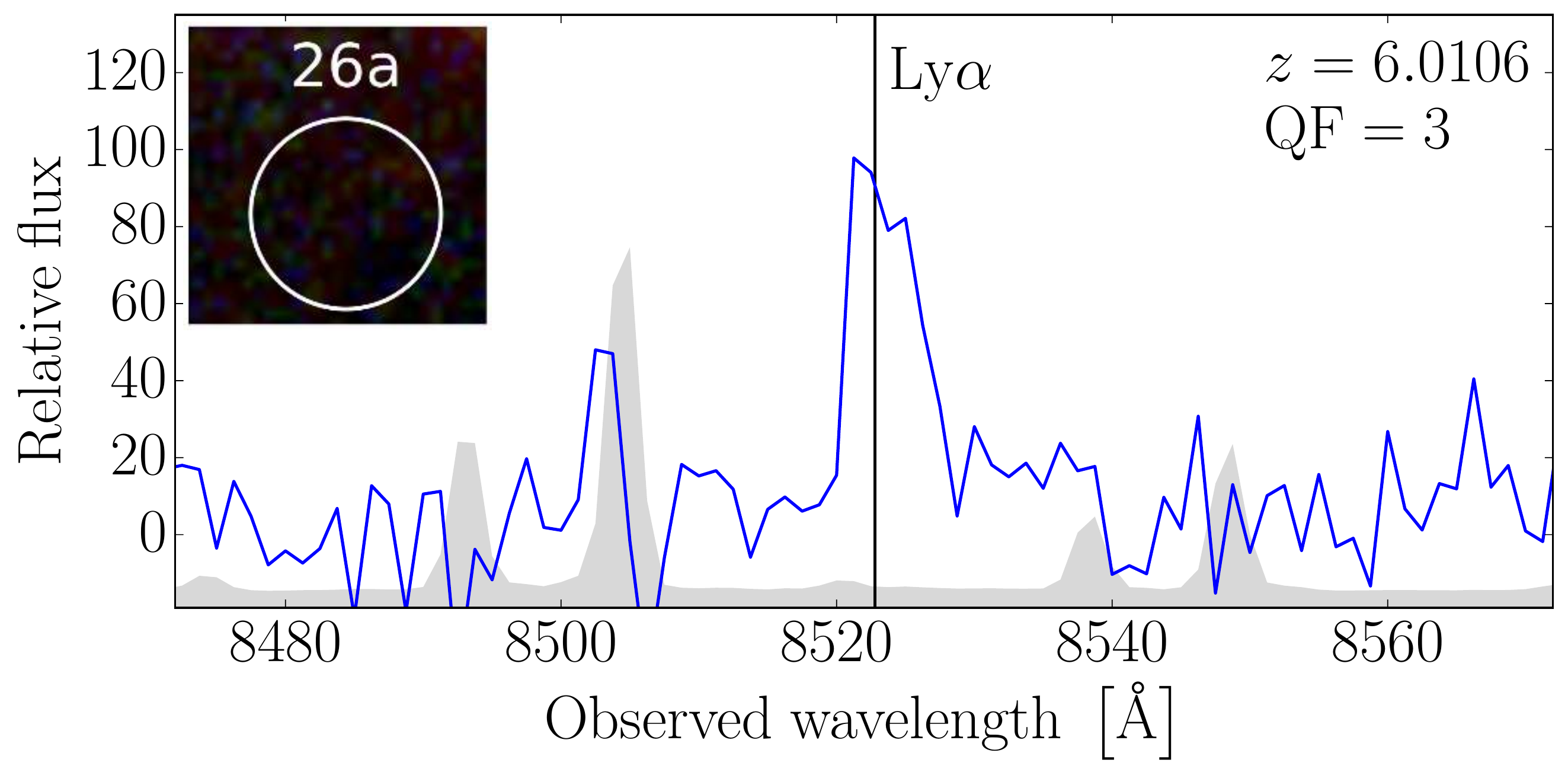}
   \includegraphics[width = 0.666\columnwidth]{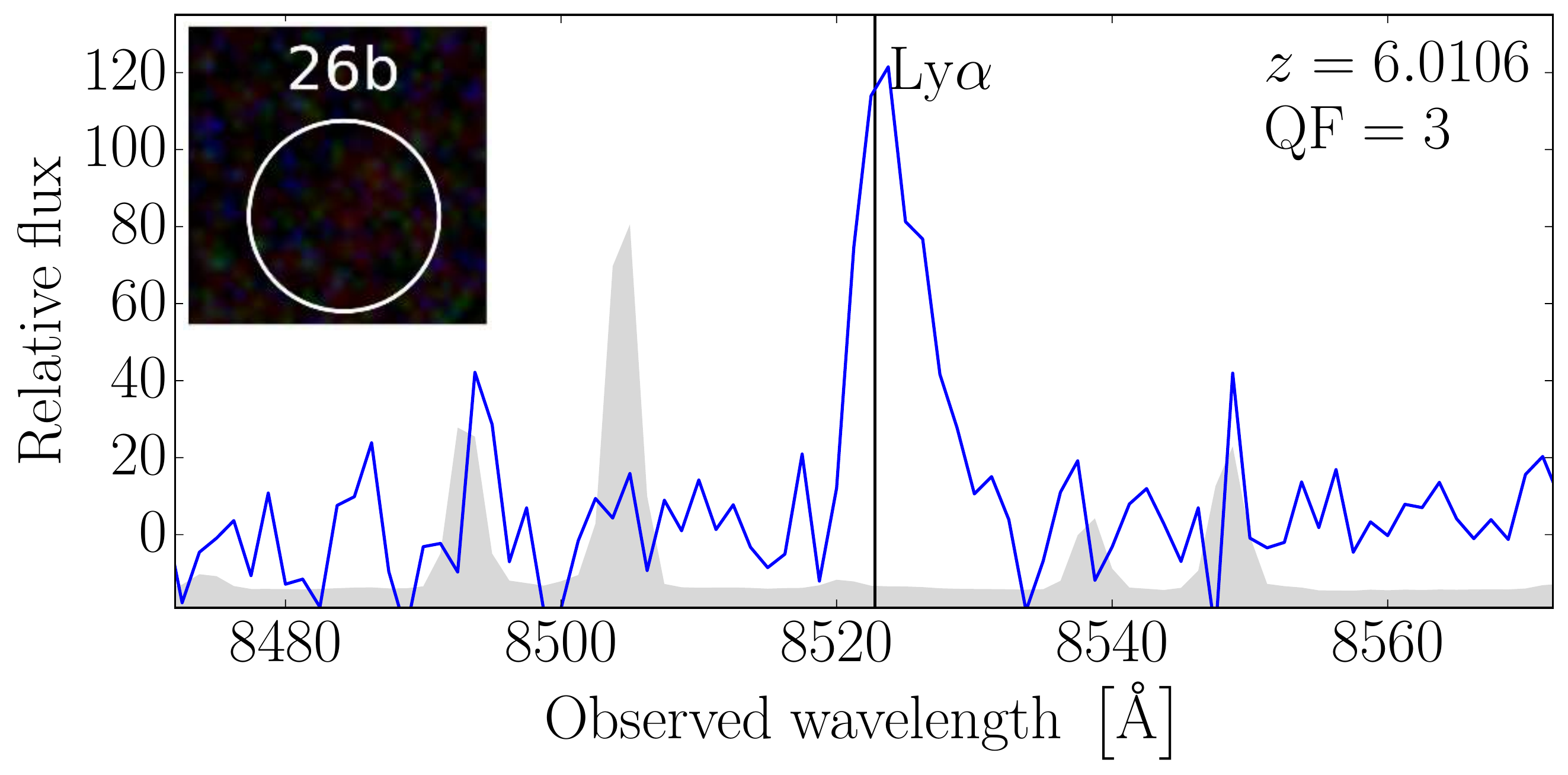}

Family 27:

   \includegraphics[width = 0.666\columnwidth]{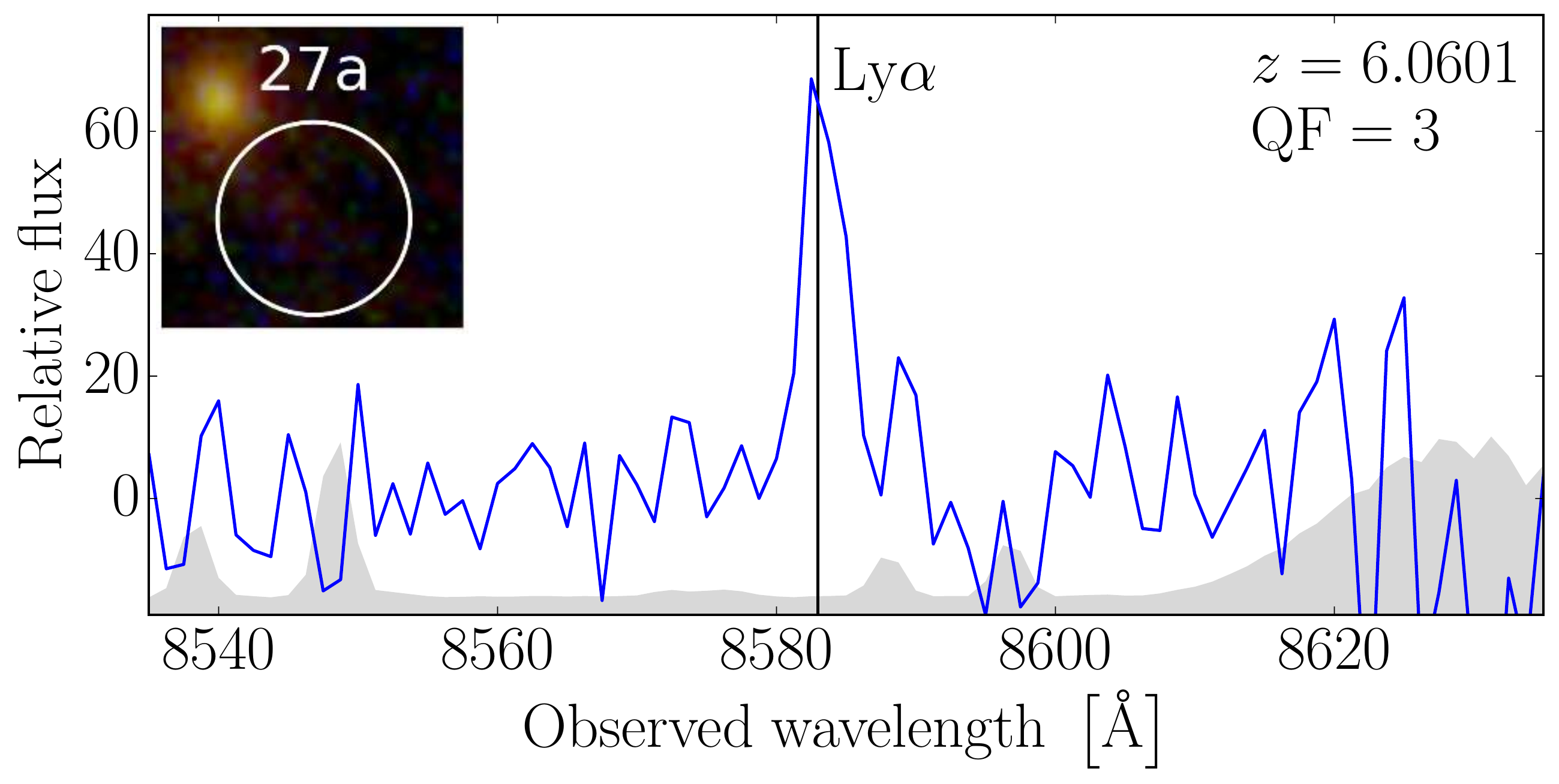}
   \includegraphics[width = 0.666\columnwidth]{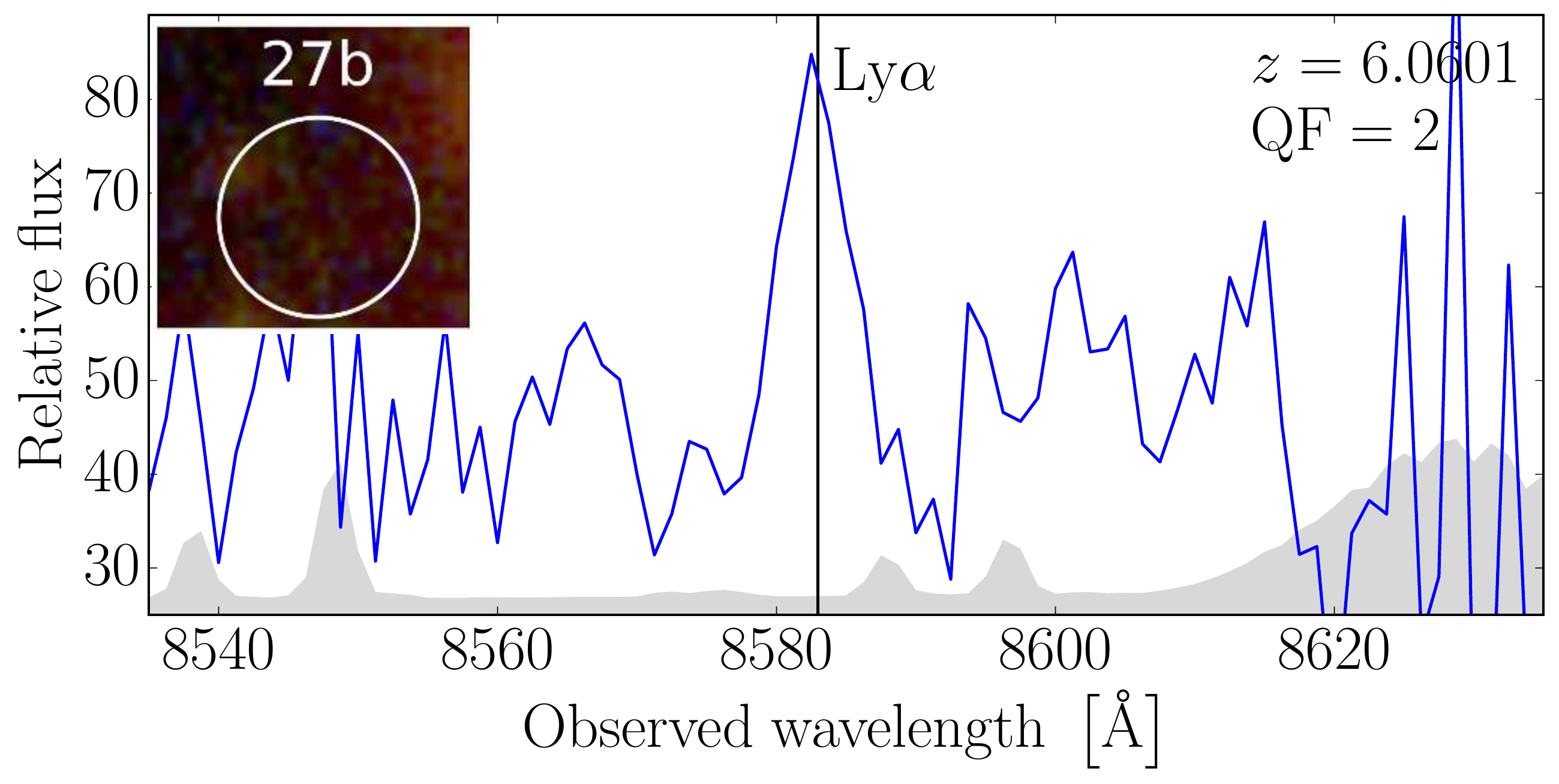}

  \caption{(Continued)}
  \label{fig:specs}
\end{figure*}

\renewcommand{\arraystretch}{1.3}
\begin{table}[h]
\setcounter{table}{1}
\centering
    \caption{Median parameter values and confidence levels for the reference lens model. Angles are measured from west to north. The values of all velocity dispersions ($\sigma_v$) are corrected by the factor $\sqrt{2/3}$ as described in the {\tt lenstool} manual (see \href{http://projets.lam.fr/projects/lenstool/wiki/PIEMD}{http://projets.lam.fr/projects/lenstool/wiki/PIEMD}).}
    \begin{tabular}{c c c c c} \hline \hline
    ~ & Median & 68\% CL & 95\% CL & 99.7\% CL \\ \hline
\hline
\multicolumn{5}{l}{Main diffuse mass component} \\ \hline
$x_1 \; (\arcsec)$ & $-0.5$ & $_{-0.6}^{+0.5}$ & $_{-1.2}^{+1.0}$ & $_{-1.8}^{+1.4}$ \\
$y_1 \; (\arcsec)$ & $0.4$ & $_{-0.3}^{+0.3}$ & $_{-0.6}^{+0.5}$ & $_{-1.0}^{+0.8}$ \\
$\varepsilon_1$ & $0.68$ & $_{-0.04}^{+0.04}$ & $_{-0.08}^{+0.09}$ & $_{-0.12}^{+0.14}$ \\
$\theta_1$ (deg) & $19.1$ & $_{-1.1}^{+1.2}$ & $_{-2.3}^{+2.6}$ & $_{-3.5}^{+4.8}$ \\
$r_{\rm core,1} \; (\arcsec)$ & $6.0$ & $_{-0.8}^{+0.7}$ & $_{-1.7}^{+1.3}$ & $_{-2.5}^{+1.9}$ \\
$\sigma_{v1} \; \rm (km\,s^{-1})$ & $951$ & $_{-79}^{+63}$ & $_{-161}^{+115.}$ & $_{-243}^{+159}$ \\
\hline
\multicolumn{5}{l}{Second diffuse mass component} \\ \hline
$x_2 \; (\arcsec)$ & $8.6$ & $_{-1.1}^{+1.2}$ & $_{-2.2}^{+2.8}$ & $_{-3.0}^{+5.1}$ \\
$y_2 \; (\arcsec)$ & $4.4$ & $_{-0.9}^{+1.0}$ & $_{-1.9}^{+2.3}$ & $_{-3.1}^{+3.7}$ \\
$\varepsilon_2$ & $0.44$ & $_{-0.15}^{+0.14}$ & $_{-0.28}^{+0.27}$ & $_{-0.37}^{+0.37}$ \\
$\theta_2$ (deg) & $113.5$ & $_{-4.0}^{+5.0}$ & $_{-7.3}^{+12.4}$ & $_{-10.4}^{+23.0}$ \\
$r_{\rm core,2} \; (\arcsec)$ & $15.7$ & $_{-1.8}^{+2.4}$ & $_{-3.0}^{+5.9}$ & $_{-4.0}^{+10.0}$ \\
$\sigma_{v2} \; \rm (km\,s^{-1})$ & $863$ & $_{-55}^{+59}$ & $_{-110}^{+121}$ & $_{-168}^{+183}$ \\
\hline
\multicolumn{5}{l}{Third diffuse mass component} \\ \hline
$x_3 \; (\arcsec)$ & $-26.3$ & $_{-1.6}^{+1.6}$ & $_{-3.4}^{+3.6}$ & $_{-6.2}^{+6.0}$ \\
$y_3 \; (\arcsec)$ & $-8.0$ & $_{-0.9}^{+0.9}$ & $_{-1.8}^{+1.9}$ & $_{-2.7}^{+3.1}$ \\
$\varepsilon_3$ & $0.27$ & $_{-0.09}^{+0.09}$ & $_{-0.18}^{+0.18}$ & $_{-0.25}^{+0.31}$ \\
$\theta_3$ (deg) & $-18.9$ & $_{-19.4}^{+14.0}$ & $_{-37.0}^{+23.0}$ & $_{-51.5}^{+30.2}$ \\
$r_{\rm core,3} \; (\arcsec)$ & $12.1$ & $_{-2.4}^{+2.9}$ & $_{-4.4}^{+6.4}$ & $_{-6.1}^{+10.6}$ \\
$\sigma_{v3} \; \rm (km\,s^{-1})$ & $667$ & $_{-55}^{+65}$ & $_{-101}^{+144}$ & $_{-140}^{+229}$ \\
\hline
\multicolumn{5}{l}{External shear} \\ \hline
$\gamma_{\rm ext}$ & $0.12$ & $_{-0.02}^{+0.02}$ & $_{-0.03}^{+0.03}$ & $_{-0.05}^{+0.05}$ \\
$\theta_{\rm ext}$ (deg) & $101.5$ & $_{-2.1}^{+2.3}$ & $_{-4.1}^{+4.9}$ & $_{-6.1}^{+8.0}$ \\
\hline
\multicolumn{5}{l}{Galaxy members} \\ \hline
$r_{\rm cut}^{\rm gals} (\arcsec)$ & $3.3$ & $_{-0.9}^{+1.3}$ & $_{-1.7}^{+3.3}$ & $_{-2.1}^{+7.1}$ \\
$\sigma_{v}^{\rm gals} (\rm km\,s^{-1})$ & $342$ & $_{-36}^{+39}$ & $_{-68}^{+86}$ & $_{-97}^{+141}$ \\
\hline\hline
    \end{tabular}
\label{tab:model_params}
\end{table}


\end{document}